\documentclass[12pt,a4paper]{report}
\usepackage{fancyhdr}
\usepackage{titlesec}
\usepackage{xcolor}
\usepackage{subcaption}
\usepackage[symbol]{footmisc}
\usepackage[hyperfootnotes=false]{hyperref}
\usepackage{graphicx}
\usepackage{bbold}
\usepackage{physics}
\graphicspath{{figures/}{./chapters/QHO1}{./chapters/QHO2}{./chapters/disc1}{./chapters/WD}}
\usepackage{amsmath}
\usepackage{fontawesome5}
\usepackage{dcolumn}% Align table columns on decimal point
\usepackage{bm}% bold math
\usepackage{verbatim}
\usepackage{xcolor}
\usepackage{epsfig}
\usepackage{bbm}
\usepackage{epstopdf}
\usepackage{enumerate}
\hypersetup{
linktoc=all,     %set to all if you want both sections and subsections linked
}
\usepackage[outline]{contour}

\usepackage{amsmath,mathtools}
\usepackage[isbn=false,style=nature,sorting=none,maxnames=50]{biblatex}%<- specify style
\AtEveryBibitem{
    \clearfield{urlyear}
    \clearfield{urlmonth}
    \clearfield{url-date}
}
\newbibmacro{string+doi}[1]{%
  \iffieldundef{doi}{#1}{\href{http://dx.doi.org/\thefield{doi}}{#1}}}
\DeclareFieldFormat{title}{\usebibmacro{string+doi}{\mkbibemph{#1}}}
\DeclareFieldFormat[article]{title}{\usebibmacro{string+doi}{\mkbibquote{#1}}}

\usepackage{epigraph}
\usepackage{dsfont}
\usepackage{multirow}
\usepackage{physics}
\usepackage{cellspace}
\usepackage{appendix}
\usepackage{chngcntr}
\usepackage{etoolbox}
\usepackage{setspace}

\renewcommand{\headrulewidth}{0pt}
\titlespacing*{\section}
{0pt}{1ex plus 1ex minus .2ex}{1.3ex minus .2ex}
\titlespacing*{\subsection}
{0pt}{1ex plus 1ex minus .2ex}{1.3ex minus .2ex}

\usepackage[T2A,T1]{fontenc}
\usepackage[utf8]{inputenc}
\usepackage[russian,english]{babel}
\numberwithin{equation}{section}

\newcommand\beq{\begin{equation}}
\newcommand\eeq{\end{equation}}
\newcommand\bea{\begin{eqnarray}}
\newcommand\eea{\end{eqnarray}}

\usepackage{geometry}
\usepackage{epstopdf}
\usepackage{framed} % or, "mdframed"
\usepackage[framed]{ntheorem}
\newframedtheorem{frm-thm}{Theorem}

\def \MR {\rm MR}
\def\NSIT{\rm NSIT}
\def\NIM{\rm NIM}
\def\Ind{\rm Ind}
\def\erf{{\rm erf}}
\def\erfi{{\rm erfi}}
\def\MRps{\rm MRps}
\def\P{{\bf P}}
\def\sgn{{\rm sgn}}
\newcommand\LG[1]{\rm LG{#1}}
\title{Tests of Macrorealism in Discrete and Continuous Variable Systems}
\author{Clement Mawby}
\date{\today}
\makeatletter
\def\@makechapterhead#1{%
  \vspace*{50\p@}%
  {\parindent \z@ \raggedright \normalfont
    \ifnum \c@secnumdepth >\m@ne
      \if@mainmatter
        \Huge\bfseries \thechapter.\space%
      \fi
    \fi
    \interlinepenalty\@M
    \Huge \bfseries #1\par\nobreak
    \vskip 40\p@
  }}
 \makeatother
\newcommand\bookepigraph[5]{
\vspace{-#5em}\hfill{}\begin{minipage}{#1}{\begin{spacing}{0.9}
\small\noindent\textit{#2}\end{spacing}
\vspace{1em}
\hfill{}{\textemdash~#3 \textsl{#4}}
}
\vspace{3em}
\end{minipage}}
\DeclareMathAlphabet{\mathbbmsl}{U}{bbm}{m}{sl}

\newsavebox{\foobox}

    \makeatletter
    \let\@fnsymbol\@arabic
    \makeatother
        \usepackage{titletoc}
        \titlecontents{chapter}
    [5.5em] %5.3
    {\bigskip \phantomsection \label{link:\thecontentslabel}}
    {\hspace*{-4em}\contentslabel[\bfseries\thecontentslabel]{1.5em}\textbf}%\thecontentslabel
   {\hspace*{-5.5em}\textbf}% unnumbered chapters
   {\titlerule*[1pc]{}\contentspage}[\smallskip]%
\begin{document}
\thispagestyle{empty}
\begin{flushleft}
\end{flushleft}
{\centering

\Huge \textbf{Tests of Macrorealism in Discrete and Continuous Variable Systems}
\\[1em]
\huge Clement Mawby
\\[2em]
\normalsize
\begin{center}
\includegraphics[height=10.4cm]{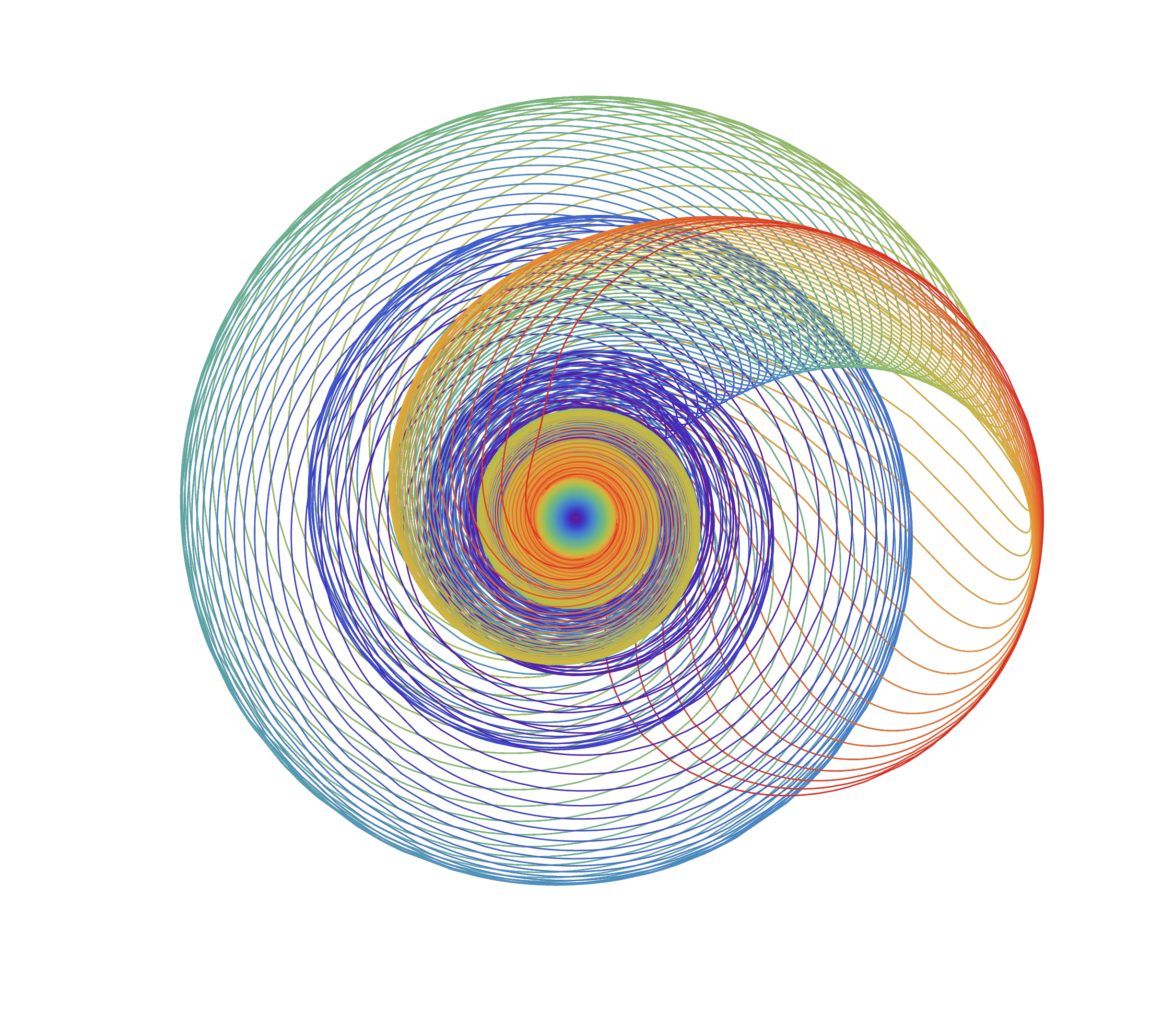}
%Department of Theoretical Physics, Imperial College
\vspace{1.3em}
\normalsize
\\
\Large Imperial College London\\
Department of Physics
\vspace{1.3em}
\normalsize

Thesis submitted for the fulfilment of the requirements for the degree of Doctor of Philosophy (PhD) in Theoretical Physics, April 2023
\end{center}
}
\clearpage
\newpage
\doublespacing
\begin{center}
{\LARGE  Abstract}
\end{center}
\phantomsection
\addcontentsline{toc}{chapter}{Abstract}
I study several aspects of tests of macrorealism (MR), which for a given data set serves to give a quantitative signal of the presence of a specific notion of non-classical behaviour.  The insufficiency of classical understanding underpins both the paradoxes of quantum mechanics, its future technological promise, and so these tests are of interest both foundationally and pragmatically.  I derive generalisations of the Leggett-Garg (LG) inequalities and Fine's theorem, which together establish the necessary and sufficient conditions for macrorealism.  First, I extend these conditions to tests involving an arbitrary number of measurement times.  Secondly, I generalise them beyond the standard dichotomic variable, to systems described by many-valued variables.  I also perform a quantum mechanical analysis examining the interplay of different conditions of MR.  I then develop the theoretical framework to support tests of macrorealism in continuous variable systems, where I define variables based on coarse-grainings of position.  I calculate temporal correlators for general bound systems, and analyse LG violations within the quantum harmonic oscillator (QHO), in its energy eigenstates and coherent states.  I analyse the precise physical mechanisms underpinning the violations in terms of probability currents, Bohm trajectories.  Staying within continuous variable systems, we outline a different approach to meeting the invasiveness requirement of LG tests.  Reasoning that we may approximately non-invasively measure whether a particle crosses the axis, we measure an object which is related to the standard correlators, and derive a set of macrorealistic inequalities for these modified correlators. We demonstrate violations of these modified LG inequalities for several states within the QHO.
%Altogether this gives the physical portrait of a Leggett-Garg violation in a system more intuitive than the canonical spin-$\frac12$ particle prevalent in the field.

\newpage
\noindent \textbf{Cover Image:}
\noindent  Looking down the $x-$axis at the time evolution of a gaussian wave function, in the scenario its particle was measured within the $x<0$ side of the axis.  This is plotted with the page face as complex plane, for a variety of time intervals post measurement, with time increasing as we move up the rainbow.  It is both a pretty picture, and a hint of the extreme complexity under the hood of one of the simplest dichotomic statements you could make about a particle.

\vspace{4em}
\begin{center}
This work was supported by an EPSRC studentship.\\
\vspace{3em}
\noindent \textbf{User Guide:}\\
\noindent Should you get lost in this thesis\\ the page numbers are hyperlinks\\ to the \hyperref[link:1]{table of contents}
\vspace{6em}
\end{center}
\noindent
\textbf{Copyright Declaration}
\begin{spacing}{1.3}
\noindent
The copyright of this thesis rests with the author. Unless otherwise indicated,
its contents are licensed under a Creative Commons Attribution-Non
Commercial 4.0 International Licence (CC BY-NC).
Under this licence, you may copy and redistribute the material in any medium
or format. You may also create and distribute modified versions of the work.
This is on the condition that: you credit the author and do not use it, or any
derivative works, for a commercial purpose.
When reusing or sharing this work, ensure you make the licence terms clear to
others by naming the licence and linking to the licence text. Where a work has
been adapted, you should indicate that the work has been changed and
describe those changes.
Please seek permission from the copyright holder for uses of this work that are
not included in this licence or permitted under UK Copyright Law.

\end{spacing}

\newpage
\begin{center}
{\LARGE  Statement of Originality}
\end{center}
\phantomsection
\addcontentsline{toc}{chapter}{Statement of Originality}

The work presented in this thesis is my own work, except where otherwise referenced.  The majority of this thesis comes from the four papers published with my supervisor Prof. Jonathan Halliwell.  Chapters \ref{chap:ntime} and \ref{chap:mlev} correspond to Phys. Rev. A, \textbf{100}, 042103 (2019) and  
Phys. Rev. A, \textbf{102}, 012209 (2020)~\cite{halliwell2019,halliwell2020}, which are extensions of earlier work \cite{halliwell2019b,halliwell2017}.  Chapters \ref{chap:QHO1} and \ref{chap:QHO2} correspond to  Phys.~Rev.~A, \textbf{105} 032216 (2022) and Phys.~Rev.~A, \textbf{107} 032216 (2023)~\cite{mawby2022,mawby2023}, and give theoretical basis to the proposal in \cite{bose2018}.  Chapter \ref{chap:wd} represents a continuous variables extension of Ref.~\cite{halliwell2016a}, and at the time of submission is unpublished.

My main contribution to \cite{halliwell2019} was to develop the generalized Fine ansatz, and the inductive proof used in developing the main results.  I also performed the statistical calculations which serve to estimate the large $n$ asymptotic behaviour of the inequalities derived in this chapter, and produced all corresponding figures.  In \cite{halliwell2020}, my main contribution was developing the framework and proofs for Fine's theorem, and the many-valued variable LGs.

In both \cite{mawby2022} and \cite{mawby2023}, I performed the very significant majority of the work for carrying out and checking the calculations, and the writing of both papers.  The development of the conceptual basis of the work and its interpretation was carried out approximately equally between the two authors.  I produced all the figures and numerical code.

For Chapter \ref{chap:wd}, conceptual development and calculations were divided approximately 50:50, and I did about 30\% of the writing here.  I produced all the figures and numerical code.

\newpage
\phantomsection
\begin{center}
{\LARGE  Acknowledgements}
\addcontentsline{toc}{chapter}{Acknowledgements}
\end{center}
\normalsize
\begin{spacing}{1.52}
I am eternally grateful for the patience, wisdom, humility, and kindness from my supervisor Prof. Jonathan Halliwell. He has been a source of limitless encouragement, and has during these few years shown me out of more than a few dark woods.  It has been an incredible experience studying this universe with him, and I look forward to unpacking the lessons shared for many years to come.

I thank Shayan Majidy, James Yearsley, Dipankar Home, Sougato Bose and Debarshi Das for for useful conversations about the LG inequalities, and Sofia Qvarfort for organising the UCL Foundations Reading Group.  I thank Prof. Allan Adams for including the obscure node-theorem in his online QM lecture series, which led to a pivotal result for this thesis.  Some of this work was carried out at two l'Agape summer schools in M\'ezeyrac, so I thank Pierre Martin-Dussaud and the other organisers for their hospitality, and I thank Alex, Robin, Jan, John, Emily,  V\'aclav, Isha, Titouan, Johannes, Celeste and Scott for the stimulating conversations and song, under starry night skies. I am grateful for the camaraderie and friendships within the Theory Group; Andreas, Antoine, Aoibheann, Ariana, Arshia, Chris, Dave, David, Eliel, emma, Fede, George, Giorgio, Graziela, Matthew, Henry, Julius, Justin, Lucas, Matt, Matthew, Nat, Rahim, Santi, Stav, Sumer, Victor, Zhenghao, and Ed whom we deeply miss.

In times surrounded by anthropogenic ecological collapse and global crises, I have needed my friends more than ever, and I could not have done this without you all.  For the emotional support, for joining in activism, for putting up with me relating any topic of conversation to quantum mechanics, for making art and music with me, and for opening doors in my mind and helping my understanding of quantum physics progress to places I could never have reached alone, I thank: Asmaa, Ben, Carol, Connor, Darije, Davide, Dom, E2R-folk, Fariha, Freya, Joe, Lisa, Lucille, Max, Otavio, Richard, Rosemary, Ruchir, Sammy, Varja, and Verity; as well the many strangers, my plants, my piano, and countless other beings.

I thank my mum for her endless support and understanding. Finally, I dedicate this work to my father, who always had a fervour for philosophical conversations, and inspired me to pursue further study at a time when that was highly doubtful -- I wish we could talk now, about some of the questions I have encountered through the course of this PhD.
\end{spacing}
\newpage 

\singlespacing
\tableofcontents
\addtocontents{lof}{\vskip -1.2cm}
\listoffigures
%\listoftables
%Introduction

\doublespacing
\fancyhf{}
\renewcommand{\headrulewidth}{0pt}
\fancyfoot{\makebox[\textwidth][c]{\hyperref[link:1]\thepage}}

\chapter{Introduction}

\fancyhfoffset[LE,RO, RE, LO]{0cm}
\renewcommand{\chaptermark}[1]{ \markboth{#1}{} }
\renewcommand{\sectionmark}[1]{ \markright{#1}{} }
\fancyhf{}
\fancyhead[L]{\textsl{\thesection~~ \rightmark}}
\fancyhead[R]{\hyperref[link:1]\thepage}
\renewcommand{\headrulewidth}{1pt}
%
%\vspace{0em}
\bookepigraph{3in}{And anyway,\\
what's wrong with \textsl{Maybe}?}{Mary Oliver,}{The World I Live In}{4.5}
\vspace{0em}
\section{Overview}
\vspace{0.5em}
Quantum mechanics is one of humanity's most empirically successful scientific theories.  Its unprecedented success in predicting the behaviours of the smallest elements of the universe -- with some predictions experimentally verified to the level of parts per trillion \cite{aoyama2018,morel2020} -- has formed the intellectual bedrock for much of the technological advancement of the last century, essential for modern computers \cite{lukasiak2010,turing2001}, MRI machines \cite{lauterbur1973}, the GPS system \cite{rabi1937, essen1955} and solar power \cite{einstein1905, chapin1954}, to name a few.  It is not however without its problems, philosophical and interpretational; further, the study of such foundational issues we will see, has a track record of bearing fruit.

There is a rift between our intuitive, classical understanding of the physical world, and the understanding demanded by quantum mechanics (QM).  The classical understanding elicits a worldview where the state of reality may in each moment be transcribed losslessly into the pages of a book, whereas in the quantum mechanical case, we find such an approach fails\footnote{Bohm refers to this as the need to study quantum \textsl{non}-mechanics, hinting that whatever underlies the theory may not be machine-like at all; the \textsl{organic} universe \cite{bohm1989}.}.  This failure results in concepts such as objective properties of systems no longer being sufficient to describe their future evolution.  In short, data in a one-to-one mapping with observables seems an inadequate description of the natural world, with the universe better thought of as fundamentally non-binary.  This divergence from the classical however represents opportunity, and is what gives quantum algorithms a theoretical edge over their classical counterparts \cite{feynman1982,shor1994,ladd2010}. 

\noindent Foundationally speaking, this failure to conform to the conceptual constraints of $0$ and $1$, and their associated logical structures, has been described as `the only mystery in QM' \cite{feynman1992}, and indeed may be what makes it so infamously difficult for us to understand.  However we note that since the times of the Ancient Greeks, mankind has written entertaining the idea that there may be more to reality than can be directly seen (and thus transcribed) of it \cite{plato1888}, and nearly two centuries ago Gauss proved that following a map as flat as our intuitive experience of our planet, will ultimately lead us astray \cite{gauss1900}.  In this vein, quantum mechanics could be stated as the discovery that Plato's Cave is in fact a lot smaller than we thought, and that when mapping the fundamentals of reality we may get similarly lost.
%\footnote{The poet Keats spoke of comfort in mystery as \textsl{negative capability} --``when a man is capable of being in uncertainties, mysteries, doubts, without any irritable reaching after fact and reason" -- as a hallmark of the greatest artists, and perhaps later the early quantum mechanists.}

Despite this damning indictment of the classical realist worldview, it does have its own realm of validity -- the realm of the macroscopic.  However with the classical worldview in such stark contrast with the highly successful non-classical worldview of QM, an unavoidable question is present -- what divides these two realms?  This division is known as the Heisenberg cut\footnote{Schr{\"o}dinger's `\textsl{What is Life}' provides an interesting discussion on why we as biological organisms seem to find ourselves on this particular side of the cut \cite{schrodinger2012}.}.  Its study is of interest both from the foundational perspective, and from the perspective of the potential technological advancements that may lie on the other side of it.  Where classical computers operate on an array of zeros and ones, quantum computers are set to exploit the further logical reach of quantum information \cite{feynman1982,hayashi2006,vedral2006,ladd2010,nielsen2010}.  The detection of quantum behaviour may even shine light on one of the great unresolved questions of modern physics, through the proposal of experiments set to interrogate the possible quantum nature of the gravitational force \cite{bose2017a, marletto2017, galley2022,matsumura2022,anastopoulos2018, anastopoulos2021}. Further, the direct application of quantum physics may unlock some of the secrets of living creatures in the field of quantum biology  \cite{schrodinger2012,abbott2008,wilde2009,lambert2013}. Led by paradoxes in our understanding of the physical world, the development of QM revealed something about the nature and texture of reality, and our relationship with it. This radical change in worldview within physics has spread to other disparate fields of research such as quantum cognition {\cite{bohm1989,trueblood2017,asano2014a}, quantum social science \cite{haven2013,obrien2016}, quantum anthropology \cite{trnka2016}, quantum international relations \cite{derian2023}, and quantum music \cite{miranda2022}. Some even suggest quantum mechanical behaviour may hold the key to the hard problem of consciousness \cite{penrose2016,hameroff2014,smart2015}.  
% \cite{spekkens2005,abramsky2011,dzhafarov2015}

Of crucial importance to many of the above pursuits, is being able to accurately discern whether a given phenomenological behaviour is acting in a way which is absent of a classical explanation.  One of the main approaches\footnote{Other approaches include the no-signalling in time conditions \cite{kofler2013}, and the recently developed Tsirelson inequalities \cite{tsirelson2006,zaw2022,plavala2023}.} to this problem is the Leggett-Garg (LG) framework, where a specific notion of realism leads to experimentally testable statements \cite{leggett1985}.  In this thesis, we explore numerous aspects of the LG framework, devoting two chapters to generalising the LG inequalities to sequences of arbitrarily many measurements, and to systems beyond the standard dichotomic variable.  We devote a further two chapters to the study of macrorealism within bound continuous variable systems, with particular focus on the quantum harmonic oscillator.  The final chapter loosely details a new approach to meeting the requirement of non-invasive measurements in LG tests.

%q cognition\cite{bohm1989,trueblood2017}, q anthropology\cite{trnka2016} to name a few. schrodingers tardigrade, q sensing
%
%Even the classial worldview creaks, when the page in the book needs to describe itself.  
\section{Realism in Quantum Physics}

\subsection[An Alluring Idea]{An Alluring Question}
\label{sec:iaq}
When introduced to quantum mechanics, one of the first concepts to be introduced is the quantum superposition principle; that when there are multiple possibilities available to a quantum system, its most general state must be described as a combination of these possibilities, weighted by complex numbers.  For a system with two orthogonal configurations available to it (a dichotomic variable)\footnote{We need not specify the system of course, but this could be considered the spin state of some electron, or a coarse-graining of position - is a particle on the left or right hand side of a well? Or something altogether more sophisticated such as the wellbeing of a cat \cite{schroedinger1935a}.  Orthogonality encodes that the two possibilities are mutually exclusive.}, the most general state is written
\begin{equation}
\label{eqn:superpos}
\ket\psi = a\ket++b\ket-	,
\end{equation}
which with normalisation, and the arbitrarity of global phases, represents two degrees of freedom, naturally represented on the surface of a (Bloch) sphere \cite{bloch1946}. Through unitary time evolution, the representation of the state simply moves around the surface of this sphere, a motion requiring two degrees of freedom to fully describe.

Next up on this whistle-stop tour, is the implementation of measurements within quantum mechanics, through the Born rule.  Broken away from the deterministic Schr{\"o}dinger evolution, we find the dice that God does and or does not play with, and see that QM only provides probabilistic predictions.  The probability for a given measurement outcome $s=\pm1$ of the dichotomic variable is given by,
\begin{equation}
p(s)=\Tr(\rho P_{s})	,
\end{equation}
where $\rho$ is the density matrix. For the state in Eq.~(\ref{eqn:superpos}) leads to the probabilities $p(+)=|a|^2$, and $p(-)=1-|a|^2$.  In the process of making the measurement, the state of the system changes, the process historically known as `collapse of the wavefunction', where in the above example the measurement leads to a collapsed state of $\ket\pm$ with probability $p(\pm)$.  This leads to  an update according to the L{\"u}ders rule \cite{lueders1950} to the mixed state
\begin{equation}
\rho\rightarrow \rho'=\sum_{s}p(s)\dyad{s},
\end{equation}
where the state has gone from being described by two degrees of freedom, to just one degree of freedom -- aligned with the classical outcome of the experiment.  This vanishing degree of freedom corresponds to the non-unitary nature of quantum measurements, and in a world presumed fully describable by quantum mechanics, is the calling card of the quantum measurement problem.

Taking stock, we have presented a physical theory, which for any measurement we throw at it, will output a well-behaved probability.  Sure things may be colourful and inhabit the complex plane in between measurements, but whenever a measurement is made, a classical black and white answer comes out.  At first (and maybe even second) glance, we could reason this theory may not be too different to other probabilistic theories.  

You could flip a coin, and watching it tumble through the air before clasping it to the back of your hand reason `I can only make a probabilistic prediction about the current state of the coin, but this is due to the incompleteness of my knowledge!  If I knew precisely the initial conditions of coin in my hand, and calculated its Newtonian trajectory, I could exactly predict the state of the coin under my hand'.

In the heady mist of electrons traversing multiple paths, cats both dead and alive, decreed mandatory shock \cite{feynman2011,merli1976,schroedinger1935a,heisenberg1971}, and in the wake of the deft, elegant, and satisfying explanatory power of statistical physics, the alluring question may arise --
\vspace*{-1em}\begin{center}\textsl{is there, perhaps obscured by our ignorance,\\ a similar, intuitive understanding of the predictions of quantum mechanics?}
\end{center}

\subsection{A History of the Real World}
Following physical intuition\footnote{Intriguingly, the faith that objects in the world have a definite existence, even when unobserved by us, is something that must be learnt by human babies, and marks a developmental stage known as \textsl{object permanence} \cite{piaget1954}.  For another connection to human perception, see Ref. \cite{asano2014a}, where LG inequalities are used to detect contextuality in our perception of optical illusions.}, leads one to the realist position, and the proposal of a hidden-variable theory of QM, the same position that Einstein took in his debates with Bohr~\cite{schilpp1949}.  This metaphysical debate, perhaps came to a head with the publishing of EPR paper, refuting the completeness of QM, by showing that the wave-function does not give a complete description of physical reality~\cite{einstein1935}. Following the successes of relativity, with its logical guardrails for causality, the notion that quantum measurements could have an effect outside of their light-cone was deeply troubling to Einstein, `spooky action at a distance' was so distressing it led him to believe a deeper, more metaphysically tasteful theory must exist.

Initially this was left as a purely philosophical argument, by some seen as an unaesthetic blemish on our rational understanding of the world, however safely hidden under the rug of the pragmatic \textsl{`shut up and calculate!'} approach.  For the few who were interested in whether a hidden variable approach were possible, a proof in the authoritative 1932 von Neumann textbook was recognised to rule it out \cite{vonneumann2018}.  There was however an error in this reasoning -- though debate persists as to whether the proof was incorrect or just widely misconstrued.  The von Neumann proof was challenged at the time, by Hermann, who was widely ignored\footnote{It does not require much speculation to imagine that sexism played a role here in ignoring the work of Grete Hermann, a female philosopher.  For an in depth discussion of the technical discrepancies, see Ref \cite{mermin2018}.}, and it would be 30 years before progress was to be made again on this topic \cite{hermann1935,seevinck2016,mermin2018}.

In 1952 Bohm published his pilot-wave theory of quantum mechanics, which successfully reproduced the predictions of QM, in terms of a hidden-variable theory \cite{bohm1952,bohm1952a}.  The success of this theory no doubt raised eyebrows in relation to the von Neumann proof, not least, those of John Stewart Bell.  
\subsection{Bell's Theorem}
\label{subsec:bineqs}
In 1964, Bell arguably transformed the study of quantum foundations forever, by proving that if any physical theory\footnote{In some sense, this is where Bell's genius lay, as his test was not specifically a test of quantum mechanics, but a test of the nature of reality more generally; a theory of theories. The approach in this thesis largely continues in this vein.} respects his precise codification of the realist worldview, then this will have direct consequences, limiting the possible scope of its experimental predictions to within the Bell inequalities (BI) \cite{bell1964,brunner2014}.  Then using a transcription by Bohm of the EPR scenario into an entangled spin pair, he was able to find a scenario where QM makes predictions which lay beyond the remit of realist theories.  The framework of a Bell test, considers the scenario where measurements are made on a pair of spatially separated particles, $A$ and $B$.  The specific notion of reality tested here, is local realism (LR), the conjunction of two ideas:
\begin{itemize}
	\item Realism: Physical properties of a system have a definite existence, independent of being measured.
	\item Locality: Space-like separated events cannot influence each other.
\end{itemize}
These two ideas largely codify the classical worldview, and any theory satisfying these two would have surely have satisfied Einstein.  Proceeding in a different manner to Bell's original proof, we note that within this classical worldview, it is possible to ascribe to any experimental outcomes a joint probability distribution  \cite{son2005, halliwell2017}.  Hence for the dichotomic quantities measured in Bell tests, for an LR theory there must exist the joint probability $\mathbbmsl{p}(s_1, s_2, s_3, s_4)$, where the two measurements made on particle $A$ have outcomes $s_1$, $s_2$, and those on particle $B$ have outcomes $s_3$, $s_4$.  Of all the possible measurements, we measure the four probabilities $p(s_1, s_3)$, $p(s_1, s_4)$, $p(s_2, s_3)$, $p(s_2, s_4)$, in four experimental runs, which we note includes no two measurements on the same particle.  Now if the theory satisfies LR, then these measurement probabilities may be considered as marginals of the underlying probability distribution, for example
\begin{equation}
	p(s_2, s_3)=\sum_{s_1, s_4}\mathbbmsl{p}(s_1, s_2, s_3, s_4),
\end{equation}
which for four marginal probabilities, ends up quite an intricately threaded set of constraints.  From each of these measurements, the dependence of one on the outcome of the other is characterised by the correlator, defined
\begin{equation}
	C_{ij}=\sum_{s_i, s_j}s_i s_j p(s_i, s_j).
\end{equation}
It is only possible to match these marginals to an underlying distribution, if the eight CHSH inequalities of the form are satisfied,
\begin{equation}
	-2 \leq C_{13}+ C_{14} + C_{23} - C_{24}\leq 2,
\end{equation}
with the other six inequalities found permuting the minus sign \cite{clauser1969,clauser1978,peres2010}.  This proof has only referred to  elementary properties of probability distributions, representative of the thread of Bell's original theory-free thinking.

A key result for Bell tests is Fine's theorem, which states that the following five statements are equivalent: (1) There is a deterministic hidden-variables model for the experiment. (2) There is a factorizable, stochastic model. (3) There is one joint distribution for all observables of the experiment, returning the experimental probabilities. (4) There are well-defined, compatible joint distributions for all pairs and triples of commuting and noncommuting observables. (5) The Bell inequalities hold \cite{fine1982,halliwell2014}.  This means the Bell inequalities are a necessary and sufficient condition for LR -- if they are satisfied we are guaranteed an LR model exists for the system at hand, and if they are not satisfied, we are certain such a model does not exist.  This makes the BIs a decisive test of LR.

As intimated before, the development of the Bell inequalities proceeds with no mention of a specific underlying theory, merely the statistical correlations of what may be observed in a lab.  We may however see how the predictions of QM fare.  With two spin-$\frac12$ particles assembled into the entangled state 
\begin{equation}
	\ket\Psi = \frac{1}{\sqrt2}(\ket\uparrow\otimes\ket\downarrow + \ket\downarrow\otimes\ket\uparrow),
\end{equation}
by taking spin measurements taken in different directions on each particle, these Bell states\footnote{While Bell states were introduced by Bohm to represent the EPR paradox, it has in fact been shown the original EPR scenario admits a locally realistic description, and so the representation as an entangled spin-pair was in fact not a direct translation \cite{bohm1989, khrennikov2002}.} lead to a violation of the Bell inequalities \cite{bell2004, nielsen2010}.  Hence their behaviour provably does not admit a local hidden-variables description, as codified by LR.  

When Bell initially published his seminal paper, it was in the new and obscure journal
Physics Physique\begin{otherlanguage*}{russian}
Физика\end{otherlanguage*}, which for its short life, would pay those who published in it -- as a travelling physicist, he dared not ask his host institution to pay journal fees on such an outlandish topic away from what he had been invited for \cite{whitaker2016, becker, bell1964}. Perhaps inspired by certain resonances between QM and New Age beliefs, Clauser of the hippie Fundamental Fysiks group \cite{kaiser2012, hofmann1959} sought out one of the few copies, and would go on to find the first experimental violations of a Bell inequality \cite{freedman1972}.  Experimental results have been many since \cite{aspect1982,aspect1982a, brunner2014}, until in 2015 a loophole-free Bell violation was reported \cite{hensen2015}.  Aspect, Clauser and Zeilinger would go on to win the Nobel Prize in physics, recognising the role Bell tests played in the development of the field of quantum information \cite{hayashi2006,vedral2006}.  A question once considered metaphysical, is hence seen to inspire new fields of science and technology.
\section{Macrorealism, and the Leggett-Garg Inequalities}
The Bell inequalities are powerful tests of the nature of reality, and are unique in giving loophole free demonstrations that in order to replicate the predictions of quantum mechanics, a theory must either violate realism, or locality.  However, to perform a Bell test, one must be able to create an entangled space-like separated pair, which limits the experimental scenarios in which the test can be performed.  To get around this restriction, we turn to Leggett-Garg tests, which are sometimes known as the `temporal Bell inequalities'.  Instead of measurements on two subsystems separated by space, in LG tests, measurements are separated by time and are performed on an individual system.

\subsection{Macrorealism}
The Leggett-Garg inequalities \cite{leggett1985, leggett1988,leggett2008} were introduced to provide a quantitative test capable of demonstrating the failure of the precise world view known as macrorealism (MR)\footnote{An extensive review of theoretical and experimental aspects of LG tests by Emary, Lambert \& Nori is found in Ref. \cite{emary2013}, with a review by Vitagliano \& Budroni \cite{vitagliano2022} covering developments in the decade since .}.  The definition of MR is an attempt to codify our everyday intuition about what it means for an object to behave classically.  The LG inequalities then test whether this specific notion of classicality has sufficient explanatory power for the observed time-evolution of a system.  Macrorealism is defined as the conjunction of three realist tenets:

\begin{itemize}
	\item Macrorealism \textsl{per se} (MRps): a system definitely exists in one of its available observable states, for all moments of time.
	\item Non-invasive measurability (NIM): it is in principle possible to determine this state, without influencing future dynamics of the system.
	\item Induction (Ind): Future measurements do not affect past states.
\end{itemize}

Of our experience of the classical world, the aspect that we only observe its elements in definite states, is reflected by MRps. The fact that we can make these observations without changing the state is reflected by NIM\footnote{There is also the implicit notion of measurements being \textsl{faithful}, that is genuinely revealing the underlying value of an observable \cite{bacciagaluppi2014,emary2013,leggett2002}.} The Ind assumption reflects our experience of causality and the arrow of time, and can even be considered a time-symmetric extension of NIM. Its satisfaction is largely unchallenged in discussions of LG tests, however discussion of retrocausality is ongoing \cite{adlam2022}.  In brief, with $\wedge$ denoting logical conjunction,
\beq
\MR = \NIM \wedge \MRps \wedge \Ind.
\label{MRdef}
\eeq
In the language of the first section of this chapter, MRps guarantees that for any experimental run, that we could write in a book the past and future behaviour of some system.  NIM then means that making measurements on this system is equivalent to accessing the pages of this book.  Hence for a system satisfying MR, any sequence of measurements made on it will correspond to a trajectory through its observable states. Through consulting the pages of the book, we could count the appearance of each trajectory, hence allowing us to construct a joint probability distribution for the sequence of measurements.  This argument of course works in reverse, we could determine a set of realistic trajectories consistent with a given joint probability.   We hence consider the existence of a joint probability distribution for all observables as equivalent to MR holding.

% Considering making two sequential measurements on the simplest possible dichotomic system as in Section \ref{sec:iaq}, each experimental run will correspond to a trajectory of either $(+,+)$, $(+,-)$, $(-,+)$ or $(-,-)$, and hence consulting the pages of the book, we could count the appearances of each trajectory, and construct a joint probability distribution for each outcome, $p(s_1, s_2)$.  

For now we proceed by taking this definition at face-value, using its logical implications to derive the LG inequalities.  We will however need to return to the sizeable nuances and intricacies present in this definition, and its implementation in experiment.

\subsection{Deriving the Leggett-Garg Inequalities}
\label{subsec:derivlg}
The Leggett-Garg inequalities are typically derived for the measurements of a dichotomic variable $Q=\pm1$ as it evolves in time, for a given system.  We will consider the value of $Q$ at a series of moments in time $t_i$, with $Q_i$ denoting the value of $Q$ at time $t_i$. The LG inequalities are defined for a particular data set, with the standard three-time data set consisting of the single time expectations $\ev{Q_i}$ at each of the three times, and the second-order correlators $\ev{Q_i Q_j}$ for each pair of times.  We now derive the three-time LG inequalities, the LG3s.

For a system upholding MRps, there must exist an underlying probability distribution, which we call $\mathbbmsl{p}(s_1, s_2, s_3)$, where we use double-struck font to indicate MRps by definition. Without making reference to measurements yet, we note that this probability distribution may be characterised by its moments, for instance the temporal correlator
\begin{equation}
\label{eq:classcorr}
\mathbbmsl{C}_{ij}=\ev{Q_i Q_j}=\sum_{s_1, s_2, s_3}s_i s_j\mathbbmsl{p}(s_1, s_2, s_3)
\end{equation}
which explicitly written out yields for example
\begin{multline}
	\mathbbmsl{C}_{12}=\mathbbmsl{p}(-, -, -) + \mathbbmsl{p}(-, -, +) - \mathbbmsl{p}(-, +, -) - \mathbbmsl{p}(-, +, +) - \\
 \mathbbmsl{p}(+, -, -) - \mathbbmsl{p}(+, -, +) + \mathbbmsl{p}(+, +, -) + \mathbbmsl{p}(+, +, +),
\end{multline}
which similar expressions for the other two temporal correlators $\mathbbmsl{C}_{23}$ and $\mathbbmsl{C}_{13}$.  Using this definition, and completeness $\sum_{s_1,s_2, s_3}\mathbbmsl{p}(s_1, s_2, s_3)=1$, with some simple algebra we find that
\begin{equation}
\mathbbmsl{L}_1=1+\mathbbmsl{C}_{12}+\mathbbmsl{C}_{23}+\mathbbmsl{C}_{13} = 4[\mathbbmsl{p}(+,+,+)+\mathbbmsl{p}(-,-,-)]\geq0,
\label{protoLG}
\end{equation} 
for instance, with similar statements found permuting two minus signs on the correlators.  Hence this particular combination of the correlation functions yields a quantity proportional to the sum of the likelihood of two trajectories, which since $\mathbbmsl{p}(s_1, s_2,s_3)$ satisfies MRps, must be positive yielding the given inequality.  

At this stage we have only used the first tenet of our definition of MR.  Indeed, for a sequence of measurements, QM has no problem assigning a \textsl{sequential} measurement probability to any given outcome, and so quantum mechanics will satisfy the preceding inequality\footnote{This is true more generally as well.  If we make a sequence of measurements, we could count the appearances of each trajectory $N_{s_1, s_2, s_3}$ in the $\mathcal{N}$ experimental trials, building a sequential measurement probability $p_{123}(s_1, s_2, s_3)=N_{s_1, s_2, s_3}/\mathcal{N}$, which is manifestly positive thus satisfying the inequality (\ref{protoLG}).}.  This reflects that it is impossible to test just for MRps, and we can only make progress if we also employ the second MR tenet, NIM\footnote{This parallels the  Bell scenario, where it is impossible to test just for realism, and we can only to test the conjunction of locality \textsl{and} realism.}.  Continuing the earlier analogy, the inequality $\mathbbmsl{L}_1\geq0$ is satisfied on the pages of the book, and we must employ NIM to let us access the pages.

We shall see that the ability to non-invasively measure plays two roles in the following. Firstly, and loosely speaking, it allows us to surgically assemble the results of multiple different experiments, into something representative of the underlying system.  Specifically, it entails relations such as
\begin{equation}
\label{eq:3nim}
\mathbbmsl{p}_{12}(s_1, s_2, s_3)=	\mathbbmsl{p}_{23}(s_1, s_2, s_3)=	\mathbbmsl{p}_{13}(s_1, s_2, s_3)=\mathbbmsl{p}(s_1, s_2, s_3),
\end{equation}
where double-struck subscripted indicates the combination of NIM and MRps, i.e. $\mathbbmsl{p}_{12}(s_1, s_2, s_3)$ is the underlying MRps distribution in the presence of non-invasive measurements at $t_1$ and $t_2$.  Now since $\mathbbmsl{p}(s_1, s_2, s_3)$ is a well-behaved probability distribution, over exclusive options, we may write its marginals 
\begin{equation}
\mathbbmsl{p}(s_2, s_3)= \sum_{s_1}\mathbbmsl{p}(s_1, s_2, s_3)= \sum_{s_1}\mathbbmsl{p}_{23}(s_1, s_2, s_3)=\mathbbmsl{p}_{23}(s_2, s_3).
\end{equation}
The second role of NIM, is it implies we can in principle actually measure the underlying distribution, that is it is possible to have
\begin{equation}
p_{12}(s_1, s_2)=\mathbbmsl{p}_{12}(s_1, s_2),
\end{equation}
where $p_{12}(s_1, s_2)$ now indicates an experimentally measured quantity.  This now means we may measure a temporal correlator in a two-time experiment,
\begin{equation}
C_{12}=\sum_{s_1, s_2}s_1 s_2 p_{12}(s_1, s_2) \overset{\mathrm{MR}}{=}\sum_{s_1, s_2, s_3} s_1 s_2 \mathbbmsl{p}(s_1, s_2, s_3)=\mathbbmsl{C}_{12},
\end{equation}
which has the quantum mechanical definition \cite{fritz2010,anastopoulos2006}
\beq
C_{12} = \frac12 \langle \hat Q_1 \hat Q_2 + \hat Q_2 \hat Q_1 \rangle.
\label{corr2}
\eeq
Hence, for a system which respects NIM as well as MRps, we may measure the temporal correlators in three different experiments, and MR holding, place these results into the inequality Eq.~(\ref{protoLG}) and its brethren, leading at last, to the LG3 inequalities:
\begin{align}
1 + C_{12} + C_{23} + C_{13} & \ge 0,
\label{LG3a}
\\
1 - C_{12} - C_{23} + C_{13} & \ge  0,
\label{LG3b}
\\
1 + C_{12} - C_{23} - C_{13} & \ge  0,
\label{LG3c}
\\
1 - C_{12} + C_{23} - C_{13} & \ge  0.
\label{LG3d}
\end{align}
%The combination of correlators in Eq.~(\ref{protoLG}) makes implications about three-time trajectories, whereas we only make two-time measurements in the test.
Whereas Eq.~(\ref{protoLG}) represented a truism by virtue of measurements in a single experimental run having definite outcomes, this inequality through the conjunction of NIM and MRps stitches together the outcomes of three distinct experimental set-ups resulting in a non-trivial bound.  For a general theory, there is no reason to expect this inequality to be satisfied, however as we have seen in this derivation, if the theory is macrorealistic, its predictions \textsl{must} satisfy this inequality, making these four inequalities a \textsl{test of macrorealism}.

We have shown for an MR system, necessarily the LG3 inequalities will hold, reflecting the existence of the underlying distribution $\mathbbmsl{p}(s_1, s_2, s_3)$.  We may also approach this problem in reverse. Here we would ask what conditions must hold on a given data set to \textsl{guarantee} they may be reproduced by an underlying MR model.  This leads to Fine's theorem for LG tests, which we introduce in Section~\ref{sec:fine}, and proceed to generalize in Chapters \ref{chap:ntime} and \ref{chap:mlev}.

\subsection[An Alluring Question given Answer]{An Alluring Question given Answer}
We are now in a position to respond to the question raised at the end of Section \ref{sec:iaq}.  We first point out that while a coin flips in the air, at every moment in time (ignoring the split seconds it is vertical), the coin possesses a definite value of either heads or tails.  Indeed, assuming some replicability in the coin toss, each experimental trajectory takes the from of a square-wave of a given frequency, between heads and tails, leaving MRps satisfied.  We could further imagine making measurements by non-invasively taking a photograph of the coin at two relevant instants of time.  We hence could calculate, or even experimentally determine the temporal correlators here, and would find them triangle waves of the same frequency of the coin's rotation.  Such linear correlation functions do indeed satisfy the LG3s Eqs.~(\ref{LG3a})--(\ref{LG3d}), verifying the LG inequalities correctly verify their coin as a classical object.

To give a concrete example of a non-classical object, we look at the behaviour of a qubit, Eq.~(\ref{eqn:superpos}), we use some time dynamics, a typical example being $\hat H = \frac12 \omega \hat \sigma_x$, the precession of a spin-$\frac12$ particle in a uniform magnetic field (in the $x$-direction) \cite{kofler2007, athalye2011}.  This Hamiltonian generates uniform precession in the $z$-direction, and so we consider $Q=\hat \sigma_z$. This results in an expression for the correlators 
\begin{equation}
C_{ij}=\ev{\sigma_z(t_i)\sigma_z(t_j)}=\cos(\omega(t_j-t_i)).
\end{equation}
  Noting that the correlator depends only on the time interval between measurements $\tau=t_j-t_i$, and is independent of the initial state, if we choose equal time spacing between measurements, we reach the simplifications $C_{12}=C_{23}=\cos\omega\tau$, and $C_{13}=\cos 2\omega\tau$.  Using these correlators in for example Eq.~(\ref{LG3a}), we find at $\omega\tau=\frac{2\pi}{3}$ a largest violation of $-\frac12$.  Hence  the simplest quantum two-level system violates the LG inequalities, and thus exhibits non-classical behaviour, which cannot be understood from a macrorealistic perspective.
%  
%More generally, $[x,p]$ encoded into time evolution. (this preempts expectation of violations in continuous systems)

This seemingly answers the question in Section~\ref{sec:iaq} in the negative, that a theory of ignorance over classical states is insufficient to explain quantum mechanical behaviour.  However with a quick modification of the coin thought experiment, this conclusion runs into difficulties.  If we now consider the coin to be a microscopic object, so small that it may not be resolved in a photograph. To be concrete, it could be a small particle oscillating in a harmonic potential, where we define the variable as which side of the potential it is on.  To make a measurement, we now must imagine interacting with it, for example by shining a beam of particles on one half of the well, and looking to see if the beam is broken.  Crucially, these measuring particles could be of a similar scale to the particle being measured. This would be akin to taking a macroscopic pendulum verified through the LGs as classical, and instead of measuring it by taking a photograph, we measure it by bombarding it with billiard balls.  We could then find that due to the inevitable changes in trajectory post measurement, the pendulum may well be capable of matching the observations on a genuinely non-classical object\footnote{For a less farfetched example of a realist model incorporating invasive measurements which violates the LG inequalities see Refs. \cite{yearsley2013a,montina2012}.}.  

Although the pendulum itself is unchanged, it now fails its test for macrorealism.  In this ad absurdum case, we can identify this as a false-positive in the test for non-classicality -- we could easily observe the trajectories being changed through measurement, which leads to the LG violations.  However, this is a luxury we do not have when it comes to systems where there is genuine doubt about their classicality -- in this situation the only method we have of probing their properties is through measurements which we cannot prove are non-invasive.  This means for any given violation, one may always argue that the underlying object is in fact classical, we just have not been able to implement a non-invasive measurement on it.  This is the problem of invasiveness in LG tests, which is persistent and without clean resolution. We will discuss it further in Section~\ref{subsec:nim}.

%\footnote{From the theoreticians perspective we can make statements about which is failing, for instance within de-Broglie Bohm theory~\cite{bohm1952,bohm1952a}However from the perspective of an interrogator of the nature of reality, intrinsic-NIM failure remains a sufficient model for any LG violations.}
\section{A Closer Look at Macrorealism}
Where we initially took the definition of MR Eq.~(\ref{MRdef}) at face-value, in this section we re-examine it, acquainting us with a wider array of tests of MR which we begin to distinguish.  We also pry a bit deeper into the NIM requirement, and look at the maximal violations of these inequalities within QM. We as well lay out an intuitive understanding of the physical implications when a system violates MR.
\label{sec:genchar}
\subsection{Necessary and Sufficient Conditions for Macrorealism}
\label{sec:fine}
We have thus far presented just the LG3 inequalities (\ref{LG3a}--{\ref{LG3d}), there is a rich, and growing set of conditions for MR \cite{avis2010,halliwell2019a,halliwell2019b}.  Unlike the BI, the LG3 inequalities are not sufficient conditions for MR, however this can be remedied.  While we give a detailed proof of Fine's theorem in the LG context in Chapter~\ref{chap:ntime}, here it is sufficient to state that its result is contingent on the existence and well-behaviour of the relevant two-time marginal probability distributions.  Where this is incorporated by design in Bell tests, in LG tests this is not necessarily true.  To test their existence, we appeal to the moment expansion (which we make much use of throughout this thesis) of these two-time probabilities \cite{klyshko1996,halliwell2013},
\begin{equation}
\mathbbmsl{p}(s_1, s_2)=\frac{1}{4}(1+s_1 \ev{Q_1} + s_2 \ev{Q_2} + s_1 s_2 C_{12}).
\label{ps2}
\end{equation}
We can imagine in three separate experiments measuring the quantities $\ev{Q_1}$, $\ev{Q_2}$ and $C_{12}$, where we must ensure the measurement of $C_{12}$ is non-invasive \cite{halliwell2019a}.  If the system is macrorealistic, then we will have $\mathbbmsl{p}(s_1, s_2)\geq0$, which translates to the four inequalities
\begin{align}
\label{LG2a}
1+\ev{Q_1}+\ev{Q_2}+C_{12}\geq0,\\
\label{LG2b}
1-\ev{Q_1}+\ev{Q_2}-C_{12}\geq0,\\
\label{LG2c}
1+\ev{Q_1}-\ev{Q_2}-C_{12}\geq0,\\
\label{LG2d}
1-\ev{Q_1}-\ev{Q_2}+C_{12}\geq0,
\end{align}
which we call the LG2 inequalities, with their violation indicating a failure of MR at the level of two times \cite{halliwell2016b,halliwell2017,halliwell2019b,halliwell2019, halliwell2020}.

Fine's theorem then provides that the complete set of necessary and sufficient conditions for MR for a dichotomic system measured three times is the four LG3 inequalities, and the twelve LG2 inequalities.  The implications of this are two-fold.  Firstly, to test if a given data set admits a macrorealistic description, the entire set of $\LG{2}$ and $\LG{3}$ inequalities must be tested.  Secondly, a violation of just one $\LG{2}$ inequality is sufficient to show a failure of MR, which involving just two measurement times, may well be logistically simpler than testing the $\LG{3}$ inequalities.  In Chapters \ref{chap:ntime} and \ref{chap:mlev} we generalise the concepts of this subsection to measurements made at $n$ times, and for systems beyond dichotomic variables.

\subsection{Weak and Strong Macrorealism}
It is important to emphasize that the definition of MR indicated by Eq.~(\ref{MRdef}) is open to a number of different interpretations due to the fact that there are a number of different and physically reasonable ways of interpreting both MRps and NIM \cite{halliwell2017,kofler2013,clemente2015,clemente2016,maroney2014a}.  MRps comes in three different types \cite{maroney2014a}, but the LG framework tests only one of them, which is essentially quantum-mechanics with an added macroscopic superselection rule.  The other two types are hidden variable theories in which the wave function itself is included in the specification of the ontic state, as is the case, for example, in de Broglie-Bohm theory, and models of this type are much harder to rule out experimentally.

In Ref.~\cite{halliwell2017}, a number of different notions of MR were identified, corresponding to different ways of implementing NIM. In this classification,
the previous characterization of MR using LG inequalities at two and three times, with the $\langle Q_i \rangle$ and $C_{ij}$ measured in six experiments, is referred to as {\it weak} MR.

The other important class of MR conditions are those involving no-signalling in time (NSIT) conditions, which, at two times entail the determination of a probability $p_{12} (s_1, s_2)$ through two sequential measurements of $Q$ in a single experiment and requiring that
\beq
\sum_{s_1} p_{12} (s_1, s_2) = p_2 (s_2),
\label{NSIT}
\eeq
where $p_2(s_2)$ is the probability for $Q$ at $t_2$ with no earlier measurement at $t_1$ \cite{kofler2013}. Analogous conditions at three or more times are readily constructed \cite{clemente2015}. Such conditions also ensure the existence of an underlying probability, but as argued in Refs.~\cite{halliwell2016b,halliwell2017} these conditions test a {\it different notion} of macrorealism which is stronger than that characterized purely by LG inequalities. Characterizations of MR which entail only NSIT conditions are therefore referred to in Ref.~\cite{halliwell2017} as {\it strong} MR. Intermediate possibilities also exist involving combinations of LG inequalities and NSIT conditions.  Furthermore, these possibilities are not the only ones -- see for example Refs.~\cite{kumari2017,saha2015, kumari2019}.  In Chapter~\ref{chap:mlev}, we analyse the interplay between the NSIT and LG inequalities, calculating the quantum mechanical interferences responsible for the violation of each, with an extended discussion of classifications of MR in Section~\ref{c3a}.  

\subsection{Nuances of Non-invasiveness} 
\label{subsec:nim}
In Section~\ref{subsec:derivlg}, we employed the logical implications of NIM to allow us to construct our MR test. We now return with a more critical eye to its meaning. This is important to understand both the meaning of LG violations, and as well to understand their experimental implementation.  
%Th
Although the LG inequalities are often thought of as the temporal equivalent of the Bell inequalities, there is significant distinction.  Recalling from Section \ref{subsec:bineqs}, that in a Bell test, measurement pairs are chosen so it is never required to measure a particle sequentially in time.  This means we can ensure that any pair of measurements are made well outside each others light-cones, and so within the LR worldview, there is no way for one wing of the experiment to alter the other.  Of course, if it were possible for some faster than light signalling to occur, implying failure of the locality condition, a violation in this scenario would faithfully show the LR understanding as insufficient.

In Leggett-Garg tests, NIM takes the logical place of locality.  Hence if in some experimental scenario an LG violation is reported, we must be able to convince ourselves that it is not the result of a disturbing measurement.  This at first glance seems like a tall order, as any measurement in QM which reveals non-trivial information will lead to disturbance of state \cite{busch2009a}.  However as in the case of the BIs just discussed, our measurements must be non-invasive, only \textsl{within} the MR worldview.  That is they must be macro-realistically non-invasive.  As a particle must reasonably be within its own light-cone at all times, the LG tests do not have a rock solid physical principle to rely upon to ensure non-disturbance, and instead NIM is something that must be argued for, on a per-experiment basis.  

In the original Leggett-Garg paper, their approach to the invasiveness problem, was with the introduction the concept of ideal negative result measurement (INRM) \cite{leggett1985,leggett1988}.  Here we imagine our experimental apparatus interacts very strongly with say, only the $Q=+1$ state, and for the $Q=-1$ state there is no interaction.  This coupled with the MRps assumption, means that in the case where the apparatus reports no measurement, we can be certain that the system possessed value $Q=-1$.  We discard the cases where the apparatus did report a measurement, as in this case the direct interaction may influence a second measurement.  The second measurement itself can be an ordinary measurement, as only two sequential measurements are needed.  Within QM, this measurement procedure \textsl{is} invasive, since even a null result in QM leads to wave-function collapse, and hence will influence future dynamics.  However from a macrorealistic perspective, the procedure satisfies NIM.

%While theoretically the procedure of INRM should meet the NIM requirement, there is not necessarily a one-to-one mapping from thought experiment to real experiment here.
While theoretically the procedure of INRM should meet the NIM requirement, as it is simple enough to talk about how we anticipate a measurement to interact (or not), when it comes to implementing the experiment, we can never in fact be certain.  This is known as the clumsiness loophole \cite{wilde2012}. It is disturbing since any time LG violations are reported, rather than them reflecting a failure of the classical worldview, they could instead just imply the rather less profound statement -- that our inadvertently invasive experimental procedure obfuscated a genuinely classical underlying data set, which perhaps a better experimentalist could have accessed.  Since NIM failures have the flexibility to represent any non-classicality, disentangling the two is challenging, and requires the devising of measurement protocols which render potential clumsiness unlikely, such as Wilde-Mizel adroit measurements \cite{wilde2012}, and continuous in time velocity measurements \cite{halliwell2016a}.

Despite the use of ingenious protocols that make it very plausible that from a macrorealistic perspective NIM is satisfied, it is hard to get away from the conclusion that it is always the conjunction of MRps and NIM that is being tested. Indeed, MRps by itself is capable of matching the predictions of QM, well demonstrated by De Broglie-Bohm theory \cite{debroglie1927, bohm1952a,bohm1952a}.  It has been proposed that it may even be fundamentally impossible to measure a system without disturbing its behaviour \cite{vitagliano2022}, which is referred to as intrinsic-NIM \cite{emary2013}.  It is hard to see how an MRps failure does not \textsl{require} an intrinsic-NIM failure -- if the system is not describable as definitely in one of its observable states, then when we measure it in one, this is necessarily a projection away from its true state space, and must influence a change of behaviour, somewhere in the system \cite{clemente2015}.

Although this may leave philosophical appetites unsatiated, the fact we may not make concrete statements about which of NIM or MRps fails, is equivalent to saying its resolution leaves no experimental trace.  That is, if at a fundamental level measurements change the answer by asking the question, this is \textsl{still} bona-fide non-classical behaviour\footnote{While this behaviour may be understood as a classical system with memory, this acts to introduce another degree of freedom beyond the observable itself.  This point is revisited in the conclusion chapter.}.

%, and it implies that if a system does not satisfy MRps, the LG violations can be seen as state change due to measurements being intrinsically and unavoidably invasive
\vspace{-1em}
\subsection{The L{\"u}ders Bound}
\label{subsec:lb}
When looking at any given LG inequality, there exists the algebraic bound, which is the largest violation that would be possible if we were freely able to change the values of correlators, so for example in the LG2s and LG3s, this limit is $-2$.  However, by writing
\begin{equation}
	\ev{(s_1 \hat Q_1 + s_2 \hat Q_2 + s_3 \hat Q_3)^2}\geq0,
\end{equation}
we see that for dichotomic variables in QM, we have
\begin{equation}
1+s_1s_2 C_{12} + s_1 s_3 C_{13} +s_2 s_3 C_{13} \geq -\frac12,	
\label{Luders}
\end{equation}
which is known as the L{\"u}ders bound \cite{barbieri2009, kofler2008, fritz2010, budroni2013, budroni2014}, which is numerically equivalent to the similar Tsirelson bound on the BIs \cite{cirelson1980}.  By writing the LG2s in the form given in Ref.~\cite{halliwell2019c},
\begin{equation}
1+s_1 \ev{Q_1} +s_2 \ev{Q_2}+s_1 s_2 C_{12}=\frac12\ev*{(1+s_1 \hat Q_1 +s_2 \hat Q_2)^2 -1},
 \end{equation}
 it is made explicit that the state achieving their maximal violation of $-\frac12$ satisfies the eigenvalue equation 
\begin{equation}
\label{eq:maxQP}
(s_1\hat Q_1 + s_2 \hat Q_2)\ket\psi=-\ket\psi. 
\end{equation}

Though there has been much progress in characterising these maximal violations, there is much debate as to what physical principle it may be upholding.  The authors of Ref.~\cite{dakic2014} claim that violation beyond the L{\"u}ders bound implies the presence of third-order interference terms\footnote{Although this is encoded by the Born rule leading to only paired interference terms, it is non-trivial that this is enough to describe our universe -- in physical terms this means a triple-slit interference experiment contains no new physics beyond the double-slit experiment.  There has been discussion about the stringent experimental requirements needed to reveal third-order interferences \cite{garner2017}.}, where quantum mechanics only has second order interference terms \cite{sinha2009}.

It is hence surprising that when we are dealing with systems described by many-valued variables, LG violations may surpass this bound. Indeed LGIs may be violated right up to the algebraic bound \cite{budroni2014, kumari2018, katiyar2017, wang2017,pan2018a}.  These violations occur under degeneracy breaking von-Neumman measurements.  This corresponds to the difference between constructing a dichotomic variable through the coarse-graining over amplitudes (L{\"u}ders \cite{lueders1950}) versus over probabilities (von-Neumann \cite{vonneumann2018}).  This surprising result, which can imply the outright logical paradox $C_{12}=C_{23}=C_{13}=-1$, motivates Chapter \ref{chap:mlev} of this thesis, where we give an in depth analysis of LG tests for $N$-levelled variables, and give an understanding of L{\"u}ders bound violations.

\subsection{Intuitive Picture of Leggett-Garg Tests}
\label{subsec:intuit}
In Section~\ref{subsec:derivlg}, when deriving the LG3s we essentially  pulled from thin air the correct string of correlators, and then showed that leads to a useful isolation of trajectory probabilities.  In this section we give a more physically motivated approach, and an interpretation of the meaning of the LG inequalities, in terms of intuitive physical concepts.

We initially derived the LGs by looking at the probabilities of individual measurement outcomes.  However, since we are in effect studying change, we could also consider representing this in the `atoms' of change, which for a dichotomic variable is a flip between its two states, which corresponds to two possibilities $+\rightarrow-$ and $-\rightarrow+$.  By explicitly writing out the correlator,
\begin{equation}
C_{ij}=\mathbbmsl{p}(+,+)+\mathbbmsl{p}(-,-) - \mathbbmsl{p}(-,+) - \mathbbmsl{p}(+,-),	
\end{equation}
we note we can write this as
\begin{equation}
\label{eq:c12f12}
C_{ij}=1-2f_{ij},
\end{equation}
where $f_{ij}$ is the likelihood of observing the variable having opposite signs\footnote{It is shown in Ref.~\cite{halliwell2016} that the quantum-mechanical histories for the two possibilities of flipping or not flipping are non-interfering, and hence consistent probabilities can always be assigned here, giving this representation an intriguingly classical air.}} at times $t_i$ and $t_j$.  Purely from our experience of classical objects, we know that inequalities such as
\begin{equation}
f_{12}+f_{23}\geq f_{13}	,
\end{equation}
must hold, which through Eq.~(\ref{eq:c12f12}) represents the LG3 inequality Eq.~(\ref{LG3b}).  To be more precise, we define the crossing variable $X_{12} = Q_1 Q_2$, which takes values $+1$ or $-1$ depending on whether or not a flip happens. Intriguingly, the LG3s may then be written
\begin{equation}
1+s_1\ev{X_{12}} +s_2\ev{X_{23}}+ s_1 s_2 \ev{X_{12}X_{23}}\geq 0,
\end{equation}
which we recognise as the set of LG2 inequalities which guarantee the existence of a joint probability distribution for the dynamical part of the data set.  Loosely, by considering change, we step down from a three-time condition to a two-time condition, at least outwardly.  This is akin to a velocity, a single number, implicitly being the difference between two numbers.  If this velocity can be measured in a single measurement, it would allow for a genuinely non-invasive measurement of correlation functions.  This was explored both theoretically \cite{halliwell2016a} and experimentally~\cite{majidy2019a} for simple spin systems, and its extension to continuous variable systems is the subject of Chapter~\ref{chap:wd}.

\section{Experimental Tests of Macrorealism}
 In this section we briefly discuss the experimental state of play, but refer to the two extensive reviews \cite{emary2013, vitagliano2022} for more details.  The first experimental LG test was conducted 25 years after their proposal, on a superconducting qubit {\cite{palacios-laloy2010}.  In the years since, there have been tests in a wide variety physical systems, including photons \cite{goggin2011,dressel2011, xu2011,suzuki2012,wang2017a, joarder2022}, silicon impurities \cite{knee2012}, nuclear magnetic resonance \cite{souza2011, majidy2019a}, neutrino oscillations \cite{formaggio2016}, solid state qubits \cite{williams2008}, superconducting flux qubits \cite{knee2016}, atoms hopping in a lattice \cite{robens2015}, the IBM Quantum platform \cite{huffman2017, ku2020, santini2022}, and others \cite{waldherr2011,athalye2011,groen2013,george2013,katiyar2013,zhou2015}.  As indicated in Section~\ref{sec:fine}, there is a large set of inequalities required to make a decisive test of MR.  All experimental tests check only a subset of this set, so test only necessity, but an experimental test of the complete set of inequalities was recently carried out \cite{majidy2019a,majidy2019b}. 
 
Great care must be taken in the implementation of measurements in any LG test, taking seriously the loopholes outlined in Section~\ref{subsec:nim}.  The LG test by Robens \textsl{et al.}  \cite{robens2015} employs the INRM procedure.  Other experiments have adopted alternative assumptions such as stationarity \cite{zhou2015}, where for neutrino oscillations such an assumption is essential, as sequential measurements are impossible for this system \cite{formaggio2016}.  The LG test on superconducting qubits by Knee \textsl{et al.} takes account of the clumsiness loophole by performing control experiments which rule out classical disturbance models.  The experiment in Ref.~\cite{joarder2022}  convincingly implements the INRM procedure on heralded single photons.  Also working with heralded single photons an ambiguous INRM procedure is employed in Ref.~\cite{wang2018}.  An alternative approach to NIM is taken in the NMR experiment by Majidy~\textsl{et al.} \cite{majidy2019a,majidy2019b}, implementing the continuous in time velocity measurement protocol given in Ref.~\cite{halliwell2016a}.
	
More generally there have also been demonstrations of  non-classicality based on interferometric experiments, with quantum interference observed in $C_{60}$ and $C_{70}$ molecules \cite{arndt1999,arndt2001}, and in large organic molecules of masses up to 6910 amu \cite{gerlich2011}. These are not specifically LG tests, but proposals have been made to adapt them for this purpose \cite{pan2020,halliwell2021}. The cutting edge experimentalists are able to create superposition states in nano-scale oscillators of $10^9$ amu \cite{oconnell2010, scala2013, wan2016}, and in mechanical oscillators up to $10^{13}$ amu \cite{kotler2021, mercierdelepinay2021}.  These experimental successes indicate LG tests within the quantum harmonic oscillator (QHO) could be a fruitful avenue for pushing LG tests closer into what could be considered the macroscopic domain. There have been some proposals for tests of non-classicality within the QHO \cite{tsirelson2006,asadian2014a,bose2018,zaw2022a,zaw2022}.

Although originally proposed within the context of testing for macroscopic quantum coherence, experimental LG tests have been nearly exclusively conducted on the discrete properties of microscopic systems \cite{emary2013}.  This could be called the problem of macroscopicity, and addressing it is one of the aims of this thesis.  There are several approaches to mathematically characterising macroscopicity, see Refs.~\cite{leggett1980,leggett2002,nimmrichter2013, frowis2018}, however we will use the concept more loosely and intuitively. Put simply, we consider systems with a higher mass as closer to macroscopicity.  In Chapters \ref{chap:QHO1} and \ref{chap:QHO2}, we develop the mathematical framework for LG tests based on the coarse-graining of position measurements, where the mass is a freely adjustable parameter.
%\newpage
\section{Summary and Overview}
\label{introsum}
To summarise, Bell's seminal work showed the metaphysical nature of a theory matters, and indeed may have measurable physical implications.  Later, the Leggett-Garg and tests of MR research programme developed to study these implications in physical scenarios beyond entangled pairs.  The research area has been fruitful since, and is becoming more relevant, as tests of quantumness play a vital role in the quantum-technologists toolbox, and the `weirdness' and further reach of non-classical statistics is found to be useful in other areas of study.  LG research is thus likely to be useful both at home and away.  We identify four key goals facing the field moving forward:
\begin{enumerate}[(a)]
	\item Macroscopicity: Most experimental and theoretical research has been conducted in the microscopic domain. There is hence unexplored experimental and theoretical territory around LG tests in systems which may broadly be considered genuinely macroscopic.\label{macrogoal}
	\item Problem of NIM: The NIM assumption of MR is very strong, and it may be impossible to avoid, so alternative approaches and convincing macrorealistic implementations are highly valuable. \label{nimgoal}
	\item \label{naturegoal} Nature of MR violations: Although indicative of non-classical behaviour, there is much scope for more understanding of the precise significance and meaning of an MR violation.
	\item Broaden tests of MR: Much theoretical research considers the simplest and restrictive scenario of a dichotomic variable, a broader exploration and understanding of MR conditions for more general systems is valuable.\label{broadgoal}
	\end{enumerate}
In this thesis we aim to make contributions focussed on each of these problems.  Firstly to address (\ref{broadgoal}), we generalize existing conditions for MR. We do so by considering extending the number of measurements made (LGn), and considering many-valued variables.  These two generalisations are compatible, so we hence extend LG tests to an $N$-level variable, measured at $n$ times.  We note these conditions may be useful more generally as LG tests become used in broader contexts.  Addressing (\ref{macrogoal}), a second aim of this thesis is to give a theoretical basis for experiments on systems which may more broadly be considered macroscopic.  We do this by studying LG tests in continuous systems, where the size of the system at hand becomes a free parameter in our calculations, and our analysis of the location and behaviour of LG violations hence persists up to the macroscopic scale.  The third aim of this thesis is to lay the groundwork of an alternative approach to tests of MR, where loosely, we swap the assumption of NIM for some assumptions motivated by ostensibly reasonable physical intuitions.  By providing an alternative to NIM, this addresses (\ref{nimgoal}), and may yet be relevant to (\ref{naturegoal}) as we in effect develop a variant of the standard MR framework.
\section{The Remainder of this Thesis}
The LG inequalities and Fine's theorem are routinely derived for measurements at three and four times \cite{busch2001,suppes1981,pitowsky2014,garg1984,zukowski2002,abramsky2011,halliwell2014}, with some individual larger $n$-time inequalities dotted around \cite{liang2011,budroni2013a, emary2013, araujo2013}.  In Chapter \ref{chap:ntime} of this thesis, we generalise these existing derivations, to the case of measurements of a dichotomic variable at arbitrarily many times.  At the level of three times, the LG inequalities involve all of the temporal correlators, however at four times or more, they involve only a subset of the possible correlators.  In this Chapter we look at some of the conditions possible when measurements of all the correlators are made.  Using statistical arguments, we establish the expected potency of these tests in the limit of a large number of measurements.

In Chapter \ref{chap:mlev} of this thesis we extend the framework of necessary and sufficient conditions for MR to many-valued variables.  We derive the necessary and sufficient conditions for MR for $N=3$ and $n=3$, with details of the procedure involved for arbitrary $n$ and $N$.  Within QM for a many-valued variable, we may consider coarse-graining at the level of amplitudes, or at the level of probabilities.  The two are equivalent under MR, but within QM are equivalent only for $N=2$, with their difference for $N>2$ resulting in a mingling of interference terms which ends up responsible for the tension mentioned in Section \ref{subsec:lb} around L{\"u}ders bound violations.  Overall, the generalisation to $N$ variables is straightforward, by virtue of there being only second-order interference terms.  However the hierarchy between MR conditions becomes much more complicated, which we detail in this chapter.

In Chapter~\ref{chap:QHO1}, we give a theoretical analysis of LG tests, where the variable tested is a coarse-graining of the position variable. We exactly calculate the temporal correlators for energy eigenstates of the QHO, and as well prescribe a method to approximate temporal correlators for any coarse-graining of space, for any system where the energy eigenspectrum is known.  We find a surprising correspondence between the QHO and the canonical spin-$\tfrac12$ example of much LG research.  We find significant LG violations in this system, and detail scenarios where the LG2 and LG3 inequalities are independently satisfied or violated, indicating the importance of performing the decisive tests detailed in Chapter~\ref{chap:ntime}.

We continue the analysis of LG tests in the QHO in Chapter~\ref{chap:QHO2},
focusing on LG violations in coherent states. LG tests on a coherent state of the QHO may be related to the results on the pure ground-state, and hence the results from the Chapter~\ref{chap:QHO1} apply.  We give details of the largest LG violations possible for coherent states, and give a survey of their location in phase-space, which may be contrasted with existing numerical work \cite{bose2018}.  Since the behaviour of a particle is simpler to imagine than that of a microscopic spin, this chapter serves to give a portrait of the underlying physical mechanisms leading to an LG violation.  We demonstrate this by analysing the probability currents involved, the corresponding Bohm trajectories, and the Wigner representation of the operators involved.

In Chapter~\ref{chap:wd}, we present an approach to tests of MR which steps away from the use of the INRM procedure.  By using the MR understanding of a particle's motion in terms of trajectories, we show that a measurement made at the origin is sufficient to measure an analogue to the standard temporal correlators from Chapter~\ref{chap:QHO1}.  We argue that this measurement protocol, consisting of a single interaction, is non-invasive, under certain reasonable and testable assumptions.  This hence proposes a useful alternative to ideal negative measurements.  We show a violation of these modified LG inequalities for a variety of states.

In Chapter~\ref{chap:summ} we summarize, and outline some speculative ideas.  For readability, technical details are relegated to a series of appendices at the ends of Chapters~\ref{chap:mlev}--\ref{chap:wd}.

\fancyhfoffset[LE,RO, RE, LO]{0cm}
\renewcommand{\chaptermark}[1]{ \markboth{#1}{} }
\renewcommand{\sectionmark}[1]{ \markright{#1}{} }
\fancyhf{}
\fancyhead[L]{\textsl{\thesection~~ \rightmark}}
\fancyhead[R]{\hyperref[link:2]\thepage}
\renewcommand{\headrulewidth}{1pt}

	\chapter{Fine's Theorem for Leggett-Garg tests with an arbitrary number of measurement times}
\label{chap:ntime}
\bookepigraph{3.5in}{At best I can skim a stone 17 steps with luck
,\\
But after that I have no control of the trajectory}{James Yorkston,}{Repetition}{1.0}
	\section{Introduction}
We stated in Section~\ref{sec:fine} that through Fine's theorem, we are able to determine the necessary and sufficient conditions for MR to hold, which at the level of three times, are the four LG3 inequalities (\ref{LG3a})--(\ref{LG3d}), and the twelve LG2 inequalities $p(s_i,s_j)\geq0$ Eqs. (\ref{LG2a})--(\ref{LG2d}), where $ij$ are the pairs $12$, $23$, $13$.   In this chapter, we will extend these conditions to measurements made at $n$ times, by considering experiments at pairs of times chosen from the set $\{ t_1, t_2 \cdots t_n\}$.  This allows us to perform a theoretical investigation of how systems which violate macrorealism behave when these tests are extended to an arbitrary number of measurements.

At first we will consider picking pairs from an $n$-cycle \cite{araujo2013}, i.e. the pairs $12$, $23$, \ldots, $(n-1)n$ and $1n$.  We will call the corresponding set of $n$-time LG inequalities the LG$n$s.  These measurements, which are assumed to be non-invasive, determine a particular set of pairwise probabilities $p(s_1, s_2)$, $p(s_2, s_3)$, \ldots, $p(s_{n-1}, s_n)$, $p (s_1, s_n)$, and in the instance that MR holds, will possess an underlying joint probability $p(s_1,s_2 \cdots s_n)$.  

For $n=3$, the LG inequalities include all possible correlators. However more generally these inequalities involve only a subset of correlators, for instance in the LG4s, $C_{13}$ and $C_{24}$ play no role.  In a departure from the $n$-cycle case, we will also consider the necessary and sufficient conditions when all possible pairwise probabilities are involved.

In Section \ref{c2s2}, we give a streamlined proof of Fine's theorem for the case of pairs of measurements taken from measurements made at three or four times, based on previously given proofs \cite{fine1982,halliwell2014} and in particular on Fine's ansatz.
In Section \ref{c2s3} we show how Fine's ansatz can be generalized to an arbitrary number of measurement times. We use this ansatz to prove Fine's theorem for LG inequalities at arbitrarily many times, using an inductive proof, and in the process, deduce the correct form of the complete set of LG inequalities in this case.

LG inequalities for $ n \ge 4$ always involve less than the complete set of two-time correlation functions (for example, in the familiar $n=4$ case there are a total of six correlators but only four appear in the LG inequalities). This raises the question of necessary and sufficient conditions for the existence of an underlying probability when all possible two-time correlators are fixed. We prove some results for this case in Section \ref{c2s4}, making contact with the ``pentagon inequality'' derived in Ref.~\cite{avis2010}.

In Section \ref{c2s5} we examine condition of both of the above types in the large $n$ limit, in which the LG inequalities appear to become easier to satisfy.
%Here we have concentrated on the most commonly-studied case of measurements of a single dichotomic variable $Q$ but a number of recent LG studies have looked at situation in which the variables of interest take three or more values at each time, raising the question of necessary and sufficient conditions in this case. We show in Section 5 that such situation are readily mapped onto the $n$-time LG inequalities for a single dichotomic variable and the appropriate conditions are then readily derived.
We summarize and conclude in Section \ref{c2s6}.

    \section{Fine's Theorem for the LG Inequalities at Three and Four Times}
\label{c2s2}
\subsection{Three-Time Case}
    
    In the three-time case, we are tasked with finding the conditions under which we can find a probability $p(s_1, s_2, s_3)$, which matches the three non-negative marginals $p(s_1, s_2)$, $p(s_2, s_3)$ and $p(s_1, s_3)$. Hence, this joint probability must be such that,       
    \beq
    p(s_1, s_2)=\sum_{s_3}p(s_1, s_2, s_3),
    \eeq
and likewise for $p(s_2,s_3)$ and $p(s_1,s_3)$.   We proceed using the moment expansion \cite{halliwell2013,klyshko1996} of the three-time probability,
    \beq
    p(s_1, s_2, s_3) = \frac{1}{8}\left(1+\sum_i B_i s_i +\sum_{i<j}C_{ij}s_i s_j + Ds_1 s_2 s_3\right),
    \label{qs2}
    \eeq
    where $i,j = 1,2,3,$ and the coefficients are defined by 
    \bea
    B_i &=& \sum_{s_1 s_2 s_3}s_i  \ p(s_1, s_2, s_3) = \langle Q_i \rangle,
    \\
    C_{ij}&=&\sum_{s_1 s_2 s_3}s_i s_j \ p(s_1, s_2, s_3) = \langle Q_i Q_j \rangle,
    \\
    D &=&\sum_{s_1 s_2 s_3}s_1 s_2 s_3 \ p(s_1, s_2, s_3) = \langle Q_1 Q_2 Q_3 \rangle. \label{EqD}
    \eea
    It is readily seen from the moment expansions of the form Eq.~(\ref{ps2}) that Eq.~(\ref{qs2}) matches the three marginals.
    %As discussed, we assume these two-time marginals are non-negative.  %This imposes restrictions on the coefficients $B_i$ and $C_{ij}$ such as
%    \beq
 %   1+B_1 - B_3 - C_{13}\geq 0.
 %   \eeq
  %  We will assume that all such restrictions are satisfied.
    
    Since the coefficients $B_i$ and $C_{ij}$ are fixed, the question is whether or not the coefficient $D$ may be chosen so that the unifying probability Eq.~(\ref{qs2}) is non-negative. We prove that a necessary and sufficient set of conditions for this are the three-time LG inequalities Eq.~(\ref{LG3a})--(\ref{LG3d}).

    Necessity is easy to establish. To prove sufficiency note that Eq.~(\ref{qs2}) is non-negative as long as,
    \beq
    A(s_1, s_2, s_3) \equiv 1 + \sum_i B_i s_i + \sum_{i < j} C_{ij} s_i s_j  \ge - D s_1 s_2 s_3.
    \label{A}
    \eeq
    For the four values of $s_1, s_2, s_3 $ for which $s_1 s_2 s_3 = - 1$, Eq.~(\ref{A}) gives
    four upper bounds on $D$,
    \beq
    A(s_1, s_2, s_3 ) \ge D,
    \eeq
    and for the values with $s_1 s_2 s_3 = 1$, this give four lower bounds on $D$
    \beq
    D \ge - A(s_1, s_2, s_3). 
    \eeq
    Hence a value of $D$ exists as long as all four upper bounds are greater that the
    all four lower bounds. This yields sixteen inequalities which are readily shown \cite{halliwell2014} to be the four three-time LG inequalities Eqs.~(\ref{LG3a})--(\ref{LG3d}),
    together with the twelve conditions $p(s_i,s_j) \ge 0 $ already assumed. 
   
   A natural question is whether the above upper and lower bounds on $D$ are compatible with the requirement $| D | \le 1 $, which follows from Eq.~(\ref{EqD}).   It is readily seen that this is the case as long as $A(s_1,s_2,s_3) \ge - 1$. It is not immediately obvious that this relation holds, but this may be shown as follows. First, the conditions $p(s_i,s_j) \ge  0$ imply that
   \beq
   p(s_1,s_2) + p(s_2,s_3) + p(s_1,s_3) \ge 0,
   \eeq
   which may be written
   \beq
   3 +  2 \sum_i B_i s_i + \sum_{i < j} C_{ij} s_i s_j \ge 0. \label{3p}
   \eeq 
We also have the three-time LG inequalities, which may be written,
\beq
1 + \sum_{i<j} C_{ij} s_i s_j \ge 0. \label{LGx}.
\eeq    
Adding Eq.~(\ref{3p}) and Eq.~(\ref{LGx}), we obtain,
\beq
2 +  \sum_i B_i s_i + \sum_{i < j} C_{ij} s_i s_j \ge 0,
\eeq    
which is precisely the condition $A(s_1,s_2,s_3) \ge -1$. Hence we find compatibility with $|D|  \le 1$.
This completes the proof.

%[CHECK: that $D$ is in range].
    
    \subsection{Four-Time Case}
    
    In the four-time case, the task is to find necessary and sufficient conditions for the existence of a joint probability $p(s_1,s_2,s_3,s_4)$ matching the four marginals $p(s_1,s_2)$, $p(s_2,s_3)$, $p(s_3,s_4)$, $p(s_1,s_4)$. As we will establish, these conditions are the eight four-time LG inequalities:
    \begin{align}
   -2  \le  C_{12} + C_{23} + C_{34} - C_{14} &\le 2,
   \label{LG4a}
   \\
   -2  \le  C_{12} + C_{23} - C_{34} + C_{14} &\le 2,
   \label{LG4b}
   \\
   -2 \le  C_{12} - C_{23} + C_{34} + C_{14} &\le 2,
   \label{LG4c}
   \\
   -2  \le  - C_{12} + C_{23} + C_{34} + C_{14} &\le 2.
   \label{LG4d}
   \end{align}
Necessity is again easy to establish.  Only four of the possible six marginals are fixed in this problem.  This means that although the four $B_i$ are fixed, the two correlators $C_{13}$ and $C_{24}$ are not. This matching problem may be solved using Fine's insightful ansatz,
    \beq
    p(s_1, s_2, s_3,s_4)=\frac{p(s_1, s_2, s_3)\ p(s_1, s_3, s_4)}{p(s_1, s_3)},
    \label{fineansatz}
    \eeq
which breaks the problem down into demonstrating the non-negativity of two three-time probabilities and a two-time probability. It is readily shown, by summing out the appropriate pairs of variables (with a judicious choice of the order in which this is done), that this ansatz matches the four marginals of interest. 

The three-time probability $p(s_1, s_2, s_3)$ is non-negative as long as its three two-time marginals $p(s_1,s_2)$, $p(s_2,s_3)$ and $p(s_1,s_3)$ are non-negative and as long as the four LG3 inequalities hold. These  inequalities may be written in the convenient form,
\beq
-1 + \left| C_{12} + C_{23} \right| \le C_{13} \le 1 - \left| C_{12} - C_{23} \right|,
\label{Cineq1}
\eeq
which puts bounds on the unfixed quantity $C_{13}$.
Similarly, the three-time probability $p(s_1, s_3, s_4)$ is non-negative as long as its three two-time marginals $p(s_1,s_3)$, $p(s_3,s_4)$ and $p(s_1,s_4)$ are non-negative and as long as the corresponding four LG3 inequalities hold, which may be written,
\beq
-1 + \left| C_{14} + C_{34} \right| \le C_{13} \le 1 - \left| C_{14} - C_{34} \right|.
\label{Cineq2}
\eeq
Note that the marginal $p(s_1,s_3)$ appears both in the demoninator of the Fine ansatz and also as a marginal of both three-time probabilities, and since it is not fixed, its non-negativity must be imposed as another restriction on $C_{13}$, which has the form,
\beq
-1 + | B_1 + B_3 | \le C_{13} \le 1 - | B_1 - B_3 |.
\label{Cineq3}
\eeq

Eq.~(\ref{Cineq1}) and Eq.~(\ref{Cineq2}) together imply that a value of $C_{13}$ may be chosen as long as the two lower bounds are less than the two upper bounds, which is equivalent to,
\beq
 \left| C_{12} \pm C_{23} \right| +  \left| C_{14} \mp C_{34} \right|\le 2.
 \eeq
These two equations are in fact a concise rewriting of the eight ${\rm LG}4$ inequalities, Eqs.~(\ref{LG4a})--(\ref{LG4d}), the desired result.

However, we must also ensure that the upper and lower bounds in Eq.~(\ref{Cineq3}) are compatible with those in     
Eq.~(\ref{Cineq1}) and Eq.~(\ref{Cineq2}) which is by no means obvious. Fortunately this is ensured by the fact that the four fixed marginals are non-negative, from which follow the inequalities
\bea
p(s_1,s_2) + p(-s_2,s_3) & \ge & 0,
\\
p(s_1,s_4) + p(s_3, -s_4) & \ge & 0.
\eea
Written out more explicitly these read,
\bea
2 + B_1 s_1 + B_3 s_3 &\ge& - C_{12} s_1 s_2 + C_{23} s_2 s_3,
\\
2+ B_1 s_1 + B_3 s_3 & \ge &- C_{14} s_1 s_4 + C_{34} s_3 s_4.
\eea
From this we see the compatibility of the bounds in Eq.~(\ref{Cineq3}) with those in the other two relations,
Eq.~(\ref{Cineq1}) and Eq.~(\ref{Cineq2}).
This completes the proof.

    \section{Generalization to an Abitrary Number of Times}
  \label{c2s3}  
    \subsection{Generalized Fine Ansatz}
    
   Given that the four-time LG inequalities ensure that a non-negative probability $p(s_1, s_2, s_3, s_4)$ may be found, we can ask about extending Fine's theorem to the case $n=5$.   We thus seek a joint probability $  p(s_1, s_2, s_3, s_4, s_5) $ matching the five pairwise probabilities $p(s_1,s_2)$, $p(s_2,s_3)$, $p(s_3,s_4)$, $p(s_4, s_5)$ and $p(s_1,s_5)$. We note that this may be solved using the generalized Fine ansatz,
    \beq
    p(s_1, s_2, s_3, s_4, s_5)=\frac { p (s_1, s_2, s_3 , s_4) \  p(s_1, s_4, s_5)} { p(s_1, s_4) }.
\label{fine5}
    \eeq
It is readily shown, by summing out triplets of values of the $s_i$ (where $i=1, 2, \cdots 5$) in a judiciously chosen order, that this ansatz matches the five fixed pairwise probabilities. The problem therefore reduces to the question of establishing the non-negativity of the four-, three- and two-time probabilities appearing in the ansatz, which will involve the four- and three-time LG inequalities and the non-negativity condition on $p(s_1,s_4)$.

We will not solve this problem explicitly, but note that it is suggestive of the Fine ansatz for the $n$-time case, which we postulate to be, 
\beq
    \label{fineansatzgen}
    p(s_1, \ldots, s_{n+1}) = \frac{p(s_1, \ldots, s_{n})\ p(s_1, s_{n}, s_{n+1})}{p(s_1, s_{n})}
    \eeq
    It is readily shown that this matches the $n$ pairwise marginals of interest, $p(s_i, s_j)$, where $(i,j)$ take the values $(1,2), (2,3), .... (n-1,n), (1,n)$. We will use this ansatz to given an inductive proof of Fine's theorem for $n$ times.

 We note in passing that through iterative application of the Fine ansatz, the $n$-time case may be reduced to a set of three-time problems, in terms of which the ansatz has the form, 
    \beq
    p(s_1, \ldots, s_{n+1})=p(s_1, s_2,s_3) \  \prod_{i=1}^{n-2}\frac{p(s_1, s_{i+2}, s_{i+3})}{p(s_1, s_{i+2})}.
    \eeq
This means that all $n$-time LG inequalities may be reduced to sets of three-time inequalities. A similar observation was noted in Ref.\cite{avis2010}.

\subsection{The LG Inequalities for An Arbitrary Number of Times}
    
    Based on the three and four-time inequalities (and the five-time inequalities given in Ref.\cite{emary2013}), we postulate that the $n$-time LG inequalities can be written as the $2^{n-1}$ relations, 
    \begin{equation}
    \label{lgansatz}
    a_1 C_{12}+a_2 C_{23}+ \ldots + a_{n-1}C_{n(n-1)}+ a_n C_{1 n} \leq n-2,
    \end{equation}
    where the coefficients $a_1, \ldots, a_{n}$ take values $\pm1$, and we constrain the product of all the coefficients $a_i$ to be negative,
    \begin{equation}
    \prod_{i=1}^n a_i = -1.
    \end{equation}
    
    This allows us to write one of the coefficients in terms of the others, for example, $a_n = -a_1 a_2 \ldots a_{n-1}$.  We see that the LG inequalities involve all possible sums of correlation functions with coefficients $\pm 1$ with an odd number of minus signs.  Some specific higher order LG inequalities have been written down previously, e.g. Refs. \cite{emary2013,athalye2011,wilde2012,barbieri2009}, and we note that one of LG5s derived here has the same mathematical form as the KCBS inequality used in contextuality studies \cite{klyachko2008}.  The inequality presented in Ref.~[\cite{athalye2011}] includes the lower bounds of $-n$ for odd $n$, and $-(n-2)$ for even $n$. For odd $n$ this bound represents the algebraic limit and is hence trivially satisfied.  For even $n$, we note our parametrisation Eq.~(\ref{lgansatz}) includes this lower bound, however expressed as an upper bound on the negative version of the kernel.

Eqs.~(\ref{lgansatz}) are readily seen to be necessary conditions for the existence of an underlying probability. The proof of this proceeds from the inequality
\beq
a_1 s_1 s_2 + a_2 s_2 s_3 + \cdots a_{n-1} s_{n-1} s_n + a_n s_n s_1 \le n-2,
\eeq
where the $s_i$ take values $\pm 1$, which can be established by choosing a fixed set of values of the $s_i$ (such as setting them all equal to $+1$) and the considering the effect of flipping their signs. Averaging both sides of this inequality using an underlying probability distribution over $s_1, s_2, \cdots, s_n$ then yields 
Eq.~(\ref{lgansatz}). We now establish sufficiency.

\subsection{Inductive Proof}
     
     Following the method of Section \ref{c2s2}, we now use the Fine ansatz Eq.~(\ref{fineansatzgen}) and the $n$-time LG inequalties Eq.~(\ref{lgansatz}) to show that the sufficient conditions for non-negativity of $p(s_1, \cdots s_{n+1})$ are the $(n+1)$-time LG inequalities.
     The probability $p(s_1, \cdots s_n)$ is non-negative as long as the LG inequalities Eq.~(\ref{lgansatz}) are satisfied. These may be written,
    \beq
    A(a_n) + a_n C_{1n}\leq n-2,
    \label{eqn:shorthandlgn}
    \eeq
    where the function $A(a_n)$ is
    \beq
    A(a_n)= a_1 C_{12}+a_2 C_{23}+ \ldots + a_{n-1}C_{n(n-1)}.
    \eeq
    Noting that the argument $a_n$ takes values $\pm1$ and also that $a_1 \ldots a_{n-1}=-a_n$, we see that $A(\pm)$ are the sums of correlators with an odd/even number of minus signs.  We can now rewrite inequality Eq.~(\ref{eqn:shorthandlgn}) as upper and a lower bound on $C_{1n}$,
    \beq
    -(n-2)+A(-)\leq C_{1n}\leq (n-2)-A(+).
\label{B1}
    \eeq
Similarly, the probability $p(s_1, s_{n}, s_{n+1})$ is non-negative if a set of three-time LG inequalities hold which if written in the general form Eq.~(\ref{lgansatz}), are
    \beq
    b_1 C_{1n}+b_2 C_{n(n+1)}-b_1b_2 C_{1(n+1)}\leq1,
    \eeq 
    where $b_1, b_2$ take values $\pm 1$. This may 
be rewritten as
    \beq
    b_1 C_{1n} + B(b_1) \leq 1,
    \eeq    
    where
    \beq
    B(b_1) = b_2 C_{n(n+1)} - b_1 b_2 C_{1(n+1)}.
    \eeq
    Note that $B(\pm)$ are the sums of $C_{n(n+1)}$ and $C_{1(n+1)}$ with an odd/even number of minus signs.  This can be rearranged to give another upper and a lower bound on $C_{1n}$:
    \beq
    -1 +B(-)\leq C_{1n}\leq 1 - B(+).
\label{B2}
    \eeq
    A value of $C_{1n}$ obeying both set of bounds Eqs.~(\ref{B1}), (\ref{B2}) may then be found as long as
\bea
B(-)+A(+)  & \leq(n-2)+1,
\\
B(+)+A(-)  & \leq(n-2)+1.
\eea
These relations may be rewritten,
\beq
A(a_n) + B( - a_n) \le (n+1) - 2
\label{res}
\eeq
It is not difficult to see that the left-hand side consists of all possible sums of correlators with an odd number of minus signs hence we have basically achieved a condition of the form Eq.~(\ref{lgansatz}) with $n$ replaced with $(n+1)$, as required.
To be more explicit, Eq.~(\ref{res}) reads,
\beq
a_1C_{12}+a_2 C_{23}+\ldots + a_{n-1} C_{n(n-1)}+b_2 C_{n(n+1)} + a_n b_2C_{1(n+1)}\leq (n+1)-2.
\eeq
This is a sum over $(n+1)$ correlators with $(n+1)$ independent coefficients taking values $\pm 1$ whose product is $ a_1 \cdots a_n b_2^2  = -1$. This is precisely of the form Eq.~(\ref{lgansatz}).

Finally, as in the four-time case, we must also confirm that the restrictions on $C_{1n}$ are compatible with the non-negativity of $p(s_1,s_n)$. Using the moment expansion of the probability $p(s_1,s_n)$, its non-negativity gives us the following upper and lower bounds on $C_{1n}$,
    \beq
-1+\left|B_1+B_{n}\right|\leq C_{1n} \leq 1- \left|B_1 - B_{n}\right|,
    \label{eqn:positivepp}
    \eeq 
    and these must be compatible with the bounds Eqs.~(\ref{B1}), (\ref{B2}).
    That is we require that we are always able to pick a $C_{1n}$ that satisfies the three sets of inequalities.  
    Since the measured pair probabilities are taken to be non-negative, we can add them and form new inequalities, 
    \beq
    p(s_1, -s_2)+p(s_2, -s_3)+\ldots+p(s_{n-1}, -s_n)\geq0,
    \eeq
    which using the moment expansion may be written explicitly as
    \beq
    (n-2)+1+B_1 s_1-B_n s_n - \sum_{i=1}^{n-1}C_{i(i+1)}s_i s_{i+1}\geq0.
    \label{eqn:positiveppn}
    \eeq
    For the case $s_1=-s_n$, the sum may be expressed as a sum of correlators with an odd amount of minus signs, e.g. a member of $A(+)$, and conversely for $s_1=s_n$, the sum may be expressed as a member of $A(-)$.  With this observation, it is simple to show that inequalities~(\ref{eqn:positivepp}) and Eq.~(\ref{B1}) are indeed compatible.
   
    To ensure compatibility with Eq.~(\ref{B2}), a similar argument may be made using the sum of pair probabilities,
    \beq
    p(s_1, -s_{n+1})+p(-s_n, s_{n+1})\geq 0,
    \eeq
    yielding
    \beq
    2+B_1s_1 - B_ns_n-C_{1(n+1)}s_1 s_{n+1}-C_{n(n+1)}s_n s_{n+1}\geq0.
    \eeq
    This inequality takes the same form as (\ref{eqn:positiveppn}), and the same arguments can be made to show that the inequalities~(\ref{eqn:positivepp}) and Eq.~(\ref{B2}) are compatible.
This completes the inductive step of the proof.

We now observe that the three-time inequalities Eqs(\ref{LG3a})--(\ref{LG3d}) we proved earlier also match the form of (\ref{lgansatz}), and hence act as the base case for the inductive proof.  We have therefore proved the $n$-time generalisation of Fine's theorem, that for any $n\geq3$, the joint $n$-time probability distribution is guaranteed to exist, as long as all $2^{n-1}$ $n$-time LG inequalities Eq.~(\ref{lgansatz}) are satisfied, together with the $4n$ two-time LG inequalities consisting of the non-negativity conditions on the fixed pairwise probabilities.

\section{Inequalities involving all of the two-time correlators}
\label{c2s4}
An interesting feature of the LG inequalities is that they in general involve only a {\it subset} of all possible two-time correlators. For the three-time case all three correlators are measured and an underlying probability sought, but for the four time case only four out of the six possible correlators are measured. In general, the LG inequalities at $n$ times involve $n$ two-time correlators out of a total possible number of $n(n-1)/2$.
This choice of using only a subset of the total set of correlators arose because LG experiments were devised by way of analogy to Bell tests, in which one carries out pairs of measurements on a pair of particles, but not two measurements on the same particle, which means that certain correlators are not relevant experimentally. However this restriction is irrelevant in LG tests since all pairs of measurements are carried out on the same particle and furthermore, there is no obvious barrier experimentally to measuring the set of {\it all} two-time correlators. {This naturally raises the question as to the form of necessary and sufficient conditions for an underlying probability in the case in which the full set of $n(n-1)/2$ two-time correlators are measured, not just the $n$ correlators measured in standard LG tests. Since far more data is specified in this case than in the usual LG case, we expect that the necessary and sufficient condtions will be stronger than any set of LG inequalities. This case is rarely considered in the LG literature (and in fact Ref.\cite{avis2010} is the only paper we are aware of that considers it). It involves some new features compared to the standard LG case which we will describe.}

\subsection{General Properties}

Avis \textsl{et al.} \cite{avis2010} consider the following condition for the case of measurements made at five possible times involving a sum of all ten possible correlators:
\beq
2 + \sum_{i<j} C_{ij} \ge 0,
\label{Av}
\eeq
where $i,j = 1, 2, \cdots 5$. 
{They argue that this condition is not a consequence of any LG inequalities and also that it may be violated by quantum mechanics. They refer to it as a ``pentagon inequality'' (out of acknowledgement for its geometric origins using the cut polytope).}

We now examine conditions of this type systematically. A general class of relations of the form Eq.~(\ref{Av}) are readily derived by noting the definition of the correlator Eq.~(\ref{eq:classcorr}) and using the simple relation,
\beq
\left\langle \left( \sum_{i=1}^n s_i Q_i \right)^2 \right\rangle \ge  
\begin{cases}
1   &\mbox{if} {\ n \  {\rm odd} },  \\ 
0  & \mbox{if} {\ n \ { \rm  even} }.
\end{cases}
\eeq
This is readily seen to be true for a macrorealistic theory since all the $Q_i$ (and the $s_i$) are $\pm 1$, and for $n$ even, all the terms in the sum may cancel, but in the odd case, there must always be one left over.
This in turn may be written,
\beq
n + 2 \sum_{i<j} s_i s_j C_{ij} \ge 
\begin{cases}
1   &\mbox{if} {\ n \  {\rm odd} },  \\ 
0  & \mbox{if} {\ n \ { \rm  even} }.
\end{cases}
\label{neweq}
\eeq
{We will refer to these conditions as ``$n$-gon inequalities''}.
For $n=3$, these are in fact just the three-time LG inequalities:
\beq
1 + s_1 s_2 C_{12} + s_2 s_3 C_{23} + s_1 s_3 C_{13} \ge 0.
\label{L123}
\eeq
For $n=4$ and $n=5$, {the $n$-gon inequalities} are in fact the same,
\beq
2 + \sum_{i<j} s_i s_j C_{ij} \ge 0,
\label{Av2}
\eeq
which is a clear generalization of Eq.~(\ref{Av}). However, {the $n$-gon inequalities} in the $n=4$ case can be written as an average of four sets of three-time LG inequalities, namely the inequalities Eq.~(\ref{L123}), averaged with the three other sets obtained by choosing the time pairs from the triples $(t_1,t_2,t_4)$, $(t_2,t_3,t_4)$ and $(t_1,t_3,t_4)$. This is not possible in the $n=5$ case, as indicated in Ref.~\cite{avis2010} and as we see explicitly below.  Hence for $n=5$ the {$n$-gon inequalities}
are stronger than the LG inequalities.

An interesting observation concerns whether the {$n$-gon inequalities} may continue to be satisfied in quantum mechanics or not. Replacing the variables $Q_i$ with their quantum operator counterparts $\hat Q_i$, we see that for all $n$ we have
\beq
\left\langle \left( \sum_{i=1}^n s_i \hat Q_i \right)^2 \right\rangle \ge  0.
\label{qc}
\eeq
These equalities therefore have the form
\beq
n + 2 \sum_{i<j} s_i s_j C_{ij} \ge  0,
\eeq
for all $n$, where here the quantum correlators are given by $C_{ij} = \frac{1}{2}\langle \hat Q_i \hat Q_j + \hat Q_j \hat Q_i \rangle$. This means that the {$n$-gon inequalities} Eq.~(\ref{neweq}) are most interesting for $n$ odd where there is clear difference between the classical and quantum cases. For $n$ even, the classical and quantum versions coincide, at least for quantum correlators given by the above formula. (There may be a difference in the quantum case if the correlators are obtained differently, by so-called degeneracy-breaking measurements \cite{budroni2014,pan2018}).

For $n=3, $ Eq.~(\ref{qc}) does in fact give the L{\"u}ders bound \cite{cirelson1980} for the three-time LG inequalities,
\beq
s_1 s_2 C_{12} + s_2 s_3 C_{23} + s_1 s_3 C_{13} \ge - \frac{3}{2}.
\eeq
Similarly for $n=5$ we have
\beq
\frac{5}{2} +  \sum_{i<j} s_i s_j C_{ij} \ge  0,
\eeq
in contrast to the classical version Eq.~(\ref{Av2}).

\subsection{Sufficient Conditions: A Conjecture}

{We suppose that measurements are made (as described in the Introduction) to determine all possible pairwise probabilities of the form $p(s_i,s_j)$, where $i<j$ and $i,j$ run over $n$ values. We will suppose that they are all non-negative, i.e. all possible two-time LG inequalities are satisfied. The question is then to determine the necessary and sufficient conditions for which there exists a joint probability $p(s_1,s_2, \cdots s_n)$ which matches all the pairwise probabilities.  The $n$-gon inequalities Eq.~(\ref{neweq}) are clearly necessary conditions and likewise all possible $n$-time LG inequalities. Since the $n$-time LG inequalities are equivalent to sets of three-time inequalities a natural conjecture is as follows.

\begin{framed}\noindent \textbf{Conjecture:}
\textsl{A set of sufficient conditions for the existence of a joint probability matching all possible pairwise probabilities consists of the $n$-gon equalities Eq.~(\ref{neweq}) together with all possible three-time LG inequalities of the form}
\beq
1 + s_i s_j C_{ij} + s_i s_k C_{ik} + s_j s_k C_{jk} \ge 0,
\label{LGcomp}
\eeq
\textsl{where $i<j<k$.}
\end{framed}

A simple combinatorial analysis shows that
there are $2^{n-1}$ $n$-gon inequalities and $2n(n-1)(n-2)/3$ three-time inequalities.
(There are also $2n(n-1)$ two-time LG inequalities but these are assumed satisfied, as stated).}

We prove this conjecture for the (essentially trivial) case $n=4$ and the non-trivial case $n=5$. Our proof is restricted to the special ``symmetric'' case in which correlators involving an odd number of variables are zero.
This is not necessarily that restrictive for at least two reasons. One is that, as seen in Ref.\cite{halliwell2014}, the symmetric case is typically enough to establish the form that sufficient conditions should take -- no significant new conditions arise when going to the general case. Secondly, one can see from a quantum-mechanical analysis that the odd correlators can be made to vanish by choice of initial state. (This is accomplished by finding a reflection operator $R$ for which $ R \hat Q R = - \hat Q$, $ R H R = H$, where $H$ is the Hamiltonian, and chosing the state such that $R | \psi \rangle = | \psi \rangle $.)
This is not a significant restriction since the time spacings between measurements are the most important adjustable parameters in experimental tests.

\subsection{Sufficient Conditions: The Case $n=4$}

Consider then the case $n=4$. We seek a probability $p(s_1,s_2,s_3,s_4)$ matching all six correlators $C_{ij}$ where $ij = 12,13,14,23,24,34$. This is conveniently approached using the moment expansion,
\beq
p(s_1,s_2,s_3,s_4) = \frac{1}{16} \left(1 + \sum_{i<j} s_i s_j C_{ij} + E s_1 s_2 s_3 s_4 \right),
\eeq
for some constant $E$, where $-1 \le E \le 1$. This is non-negative as long as $E$ may be chosen so that
\beq
f(s_1,s_2,s_3,s_4) \equiv 1 + \sum_{i<j} s_i s_j C_{ij} \ge - E s_1 s_2 s_3 s_4.
\eeq
This reads
\beq
f(s_1,s_2,s_3,s_4) \ge  E,
\label{E1}
\eeq
for all values of the $s_i$ for which $ s_1 s_2 s_3 s_4 = - 1$ and reads
\beq
E \ge - f(s_1',s_2',s_3',s_4'),
\eeq
for all values for which $ s_1' s_2' s_3' s_4' = +1 $. Hence and $E$ may be found as long as all the lower bounds on it are less than all its upper bounds. This clearly leads to a set of conditions of the form,
\beq
f(s_1,s_2,s_3,s_4) + f(s_1,s_2,s_3,- s_4) \ge 0,
\eeq
plus three more with the minus sign in the other three places. (In general there would also be conditions with three sign flips but since $f(s_1,s_2,s_3,s_4)$ is symmetric under flipping all four signs this is equivalent to just one sign flip.) It is readily seen that these conditions are equivalent to the conditions that all four possible three-time probabilities $p(s_i,s_j,s_k)$  with $i<j<k$ are non-negative, which is guaranteed if all sets of three-time LG inequalities are satisfied, {as we have assumed}.
This proves the conjecture for $n=4$ in the symmetric case.

\subsection{Sufficient Conditions: The Case $n=5$}

For the $n=5$ case we seek a probability $p(s_1,s_2,s_3,s_4,s_5)$ matching all ten correlators $C_{ij}$. The moment expansion in the symmetric case is,
\beq
\begin{split}
p(s_1,s_2,s_3,s_4,s_5) = \frac{1}{32} \bigg(1 +& \sum_{i<j} s_i s_j C_{ij}  
+ E_1  s_2 s_3 s_4 s_5 
+ E_2 s_1 s_3 s_4 s_5
\\ 
+& E_3 s_1 s_2 s_4 s_5
+ E_4 s_1 s_2 s_3 s_5
+E_5 s_1 s_2 s_3 s_ 4
\bigg),
\end{split}
\eeq
for some constants $E_1, E_2, E_3, E_4, E_5$. We seek the conditions under which these constants may be chosen to ensure that the probability is non-negative. Following the method used in the $n=4$ case, we proceed to eliminate $E_1$ by
noting that the condition $p(s_1,s_2,s_3,s_4,s_5) \ge 0 $ may be written,
\beq
F(s_1,s_2,s_3,s_4,s_5) + E_1 s_2 s_3 s_4 s_5 \ge 0,
\eeq
where
\beq
\begin{split}
F(s_1,s_2,s_3,s_4,s_5) = 1 +& \sum_{i<j} s_i s_j C_{ij}  
+ E_2 s_1 s_3 s_4 s_5
+ E_3 s_1 s_2 s_4 s_5
\\
+& E_4 s_1 s_2 s_3 s_5
+E_5 s_1 s_2 s_3 s_ 4.
\end{split}
\eeq
In close analogy to the $n=4$ case, we readily find that the upper bounds on $E_1$ are greater than the lower bounds, and so a suitable value of $E_1$ may be chosen, as long as firstly,
\beq
F(s_1,s_2,s_3,s_4,s_5) + F(s_1,s_2,s_3,s_4,-s_5) \ge 0,
\label{F1}
\eeq
plus three more similar conditions in which $s_2$, $s_3$ or $s_4$ have their sign flipped; and secondly,
\beq
F(s_1,s_2,s_3,s_4,s_5) + F(s_1,s_2,-s_3,-s_4,-s_5) \ge 0,
\label{F2}
\eeq
plus three more conditions in which the triples $(s_2,s_3,s_4)$, $(s_2,s_4,s_5)$ or $(s_2,s_3,s_5)$ have all three signs flipped.

Eq.~(\ref{F1}) has the form,
\beq
1 + \sum_{{i<j} \atop {i,j \ne 5}} s_i s_j C_{ij} + E_5 s_1 s_2 s_3 s_4  \ge 0,
\label{F3}
\eeq
which is precisely the $n=4$ case. It follows that
$E_5$ can be chosen to ensure that these inequalities are satisfied as long as a set of three-time LG inequalities holds. There will in addition be three more variants involving $E_2, E_3, E_4$. {It is readily shown that the three-time LG inequalities that need to be satisfied for Eq.~(\ref{F3}) and its three variants to be satisfied is the complete set of three-time LG inequalities, Eq.~(\ref{LGcomp}), with $ijk$ taking values $123, 124, 125, 134, 135, 145, 234, 235, 245, 345$.}

However, the upper and lower bounds on $E_5$ implied by Eq.~(\ref{F3}) must be also compatible with the further upper and lower bounds on $E_5$ derived below, and likewise for the three variants of this inequality.

Similarly, Eq.~(\ref{F2}) has the form,
\beq
\begin{split}
1 &+ s_1 s_2  C_{12} + s_3 s_4 C_{34} + s_3 s_5 C_{35} + s_4 s_5 C_{45}
\\
&+ E_3 s_1 s_2 s_4 s_5 + E_4 s_1 s_2 s_3 s_5 + E_5 s_1 s_2 s_3 s_4 \ge 0.
\end{split}
\label{F4}
\eeq
There will be three similar relationships for the three variants of Eq.~(\ref{F2}), but will clearly be qualitatively the same by relabelling. We can solve this inequality by eliminating, say, $E_4$, by writing,
\beq
G(s_1,s_2,s_3,s_4,s_5) + E_4 s_1 s_2 s_3 s_5 \ge 0
\eeq
where
\beq
\begin{split}
G(s_1,s_2,s_3,s_4,s_5) =
1 &+ s_1 s_2  C_{12} + s_3 s_4 C_{34} + s_3 s_5 C_{35} + s_4 s_5 C_{45}
\\
&+ E_3 s_1 s_2 s_4 s_5 + E_5 s_1 s_2 s_3 s_4 
\end{split}
\label{G2}
\eeq
Proceeding as above, we readily find that a value of $E_4$ may be chosen as long as the following two sets of conditions hold. Firstly,
\beq
G(s_1,s_2,s_3,s_4,s_5) + G(s_1,s_2,s_3,s_4,-s_5) \ge 0,
\label{G3}
\eeq
plus three more similar conditions with $s_1$, $s_2$ or $s_3$ having their sign flpped; and second,
\beq
G(s_1,s_2,s_3,s_4,s_5) + G(-s_1,-s_2,-s_3,s_4,s_5) \ge 0,
\label{G4}
\eeq
plus three more conditions with the triples $(s_1,s_2,s_5)$, $(s_1,s_3,s_5)$ and $(s_2,s_3,s_5)$ having all three signs flipped.

Eq.~(\ref{G3}) has the form,
\beq
1 + s_1 s_2 C_{12} + s_3 s_4 C_{34} + E_5 s_1 s_2 s_3 s_4 \ge 0.
\label{G5}
\eeq
It is easily shown that a value of $E_5$ may be chosen so that these inequalities are satisfied, relying only on the property $ |C_{ij} | \le 1 $ of the correlators. We must also confirm that Eq.~(\ref{G5}) is compatible with other bounds on $E_5$ and in particular the bounds implied by Eq.~(\ref{F3}). It is readily seen that this is the case subject only to a set of three-time LG inequalities. Similar statements apply to the three variants of Eq.~(\ref{G3}).

Consider now the four inequalities of the form Eq.~(\ref{G4}). Two of them are,
\bea
1 + s_1 s_2 C_{12} + s_3 s_4 C_{34} + E_5 s_1 s_2 s_3 s_4 &\ge& 0,
\\
1 + s_1 s_2 C_{12} + s_4 s_5 C_{45} + E_3 s_1 s_2 s_4 s_5 &\ge& 0,
\eea
which are straighforwardly handled as above. The other two inequalities turn out in fact to be the same and have the form
\beq
1 + s_3 s_5 C_{35} + E_3 s_ 1 s_2 s_4 s_5 + E_5 s_1 s_2 s_3 s_4 \ge 0.
\label{G8}
\eeq
This set of inequalities taken on its own can be satisfied for some $E_3$ and $E_5$ using $|C_{35}| \ge 1$.
However, the bounds on $E_3$ and $E_5$ implied by Eq.~(\ref{G8}) must be compared with all other inequalities involving $E_3$ and $E_5$. A potentially lengthy search through all the possibilities is required to check all the cases, but it is not difficult to see that 
the only ones that give non-trivial new conditions are Eq.~(\ref{F3}) and its variant involving $E_3$. These two inequalities may be written,
\bea
E_5 s_1 s_2 s_3 s_4 & \ge& - 1 - \sum_{{i<j} \atop {i,j \ne 5}} s_i s_j C_{ij},
\\
E_3 s_1 s_2 s_4 s_5  & \ge & - 1 - \sum_{{i<j} \atop {i,j \ne 3}} s_i s_j C_{ij}.
\eea
We compare these two inequalities to Eq.~(\ref{G8}) with the signs of $s_3$ and $s_5$ reversed, which reads,
\beq
1 + s_3 s_5 C_{35} \ge  E_3 s_ 1 s_2 s_4 s_5 + E_5 s_1 s_2 s_3 s_4.
\eeq
The last three inequalities are compatible as long as,
\beq
1 + s_3 s_5 C_{35} \ge -2 - \sum_{{i<j} \atop {i,j \ne 3}} s_i s_j C_{ij}
- \sum_{{i<j} \atop {i,j \ne 5}} s_i s_j C_{ij}.
\eeq
Written out in full this inequality is conveniently written
\beq
\left( 1 + s_1 s_2 C_{12} + s_1 s_4 C_{14} + s_2 s_4 C_{24} \right) + 2 + \sum_{i<j} s_i s_j C_{ij} \ge 0,
\eeq
where the sum is over all $ij$. Noting that the first term in parantheses is a LG inequality so is non-negative, the total inequality is satisfied as long as
\beq
2 + \sum_{i<j} s_i s_j C_{ij} \ge 0.
\eeq
This is the desired result, Eq.~(\ref{Av2}), and establishes the conjecture for $n=5$ in the symmetric case.

\subsection{Other Approaches}

Although as argued the symmetric case is not as restrictive as it might seem, it would clearly be desirable to extend the proof to the general case. Here we have focused on a purely algebraic proof but a geometric one involving the cut polytope geometry discussed in Ref.~\cite{avis2010} is clearly a natural place to look and may get
away from the restriction to the symmetric case considered here. 

Another very different approach would be to do numerical experiments which confirm the conjecture.
We start with the moment expansion in the general case:
\beq
\begin{split}
p(s_1, s_2, s_3, s_4, s_5) = \frac {1} {32} \bigg( 1 +&
\sum_i B_i s_i + \sum_{i<j} C_{ij} s_i s_j + \sum_{i<j<k} D_{ijk} s_i s_j s_k
\\
+& \sum_{i<j<k<\ell} E_{i j k \ell}  s_i s_j s_k s_\ell 
+ F s_1 s_2 s_3 s_4 s_5 \bigg).
\end{split}
\label{qs}
\eeq
We then suppose that values of the fixed quantities $B_i$, $C_{ij}$ are chosen at random, which may or may not satisfy the requisite conditions (i.e. the full set of two and three-time LG inequalities and the generalized pentagon inequalities Eq.~(\ref{Av2})). For each set of values of the fixed quantities, a numerical search is carried out to find
values of the coefficients $D_{ijk}$, $E_{ijk\ell}$ and $F$ for which the $32$ inequalities $p(s_1, s_2, s_3, s_4, s_5) \ge 0 $ are satisfied. 
The conjecture is upheld if the success or failure of the search is found to be in direct correspondence to whether the fixed values do or do not satisfy the requisite conditions.
%This sort of numerical experiment supplies grounds for confidence in the conjecture if no counter-examples are found after a large number of trials.
{We have carried out a series of such numerical experiments using Mathematica and found no counter-examples to the conjecture. However, we find that the interpretation of such numerical experiments is subtle and it is difficult to see how to use them to make reliable claims.}

\subsection{Further Generalizations}

A further natural generalization is to consider situations in which third and higher-order correlators are measured. The appropriate necessary and sufficient conditions that these must satisfy have been given in Refs.~\cite{halliwell2019a,pan2018}.  It has been shown for the canonical spin system, that violations of MR may be concealed in such higher order correlations~\cite{majidy2021a}. These conditions are of experimental interest because
measurement of some of these higher-order correlators has been carried out \cite{bechtold2016}.

\section{The Limit of a Large Number of Measurements}
\label{c2s5}
We now make some observations about the simplifications and general features that arise for large $n$, for the ${\rm LG}n$ inequalities Eq.~(\ref{lgansatz}), and the $n$-gon inequalities Eq.~(\ref{neweq}).

\subsection{Algebraic Argument}
If we assume that the $C_{ij}$ depend solely on the time difference $t_j - t_i$, and also make the common assumption of equal time spacing $\tau$ between the times $t_i$,
then we have $C_{12}=C_{23}=\ldots=C_{n(n+1)}$.  The general $n$-time LG inequalities may then be written as
\beq
C_\tau\left(a_1+\ldots+a_{n-1}\right) \pm C_{n\tau} \leq n-2,
\eeq
Now note that the sum of the $a_i$ coefficents takes extremal values of $\pm(n-1)$, so we may write the inequality in both worst cases as
\beq
-\frac{n-2}{n-1}\leq C_\tau + \frac{C_{n\tau}}{n-1}\leq \frac{n-2}{n-1}
\eeq
In the limiting case of large $n$, this inequality tends toward
\beq
-1\leq C_{\tau}\leq 1
\eeq
which is a condition the correlators satisfy already. Hence the ${\rm LG}n$ become easier to satisfy for large $n$, becoming identically satisfied as $n$ goes to infinity. 
Mathematically, there is an intuitively clear reason why it might become easier to satisfy ${\rm LG}n$ for large $n$. This is that the LG inequalities involve just $n$ correlators, but the probabilities for $n$ times have $2^n-1$ independent components so the requirement to match $n$ correlators becomes less restrictive, relatively speaking, as $n$ increases. However, this is perhaps counter to physical intuition, which suggests that it should be harder to assign probabilities to histories in which more variables are specified, especially in quantum mechanics.

There are two cases to consider. One is the case in which we have a fixed time interval $[0,T]$ and just add progressively more measurements in that interval. However, in that case, $C_{12}$ will tend to $1$ for large $n$ (since $t_2-t_1$ goes to zero) and then ${\rm LG}n$ can only be satisfied for $C_{1n}$ = 1. 

The second case is that which we keep adding extra measurements at later times, so prolonging the total time interval, and in that case it does indeed seem that, as indicated, ${\rm LG}n$ gets easier to satisfy as $n$ increases.
In LG tests we are trying to determine whether or not there exists a probability $p(s_1, s_2 \cdots s_n)$ corresponding to certain sets of measurements taken at $n$ times. In general, 
we would expect it to be harder to find a probability for $(n+1)$ times,  $p(s_1, s_2 \cdots s_n, s_{n+1})$, since that would be a fine graining, and easier to find a probability for $(n-1)$ times, $p(s_1, s_2, \cdots s_{n-1})$ since that is a coarse-graining.

However, LG tests refer to a very specific set of measurements in which only certain correlators are measured and as a consequence the probabilities it seeks for $(n-1)$, $n$, and $(n+1)$ times are not simply related. The $n$-time LG inequalities refer to the correlators $C_{12} \cdots C_{(n-1)n}$ and $ C_{1n}$ but the $(n+1)$-time inequalities refer to the correlators $C_{12} \cdots C_{n(n+1)}$ and $ C_{1(n+1)}$, which does not involve $C_{1n}$ but does involve two new correlators that do not appear in the $n$-time case. This means that the probabilities for $n$ times, and $(n+1)$ times, that these LG inequalities test for are not simple fine or coarse grainings of each other -- the fixed quantities in the $n$-time case are not a subset of the fixed quantities of the $(n+1)$-time case.

Clearly a case in which the set of fixed quantities {\it are} simply related subsets for different values of $n$ 
is that in which {\it all} possible two-time correlators are fixed.  There, as demonstrated, we require not just the LG inequalities to hold but also the conditions 
Eq.~(\ref{neweq}). Since the total number of two-time correlators is  $ n (n-1)/2$, Eq.~(\ref{neweq}) will become {\it harder} to satisfy as $n$ increases, even if the LG inequalities become easier. This is the intuitively expected result.

\subsection{Approximate Measure of Violation from the Central Limit Theorem}

By looking at the case where measurements are spaced equally in time, we have provided a simple argument for the behaviour of the $n$-time LG inequalities in the large $n$ regime.  We now make a more sophisticated and general argument, which includes the case of arbitrarily spaced measurements.  This argument is applicable to both the ${\rm LG}n$ and $n$-gon inequalities Eqs.~(\ref{lgansatz}), (\ref{neweq}), and more generally to any inequality formed as a sum of correlators.

We aim to establish for a given $n$, a measure of how easy, or hard it is to violate a given inequality from the family of ${\rm LG}n$ or $n$-gon inequalities.  A crude approach is to ask what volume of the parameter space of $C_{12},\ldots,C_{1n}$ leads to a violation of the given inequality.  This should be viewed as a purely geometric estimate, i.e. estimating the volume of the polytope defined by the given inequalities.  

Owing to the equivalence between volume in parameter space, and the probability of a uniformly distributed event happening, this question is profitably reframed as a question of probabilities.  If we pick random values for $C_{12},\ldots,C_{1n}$ from a uniform distribution, then the probability of a random point leading to a violation of the inequality will exactly correspond to the volume in parameter space capable of supporting a violation.  Hence calculating this probability will give us the desired measure of the ease or difficulty of violation.

\begin{figure}
	\begin{center}
		\includegraphics[trim={0cm 4.5cm 0cm 5cm},clip,width=0.65\textwidth]{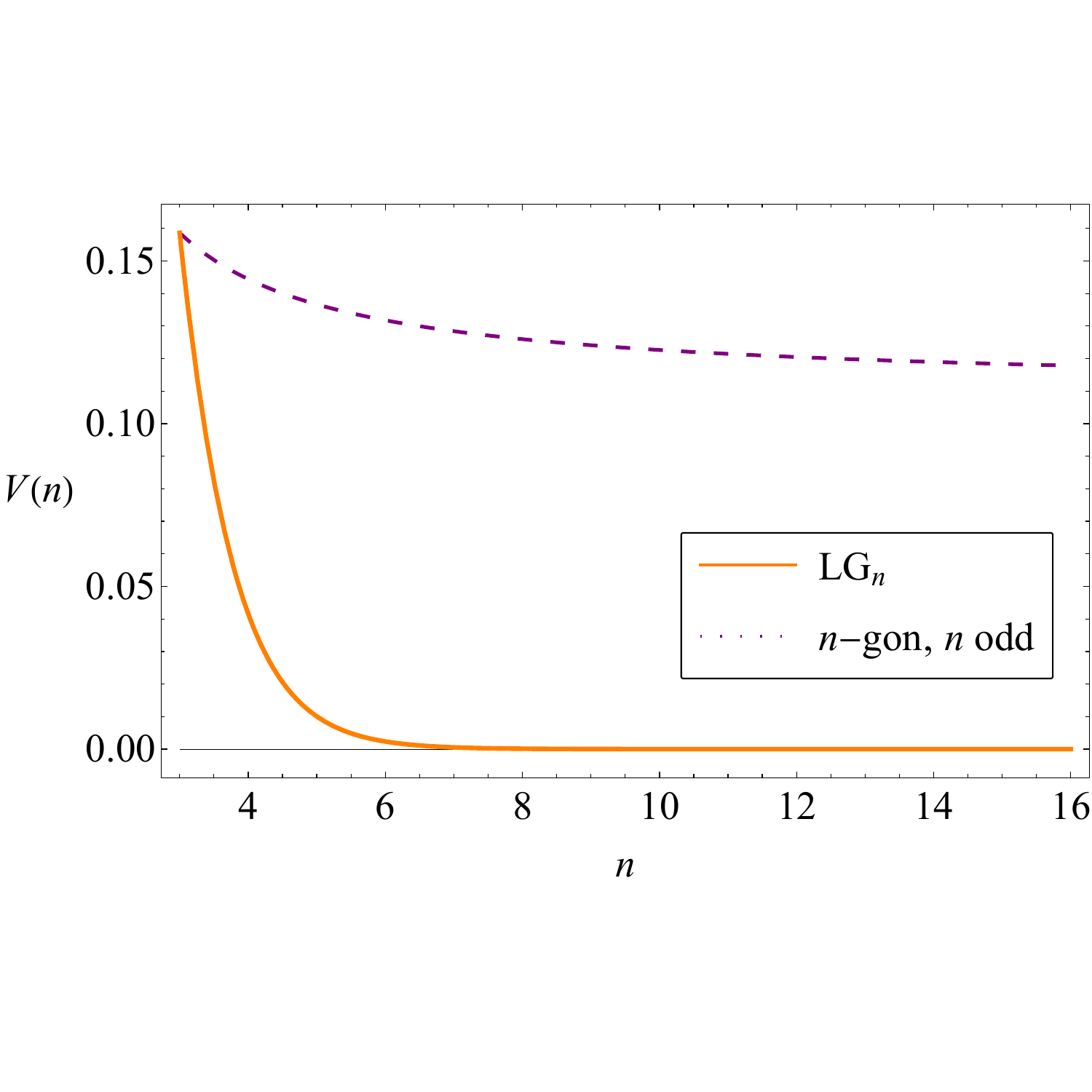}
	\end{center}
	\caption[Estimate of the potency of $n$-time MR conditions]{
		\label{fig:clt}
		The fraction of parameter space which leads to the violation of a given ${\rm LG}n$ or $n$-gon inequality, as estimated by the central limit theorem, Eq.~(\ref{eq:vlgn}), Eq.~(\ref{eq:vngon}).}
\end{figure}

Under this formulation of the problem, we treat the correlators as uniformly distributed random variables, taking values from $-1$ to $1$.  We now note that since the distribution is uniform, the distributions of $C_{ij}$ and $-C_{ij}$ are identical.  Hence when we consider the contribution from a given correlator, we see that probabilistically speaking, it does not matter whether it appears with a plus or a minus sign.  In the inequalities Eq.~(\ref{lgansatz}) and Eq.~(\ref{neweq}) where we see sums of correlators with different signs, we may instead consider just a single sum of correlators with all pluses.  Within the probabilistic formulation, the analysis of this one case corresponds to exploring the set of all possible sign permutations of correlators in the original inequality.  Then since the specific choice of signs appearing in the LG inequalities, or $n$-gon inequalities, will be contained within this set, we see the analysis of this one case is sufficient. The inequalities in which we are interested may all be analysed using the general form
\beq
\label{eq:probineq}
S_{j(n)}\leq b(n),
\eeq
where $S_j$ represents the sum of the $j$ correlators, represented by random variables, $j(n)$ is the number of correlators involved in the $n$-th order inequality, and $b(n)$ is the upper bound on the inequality.  We now invoke the central limit theorem (CLT) \cite{billingsley2012} for a uniform distribution, which states that the distribution of $\frac{S_j\sqrt{3}}{\sqrt{j}}$ for the sum of $j$ independent uniformly distributed variables from $-1$ to $1$, will tend towards the standard normal distribution.  Comparing this result to the generalised inequality Eq.~(\ref{eq:probineq}), we can estimate the region of violation as
\beq
V(n)=\frac{1}{\sqrt{2\pi}}\int_{\frac{b(n)\sqrt{3}}{ \sqrt{j(n)}}}^{\infty}e^{-\frac{x^2}{2}}\mathop{dx}.
\eeq
This integral is simply evaluated as
\beq
\label{eq:violation}
V(n)=\frac{1}{2}\left(1-\text{erf}\left(\sqrt{\frac{3}{2}}\frac{b(n)}{\sqrt{j(n)}}\right)\right).
\eeq
This result says that the difficulty or ease of violating any $n$-th order inequality formed of the sum of $j(n)$ correlators (with arbitrary signs on each correlator), with an upper bound $b(n)$, is determined solely by the functional form of the ratio of $\frac{b(n)}{\sqrt{(j(n))}}$.

Comparing Eq.~(\ref{eq:probineq}) to the general LG inequality Eq.~(\ref{lgansatz}), we see that for the LG case, we have $b(n)=n-2$, and $j(n)=n$.  Using these values in Eq.~(\ref{eq:violation}), we find
\beq
\label{eq:vlgn}
V_{\text{LG}}(n)=\frac{1}{2}\left(1-\text{erf}\left(\sqrt{\frac32}\frac{n-2}{\sqrt{n}}\right)\right).
\eeq
For large $n$, the argument of the error function behaves as $\sqrt{n}$, and hence the violating region of parameter space vanishes with increasing $n$.  This behaviour is shown in Fig.~\ref{fig:clt}.

Comparing the form of the $n$-gon inequality Eq.~(\ref{neweq}) to Eq.~(\ref{eq:probineq}), we have $b(n)=\frac{n}{2}$ for $n$ even, $b(n)=\frac{n-1}{2}$ for $n$ odd, and $j(n)=\frac12n(n-1)$. In the case of even $n$, this leads to a violation region of
\beq
\label{eq:vngon}
V_{n-\text{gon}}(n)=\frac{1}{2}\left(1-\text{erf}\left(\frac{\sqrt{3}}{2}\frac{n}{\sqrt{n(n-1)}}\right)\right)
\eeq
In both the even and odd cases, for large $n$, the argument of the error function tends toward $\frac{\sqrt{3}}{2}$, and hence the violation asymptotes to $V\approx0.11$, as depicted in Fig.~\ref{fig:clt}.   

%We note that included in this result is the case of very rapid measurements, i.e. $t$ goes to 0.  As mentioned before, this case implies that the correlators tend towards 1, and hence we should no longer consider the entirety of the parameter space as being accessible or relevant. Equivalently, the CLT breaks down owing to the correlators no longer uniformly taking values from $-1$ to 1.

This CLT approximation gives a geometric understanding of the earlier arguments that ${\rm LG}n$ inequalities become easier to satisfy for large $n$, and illuminates that the generalised $n$-gon inequalities maintain significant regions of violation for all $n$.  This analysis could be improved, since in QM the L{\"u}ders bound places a limit on observed correlations, and so for these inequalities in the context of QM, we should consider the fraction of violation compared to the volume of space in which the L{\"u}ders bound is satisfied.  One could as well perform a QM-friendly analysis by sampling with the Haar measure from correlation functions defined over unitaries.

%Note this result applies to only a single inequality from the entire $n$-gon set, in isolation.  That is it tells us the magnitude of the volume in parameter space which leads to a violation of any one of the inequalities, however it says nothing about how these regions overlap.

%wever for ${\rm LG}n$, we may place an upper bound on $V(n)$ as $V(n)\leq\exp(-\frac{3n}{2}+6)$ by approximating the error function.  Hence if we assume each individual violation space is entirely distinct from any other, the upper bound on the total region of parameter space leading to any violation is given by $2^n\exp(-\frac{3n}{2}+6)$.  This upper bound can be shown to decrease exponentially quicklyshko1996, and so the upper bound on the total region of violation becomes negligibly small.

\subsection{Illustration in a Simple Spin Model}
\begin{figure}
	\subfloat[]{{\includegraphics[height=5.3cm]{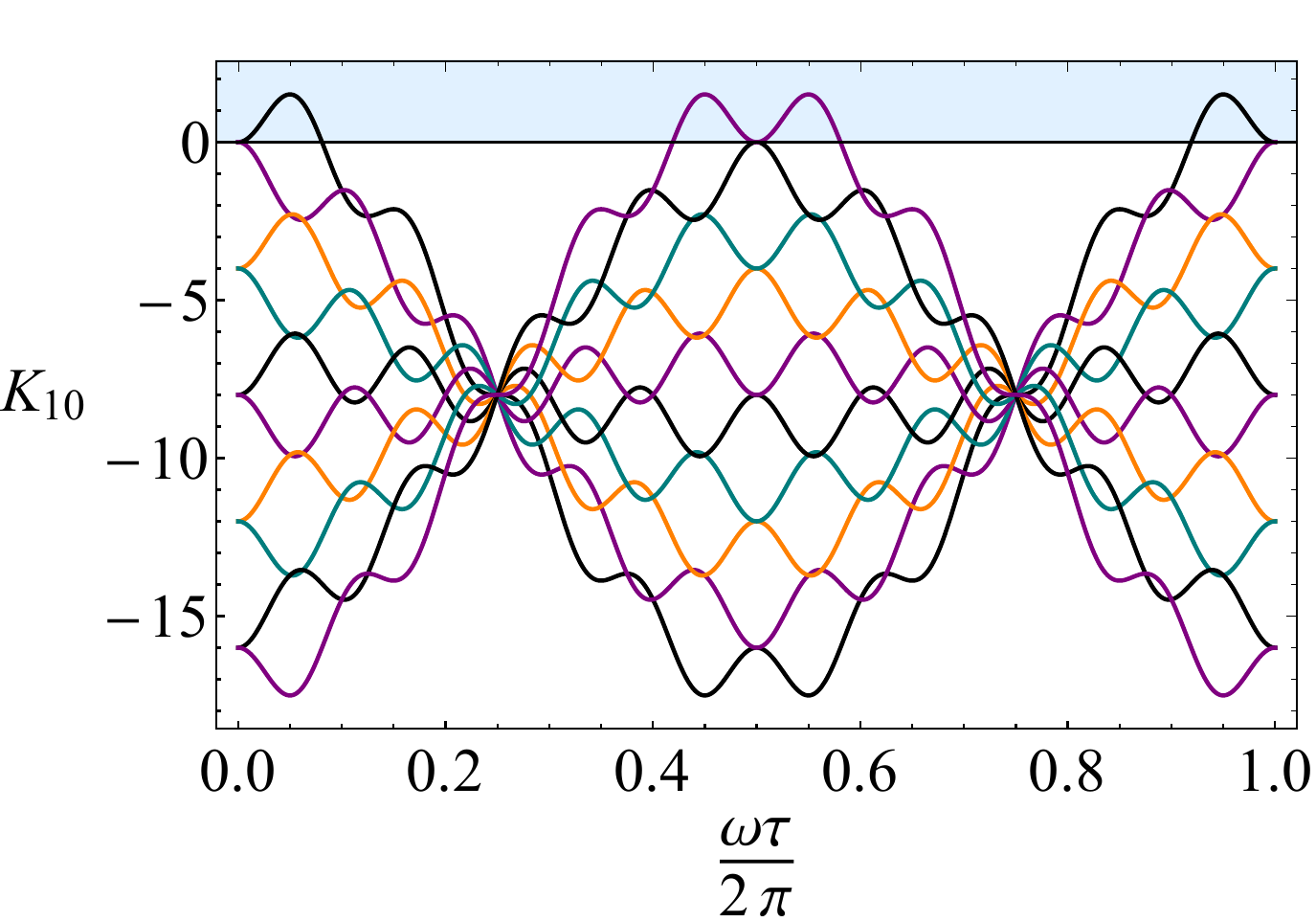}}}%
	\qquad
	\subfloat[]{{\includegraphics[height=5.05cm]{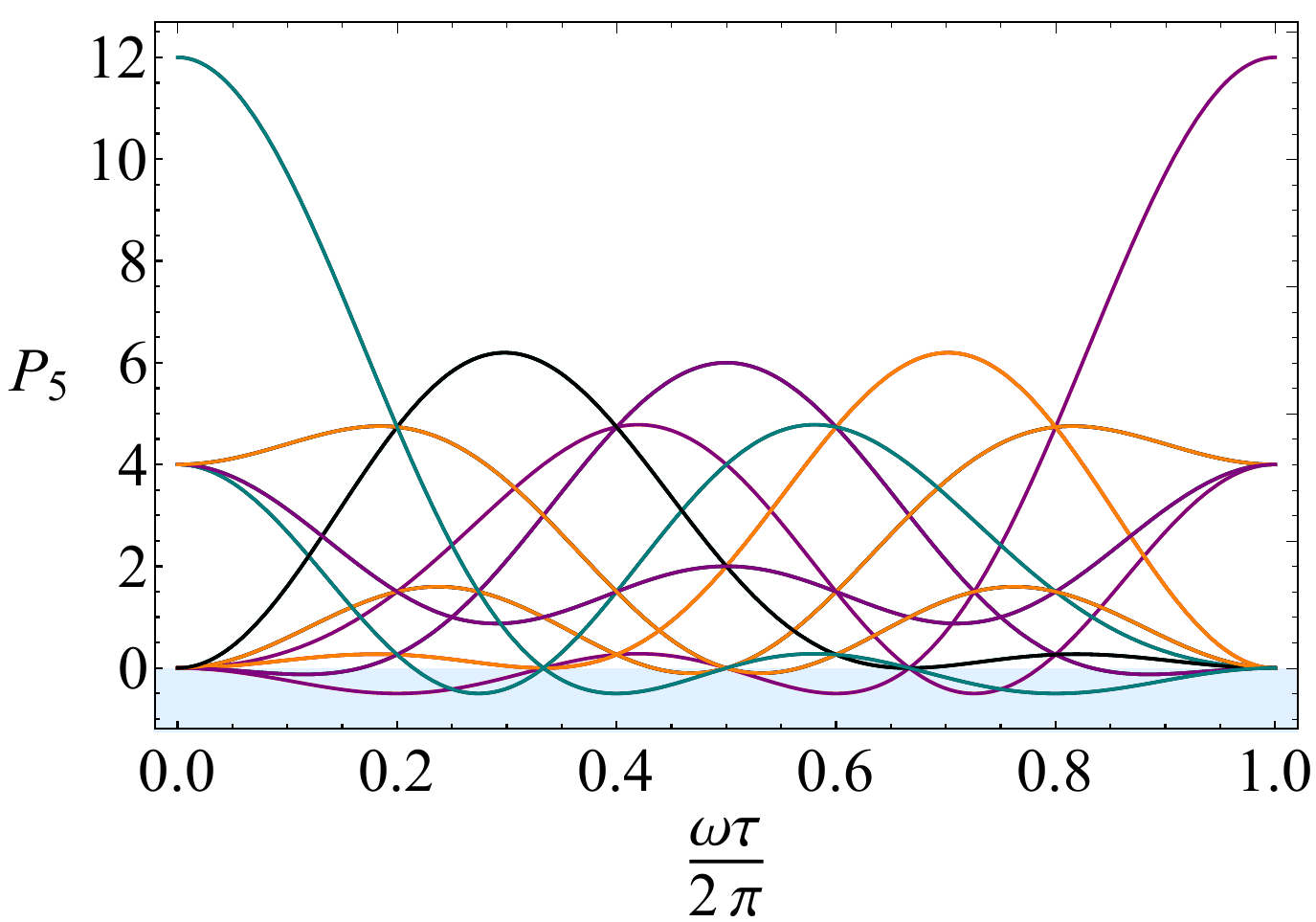}}}
	\caption[Behaviour of $n$-time MR conditions in a simple spin model]{In (a), the 10 distinct inequalities from the $n=10$ LG case are plotted, for the simple spin system.  The inequalities are violated in the shaded region,  $K_{10}\geq0$.  In (b), the 10 distinct inequalities from the $n=5$ $n$-gon case are plotted.  In this case, at least one of the inequalities is violated for {each} $\tau$, except at points of measure zero.}%
	\label{fig:simplespin}%
\end{figure}
We now illustrate the above results using plots in the commonly-studied simple  spin model \cite{kofler2007, athalye2011,emary2013}
in which the correlators are given by
%We look at the behaviour of these inequalities for a simple quantum mechanical system.  We consider the time %evolution of a simple spin system, where the correlators are given by 
\beq
\label{eq:corr}
C_{ij}=\cos\left(\omega(t_j-t_i)\right).
\eeq
We also make the common assumption of equal time spacing $\tau$.
%We will then only consider the case of measurements made at equal time spacings $\tau$ for simplicity.  The behaviour of the lower order LG inequalities within this scheme is well studied, see Ref. \cite{Ema}, so we focus on their behaviour for intermediate and large $n$.  
Purely for convenience of plotting, we rearrange the LG inequalities, introducing the notation $K_n$, defined as
\beq
\label{eq:kernel}
K_n =  a_1 C_{12}+a_2 C_{23}+ \ldots + a_{n-1}C_{n(n-1)}+ a_n C_{1 n} - (n-2),
\eeq
and so violations of the ${\rm LG}n$ inequalities Eq.~(\ref{lgansatz}) are delineated by $K_n\geq0$.  We introduce a similar notation $P_n$ for the $n$-gon inequalities Eq.~(\ref{neweq}), where $P_n\leq0$ signifies a violation.

Recall that in Eq.~(\ref{eq:kernel}) $a_1, \ldots, a_n$ take all values of $\pm1$ such that their overall product is $-1$, and so we must study the behaviour of each member of the family of inequalities for a given $n$.  For example, for the $n=10$ case of the LG inequalities, there are a total of 512 inequalities.  When working with equally spaced measurements, it turns out that only 10 of these inequalities are distinct.  
These 10 inequalities are plotted for the simple spin case in Fig.~\ref{fig:simplespin}(a).  This plot can be considered representative behaviour of the ${\rm{LG}}_n$ inequalities in the simple spin case, where the bulk of the inequalities are always satisfied.  Furthermore, the only two that are not satisfied are violated over only a small region of measurement times.

For the $n$-gon case for $n=10$ there are 1024 inequalities, of which 272 are distinct.  This large number of inequalities is not particularly enlightening to plot so we plot instead the simplest non-trivial case, $n=5$,
%It is clear that by considering all two-time correlators, we reach a much richer set of conditions for macrorealism.  
in Fig.~\ref{fig:simplespin}(b).

\begin{figure}
	\subfloat[]{{\includegraphics[height=5.3cm]{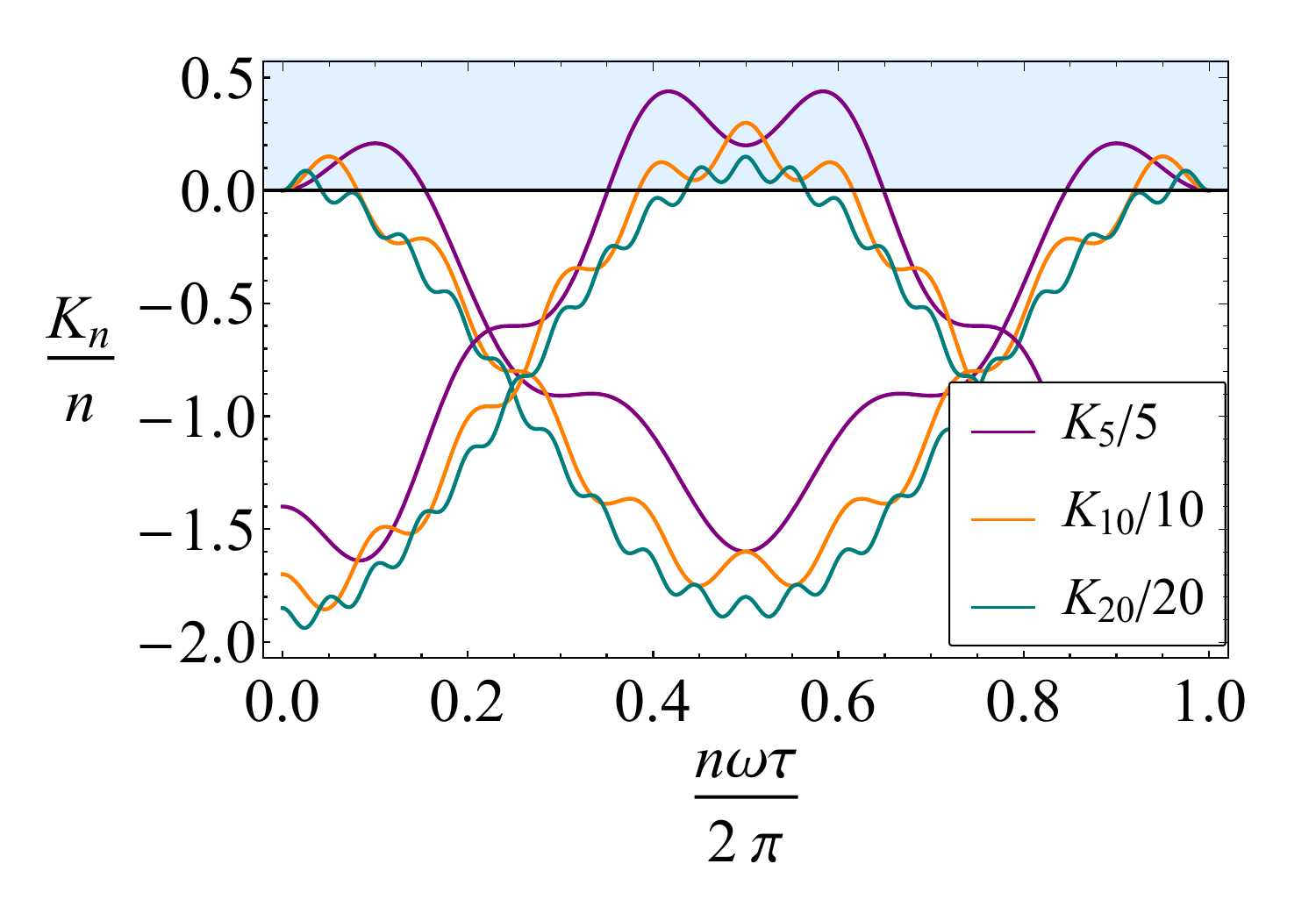}}}%
	\qquad
	\subfloat[]{{\includegraphics[height=5.3cm]{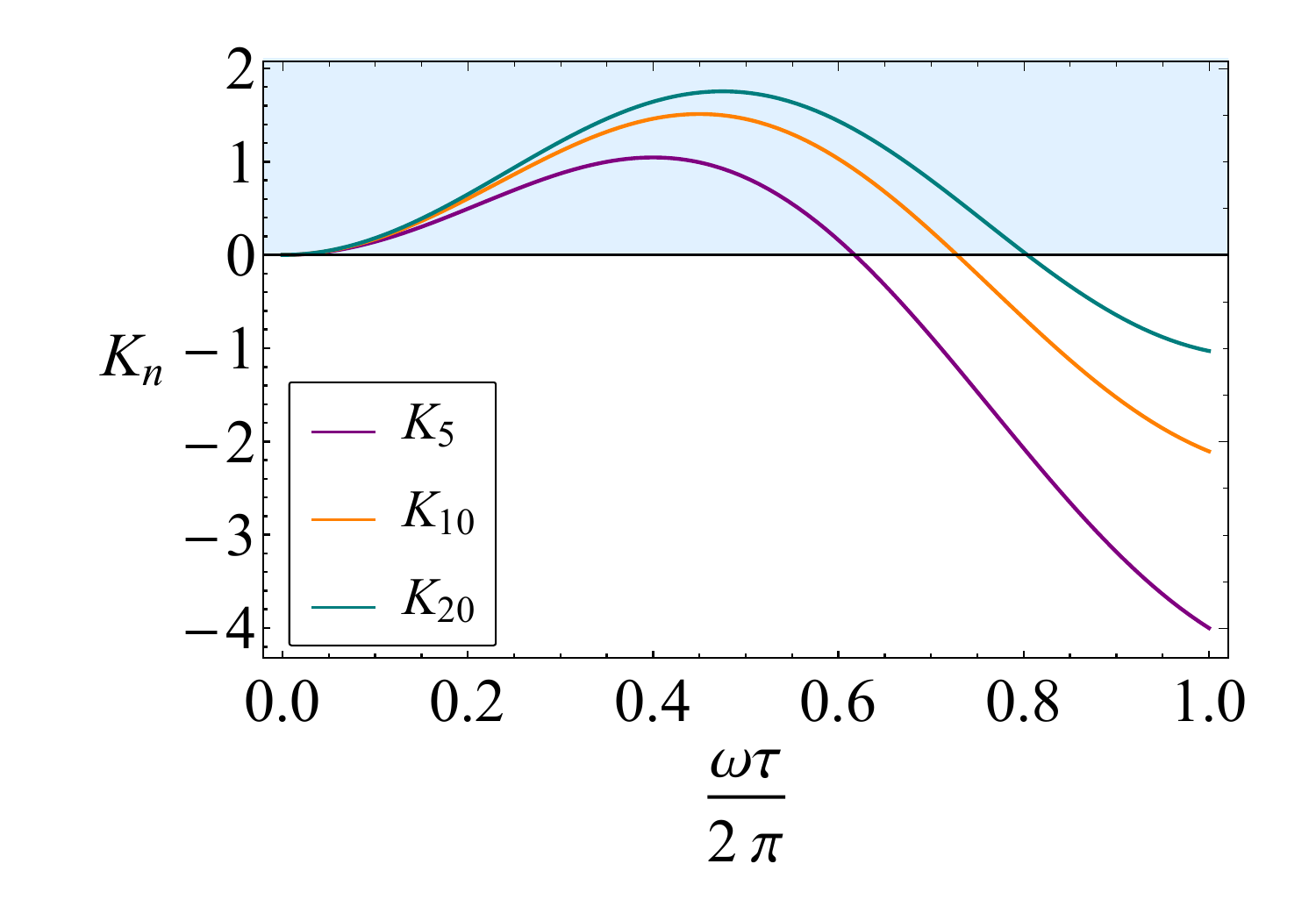}}}
	\caption[Asymptotic behaviour of $n$-time MR conditions in spin system]{The violations of the ${\rm LG}n$ inequalities within a simple quantum mechanical spin model.  We plot only the inequalities leading to the greatest violation, where violations are depicted by the shaded region.  In (a), measurements are performed in a way which extends the experimental time period, and in this regime the ${\rm LG}n$ inequalities become easier to satisfy for higher $n$.  Note that although the violations appear smaller in magnitude, this is an effect of plotting the inequalities normalised by $n$.  In (b), more measurements are performed within the same experimental time period, and in this regime, the inequalities become harder to satisfy with increasing $n$.}%
	\label{fig:largen}%
\end{figure}

LG violations as a function of $n$ are shown in Fig.~\ref{fig:largen}.
The case in which increasing $n$ extends the time interval is plotted in Fig.~\ref{fig:largen}(a).  In the simple spin model, only two of the ${\rm LG}n$ inequalities are violated for any $n$, both of which are plotted.  It can be seen that the region over which $K_n$ is violated decreases with $n$.  For the case in which the total time interval remains fixed as $n$ increases, only one of the ${\rm LG}n$ inequalities is violated for any $n>4$, which is plotted in Fig.~\ref{fig:largen}(b).  Here the opposite effect is observed, where with increasing $n$, the region of violation {\it increases}

\begin{figure}
	\begin{center}
		\includegraphics[width=0.60\textwidth]{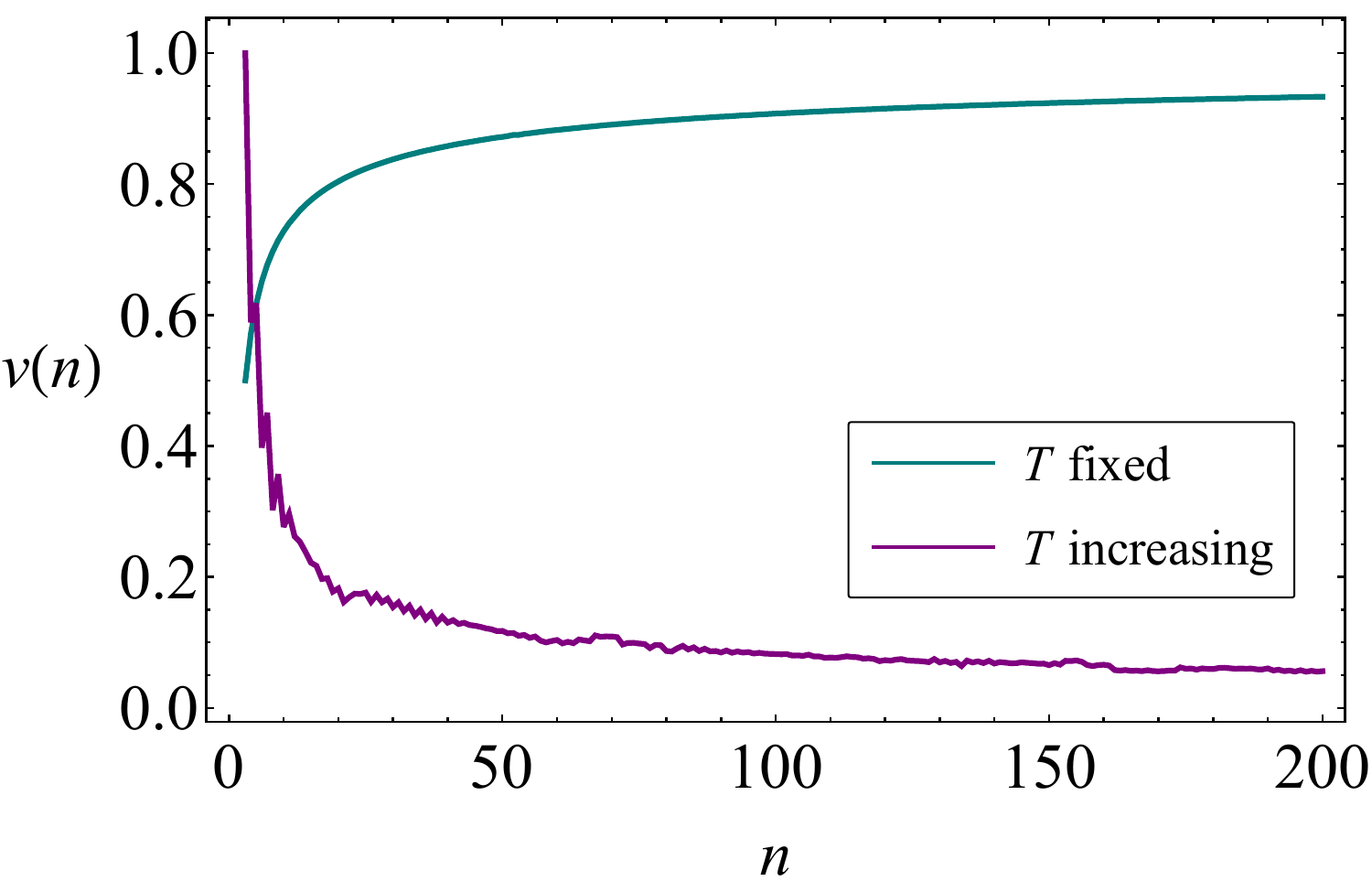}
	\end{center}
	\caption[Numerical measure of large $n$ behaviour]{
		\label{fig:nint}
		The fraction $\nu (n)$ of measurement times which lead to the violation of any of the ${\rm LG}n$ inequalities  Eq.~(\ref{lgansatz}) for the simple spin model.  Both the case where subsequent measurements subdivide the measurement window $[0,T]$ and where subsequent measurements increase the window $[0,T]$ are shown.}
\end{figure}

We have also numerically calculated the fraction of measurement times which lead to {\it at least one} of the ${\rm LG}n$ inequalities being violated, as a function of $n$.  This is plotted in Fig.~\ref{fig:nint}, for both cases, of either increasing or fixed time period with increasing $n$.  
This confirms the behaviour hinted at in Fig.~\ref{fig:largen}, that for the simple spin model, in the case of an increasing experimental time period the ${\rm LG}n$ inequalities become easier to satisfy, whereas for the case of a fixed overall time period, the inequalities become harder to satisfy.
\section{Summary and Conclusions}
\label{c2s6}
We have proved Fine's theorem for the case of $n$ measurements and in doing so established the general form
Eq.~(\ref{lgansatz}) of a complete set of LG inequalities at $n$ times.
We considered generalizations of the usual LG approach to the situation in which all possible two-time correlators are measured and conjectured that necessary and sufficient conditions for the existence of an underlying probability consist of {the set of all three-time LG inequalities together with the $n$-gon inequalities on all $n$ correlators Eq.~(\ref{neweq}).} We proved this conjecture for the cases $n=4$ and $n=5$ in the symmetric case.
This fills a gap in the previous work Ref.\cite{halliwell2019a} which explored a generalized LG approach in which higher order correlators are measured but omitted the extended two-time correlator case.
We explored both of the above sets of inequalities in the large $n$ limit and confirmed agreement with intuitive expectations.

The {$n$-gon inequalities} Eq.~(\ref{neweq}) are clearly of interest to explore experimentally since they are qualitatively different to the usual LG inequalities.  Hence certain failures of MR may only be detected using the $n$-gon inequalities, in systems reported by the LGns to be classical. The $n$-time LG inequalities are of less interest experimentally, in comparison to experimental tests carried out already, since, as we saw here, they are logically equivalent to sets of three-time LG inequalities. 

However, the $n$-time LG inequalities could find another role since there are related problems that can be mapped onto the LGn situation. LG inequalities usually concern a single dichotomic variable at each time but some recent works have explored situations involving variables taking three or more values (see for example, Ref.~\cite{emary2017}) which could easily be mapped onto the dichotomic case by adding extra times. Another possible application is in the contextuality by default approach to assessing the effects of signaling in LG inequalities
\cite{dzhafarov2015} in which the original three variables $Q_1$, $Q_2$, $Q_3$ in LG3 are adjoined with three more variables representing the values of the $Q_i$ in different contexts. This situation is equivalent to a $6$-time LG situation and can be analyzed using the inequalities derived here.

\def\id{\mathbb{1}}
\def\half{\frac {1} {2}}
\def \MR {\rm MR}
\def\LG{\rm LG}
\def\NSIT{\rm NSIT}
\def\NIM{\rm NIM}
\def\Ind{\rm Ind}
\def\MRps{\rm MRps}
\def\P{{\mathbb E}}

\fancyhf{}
\renewcommand{\headrulewidth}{0pt}
\fancyfoot{\makebox[\textwidth][c]{\hyperref[link:4]\thepage}}

\fancyhfoffset[LE,RO, RE, LO]{0cm}
\renewcommand{\chaptermark}[1]{ \markboth{#1}{} }
\renewcommand{\sectionmark}[1]{ \markright{#1}{} }
\fancyhf{}
\fancyhead[L]{\textsl{\thesection~~ \rightmark}}
\fancyhead[R]{\hyperref[link:4]\thepage}
\renewcommand{\headrulewidth}{1pt}

\chapter{Conditions for macrorealism for systems described by many-valued variables}
\label{chap:mlev}
\bookepigraph{3in}{You don't worship the gate
go into the inner temple}{Ram Dass}{}{4.5}
\label{chap:summ}
\vspace{-2em}
%\epigraph{There are more things in heaven and earth, Horatio, than are dreamt of in your philosophy.}{Shakespeare, Hamlet}
	\section{Introduction}

Almost all experimental studies of the LG framework concern microscopic systems and the original goal of using the framework to study coherence in macroscopic systems is still on the far horizon \cite{emary2013}. (Although see the recent proposal Ref.~\cite{bose2018a}). However, as the systems studied become progressively larger, it becomes of interest to extend the above frameworks to embrace the new features of larger systems not present, for example, in simple two-state systems, or for systems described by a single dichotomic variable.

The purpose of this chapter is to develop generalizations of the necessary and sufficient tests of macrorealism for the case of many-valued variables.
We thus imagine doing a set of experiments each of which may involve one, two or three times, and which determine certain averages and correlators of an $N$-valued variable. 
This includes both the case of fine-grained measurements on an $N$-level system, or coarse-grained measurements on systems  with dimension greater than $N$.  The underlying strategy in this chapter is quite simple: in each case we will show how to express measurements on a many-valued variable in terms of measurements on a related set of dichotomic variables, for which the MR conditions are closely related to those studied in Chapter~\ref{chap:ntime}.

Taking first the case of weak MR,
these measurements provide a determination, indirectly, of a set of candidate two-time probabilities of the form $p(n_1, n_2)$,
$p(n_2,n_3)$, $p(n_1,n_3)$ for the three time pairs $(t_1,t_2)$, $(t_2,t_3)$, $(t_1,t_3)$, where $n$ runs from $1$ to $N$ and $N \ge 3$. The question as to whether these candidate two-time probabilities are non-negative leads to the desired generalization of the two-time LG inequalities, Eqs.~(\ref{LG2a}--\ref{LG2d}).
From there one can ask for the conditions under which these three pairwise probabilities are the marginals of a single three-time probability $p(n_1,n_2,n_3)$. We will thus develop the appropriate generalizations of the three-time LG inequalities and Fine's theorem. We also develop definitions of strong MR by exploring the generalizations of the NSIT conditions.
We will use our analysis of LG inequalities and NSIT conditions to explore some of the novel features of MR conditions for many-valued variables not present in the dichotomic case.

In Section \ref{c3s2}, we review commonly-used conditions for MR for measurements of a single dichotomic variable as outlined above. 
The discussion is primarily from a 
quantum-mechanical perspective, from which such conditions are seen to be restrictions on the degree of interference between different histories of the system. This helps in determining the degree to which different MR conditions are independent of each other and also illustrates the link between LG inequalities and NSIT conditions, preparing the way for the many-valued case.  At the end of this section, we give a detailed account of the different types of definition of MR, where for dichotomic variables a simple hierarchy of conditions is present \cite{maroney2014a,clemente2015,clemente2016,halliwell2016b,halliwell2017}.

The extension of the standard approach involving LG inequalities for dichotomic variables to many-valued variables is presented in Sections \ref{c3s3} and \ref{c3s4}. We first, in Section \ref{c3s3}, consider the simpler case of MR conditions at two times and determine the form of the LG inequalities for variables taking $N$ values at each time.
In Section \ref{c3s4} we extend our results to the standard LG situation in which measurements are made at three pairs of times. We prove the generalization of Fine's theorem required for this case, i.e. we determine necessary and sufficient conditions for the existence of an underlying joint probability matching the measured two-time marginals for $N$-valued variables.
The conditions turn out to be a set of LG inequalities which can be written quite simply in the general case.
We note that our results are readily extended to measurements at many times by taking advantage of the generalized Fine ansatz presented in Chapter~\ref{chap:ntime}. 

In Section \ref{c3s5} we consider the different and stronger MR conditions involving NSIT conditions, primarily focusing on the two-time case and we discuss the logical relationships between such conditions and the weaker LG inequalities. We find that it is considerably less simple than the dichotomic case and we use this understanding to analyze the results of some recent experiments in which two-time LG violations are observed even though certain NSIT conditions are satisfied. 

In Section \ref{c3s6}, we use the methods developed above to examine the surprising fact that the LG framework sometimes permits violations of the LG inequalities to a degree beyond the L\"uders bound, using particular types of measurements available only for $N \ge 3$ \cite{dakic2014,budroni2014,wang2017,pan2018}.
This bound coincides numerically with the Tsirelson bound for the correlators in Bell experiments \cite{cirelson1980} but for LG tests the L\"uders bound is not in fact the maximum permitted by quantum mechanics. 
We show that
the violation naturally separates into a conventional LG violation (respecting the L\"uders bound) together with a violation of a two-time NSIT condition and discuss the consequences of this in terms of MR conditions.
% so it does say anything new about interference effects at three times. 
%However, we write down an analogous two-time situation in which the interpretation of the L\"uders bound has a clearer interpretation. 
%We give an explicit example using a spin model. 
We summarize and conclude in Section \ref{c3s7}.

Some useful technical results from the decoherent histories approach to quantum theory are given in Appendix \ref{c3b}.  
In Appendix \ref{c3c} we record for future use the full list of LG inequalities for the $N=3$ case at three times, a list too lengthy for the main text.

\section{Conditions for Macrorealism for Dichotomic Variables}
\label{c3s2}
%\subsection{Standard LG Formalism}

The  LG framework is usually described from a purely macrorealistic point of view in which the experimental situation is a black box, about which little is assumed in terms of the dynamics, initial state etc.  Although we will follow this approach where possible,
in this chapter we find that a quantum-mechanical analysis is often most convenient, since the LG inequalities and NSIT conditions that we seek to derive have a simple interpretation in quantum mechanics as restrictions on the size of certain interference terms, as will become clear in what follows (and see also Ref. \cite{halliwell2016b}). 
The resulting understanding may subsequently be re-expressed in purely macrorealistic terms but we do not always spell this out explicitly.

In this section we describe the quantum-mechanical description of MR conditions for a single dichotomic variable at two and three times. Section \ref{subsec:recap} is primarily a summary of earlier work \cite{halliwell2016b}, with some additional commentary,
but the details of this work are important for the generalizations considered in later sections.

%\blue{[Re-expression in MR terms?]}. 

\subsection{Two-Time LG Inequalities, NSIT Conditions and Interferences}
\label{subsec:recap}
We imagine a system with Hamiltonian $H$ subject to measurements at either single times in each experiment or pairs of times. 
Measurements of the dichotomic variable $Q$ are described by the projection operator
\beq
P_s = \frac{1}{2} \left( 1 + s \hat Q \right),
\label{projs}
\eeq
where $s = \pm 1$. (Hats are used to denote operators only when necessary to distinguish from a classical counterpart).
For a system in initial state $\rho$ the probability for a single time measurement at time $t_1$  is
\beq
p_1(s) = {\rm Tr} \left( P_s (t_1) \rho \right),
\label{single}
\eeq
where $P_s (t)= e^{iHt} P_s e^{-iHt} $ is the projector in the Heisenberg picture and we use units in which $\hbar = 1$.
The probability for two sequential projective measurements at times $t_1, t_2$ is
\beq
p_{12} (s_1, s_2) = {\rm Tr} \left( P_{s_2} (t_2) P_{s_1} (t_1) \rho P_{s_1} (t_1) \right).
\label{p12}
\eeq
This matches $p_1 (s_1) $ when summed over $s_2$, but does not match the single time result
%summing over the initial measurement we find
%\beq
%\sum_{s_1} p_{12} (s_1, s_2 ) = {\rm Tr} \left( P_{s_2} (t_2)  \rho_M (t_1) \right),
%\eeq
%where $\rho_M (t_1) $ denotes the measured density operator,
%\beq
%\rho_M (t_1) = \sum_{s_1} P_{s_1} (t_1) \rho P_{s_1} ,
%\label{rhoM}
%\eeq
$ {\rm Tr} (P_{s_2} (t_2) \rho ) $ when summed over $s_1$.
Hence
the NSIT condition Eq.~(\ref{NSIT}) is not in general satisfied in quantum mechanics, except for initial density operators diagonal in $\hat Q$ at time $t_1$.

It is also of interest to consider the two-time quasi-probability
\beq
q(s_1, s_2) = {\rm Re} {\rm Tr} \left( P_{s_2} (t_2) P_{s_1} (t_1) \rho \right),
\label{quasi}
\eeq
which is real and sums to $1$, but can be negative \cite{halliwell2016b,goldstein1995}. It matches both single-time marginals $p_1 (s_1)$ and $p_2 (s_2)$ when summed over $s_2$ and $s_1$ respectively.
Following the useful general results of Refs.~\cite{halliwell2013,klyshko1996}, it has a very useful moment expansion,
\beq
q(s_1,s_2) = \frac {1}{4} \left(1 + \langle \hat Q_1 \rangle  s_1 +  \langle \hat Q_2 \rangle s_2  + C_{12} s_1 s_2 \right),
\label{mom}
\eeq
where the correlator is,
\beq
C_{12} = \half \langle \hat Q_1 \hat Q_2 + \hat Q_2 \hat Q_1 \rangle,
\label{corr2}
\eeq
and the averages are quantum-mechanical ones in these expressions.
Here, $\hat Q_i$ denotes the operator $\hat Q$ at time $t_i$, in parallel with the macrorealistic notation used in the Introduction.
(We will not use the notation $\hat Q(t)$ to denote time-dependence in order to avoid confusion with a notation used later in the many-valued case).
We see from Eq.~(\ref{mom}) that, for a quantum-mechanical system, the two-time LG inequalities, Eqs.~(\ref{LG2a}--\ref{LG2d}), are equivalent to the conditions,
\beq
q(s_i, s_j) \ge 0. 
\eeq

The two-time measurement probability has a similar moment expansion
\beq
p_{12}(s_1,s_2) = \frac {1}{4} \left(1 + \langle \hat Q_1 \rangle  s_1 +   \langle \hat Q_2^{(1)}  \rangle s_2 
+ C_{12} s_1 s_2 \right).
\label{p12mom}
\eeq
Here,
\beq
\langle \hat Q_2^{(1)} \rangle = \langle \hat Q_2 \rangle + \half \langle [ \hat Q_1, \hat Q_2] \hat Q_1 \rangle
\label{Q2(1)}
\eeq
and has the interpretation as the average of $\hat Q_2$ in the presence of an earlier measurement at $t_1$ with the result summed out. The difference between Eq.~(\ref{p12mom}) and the quasi-probability therefore vanishes when $ \langle \hat Q_2^{(1)} \rangle = \langle \hat Q_2 \rangle $
(for example, when $ \hat Q_1 $ and $ \hat Q_2 $ commute, but this is clearly not the only way). 
Note that the quasi-probability and the two-time measurement probability have the same correlation function \cite{fritz2010}.

We now introduce the description of interferences in relation to 
the quasi-probability Eq.~(\ref{quasi}) and the two-time probability Eq.~(\ref{p12}). This may be seen from the simple relationship (proved in greater generality in Appendix \ref{c3b}, which is,
\beq
q(s_1, s_2) = p_{12} (s_1, s_2)  +\sum_{{s_1'} \atop {s_1' \ne s_1}}  \ {\rm  Re} D (s_1,s_2 |s_1',s_2),
\label{qp}
\eeq
where
\beq
D (s_1,s_2 |s_1',s_2') = {\rm Tr} \left(  P_{s_2} (t_2) P_{s_1} (t_1) \rho P_{s_1'} (t_1) P_{s_2'}(t_2) \right),
\label{DF1}
\eeq
is the so-called {\it decoherence functional}~\cite{halliwell2016b,griffiths1984,griffiths1993,griffiths1996,griffiths1998,omnes1988,omnes1988a,omnes1988b,omnes1989,omnes1990,omnes1992,gellmann1993,halliwell2009}, and is a very useful quantity in what follows. The off-diagonal terms of the decoherence functional are measures of interference between the two different quantum histories represented by sequential pairs of projectors, $P_{s_2} (t_2) P_{s_1} (t_1)$ and $P_{s_2'} (t_2) P_{s_1'} (t_1)$. For the dichotomic case considered here, there is in fact only one term in the sum over $s_1'$, namely $s_1' = - s_1$, so the interference terms, which for conciseness we denote $I (s_2)$, are simply
\beq
I(s_2) = {\rm Re} D(s_1,s_2|-s_1, s_2),
\eeq
and are independent of $s_1$.
Note that by inserting the moment expansions Eqs.~(\ref{mom}), (\ref{p12mom}) in Eq.~(\ref{qp}), we may make the identification $I(s_2) = \frac{1}{4}  ( \langle \hat Q_2 \rangle - \langle \hat Q_2^{(1)} ) s_2$.

When the off-diagonal terms are zero
there is no interference and we have $q(s_1,s_2) = p_{12} (s_1,s_2) $, and the NSIT condition Eq.~(\ref{NSIT})
is satisfied exactly. However, noting that $p_{12}(s_1,s_2) $ is always non-negative, we see from Eq.~(\ref{qp}) that 
the requirement that $q(s_1, s_2)$ is non-negative is equivalent to the following bounds on the off-diagonal terms of the decoherence functional:
\beq
- I(s_2)  \le p_{12} (s_1, s_2).
\label{ReDp}
\eeq
Hence, the two-time LG inequalities represent bounds on the degree of interference. 

%(We note in passing that the sign of the interference term is important in Eq.~(\ref{ReDp}) -- it is trivally satisfied when $I(s_2) \ge 0 $ so bounds only how negative $I(s_2)$ can be. In the context of the double slit experiment this corresponds to the difference between destructive and constructive interference. The former is required to get LG violations \cite{HalDS}). 

%\blue{[Links to double-slit. Positive and negative interference terms].}

From Eq.~(\ref{DF1}), it is easy to see that the decoherence functional summed over $s_2$ gives zero for $s_1 \ne s_1'$, which means that $\sum_{s_2} I(s_2) = 0$. Hence 
the interference terms in Eq.~(\ref{ReDp}) are fixed by just one independent quantity, which could be taken for example to be $I(+) = - I(-)$.
However, this single interference term has four upper bounds in Eq.~(\ref{ReDp}), which is why there are, correspondingly, four two-time LG inequalities.
By contrast, the stronger NSIT condition Eq.~(\ref{NSIT}) simply requires that all interference terms are zero. Since there is just one independent interference term the NSIT condition amounts to a single condition, $I(+) = 0$.

There is also a relationship between the interference terms $I(s_2)$ and the so-called ``coherence witness'' \cite{li2012,schild2015,wang2017b}, which measures the degree to which the NSIT condition Eq.~(\ref{NSIT}) is violated. We have
\bea
p_2 (s_2) - \sum_{s_1} p_{12} (s_1, s_2) &=& \sum_{{s_1, s_1'} \atop{s_1 \ne s_1'}}  \ {\rm  Re} D (s_1,s_2 |s_1',s_2),
\nonumber \\
&=& 2 I(s_2).
\label{Wit1}
\eea
Hence we identify the coherence witness with $2 I(s_2)$. (It is often defined with a modulus sign but the signed definition used here is convenient). 
% ambiguous measurements

In terms of measurements to check the two-time LG inequalities (or equivalently measure the quasi-probability Eq.~(\ref{quasi})), the procedure is to carry out measurements of $ \langle \hat Q_1 \rangle$, $  \langle \hat Q_2 \rangle $ and $C_{12}$ in three 
separate experiments \cite{halliwell2016b,halliwell2017}.
These must be non-invasive (from a macrorealistic perspective) to meet the NIM requirement. This is trivial for the two averages (since only a single measurement is made in those experiments). For the case of $C_{12}$ this is accomplished using a pair of sequential measurements where the first one is an ideal negative measurement (in which the detector is coupled to, say, $Q_1=+1$ and a null result is taken to imply that $Q_1=-1$). This determines the sequential probability $p_{12} (s_1, s_2)$ from which the correlator is obtained (and as noted, coincides with the correlator in the quasi-probability $q(s_1,s_2))$.  
It should be added that the extent to which ideal negative measurements really meet the NIM requirement has been a matter of debate and alternative non-invasive measurement measurement protocols have been explored. (See for example Refs.~\cite{halliwell2019a,halliwell2016a} and references therein).

\subsection{LG Inequalities for Three Times}

A discussion of the relationship between the LG inequalities and bounds on the interferences following from Eq.~(\ref{qp}) in the two-time case may also
be given for the three-time LG inequalities using the generalizations of the quasi-probability and two-time probability given in Appendix B. Following that discussion, we first consider the three-time quasi-probability
\beq
q(s_1, s_2, s_3) = {\rm Re} {\rm Tr} \left( C_{s_1 s_2 s_3} \rho \right)
\eeq
where 
\beq
C_{s_1 s_2 s_3} = P_{s_3} (t_3) P_{s_2} (t_2) P_{s_1} (t_1) .
\label{Csss}
\eeq
It has moment expansion,
\begin{equation}
q(s_1,s_2,s_3) = \frac{1}{8} ( 1 + s_1 \langle \hat Q_1 \rangle + s_2 \langle \hat Q_2 \rangle +s_3 \langle \hat Q_3 \rangle + s_1 s_2 C_{12} + s_2 s_3 C_{23} + s_1 s_3 C_{13}  + s_1 s_2 s_3 {D}),
\label{q123}
\end{equation}
where $D$ is a triple correlator whose form is not needed here, and is not normally measured in LG experiments \cite{halliwell2017,halliwell2013}.
Now we note that the quantum-mechanical expression for the correlators, Eq.~(\ref{corr2}), are equivalently written,
\beq
C_{12} = \sum_{s_1 s_2 s_3} s_1 s_2 \ q (s_1, s_2, s_3),
\eeq
and likewise for $C_{23}$ and $C_{13}$.  We focus on the LG inequality Eq.~(\ref{LG3a}), which we write here as $L_1 \ge 0 $, where
\beq
L_1 = 1 + C_{12} + C_{13} + C_{23}.
\eeq
This can be written in the quantum case as
\beq
L_1 = \sum_{s_1 s_2 s_3} \left[  1 + s_1 s_2 + s_1 s_3 + s_2 s_3 \right] \ q( s_1, s_2, s_3).
\label{L1}
\eeq
Since the term in square brackets is always non-negative $L_1$ can only be negative if $q(s_1,s_2,s_3) < 0 $.

Eq.~(\ref{pqd}) shows that
the quasi-probability $q(s_1,s_2,s_3)$ is related to the three-time sequential measurement probability $p_{123}(s_1,s_2,s_3)$ by
\beq
q(s_1,s_2,s_3) = p_{123} (s_1,s_2,s_3) + {\rm Re} {\rm Tr} \left(C_{s_1 s_2 s_3}\ \rho  \ \overline{C}_{s_1 s_2 s_3}^\dag \right),
\label{qpi}
\eeq
where $ \overline{C}_{s_1 s_2 s_3} = 1 - C_{s_1 s_2 s_3}$.
This means the interference terms need to be sufficiently large and negative in order for $q(s_1,s_2,s_3)$ to become negative, in parallel with the two-time case. However, unlike the two-time case, the LG inequality $L_1$ is a coarse-graining of the three-time quasi-probability and we need to be more specific in identifying the interference terms responsible for a LG violation.

A very convenient way of identifying the specific form of the interference terms in Eq.~(\ref{qpi}) is to note that $p_{123} (s_1, s_2, s_3) $ also has a moment expansion, which is
\bea
p_{123} (s_1,s_2,s_3) = \frac{1}{8} ( 1 \! &+& s_1 \langle \hat Q_1 \rangle + s_2 \langle \hat Q_2^{(1)} \rangle +s_3 \langle \hat Q_3^{(12)} \rangle  \
\nonumber \\
&+&  s_1 s_2 C_{12} + s_2 s_3 C_{23}^{(1)} + s_1 s_3 C_{13}^{(2)} + s_1 s_2 s_3 {D}
).
\label{p123}
\eea
Here, as in Eq.~(\ref{Q2(1)}), the superscripts on the averages and two-time correlators denote the presence of earlier or intermediate measurements whose results are summed over \cite{halliwell2017}.
Although we need only the general form of the interference terms, their exact forms are given by
 \begin{align}
 	\ev*{\hat Q^{(1)}_2}&=\ev*{\hat Q_2} - \frac12 \ev*{[\hat Q_1, \hat Q_2]\hat Q_1},\\
 	\ev*{\hat Q^{(12)}_3}&=\ev*{\hat Q_3}-\frac12 \ev*{[\hat Q_1 \hat Q_2, \hat Q_3]\hat Q_2 \hat Q_1},\\
    C_{23}^{(1)}&=  C_{23} - \frac14 \ev*{[\hat Q_1, \{\hat Q_2, \hat Q_3\}]\hat Q_1},\\
        C_{13}^{(2)}&=  C_{13} - \frac14 \ev*{[\hat Q_2, \{\hat Q_1 ,\hat Q_3\}]\hat Q_2}.
 \end{align}
 So for example, $C_{13}^{(2)} $ is the correlator for the time pair $t_1, t_3$ in the presence of an intermediate measurement at $t_2$ which has been summed out.
Taking the difference between Eq.~(\ref{q123}) and Eq.~(\ref{p123}), we find that the interference terms have the form
\bea
{\rm Re} {\rm Tr} \left(C_{s_1 s_2 s_3}\ \rho  \ \overline{C}_{s_1 s_2 s_3}^\dag \right)
= \frac{1}{8} \left(  s_2   (\langle \hat Q_2 \rangle - \langle \hat Q_2^{(1)} \rangle )  
+  s_3  (\langle \hat Q_3 \rangle - \langle \hat Q_3^{(12)} \rangle )  \right.
\nonumber \\
\left. + s_2 s_3 (C_{23} - C_{23}^{(1)}) + s_1 s_3 ( C_{13} - C_{13}^{(2)}) \right)
\eea
Inserting this in Eq.~(\ref{qpi}) and Eq.~(\ref{L1}), we find
\beq
\label{corrints}
L_1 =  \sum_{s_1 s_2 s_3} \left[  1 + s_1 s_2 + s_1 s_3 + s_2 s_3 \right] \ p_{123}( s_1, s_2, s_3)
+  (C_{23} - C_{23}^{(1)}) +  ( C_{13} - C_{13}^{(2)}),
\eeq
and note that the first order moment terms have dropped out entirely. The first term on the right-hand side is clearly non-negative, and hence the interference terms responsible for a three-time LG violation are the remaining terms on the right, involving the difference between two-time correlators.  These terms are completely independent of the quantities controlling the two-time LG violations, which as we saw above, are the quantities  $(\langle \hat Q_2 \rangle - \langle \hat Q_2^{(1)} \rangle ) s_2  $ (and similarly for the other two-time pairs). Hence the two-time and three-time LG inequalities test for the presence of completely independent types of interference terms.
This observation is relevant to the study of L\"uders bound violations in Section \ref{c3s6}.

We briefly note that the three-time LG inequalities Eqs.~(\ref{LG3a})--(\ref{LG3d}) may be written in terms of the quasi-probability Eq.~(\ref{quasi}), as
\beq
q(s_1,-s_2) + q(s_2,-s_3) + q(-s_1,s_3) \le 1.
\label{LG3q}
\eeq
We will see that this is a convenient starting point for generalizations. 

\subsection{Detailed Characterization of Different Types of Macrorealism}
\label{c3a}
%\blue{Moved from Section 1}

So far we have discussed conditions for MR based on LG inequalities and NSIT conditions, which, as indicated in Section \ref{sec:genchar}, are different types of conditions for MR. For clarity, we now make these different definitions more precise, following Refs.~\cite{halliwell2016b,halliwell2017}.
As stated already, Eq.~(\ref{MRdef}), macrorealism is the logical conjunction of NIM, MRps and induction. 
The last requirement, induction, is rarely contested. The first two can be interpreted in a number of different ways, so elaboration is required.  As stated in Section \ref{sec:genchar}, the LG inequalities rule out just one type of MRps \cite{maroney2014a}, which corresponds essentially to hidden variable theories of the GRW type \cite{ghirardi1986, pearle1989,penrose1996,bassi2013}.   We now consider the different ways of interpreting the NIM requirement, which we illustrate for MR tests at both two and three times.
%
%First of all, MRps comes in three different types \cite{maroney2014a}, but the LG framework tests only one of them, which are essentially hidden variable theories of the GRW type \cite{ghirardi1986, pearle1989,penrose1996,bassi2013}. The other two types are hidden variable theories in which the wave function itself is included in the specification of the ontic state, as is the case, for example, in de Broglie-Bohm theory, and models of this type are much harder to rule out experimentally.  Secondly, the NIM requirement may also be interpreted in a number of different ways depending on how many quantities are determined in each individual experiment. We illustrate this for MR tests at both two and three times.

At two times, there are two natural ways to proceed. One is to measure the two-time probability $p_{12} (s_1,s_2)$ using sequential measurements in a single experiment and then require that it satisfies the NSIT condition Eq.~(\ref{NSIT}), which we denote $\NSIT_{(1)2}$ \cite{kofler2013}. 
We refer to the version of NIM involved in such a test as {\it sequential} NIM, denoted $\NIM_{seq}$, and refer to the definition of MR tested in this way as strong MR:
\beq
\MR_{strong} = \NSIT_{(1)2} \wedge \Ind.
\label{MRs}
\eeq
Note that the NSIT condition embraces both NIM and MRps in this approach, since it both shows that the first of the two measurements does not disturb the second and supplies the joint probability for the pair of measurements.

The other way is to measure $\langle Q_1 \rangle$, $\langle Q_2 \rangle$ and $C_{12}$ in three different experiments, requiring non-invasiveness in each individual experiment. The only non-trivial issue in terms of invasiveness is the measurement of $C_{12}$ which is carried out using standard methods in LG experiments, as described in Section \ref{subsec:recap}. Crucially, here we do not demand that that the measurements in different experiments would still be non-invasive if combined. That is, non-invasiveness is maintained only in each separate piece of the data set, but not the whole.
(This is a direct analogy to what is done in Bell experiments). We therefore refer to this version of NIM as {\it piecewise}, and denote it $\NIM_{pw}$.
Since $\NIM_{pw}$ is clearly weaker than $\NIM_{seq}$, the corresponding version of MR is much weaker, and we denote it,
\beq
\MR_{weak} = \NIM_{pw} \wedge \LG_{12} \wedge \Ind.
\label{MRw}
\eeq
%Most LG experiments test a definition of MR of precisely this type (even though this is rarely made clear).

At three times one may proceed similarly. One may use $\NIM_{seq}$ and measure the three-time probability $p_{123} (s_1,s_2,s_3)$ directly using three sequential measurements in a single experiment
and then require that it satisfies a set of NSIT conditions \cite{kofler2013,clemente2015,clemente2016}. A suitable set are the conditions,
\bea
\sum_{s_2} p_{23} (s_2,s_3) &=& p_3( s_3),
\label{NSIT23}
\\
\sum_{s_1} p_{123} (s_1,s_2,s_3) &=& p_{23} (s_2,s_3),
\label{NSIT123a}
\\
\sum_{s_2} p_{123} (s_1,s_2,s_3) &=& p_{13} (s_1,s_3), 
\label{NSIT123b}
\eea
which are denoted $\NSIT_{(2)3}$, $\NSIT_{(1)23}$ and $\NSIT_{1(2)3}$ respectively.
The corresponding definition of strong MR is
\beq
\MR_{strong} = \NSIT_{(2)3} \wedge \NSIT_{(1)23} \wedge \NSIT_{1(2)3} \wedge \Ind.
\eeq
Alternatively, one can work with $\NIM_{pw}$ in which the results of a number of different experiments are combined. In particular, six experiments are carried out to determine the three $\langle Q_i \rangle$ and three $C_{ij}$, so no more than two measurements are made in each individual experiment. Weak MR is then defined using a combination of two and three time LG inequalities:
\beq
\MR_{weak} = \NIM_{pw} \wedge \LG_{12} \wedge  \LG_{23} \wedge \ LG_{13}  \wedge \LG_{123} \wedge \Ind.
\eeq
An intermediate possibility is to use $\NIM_{seq}$ for the two-time measurements, but use $\NIM_{pw}$ in assembling the two-time probabilities into a three time probability, leading to the definition of MR:
\beq
\MR_{int} = \NSIT_{(1)2} \wedge \NSIT_{(1)3} \wedge \NSIT_{(2)3} \wedge \LG_{123} \wedge \Ind
\eeq
Most LG experiments to date appear to be testing either $\MR_{weak}$ or $\MR_{int}$ although this is not necessarily made clear in many previous works.

There is a clear hiearchy in these conditions, namely
\beq
\MR_{strong} \implies \MR_{int} \implies \MR_{weak}
\eeq
which follows because the NSIT conditions imply that LG inequalities must hold but the converse is not true. However, this hierarchy is specifically for the case of measurements of a single dichotomic variable. 
For the case of many-valued variables we consider in this chapter there is a richer variety of NSIT conditions and LG inequalities and consequently a richer variety of intermediate MR conditions which do not have a straightforward hierarchical relationship.

%Finally, as noted in the Introduction, there is not just one definition of MR -- there are a number of different ones, depending on whether LG inequalities or the stronger NSIT conditions are used to characterize them \cite{halliwell2017}. For convenience the detailed definition of these different types of MR is summarized in Appendix A, since this background is important for the generalizations considered next.

%\blue{[More detailed discussion of 2 and 3 time interferences. NSIT at three times. Definitions of MR.]}

%\blue{kumari2017,saha2015 inequalities?}

%Also, from a purely macrorealistic perspective, there is no reason to favour the quasi-probability Eq.~(\ref{quasi}) over the two-time measurement probability, Eq.~(\ref{p12}), and one would therefore expect a relation of the form,
%\beq
% p_{12} (s_1,-s_2) + p_{23} (s_2,-s_3) + p_{13} (-s_1,s_3) \le 1,
%\eeq
%to hold. These inequalities are known as the Wigner-Leggett-Garg (kumari2017,saha2015) inequalities, since they are natural generalizations of Wigner's inequalities for Bell-type systems.  However, from the above, it can be readily seen that they consist of the usual LG inequalities plus a set of two-time interference terms.

\section{Conditions for Macrorealism using Leggett-Garg Inequalities at Two Times}
\label{c3s3}
We now come to the main work of this chapter which is to establish MR conditions for many-valued variables. In this Section and the next we do so using LG inequalities, which, in the language of Appendix A, characterise weak MR. In this Section we focus on two times and consider three and more times in the next Section.

\subsection{Projectors for Many-Valued Variables}

We suppose that measurements on our system may be described by a set of projection operators $E_n$, where $n=1, 2, \cdots N$ and $\sum_n E_n = \id$, where $\id$ denotes the identity operator.
(To avoid confusion, we use the notation $P_s$ with $s= \pm 1$, for projectors only in the dichotomic case).
These could be fine-grained measurements $E_n = |n \rangle \langle n |$ on an $N$-level system, or coarse-grained measurements on a system of dimension greater than $N$. These projections may be regarded as measurements of any hermitian operator which has a spectral expansion in terms of the $E_n$,
%\beq
%A = \sum_n \lambda_n E_n,
%\label{AP}
%\eeq
%where the $\lambda_n$ are real, 
and this is the sort  of ``many-valued variable" we have in mind, but we will not make explicit use of such an operator in what follows.
%In practice, in what follows it is sufficient to work directly with the projection operators $E_n$ without reference to the operator $A$ (although the conditions we derive in terms of the $E_n$ may equivalently be expressed in terms of powers of $A$, by inverting Eq.~(\ref{AP})).
Some of the MR conditions we develop in what follows may be extended to the case in which the $E_n$ describe ambiguous measurements (see Ref.~\cite{emary2017} for example) but we will not spell this out explicitly here.
As stated in the Introduction, our strategy for deriving MR conditions is to link the $E_n$ with a set of dichotomic variables.
%For the purposes of both actual measurements and the mathematical task of establishing necessary and sufficient conditions for macrorealism, it is very convenient to relate the many-valued case to the dichotomic case, by defining a class of dichotomic variables in terms of the projectors $E_n$. 
The widest class of such variables have the form,
\beq
\hat Q = \sum_{n=1}^N \epsilon (n) E_n,
\label{Qdef}
\eeq
where $\epsilon (n) $ takes values $\pm 1$ with at least one $+1$ and one $-1$. However,
for our purposes we have found that it is sufficient in almost all cases to focus on a more restricted class, consisting of the $N$ dichotomic variables in which $\epsilon (n)$ has a single $+1$ and the rest of the values are $-1$. (The only exception we have encountered concerns NSIT conditions for $N > 3$ briefly discussed later on).
Each $Q$ therefore has the form, for fixed $n$,
\beq
\hat Q(n) = E_n - \bar E_n.
\eeq
where $\bar E_n = \id - E_n$, the negation of $E_n$.  This variable simply discerns whether a system is occupying a given level $n$, for example
\begin{align}
\hat Q(n) \ket {E_n}&=+\ket{E_n}\\
\hat Q(n) \ket {E_m} &=-\ket{E_m},	
\end{align}
where $m\ne n$.  Each projector $E_n$ may be written,
\beq
E_n = \frac{1}{2} \left( \id + \hat Q(n) \right)
\label{PQn}
\eeq
Since the $E_n$ sum to the identity, the $N$ dichotomic variables $\hat Q(n)$ satisfy
\beq
\sum_{n=1}^N \hat Q(n) = (2 - N) \id
\label{QN}
\eeq
This means that they are still more than the minimum needed to uniquely fix the state of the system. This non-minimal set of variables is convenient to use but it will sometimes be convenient to revert to a set of $N-1$ independent variables.

So far the description is quantum-mechanical but since LG conditions are best formulated in a macrorealistic setting, we introduce the classical analogues of the projectors $E_n$, denoted $\P (n)$, which take values $0$ or $1$, and a corresponding classical object $Q(n)$, with the two related by
\beq 
\P (n) = \frac{1}{2} \left( 1 + Q(n) \right).
\eeq
%\blue{NOTE: May need a more general $\epsilon(n)$.}

In what follows we are interested in the values of $Q(n)$ and $\hat Q(n)$ at times $t_i$ and, following the notation introduced earlier, we denote those values by $Q_i (n)$ and $\hat Q_i (n)$.
The label $n$ always denotes one of the $N$ dichotomic variables and times are always denoted using subscripts. (This is why we avoid the commonly-used notation $Q(t)$ to denote time-dependence, as noted earlier).

\subsection{Two-Time LG Inequalities}

We now consider conditions for weak MR at two-times for many-valued variables, using generalizations of the two-time LG inequalities for the dichotomic case Eqs.~(\ref{LG2a}--\ref{LG2d}). We suppose that non-invasive measurements are made, as described in Section 2, on the $N$ dichotomic variables $Q(n)$ which determine the averages $\langle Q_1 (n_1) \rangle $, $\langle Q_2 (n_2) \rangle$ and the correlator $\langle Q_1 (n_1) Q_2 (n_2) \rangle $. From these quantities we seek to construct a candidate probability $p(n_1,n_2)$. It is reasonably clear that the desired probability is
\beq
p(n_1, n_2) =\langle \P_2 (n_2) \P_1 (n_1) \rangle,
\label{LGN2}
\eeq
where again the subscripts are time labels.
This is clearly the classical analogue of  the quantum-mechanical quasi-probability,
\beq
q(n_1, n_2) = {\rm Re} {\rm Tr} \left( E_{n_2} (t_2) E_{n_1} (t_1) \rho \right),
\label{quasin}
\eeq
which is the natural generalization of Eq.~(\ref{quasi}). The desired two-time LG inequalities are then simply the requirement that the two-time probability is non-negative, which, written out in terms of the $Q(n)$ variables, read:
\beq
1 + \langle Q_1 (n_1) \rangle + \langle Q_2 (n_2) \rangle + \langle Q_1 (n_1) Q_2 (n_2) \rangle \ge 0.
\label{LGN2Q}
\eeq
These $N^2$ LG inequalities are necessary and sufficient conditions for MR at two times. Necessity is trivially established and sufficiency follows since the probabilities themselves, $p(n_1,n_2)$, are proportional to the inequalities.
The LG inequalities involve a set of averages and correlators of the $N$ variables $Q(n)$. 
However, as indicated, this is a non-minimal set since they satisfy Eq.~(\ref{QN}). This means that in practice it is only necessary to make measurements on $N-1$ of the $Q(n)$, and the averages and correlators involving the unmeasured variable readily obtained using Eq.~(\ref{QN}). We will see this explicitly below for the $N=3$ case.

%This does not appear, in general, to reduce the number of conditions, but does reduce the number of correlators and averages that need to be measured in any experiment.

It may seem unusual that Eq.~(\ref{LGN2Q}) contains only plus signs, in contrast to the usual two-time LG inequalities Eqs.~(\ref{LG2a}--\ref{LG2d}), which contain plus and minus signs. However, this relates to the non-minimal nature of the set of variables $Q(n)$ and different forms of the LG inequalities may be obtained by taking linear combinations. In particular, suppose Eq.~(\ref{LGN2}) is summed over all $n_1 \ne n_1'$ for some $n_1'$, and we use the fact that
\beq
\sum_{n_1 \ne n_1'} \P_1 (n_1) = 1 - \P_1 (n_1') \equiv \overline{ \P}_1 (n_1').
\label{PPbar}
\eeq
This is easily seen to have the effect of replacing $Q_1(n_1)$ with $ -Q_1(n_1')$, so we get,
\beq
1 - \langle Q_1 (n_1') \rangle + \langle Q_2 (n_2) \rangle - \langle Q_1 (n_1') Q_2 (n_2) \rangle \ge 0.
\label{minus}
\eeq
This means that the set of inequalities Eq.~(\ref{LGN2Q}) is equivalent to any other set in which any of the $Q(n)$'s have reversed sign. It is therefore sufficient to work with any one set with a fixed set of signs. 
Note also that for the dichotomic case, $N=2$, Eq.~(\ref{QN}) implies that the two dichotomic variables $Q(1)$, $Q(2)$ satisfy $Q(1) = - Q(2)$,  and the four LG inequalities Eq.~(\ref{LGN2Q}) coincide with the two-time LG inequalities Eqs.~(\ref{LG2a}--\ref{LG2d}), as required.

%\blue{[More discussion here?].}

\subsection{Two-Time LG Inequalities for the $N=3$ Case}

The LG inequalities Eq.~(\ref{LGN2Q}) are simplest in form when written in terms of the non-minimal set of $N$ variables $Q(n)$, but in terms of measurements to check them, it is actually only necessary to measure $N-1$ variables at each time and use Eq.~(\ref{QN}) to determine the correlators and averages for the remaining unmeasured variable. We write this out explicitly for the case $N=3$. It is notationally more convenient in the $N=3$ case to take $n=A,B,C$ (following common usage \cite{emary2017}).  Instead of the three variables  $Q(n)$ for $n=A,B,C$, for notational convenience we use the three dichotomic variables $Q$, $R$, $S$ and use the classical projectors 
\bea
\P (A) & =& \frac{1}{2}(1+Q), \\ 
\P(B) &=& \frac{1}{2}(1+R), \\
\P(C) &=& \frac{1}{2} (1+S). 
\eea
The requirement that these sum to $1$ is equivalent to the statement $ Q+R+S = -1$.
%$\P (C)$ has a non-standard appearance and may look like it can take the value $-1$ (as well as $0$ and $1$). However, in this representation of the projectors the pair $(Q,R)$ may only take the three pairs of values $(-1,-1), (-1,1), (1,-1)$ but not $(+1,+1)$, since the latter corresponds to the contradictory statement that the system is in both the $A$ state and the $B$ state.
This relation between $Q$, $R$ and $S$ may seem to unduly constrain the system (because for example we cannot have $Q=R=+1$) but this does not matter since at each time only one or the other is measured.

In terms of these variables the nine two-time LG inequalities have the form Eq.~(\ref{LGN2Q}) and  four of those relations are:
\bea
1 +   \langle Q_1 \rangle +  \langle Q_2 \rangle +  \langle Q_1 Q_2 \rangle  & \ge & 0,
\label{LGNa}
\\
1 +   \langle R_1 \rangle +  \langle Q_2 \rangle +  \langle R_1 Q_2 \rangle  & \ge & 0, 
\\
1 +    \langle Q_1 \rangle +  \langle R_2 \rangle +  \langle Q_1 R_2 \rangle  & \ge & 0, 
\\
1 +    \langle R_1 \rangle +  \langle R_2 \rangle +  \langle R_1 R_2 \rangle  & \ge & 0.
\label{LGNd}
\eea
The other five are similar in form and all involve the variable $S$. However, the relation $Q+R+S=-1$ means that all averages and correlators involving $S$ may be expressed in terms of averages and correlators involving $Q$ and $R$. Carrying this out explicitly yields:
\bea
\langle Q_1 Q_2 \rangle + \langle Q_1 R_2 \rangle + \langle R_1 Q_2 \rangle + \langle R_1 R_2 \rangle & \ge & 0,
\\
\langle Q_1 \rangle + \langle R_1 \rangle + \langle Q_1 Q_2 \rangle + \langle R_1 Q_2 \rangle & \le & 0,
\\
\langle Q_1 \rangle + \langle R_1 \rangle + \langle Q_1 R_2 \rangle + \langle R_1 R_2 \rangle & \le & 0,
\\
\langle Q_2 \rangle + \langle R_2 \rangle + \langle Q_1 Q_2 \rangle + \langle Q_1 R_2 \rangle & \le & 0,
\\
\langle Q_2 \rangle + \langle R_2 \rangle + \langle R_1 Q_2 \rangle + \langle R_1 R_2 \rangle & \le & 0.
\label{LGNi}
\eea
Note that the last four inequalities are upper, not lower bounds. This form is the most useful form to use for experimental tests since no measurements on $S$ are required, only on $Q$ and $R$. This will clearly be true more generally -- measurements on only $N-1$ of the $Q(n)$ variables are required.

An experiment to test inequalities of this type on a three-level system was in fact carried out recently \cite{emary2017,wang2018}. The experiment used ambiguous measurements to determine all nine components of the quasi-probability Eq.~(\ref{quasin}) and observed violations. It therefore represents a test of a complete set of necessary and sufficient conditions for MR at two times for a three-level system. (This experiment also involved a NSIT condition, discussed further below).

%\blue{[More here on alternative forms?]}

%Depending on exactly what the experimental arrangement can measure, it can be of interest to explore
%different ways of representing the projectors. These will cast the LG inequalities in different forms. For example, we could consider a three-valued quantity $A $ taking values $n=-1,0,1$. In the $N=3$ case we can then represent the projectors using $\P(s) = ( A^2 + s A)/2 $, with $s=\pm 1$, and $\P(0) = 1 - A^2 $. This has the feature that it readily collapses down to the dichotomic case if the $0$ value is excluded since then $A^2 = 1$.

%\blue{[More on measurment. INMs. The importance of using dichotomic variables in the measurements. ]}.

%[General result of this form?].

\section[Conditions for MR using LG Inequalities at Three and More Times]{Conditions for Macrorealism using Leggett-Garg Inequalities at Three and More Times}
\label{c3s4}
We turn now to the three-time case and look for conditions for weak MR involving set of three-time LG inequalities for many-valued variables in conjunction with the two-time LG inequalities already derived. We thus generalize the results of Refs.~\cite{halliwell2014,halliwell2017,halliwell2019}.

\subsection{Three-Time LG Inequalities and Fine's Theorem}

We seek necessary and sufficient conditions for the existence of a joint probability $p(n_1,n_2,n_3)$ matching the three pairwise probabilities  $p(n_1,n_2)$, $p(n_2,n_3)$ and $p(n_1,n_3)$ which take the form $ \langle \P_i(n_i) \P_j(n_j) \rangle$. 
The dichotomic case expressed in the form Eq.~(\ref{LG3q}) suggests
the following proposition: the necessary and sufficient conditions are a set of two-time LG inequalities of the form $ \langle \P_i(n_i) \P_j(n_j) \rangle \ge 0 $ ensuring the non-negativity of the pairwise probabilities, together with 
the $N^3$ inequalities,
\beq
\langle \P_1 (n_1) \overline{\P}_2 (n_2) \rangle
+ \langle \P_2 (n_2) \overline{\P}_3 (n_3) \rangle
+ \langle \overline{\P}_1 (n_1) \P_3 (n_3) \rangle \le 1.
\eeq
Writing $ \P (n) = \frac{1}{2} (1 + Q(n) ) $, these inequalities read
\beq
1+  \langle Q_1(n_1) Q_2(n_2)  \rangle + \langle Q_2(n_2) Q_3(n_3)  \rangle + \langle Q_1(n_1) Q_3(n_3) \rangle \ge 0.
\label{LGN3Q}
\eeq
We will prove the proposition below.
Although this is a complete set of LG inequalities as it stands, like the two-time case it appears to involve only one set of signs in front of the correlators. However, again it is possible to demonstrate equivalence with LG inequalities with another set of signs. This can be accomplished for example, by summing Eq.~(\ref{LGN3Q}) over all $n_1 \ne n_1'$ (as in Eqs.~(\ref{PPbar}), (\ref{minus})), and using Eq.~(\ref{QN}). This yields a sum of a three-time LG inequalities of the form Eq.~(\ref{LGN3Q}) with $Q_1 (n_1) $ replaced with $ - Q_1 (n_1) $ and a two-time LG inequality. Hence taken together with the two-time LG inequalities (which we assume hold), the three-time LG inequalities may be written in a number of different forms.

For $N=2$, we may write $Q(1) = Q$ and $Q(2) = - Q$, and we see that the eight inequalities Eq.~(\ref{LGN3Q}) boil down to the four familiar LG inequalities Eqs.~(\ref{LG3a})--(\ref{LG3d}). This simplification does not seem to happen for $N \ge 3$. For $N=3$, for example, we have twenty-seven three-time LG inequalities in terms of the $Q(n)$. However, the fact that the $Q(n)$ are a non-minimal set of dichotomic variables, obeying Eq.~(\ref{QN}), reduces the number of correlators that have to be measured, as we saw in the two-time case already. In particular, we may write out the twenty-seven LG inequalities in the $N=3$ case in terms of the variables $Q$, $R$, $S$ introduced in Section 3(C), satisfying $Q+R+S=-1$. Any correlators or averages involving $S$ may then be expressed in terms of those involving only $Q$ and $R$. We thus obtain a set of three-time LG inequalities analogous to the two-time set, Eqs.~(\ref{LGNa})--(\ref{LGNi}). These are written out in full in Appendix C.

%Similarly, it is true for $N \ge 3$ the $N^3$ inequalities are not a minimal set since at each time there are $N-1$ independent variables as a result of Eq.~(\ref{QN}). 
%This means that there will in general be $(N-1)^3$ {\it independent} three-time LG inequalities for $N \ge 3$. \blue{[CHECK THIS]}.

\subsection{Proof of the Generalized Fine's Theorem}

We now prove the above proposition. Necessity trivially follows from the simple identity,
\beq
\langle ( Q(n_1) + Q(n_2) + Q(n_3) )^2  \rangle \ge 1.
\eeq
To prove sufficiency,
we use the moment expansion of $p(n_1,n_2,n_3)$, making use of Eq.~(\ref{PQn}),
defined by,
\beq
\begin{split}
p(n_1,n_2,n_3) = \langle  \P_1 ( & n_1)   \P_2 (n_2) \P_3 (n_3) \rangle
\\
= \frac{1}{8} \bigg( 1 &+ \langle Q_1(n_1) \rangle + \langle Q_2(n_2) \rangle + \langle Q_3(n_3) \rangle
\\
&+  \langle Q_1(n_1) Q_2(n_2)  \rangle + \langle Q_2(n_2) Q_3(n_3)  \rangle + \langle Q_1(n_1) Q_3(n_3) \rangle
\\
&+ \langle Q_1(n_1) Q_2(n_2) Q_3(n_3) \rangle
\bigg).
\end{split}
\label{pn3}
\eeq
This clearly matches the pairwise marginals as required. It involves a set of unfixed triple correlators $ \langle Q_1(n_1) Q_2(n_2) Q_3(n_3) \rangle$ so the aim is to determine the conditions under which these may be chosen to ensure that 
\beq
0 \le p(n_1, n_2, n_3) \le 1.
\label{UL}
\eeq
Since the probability $p(n_1,n_2,n_3)$ sums to $1$ by construction, the upper bound of $1$ is guaranteed as long as we can show that all $N^3$ terms are non-negative. 
%\blue{[More explanation here]}.
However, it turns out to be more convenient to focus on fixed values of $n_1, n_2, n_3$ and prove that $p(n_1,n_2,n_3)$ satisfies Eq.~(\ref{UL}). The lower bound is clearly satisfied by suitable choice of the triple correlator.
It turns out that the upper bound is most easily handled by re-expressing it as a set of lower bounds on related probabilities, using the identity,
\beq
\langle \left( \P_1 (n_1) + \overline{\P}_1 (n_1) \right)  \left( \P_2 (n_2) + \overline{\P}_2 (n_2) \right) 
 \left( \P_3 (n_3) + \overline{\P}_3 (n_3) \right) 
\rangle = 1.
\eeq
When expanded out, this is readily seen to imply that,
\bea
1 - p(n_1, n_2, n_3) &=&
\langle   \P_1 ( n_1)   \P_2 (n_2) \overline{\P}_3 (n_3) \rangle
+\langle  \P_1 ( n_1)  \overline{ \P}_2 (n_2) \P_3 (n_3) \rangle
\nonumber \\
&+&\langle  \overline{\P}_1 (n _1)   \P_2 (n_2) \P_3 (n_3) \rangle
+\langle  \overline{\P}_1 ( n_1)   \overline{\P}_2 (n_2) \overline{\P}_3 (n_3) \rangle
\nonumber \\
&+&\langle  \overline{\P}_1 ( n_1)   \overline{\P}_2 (n_2) \P_3 (n_3) \rangle
+\langle  \overline{\P}_1 ( n_1)   \P_2 (n_2)\overline{\P}_3 (n_3) \rangle
\nonumber \\
&+&\langle  \P_1 ( n_1)   \overline{\P}_2 (n_2) \overline{\P}_3 (n_3) \rangle.
\label{negp}
\eea
It follows that $p(n_1,n_2,n_3) \le 1$ as long as the seven probabilities on the right-hand side are non-negative.
Each of these probabilities has a moment expansion of the form Eq.~(\ref{pn3}), in which one, two or three of the $Q(n)$'s have their sign flipped, and which are readily seen to yield more upper and lower bounds on the triple correlator.

The remaining steps in the proof are very similar to the dichotomic case covered in Refs.~\cite{halliwell2014,halliwell2019}.
For simplicity we write $p(n_1,n_2,n_3) \ge 0 $ as,
\beq
F (Q_1 (n_1), Q_2 (n_2), Q_3 (n_3) )  + \langle Q_1(n_1) Q_2(n_2) Q_3(n_3) \rangle  \ge 0,
\eeq
where $ F$ is read off from Eq.~(\ref{pn3}). This clearly gives a lower bound on the triple correlator. 
The triple correlator has the same sign for the moment expansion of the last three probabilities on the right-hand side of Eq.~(\ref{negp}). We thus find that non-negativity of four of the probabilities is ensured if
the triple correlator has a total of four lower bounds defined  by
\beq
 \langle Q_1(n_1) Q_2(n_2) Q_3(n_3) \rangle \ \ge \ -F ( s_1 Q_1(n_1), s_2 Q_2(n_2), s_3 Q_3(n_3))  \bigg|_{s_1 s_2 s_3 = +1},
\eeq
where the $s_i$ take values $\pm 1$.
Similarly, the first four probabilities on the right-hand side of Eq.~(\ref{negp}), have a minus sign in front of the triple correlator, and we thus obtain the four upper bounds defined by,
\beq
F ( s_1 Q_1(n_1), s_2  Q_2(n_2), s_3  Q_3(n_3))  \bigg|_{s_1 s_2 s_3  = -1}\  \ge \ \langle Q_1(n_1) Q_2(n_2) Q_3(n_3) \rangle.
\eeq
A triple correlator ensuring that Eq.~(\ref{UL}) holds
may therefore be found as long as the four lower bounds are less than the four upper bounds, i.e.
\beq
F ( s_1' Q_1(n_1), s_2'  Q_2(n_2), s_3'  Q_3(n_3))  \bigg|_{s_1' s_2' s_3'  = -1}
+ F ( s_1 Q_1(n_1), s_2 Q_2(n_2), s_3 Q_3(n_3))  \bigg|_{s_1 s_2 s_3 = +1} \ge 0,
\eeq
for all possible choices of $s_i, s_i'$ satisfying the stated restrictions.
Written out in full, these inequalities are readily seen to coincide with  the two-time LG inequalities Eq.~(\ref{LGN2Q}) and the three-time LG inequalities Eq.~(\ref{LGN3Q}) (and their variants obtained under sign changes). 
One also needs to check that the choice of triple correlator satisfying these upper and lower bounds lies in the correct range, $[-1,1]$, however this is also ensured by the two and three-time LG inequalities (and this is readily seen from the proof in Ref.~\cite{halliwell2019}). This proves sufficiency.

%For the purpose of experimental checks, the LG inequalities Eq.~(\ref{LGN3Q}) are most conveniently written out in terms of a minimal set of variables. For the $N=3$ case there are twenty seven inequalities which are readily written out
%explicitly in terms of the two dichotomic variables $Q$ and $R$ introduced in Section 2(C), and we get a set of three-time inequalities analogous to the two-time inequalities Eqs.~(\ref{LGNa})--(\ref{LGNi}).
%\blue{How many INDEPENDENT LG inequalities? Maybe just $8$ for the $N=3$ case. (Two independent variables at each time).}

% These are given in Appendix C.

%\blue{[Work out the full set of LG inequalities in the $N=3$ case using $Q$, $R$.]}

%\blue{[MORE WORK HERE]
%[N=3 case. NSIT conditions.]

\subsection{Four and More Times}

The dichotomic case is usually formulated in terms of measurements at both three and four times. It was recently generalized to measurements made at an arbitrary number of times and the corresponding Fine's theorem derived \cite{halliwell2019}. This extension from three to many times made use of a generalization of  a famous ansatz of Fine, which expresses the four-time problem entirely in terms of three-time probabilities. This ansatz was originally given in terms of dichotomic variables, but we make the simple observation that the ansatz still works for many-valued variables and, for the four-time case is:
    \beq
    p(n_1, n_2, n_3,n_4)=\frac{p(n_1, n_2, n_3)\ p(n_1, n_3, n_4)}{p(n_1, n_3)}.
    \label{fineansatz}
    \eeq
This is the solution to the matching problem in which we seek a probability $   p(n_1, n_2, n_3,n_4)$ matching the four pairwise marginals, $p(n_1,n_2)$, $p(n_2,n_3)$, $p(n_3,n_4)$ and $p(n_1,n_4)$. It reduces the problem of showing that the four-time probability is non-negative to that of showing that two three-time probabilities are non-negative, which is guaranteed if two appropriate sets of three-time LG inequalities are satisfied, which we may choose to be,
\bea
1 + \langle Q(n_1) Q(n_2) \rangle + \langle Q(n_2) Q(n_3) \rangle + \langle Q(n_1) Q(n_3) \rangle  & \ge & 0, 
\label{QQ123}
\\
1 - \langle Q(n_1) Q(n_3) \rangle + \langle Q(n_3) Q(n_4) \rangle - \langle Q(n_1) Q(n_4) \rangle  & \ge & 0.
\label{QQ134}
\eea
The choice of signs here is purely for convenience and exploits the fact equivalence between different sets of three-time LG inequalities under sign flips of the $Q(n)$. Eliminating the unfixed correlator $ \langle Q(n_1) Q(n_3) \rangle $ between these two inequalities we obtain the $N^4$ inequalities:
\beq
\langle Q(n_1) Q(n_2) \rangle +  \langle Q(n_2) Q(n_3) \rangle + \langle Q(n_3) Q(n_4) \rangle -\langle Q(n_1) Q(n_4) \rangle \ge -2
\label{LG41}
\eeq
We compare this with the set of CHSH-type inequalities Eqs.~(\ref{LG4a})--(\ref{LG4d}), which for fixed $n_i$ consists of eight inequalities, the first two of which are,
\beq
-2 \le   \langle Q(n_1) Q(n_2) \rangle 
 + \langle Q(n_2) Q(n_3) \rangle 
 + \langle Q(n_3) Q(n_4) \rangle 
  -\langle Q(n_1) Q(n_4) \rangle \le 2,
\label{CHSH}
\eeq
and the remaining three pairs are obtained by moving the minus sign to the other three possible locations. The condition Eq.~(\ref{LG41}) just derived is clearly just one of these. However, again using the non-minimal property of the set of $Q(n)$ and the consequent possibility of 
sign flips described in Eqs.~(\ref{PPbar}), (\ref{minus}),
any one of these eight inequalities may be transformed into any other. Hence the necessary and sufficient conditions for (weak) MR at four times, consist of the non-negativity of the pairwise marginals together with any one of the CHSH-type inequalities 
Eq.~(\ref{CHSH}). As expected these conditions reduce to the standard set of eight four-time LG inequalities in the dichotomic case with $Q(1) = - Q(2) =  Q$.

The complete set of LG inequalities for arbitrarily many times in the dichotomic case was given in Ref.~\cite{halliwell2019}. This may be generalized to the case of many-valued variables at many times by proceeding along the same lines as the four-time case above.

\section{No-Signaling in Time Conditions}
\label{c3s5}
We now consider stronger conditions for MR characterized by NSIT conditions.
We focus primarily on the two-time case, with brief mention of three-time MR conditions.

\subsection{Two-Time NSIT Conditions}

The NSIT condition Eq.~(\ref{NSIT}) natural generalizes to the $N$ conditions,
\beq
\sum_{n_1 = 1}^N p_{12} (n_1, n_2) = p_2 (n_2).
\label{NSITn}
\eeq
Since both sides sum to $1$, the number of independent conditions is $N-1$.
However, for $N \ge 3$, this is not the only type of NSIT condition. 
One can instead measure any one of a number of dichotomic variables $Q$ at the first time to determine a probability $p^Q_{12} (s_1, n_2) $ for $s_1 = \pm 1 $, to which there corresponds a NSIT condition,
\beq
\sum_{s_1 } p^Q_{12} (s_1, n_2) = p_2 (n_2).
\label{NSITs}
\eeq
There will in general be a number of conditions of this type, depending on how $Q$ is defined. Recalling that the NSIT condition Eq.~(\ref{NSIT}) is only satisfied for zero interference, the natural way to determine the most complete set of NSIT conditions in the $N \ge 3$ case is to look at the interferences, generalizing the analysis of Section \ref{c3s2}, and derive a set of NSIT conditions which ensure that they are all zero. (Unlike the analysis of LG inequalities presented above, the analysis of NSIT conditions is quantum-mechanical in nature. A macrorealistic presentation is still possible here but we find the quantum-mechanical ones most convenient since it can take advantage
of the machinery of the decoherent histories approach briefly outlined in Appendix B).

The probability $p^Q_{12} (s_1, n_2)$ cannot simply be obtained from $p_{12} (n_1, n_2)$ by coarse graining over $n_1$ because of interferences. As described in Appendix B, both probabilities and interference terms are described by the decoherence functional,
\beq
D (n_1,n_2 |n_1',n_2) = {\rm Tr} \left(  E_{n_2} (t_2) E_{n_1} (t_1) \rho E_{n_1'} (t_1)  \right),
\label{DFN}
\eeq
and both of the probabilities $p_{12}(n_1,n_2)$ and $p_{12}^Q (s_1, n_2)$ may be obtained from it, as we now show.
The interferences themselves are represented by the off-diagonal terms of the decoherence functional.
The probability $p_{12} (n_1, n_2)$ is simply $D(n_1,n_2|n_1,n_2)$. The probability $p^Q_{12} (s_1, n_2)$  is the diagonal part of the decoherence functional,
\beq
D (s_1,n_2 |s_1',n_2) = {\rm Tr} \left(  E_{n_2} (t_2) P_{s_1} (t_1) \rho P_{s_1'} (t_1)  \right),
\label{DFS}
\eeq
where $P_s$ is the projector onto the values of $Q$, Eq.~(\ref{projs}), and may be written in terms of the projectors $E_n$ as,
\beq
P_s = \sum_{n} c_{sn} E_n.
\label{csn}
\eeq
Here, the coefficients $c_{sn}$ are $0$ or $1$ and depend in a simple way on how $Q$ is defined. We thus see that these two decoherence functionals are related by
\beq
D (s_1,n_2 |s_1',n_2) = \sum_{n_1, n_1'} c_{s_1 n_1} c_{s_1' n_1'} D (n_1,n_2 |n_1',n_2).
\eeq
This, via Eq.~(\ref{DFS}), gives an expression for  $p^Q_{12} (s_1, n_2)$ in terms of Eq.~(\ref{DFN}).

The NSIT conditions in a quantum-mechanical setting may be checked by writing each side in terms of the decoherence functional.
Consider first Eq.~(\ref{NSITn}). We have
\bea
p_2 (n_2) &=&  \sum_{n_1, n_1'} D (n_1,n_2 |n_1',n_2),
\nonumber \\
&=& \sum_{n_1} p_{12} (n_1, n_2) + \sum_{{n_1, n_1'} \atop {n_1 \ne n_1'}}  {\rm Re} D (n_1,n_2 |n_1',n_2).
\nonumber \\
\label{inter}
\eea
We again introduce a slightly more condensed notation for the interference terms,
\beq
I_{n_1 n_1'} (n_2)  = {\rm Re} D (n_1,n_2 |n_1',n_2),
\label{Idef}
\eeq
for $n_1 \ne n_1'$.
The coherence witness for the NSIT condition Eq.~(\ref{NSITn}) is a sum terms of this form, so
the NSIT condition is violated unless the interference terms are zero.

Similarly, the violation of Eq.~(\ref{NSITs}) is seen to be,
\bea
p_2 (n_2) &=&  \sum_{s_1 s_1'} D (s_1,n_2 |s_1',n_2),
\nonumber \\
&=& \sum_{s_1} p^Q_{12} (s_1, n_2) + \sum_{ {s_1,s_1'} \atop {s_1 \ne s_1'}} {\rm Re} D (s_1,n_2 |s_1',n_2),
\nonumber \\
&=&  \sum_{s_1} p^Q_{12} (s_1, n_2) + \sum_{{s_1,s_1'} \atop {s_1 \ne s_1'}}  \sum_{n_1, n_1'} c_{s_1 n_1} c_{s_1' n_1'} I_{n_1 n_1'} (n_2).
\label{NSITsV}
\eea
This shows that the two different types of NSIT conditions, Eq.~(\ref{NSITn}) and Eq.~(\ref{NSITs}), imply that different combinations of interference terms are zero. It then seems reasonably clear that we can make all the interference terms zero, as required by choosing suitable combinations of NSIT conditions. This is most easily seen in specific examples.

%\blue{[A reasonable conjecture is that Eq.~(\ref{NSITs}) for all choices of $Q$ does the job]}.

For general $N$ there are $ N(N-1)/2$ ways in which $n_1 \ne n_1'$. The interferences terms sum to zero when summed over $n_2$ which means that are $N-1$ interferences terms for fixed $n_1 \ne n_1'$, and therefore a total of $ N (N-1)^2/2$ independent interference terms. This means there is just one for the $N=2$ case, as we know already, but this jumps up to six for the $N=3$ case, which we now study in detail.

\subsection{Two-Time NSIT Conditions for the $N=3$ Case}
\label{subsec:nsitn3}
For the $N=3$ case the violation of the NSIT condition Eq.~(\ref{NSITn}) reads,
\beq
p_2 (n_2) = \sum_{n_1} p_{12} (n_1, n_2)  + 2 I_{AB} (n_2) + 2  I_{AC} (n_2) + 2 I_{BC} (n_2),
\label{conA}
\eeq
where we use $I_{n_1 n_1'} (n_2)$ defined in Eq.~(\ref{Idef}) and again use the labelling $n=A,B,C$ for the $N=3$ case.
Each interference term sums to zero when summed over $n_2$, so each term $I_{n_1 n_1'} (n_2)$ has two independent values for fixed $n_1, n_1'$. Hence there are a total of six interference terms as expected. There are three different choices for the NSIT condition Eq.~(\ref{NSITs}) depending on how the dichotomic variables $Q(n)$ are defined.
We denote the three choices $Q,R,S$, where $ Q = E_A - \bar E_A$, $ R = E_B - \bar E_B$ and $S = E_C - \bar E_C $,
So for example, for $Q$, the values of $c_{sn}$ defined in Eq.~(\ref{csn}) are $c_{+,A} = c_{-,B} = c_{-,C} = 1$ and the rest zero.
From Eq.~(\ref{NSITsV}), this yields the NSIT violations,
\bea
p_2 (n_2) &=& \sum_{s_1} p^Q_{12} (s_1, n_2)  + 2  I_{AB} (n_2) + 2 I_{AC} (n_2),
\label{conB}
\\
p_2 (n_2) &=& \sum_{s_1} p^R_{12} (s_1, n_2)  + 2 I_{AB} (n_2) + 2 I_{BC} (n_2),
\label{conC}
\\
p_2 (n_2) &=& \sum_{s_1} p^S_{12} (s_1, n_2)  + 2 I_{AC} (n_2) + 2 I_{BC} (n_2).
\label{conD}
\eea
%(More explanation).
Eqs.~(\ref{conA})--(\ref{conD}) is a set of eight conditions (since each one involves two independent conditions for $N=3$), but since there are only six interference terms to be constrained to zero, clearly only six of the eight NSIT conditions are required. For example, we could require that the interference terms in Eqs.~(\ref{conB})--(\ref{conD}) vanish, which implies that all of the $I_{n_1 n_1'} (n_2)$ are zero. Or, we could require that the interference terms in Eq.~(\ref{conA}) vanish along with the interferences terms in any pair of Eqs.~(\ref{conB})--(\ref{conD}), with the same consequence.

%\blue{[Is there a similar statement for general $N$?]}

We now note a significant new  and general feature that arises for $N \ge 3$, which is that the hierarchical relationship between two-time LG inequalities and NSIT conditions outlined in Appendix \ref{c3a} for the dichotomic case becomes more complicated.
In the dichotomic case, the relationship between these two different types of conditions is simple. As we can see from Eq.~(\ref{qp}), if $p_{12} (s_1, s_2)$ satisfies NSIT, then the interference term vanishes and $q(s_1, s_2) = p_{12} (s_1, s_2)$ and is therefore non-negative, i.e. the two-time LG inequalities hold. Equivalently, if the LG inequalities fail, NSIT must be violated. (However, the LG inequalities may still hold even if NSIT is violated). 

In the many-valued case, if a suitably large set of independent NSIT conditions are satisfied then all possible interference terms vanish. And since
Eq.~(\ref{pqd}) for the many-valued case reads,
\beq
q(n_1, n_2) = p_{12} (n_1, n_2)  + \sum_{{n_1'} \atop {n_1' \ne n_1}} \ I_{n_1 n_1'} (n_2),
\label{qpN}
\eeq
this means that $q(n_1,n_2) \ge 0 $, i.e. the two-time LG inequalities hold. So far this hierarchical relationship 
is the same as the dichotomic case, as outlined in Appendix A.
However, we can have a situation in which some, but not all, combinations of the interference terms vanish. This would mean that the LG inequalities could be violated but some of the NSIT conditions are still satisfied.

For example in the $N=3$ case, consider the NSIT conditions Eq.~(\ref{NSITn}) along with the LG inequalities for the dichotomic variable $Q = E_A - \bar E_A $. It is then readily shown that the quasi-probability (i.e. the set of two-time LG inequalities) is,
\bea
q(s_1, s_2) = p_{12}^Q (s_1, s_2) +  I_{AB} (s_2) +  I_{AC} (s_2),
\label{LGqpI}
\eea
where $I_{AB} (+) = I_{AB} (A)$ and $I_{AB}(-) = I_{AB}(B) + I_{AB} (C) $ and similarly for $I_{AC} (s_2)$. It is then clearly  possible for these interference terms to be non-zero and sufficiently negative that $q(s_1, s_2) \le 0 $ (i.e. the two-time LG inequalities fail), but with the sum over interference terms in Eq.~(\ref{conA}) equal to zero, so that the two NSIT conditions Eq.~(\ref{NSITn}) are satisfied.
%\blue{[More detail here?]}.

Precisely such a situation was observed in two recent experiments which both note two-time LG violations when NSIT conditions are satisfied in a three-level system \cite{emary2017,wang2018,george2013}. 
This seems surprising on the face of it, but these papers consider only NSIT conditions of the form Eq.~(\ref{NSITn}). If a complete set of all six NSIT conditions is imposed then all interferences are zero and all two-time LG inequalities must be satisfied.
A similar feature was observed in a recent LG analysis of the triple slit experiment \cite{halliwell2021a}.

Note that there are still {\it some} hierarchical relationships between NSIT conditions and LG inequalities. For example, if the NSIT time condition involving $Q$ is satisfied, i.e. the interference terms in Eq.~(\ref{conB}) vanish, then the quasi-probability Eq.~(\ref{LGqpI}) is non-negative, so the LG inequality holds. However, the general point here is that there is a clear logical relation between NSIT conditions and LG inequalities only if all possible NSIT conditions are satisfied, and if only some are, then the two types of conditions are no longer simply related.

We may also phrase this all in terms of the characterizations of MR for dichotomic variables outlined in Appendix A. For many-valued variables, a natural definition of strong MR at two times is to require that a suitably large set of NSIT conditions are satisfied (large enough to ensure that all interference terms vanish). Similarly weak MR at two times is the requirement that the full set of $N^2$ two-time LG inequalities hold. Clearly the former implies the latter but for many-valued variables there are then a variety of intermediate possibilities which are not simply related to each other.

The relationship between NSIT conditions and interference terms becomes more complicated for $N>3$, since it is not sufficient
to work with dichotomic variables of the form $\hat Q(n) = E_n - {\bar E}_n $. For example, for $N=4$,  there are eighteen interference terms in the decoherence functional. There are four choices of $Q(n)$ in the form defined above which means that there are twelve NSIT conditions involving $p_{12}^Q (s_1, n_2)$ which, unlike the $N=3$ case, is not enough to ensure that all interference terms are zero. What is required is a more general dichotomic variable, for example, of the form $Q=P_1 + P_2 - P_3 - P_4$.
By including two variables of this type we can bring the number of NSIT conditions up to eighteen which is enough to kill all the interference terms.

%\subsection{Consequences for Definitions of MR}

%\blue{[More discussion here about which versions of MR are being tested. The hierarchy of MR conditions seen in the dichotomic case is more complicated here]}.

\subsection{Three-Time NSIT Conditions}

We now briefly consider three-time NSIT conditions  for the many-valued case, the generalizations of the relations, Eqs.~(\ref{NSIT23})--(\ref{NSIT123b}), used in the dichotomic case. We saw in the two-time case in Section \ref{subsec:nsitn3} that, for the $N=3$ case, it is possible to write down a set of NSIT conditions involving three dichotomic variables $Q$, $R$, $S$, which ensures that all interference terms are zero. Hence by analogy, we anticipate that in the three-time case for $N=3$, a complete set of NSIT conditions consists of Eqs.~(\ref{NSIT23})--(\ref{NSIT123b}) in which $Q$, $R$ or $S$ is measured at the first pair of times (with a measurement $E_{n_3}$ at the final time). We will not give any more details here. For $N>3$ a more judicious choice of dichotomic variables may be required as we saw for the $N=4$ case for two times.

\section{Violations of the L\"uders Bound}
\label{c3s6}
%In quantum mechanics, the LG inequalities Eqs.~(\ref{LG3a})--(\ref{LG3d}) with correlation functions given by Eq.~(\ref{corr2}) have a maximum violation of $- \frac{1}{2} $ on the right-hand side. This follows from the inequality
%\beq
%\biggl< \left( s_1 \hat Q_1 + s_2 \hat Q_2 + s_3 \hat Q_3 \right)^2 \biggr> \ge 0,
%\eeq
%which is readily seen to imply that

%This relation has the same mathematical form as the Tsirelson bound for the correlators in Bell-type experiments \cite{cirelson1980} and in the LG context this is often known  as the L\"uders bound \cite{dakic2014,budroni2014,wang2017,pan2018}. 
%However, unlike the Tsirelson bound for measurements on entangled pairs which represents the maximal violation permitted by quantum mechanics, 
%the L\"uders bound can in fact be violated under certain circumstances. 
%These violations can in principle go right up to the algebraic maximum of $-2$ on the right-hand side, accomplished for example when $C_{12} = C_{23} = C_{13} = -1 $, an outright logical paradox from the classical point of view (rather than just the statistical paradox implied by standard LG violations).
In Section~\ref{subsec:lb} we introduced the L\"uders bound, which implies the lower bound of $-\frac12$ to LG3 violations Eq.~(\ref{Luders}).  We also mentioned the scenarios in which it is seen to be violated.  The L\"uders bound violation is possible for systems with $N \ge 3$ in which the correlator is measured in a different way.
In the usual method, one measures the dichotomic variable
\beq
\hat Q = \sum_{n} \epsilon (n) E_n,
\eeq
for some coefficients $\epsilon (n) = \pm 1 $ using a projector $P_s$ onto $\hat Q$. This is often referred to as 
a L\"uders measurement \cite{lueders1950} . We will refer to the resulting correlation function Eq.~(\ref{corr2}) as the L\"uders correlator, $C^L_{12}$, and it has an equivalent expression in terms of the quasi-probability,
\beq 
C^L_{12} = \sum_{n_1, n_2} \epsilon (n_1) \epsilon (n_2) \ q (n_1, n_2).
\eeq
However, there is a macrorealistically equivalent method, which is to determine the two-time sequential measurement probability $p_{12} (n_1, n_2) $ using von Neumann (vN) measurements (sometimes also called ``degeneracy-breaking''), modeled by the $E_n$, and related to the L\"uders measurements by Eq.~(\ref{csn}). We then construct the von Neumann correlator
\beq
C^{vN}_{12} = \sum_{n_1, n_2} \epsilon (n_1) \epsilon (n_2) \ p_{12} (n_1, n_2).
\eeq
%Such measurements are sometimes called von Neumann measurements so we will refer to this as the von Neumann (vN)  correlator. 
Unlike the L\"uders correlator, LG inequalities constructed from $C^{vN}_{12}$ need not satisfy the L\"uders bound, Eq.~(\ref{Luders}). 

However, we would argue, using the understanding of earlier sections, that such a violation is different in character to the usual three-time LG violations.
Using Eq.~(\ref{qpN}), it is readily seen that
\beq
C^{vN}_{12} = C^L_{12} -   \sum_{n_1 \ne n_1' } \sum_{n_2} \epsilon (n_1) \epsilon (n_2)  \ I_{n_1 n_1'} (n_2).
\label{vNI}
\eeq
We thus see that the difference between the von Neumann and L\"uders correlators depends on the two-time interference terms, i.e. on the degree to which the various two-time NSIT conditions are  violated. This difference vanishes in the dichotomic case, as is readily shown.

The significance of this is as follows. As argued in Section \ref{c3s2}, violations of the three-time LG inequalities with the usual L\"uders correlators arise entirely due to interference terms present at three times but not present at two times. Violation of the L\"uders bound using the von Neumann correlators therefore arises due to a combination of the usual three-time interferences plus some additional two-time interference terms for each time pair -- it is not due to any new kind of three-time interference term. 

This effect can then be regarded in two different ways. One attitude would be to say that since this new effect (compared to the more usual LG violations)  comes solely from two-time interference, in a systematic exploration of various MR conditions at two and three times, it might be more natural to identify these effects using two-time NSIT conditions, not the three-time inequalities Eq.~(\ref{Luders}). For example, one could first explore MR conditions for measurements on all pairs of times and look for parameter ranges in which such conditions are violated or satisfied. On proceeding to the three-time case, it would then be natural to restrict only to those parameter ranges for which all two-time MR conditions hold, in order to clearly distinguish between MR conditions at two and three times. The interferences producing the L\"uders bound violation
would disappear if all two-time NSIT conditions are enforced, but a three-time LG violation up to the L\"uders bound is still possible. 

%(These comments also relate to the hierarchy of MR conditions outlined in Appendix B).

The second attitude would be to note that a L\"uders violation requires violation of both
a three-time LG inequality and a two-time NSIT condition. Hence it represents a certain economy since tests the violation of two conditions in a single experiment. In the language of Appendix A, the two conditions are strong MR at two times (which for the many-valued case we take to mean that all two-time NSIT conditions hold) and weak MR at three times.

%Differently put, in exploring MR for experiments involving measurements at three times, a natural way to to proceed might be to first consider only measurements at pairs of times and look for possible MR violations there. One can then identify parameters for which MR at two times is completely satisfied. This then measn

%there is a link here to the hierarchical relationship between MR conditions described in Appendix B for the dichotomic case, which does not hold in general for many-valued systems. However, it does if one adopts a definition of intermediate MR for many-valued variables in which all possible two-time NSIT conditions hold. 

%Without this, a L\"uders bound violation of the three-time inequalities is testing a combination of two and three-time MR conditions.

%\blue{[MORE WORK HERE TO CLARIFY]}

L\"uders bounds violations are also possible with two-time LG conditions and similar comments apply in terms of the MR conditions affected. For example,
the two-time LG inequality,
\beq
1 + \langle Q_1 \rangle + \langle Q_2 \rangle + C_{12} \ge 0,
\label{LGQ++}
\eeq
and has L\"uders bound $ - \frac{1}{2}$ on the right-hand side if the correlator is the usual L\"uders one. This LG inequality can be violated if certain two-time interference terms are sufficiently large, as we have seen, for example, in Eq.~(\ref{LGqpI}).
From a macrorealistic point of view
this inequality still holds if the correlator is measured using degeneracy breaking measurements. 
The correlator is then taken to be the von Neumann one, $C^{vN}_{12}$ and a violation of the two-time L\"uders bound is then possible, due to the presence of additional two-time interference terms. Hence here a L\"uders violation tests the presence of a sum of two-time interference terms see in Eq.~(\ref{vNI}).

However, as stressed already, there are typically many interference terms for systems with $N \ge 3$ which can all be constrained to various degrees by various LG inequalities and NSIT conditions.
A L\"uders bound violation in this case signals a violation of a combination of certain two-time LG inequalities and certain two-time NSIT conditions.

To see this in more detail, consider the $N=3$ case discussed in detail in Section \ref{c2s5}. We have three dichotomic variables $Q,R,S$ and we note that the LG inequality Eq.~(\ref{LGQ++}) is proportional (up to a factor of $1/4$) to the quasi-probability $q(+,+)$ in Eq.~(\ref{LGqpI}), which we write out explicitly:
\beq
q(+,+) = p_{12}^Q (+,+) + I_{AB}(A) + I_{AC}(A)
\eeq
The see the extra interference arising in a L\"uders violation, we compute Eq.~(\ref{vNI}). We have $\epsilon (A) = 1$ and $\epsilon(B) = \epsilon (C) = -1 $ and noting that $\sum_{n_2} I_{n_1 n_1'} (n_2) = 0$, we readily find
\beq
C_{12}^{vN} = C_{12}^L + 4 I_{BC} (A)
\eeq
These relations together imply that
\beq
1 + \langle Q_1 \rangle + \langle Q_2 \rangle + C_{12}^{vN} = 4 \left( 
p_{12}^Q (+,+) + I_{AB}(A) + I_{AC}(A) + I_{BC} (A) \right)
\label{LGvn}
\eeq
and it is the presence of the term $I_{BC}(A)$ which makes a L\"uders violation possible.
%As emphasized in this paper, a comprehensive set of MR tests at two times will consist of a complete set of LG inequalities, or of NSIT conditions, or a combination thereof. 
The LG and NSIT conditions described in Section \ref{subsec:nsitn3} will imply restrictions on all six of the interference terms in the $N=3$ case.
In particular, it is not hard to see that non-trivial values of $I_{BC}(A)$ can be identified with NSIT or conventional LG violations without having to appeal to a L\"uders violation.

Eq.~(\ref{LGvn}) has another interesting feature which is that it may be negative even when the operators $\hat Q_1 $ and $\hat Q_2$ commute. The NSIT conditions Eq.~(\ref{conB}) are satisfied under those conditions and so the interference terms $I_{AB}(A)$ and $I_{AC}(A)$ are zero, but since $I_{BC}(A) \ne 0$ Eq.~(\ref{LGvn}) indicates that a LG violation is still possible.
This conclusion is in agreement with the earlier work Ref. \cite{kumari2018} in which this phenomenon was examined more generally (and also investigated the conceptual relevance of L\"uders violations, as we do here).

To be clear, none of the above remarks undermine the significance of L\"uders bound violations, which are truly striking non-classical effects, even more so than conventional LG violations. Here, we have argued here that the presence of the underlying interference terms producing them can be detected by less striking means and also identified the particular type of MR conditions that L\"uders violations test.

%\blue{[ What is the max violation if LG2 satisfied? Work out details in $N=3$ case.  Which version of MR is being tested?
%When do new 3-time interferences come into play? ].}

% Differently put, a natural view is to take a quantum-mechanical perspective from which the role of LG and NSIT conditions is to constrain the size of interference terms.

% None of these observations undermine the signifcance of Luders bounds violations, which are truly striking quantum effects. For example,

%\blue{
%\section{{An Example for $N=3$}}
%}

%\blue{Here we may also need a reasonably simple example that illustrates most of the points made throughout the whole paper, especially Section 3D and the Luders violation. A simple one may be found in the appendix of Emary Ref.[38] for a three-level system. The spin $j$ example may also be useful.}].

\section{Summary and Conclusion}
\label{c3s7}
We have shown how to extend the standard conditions for macrorealism for dichotomic variables, namely two and three-time LG inequalities and NSIT conditions, to situations described by $N$-valued variables. We have in addition explored the various new features of these conditions and their relationships that do not arise in the dichotomic case.

To prepare the ground, we carried out a detailed quantum-mechanical analysis of the dichotomic case in Section \ref{c3s2}. This highlighted the fact that conditions for MR act as constraints on the degree of interference.

In Section \ref{c3s3} we considered conditions for MR using LG inequalities at two times. We established a complete set of two-time LG inequalities, Eq.~(\ref{LGN2Q}), in terms of the dichotomic variables $Q(n)$, which are necessary and sufficient conditions for the existence of a pairwise joint probability $p(n_1,n_2)$. The $N$ variables $Q(n)$ are however not a minimal set and we exhibited a minimal set for the case $N=3$ involving just two dichotomic variables.

MR conditions using LG inequalities at three or more times were considered in Section \ref{c3s4}. We proved a generalization of Fine's theorem, i.e. established the necessary and sufficient conditions under which a set of three pairwise probabilities of the form $p(n_i,n_j)$ could be matched to an underlying joint probability. The conditions in question turned out to be a natural generalization of the familiar three-time LG inequalities for the dichotomic case.
We then generalized this treatment to four or more times.
We noted that Fine's ansatz for the $N$-valued variable case readily extends to the case of four or more times (as studied in the dichotomic case in Ref.~\cite{halliwell2019}), which indicates that, like the dichotomic case, MR conditions involving LG inequalities boil down to the three-time case

In Section \ref{c3s5}, we considered the stronger MR conditions characterized by NSIT conditions for the $N \ge 3$ case and elucidated their connection to the vanishing of certain interference terms.
We noted that  the $N  \ge 3$ case has a much richer set of NSIT conditions than the dichotomic case. In particular, Eq.~(\ref{NSITn}) is not the only type of NSIT condition and many new conditions can be generated by considering measurements of different choices of dichotomic variables $Q(n)$ at the first time.
We derived the conditions in detail in the $N=3$ case and applied this understanding to some recent experiments, in which two-time LG inequality violations were observed, even though the NSIT conditions Eq.~(\ref{NSITn}) were satisfied. This illustrates an important general feature: the
set of NSIT conditions and LG inequalities for many-valued variables do not have the simple hierarchical relationship enjoyed by the dichotomic case (summarized in Appendix \ref{c3a}).

In Section \ref{c3s6}, we used the understanding gained earlier to examine violations of the L\"uders bound for three-time LG inequalities. We noted that these entail violations of two distinct MR conditions -- a two-time NSIT condition and a three-time conventional LG violation. The effect would therefore be absent in an approach which requires all two-time MR conditions to be satisfied before proceeding to three times, but could also be regarded as an economical way of identifying two different MR condition violations in a single experiment. Similar observations were also made for the case of two-time L\"uders violations.

%We noted that the violations are entirely due to interference terms operating at two times and not due to a new three-time interference effect and we discussed the consequences of this in terms of the particular MR conditions involved.

%We also noted that for two-time L\"uders violations, the interference terms responsible can be identified by other conditions of the LG or NSIT type.

The present work suggests a number of new possibilities for experimental tests of macrorealism. The main one would be to test a complete set of MR conditions for a system with $N \ge 3$, as was recently carried out for the dichotomic case \cite{majidy2019b}. This would involve making measurements which are able to check all of the two-time and three-time LG inequalities, Eqs.~(\ref{LGN2Q}), (\ref{LGN3Q}), thereby checking for violations of MR at both two times and three times. A limited number of LG tests involving three-level systems have been carried out  but none test a complete set of two and three time MR conditions (although the one recent experiment discussed in Section \ref{c3s3} which tested a complete set of two-time conditions \cite{wang2018} is an important step forwards here).
The proposed new experiments could readily be carried out by simple extensions of existing approaches. 
We also note that the recent proposal to test MR in the context of the harmonic oscillator, involving a dichotomic variable equal to the sign of the position operator \cite{bose2018a}, could readily be extended to many-valued variables by partitioning the position into more than two values.

%
%%[Experimental consequences].
%
%\section{Acknowledgements}
%
%We are grateful to Sougato Bose, Clive Emary, Dipankar Home, George Knee, Johannes Kofler, Raymond Laflamme, Shayan Majidy, Owen Maroney, Alok Pan, Stephen Parrott and James Yearsley for many useful discussions and email exchanges about the Leggett-Garg inequalities over a long period of time. We also thank Shayan Majidy for a critical reading of the manuscript. In addition, we thank two anonymous referees for helpful comments.

\begin{subappendices}

\section{Some results from the decoherent histories approach to quantum mechanics}
\label{c3b}
Here  we briefly outline some properties of the decoherence functional and its relation to the quasi-probability and sequential measurement formula. These results are standard mathematical ones from the decoherent histories approach to quantum theory
\cite{halliwell2016b,griffiths1984,griffiths1993,griffiths1996,griffiths1998,omnes1988,omnes1988a,omnes1988b,omnes1989,omnes1990,omnes1992,gellmann1993,halliwell2009} although this is not a decoherent histories analysis. 

Histories consisting of  measurements at $n$ times are represented by ``class operators'',
\beq
C_{\alpha} = P_{s_n} (t_n) \dots P_{s_2} (t_2) P_{s_1} (t_1),
\eeq
where $\alpha$ denotes the string $(s_1,s_2, \cdots s_n)$. They sum to the identity. Each class operator has negation defined by
\beq 
\overline{C}_{\alpha} = 1 - C_{\alpha} = \sum_{\alpha' \ne \alpha} C_{\alpha'},
\eeq
which therefore represents all the histories not corresponding to the string $(s_1,s_2, \cdots s_n)$.
The probability for a sequence of measurements at $n$ times is
\beq
p(\alpha) = {\rm Tr} \left( C_{\alpha} \rho C_{\alpha}^\dag\right),
\eeq
and the associated quasi-probability is
\beq
q(\alpha) ={\rm Re} {\rm Tr} \left( C_{\alpha} \rho  \right).
\eeq
We also introduce the decoherence functional
\beq
D(\alpha, \alpha') = {\rm Tr} \left( C_{\alpha} \rho C_{\alpha'}^\dag\right),
\eeq
describing interference between the history $C_{\alpha}$ and history $C_{\alpha'}$.
Simple algebra then shows that
\beq
q(\alpha) = p(\alpha) + {\rm Re} D(\alpha, \bar \alpha),
\label{pqd}
\eeq 
where 
\beq
D(\alpha, \bar \alpha) = {\rm Tr} \left(C_{\alpha}\rho  \overline{C}_{\alpha}^\dag \right)
\eeq
is the decoherence functional describing the interference between history $C_{\alpha}$ and its negation $ \overline{C}_{\alpha} $.
Eq.~(\ref{pqd}) may also be written,
\beq
q(\alpha) = p(\alpha) + \sum_{{\alpha'} \atop {\alpha' \ne \alpha}} {\rm Re} D(\alpha, \alpha').
\label{pqd2}
\eeq 
\section{LG Inequalities at Three Times for the $N=3$ Case}
\label{c3c}
We give here the explicit form for the three-time LG inequalities Eq.~(\ref{LGN3Q}) in the $N=3$ case, in terms of the three dichotomic variables $Q$, $R$ and $S$, satisfying $Q+R+S=-1$.
%%try fit them all onto one page somehow
\begin{align}
1+\expval{Q_1 Q_2}+\expval{Q_2 Q_3}+\expval{Q_1 Q_3}&\geq 0,\\
1+\expval{R_1 Q_2}+\expval{Q_2 Q_3}+\expval{R_1 Q_3}&\geq 0,\\
1+\expval{Q_1 R_2}+\expval{R_2 Q_3}+\expval{Q_1 Q_3}&\geq 0,\\
1+\expval{R_1 R_2}+\expval{R_2 Q_3}+\expval{R_1 Q_3}&\geq 0,\\
1+\expval{Q_1 Q_2}+\expval{Q_2 R_3}+\expval{Q_1 R_3}&\geq 0,\\
1+\expval{R_1 Q_2}+\expval{Q_2 R_3}+\expval{R_1 R_3}&\geq 0,\\
1+\expval{Q_1 R_2}+\expval{R_2 R_3}+\expval{Q_1 R_3}&\geq 0,\\
1+\expval{R_1 R_2}+\expval{R_2 R_3}+\expval{R_1 R_3}&\geq 0,\\
1+\expval{S_1 Q_2}+\expval{Q_2 Q_3}+\expval{S_1 Q_3}&\geq 0,\\
1+\expval{S_1 R_2}+\expval{R_2 Q_3}+\expval{S_1 Q_3}&\geq 0,\\
1+\expval{S_1 Q_2}+\expval{Q_2 R_3}+\expval{S_1 R_3}&\geq 0,\\
1+\expval{S_1 R_2}+\expval{R_2 R_3}+\expval{S_1 R_3}&\geq 0,\\
1+\expval{Q_1 S_2}+\expval{S_2 Q_3}+\expval{Q_1 Q_3}&\geq 0,\\
1+\expval{R_1 S_2}+\expval{S_2 Q_3}+\expval{R_1 Q_3}&\geq 0,\\
1+\expval{Q_1 S_2}+\expval{S_2 R_3}+\expval{Q_1 R_3}&\geq 0,\\
1+\expval{R_1 S_2}+\expval{S_2 R_3}+\expval{R_1 R_3}&\geq 0,\\
1+\expval{Q_1 Q_2}+\expval{Q_2 S_3}+\expval{Q_1 S_3}&\geq 0,\\
1+\expval{R_1 Q_2}+\expval{Q_2 S_3}+\expval{R_1 S_3}&\geq 0,\\
1+\expval{Q_1 R_2}+\expval{R_2 S_3}+\expval{Q_1 S_3}&\geq 0,
\end{align}
\begin{align}
1+\expval{R_1 R_2}+\expval{R_2 S_3}+\expval{R_1 S_3}&\geq 0,\\
1+\expval{S_1 S_2}+\expval{S_2 R_3}+\expval{S_1 R_3}&\geq 0,\\
1+\expval{S_1 S_2}+\expval{S_2 Q_3}+\expval{S_1 Q_3}&\geq 0,\\
1+\expval{Q_1 S_2}+\expval{S_2 S_3}+\expval{Q_1 S_3}&\geq 0,\\
1+\expval{R_1 S_2}+\expval{S_2 S_3}+\expval{R_1 S_3}&\geq 0,\\
1+\expval{S_1 Q_2}+\expval{Q_2 S_3}+\expval{S_1 S_3}&\geq 0,\\
1+\expval{S_1 R_2}+\expval{R_2 S_3}+\expval{S_1 S_3}&\geq 0,\\
1+\expval{S_1 S_2}+\expval{S_2 S_3}+\expval{S_1 S_3}&\geq 0.
\end{align}
Due to the non-minimal nature of the set $Q, R, S$, we in fact do not need to measure all 
twenty-seven correlators, since all averages and correlators involving $S$ may be expressed in terms of $Q$ and $R$. So we have for example,
\beq
\expval{S_1 Q_2} = - \expval{Q_2} - \expval{Q_1 Q_2} - \expval{R_1 Q_2},
\eeq
and also
\begin{multline}
	\expval{S_1 S_2}=1 + \expval{Q_1}+\expval{Q_2}+\expval{R_1}+\expval{R_2}
+\expval{Q_1 Q_2}+\expval{Q_1 R_2}+\expval{R_1 Q_2}+\expval{R_1 R_2}.
\end{multline}
%\expval{S_1 S_2}=1 &+& \expval{Q_1}+\expval{Q_2}+\expval{R_1}+\expval{R_2}
%\nonumber \\
%&+& \expval{Q_1 Q_2}+\expval{Q_1 R_2}+\expval{R_1 Q_2}+\expval{R_1 R_2}.
%\eea
All other cases have this general form and we will not write them out here. The set of quantities to be measured then consists of the twelve correlators of the form $\expval{Q_i R_j}$, $\expval{Q_i Q_j}$, $\expval{R_i R_j}$ (the last two with $i<j$), along with the six averages $\expval{Q_i}$ and $\expval{R_i}$, for $i,j =1,2,3$.

\end{subappendices}

\fancyhf{}
\renewcommand{\headrulewidth}{0pt}
\fancyfoot{\makebox[\textwidth][c]{\hyperref[link:5]\thepage}}

\fancyhfoffset[LE,RO, RE, LO]{0cm}
\renewcommand{\chaptermark}[1]{ \markboth{#1}{} }
\renewcommand{\sectionmark}[1]{ \markright{#1}{} }
\fancyhf{}
\fancyhead[L]{\textsl{\thesection~~ \rightmark}}
\fancyhead[R]{\hyperref[link:5]\thepage}
\renewcommand{\headrulewidth}{1pt}

\chapter{Leggett-Garg tests for macrorealism in the quantum harmonic oscillator and more general bound systems}	
	\label{chap:QHO1}
	\bookepigraph{3.5in}{There are more things in heaven and earth, Horatio,\\ Than are dreamt of in your philosophy.}{Shakespeare,}{Hamlet}{0}
	\vspace{-3em}
	\section{Introduction}

Much LG research has been conducted on the discrete properties (often non-classical spin) of microscopic systems, e.g. the properties of single atoms, or photons.  That these systems are so far removed from the domain of what we may consider macroscopic objects is what we call the macroscopicity problem.  While there are several approaches to precisely characterising `macroscopicity', see Refs~\cite{leggett1980, frowis2018, nimmrichter2013}, we will take a more pragmatic approach to the term here.  For a system with freely adjustable mass, it is capable of describing an arbitrarily large particle, and we hence consider it a (potentially) macroscopic system.

Most macroscopic systems that have been investigated with a view to exhibiting quantum coherence effects have the feature that they are described by continuous variables. Hence to approach LG tests in the macroscopic domain, it is natural to develop such tests for continuous variables, using, for example, variables defined by a coarse graining of position.  In this chapter, we therefore pursue an investigation into one of the most ubiquitous continuous variable systems -- the quantum harmonic oscillator, and bound systems more generally.  %The results we derive are in a sense invariant 
	
	%with the aim that it can act as a bridge from the microscopic smacroscopic world for experimental LG tests.
	
	Our theoretical investigation is in part inspired by the harmonic oscillator LG experiment proposed by Bose \textsl{et al}, where an ideal negative result measurement procedure is used to non-invasively measure variables defined through the position coarse-graining $Q={\rm sgn}(x)$~\cite{bose2018a, das2022a}.  This then yields the LG standard data-set of single time-averages and the temporal correlators.  By contrast with other tests for non-classicality, this approach does not require the fabrication of non-Gaussian states \cite{romero-isart2011}, nor the coupling to an ancillary quantum system \cite{scala2013,asadian2014}.
	
%	
%	  The work of Bose et al suggests that these quantum correlations persist, and remain experimentally feasible to detect, at scales of up to $10^6$--$10^9$ amu, meaning LG tests on the QHO are likely to push the limits of our observations of macroscopic coherence.  
	
	In this chapter we plug a gap in the literature by performing a thorough theoretical analysis of LG tests in continuous variable systems.  We derive analytical results for the temporal correlators for the QHO in a variety of experimentally accessible states, primarily energy eigenstates and superpositions thereof.
	We also derive an approximation for the temporal correlation functions which is applicable to any quantum system for which the energy eigenspectrum is (exactly or numerically) known.
	We determine substantial areas in the various parameter spaces in which combinations of LG2, LG3 and LG4 inequalities are satisfied or violated, thereby preparing the ground for experimental tests.
	
	In Section \ref{sec:calc}, we set up the problem and calculate temporal correlators working within the energy eigenbasis.  We develop a powerful technique for studying the partial overlap of energy eigenfunctions, which enables us to calculate the temporal correlators for any system with known energy eigenspectrum.  We apply this to the QHO, which yields a simple and useful approximation for the temporal correlator in the first excited state as a simple a cosine (just like the simple spin model used in LG tests).  We also demonstrate a procedure to calculate the exact temporal correlators for any energy eigenstate of the QHO.
	
	In Section \ref{sec:1and0}, we  explore the LG violations present in the QHO.  We find significant violations in the first excited state, and see that by creating superpositions of the ground state and the first excited state, we can find substantial regimes where the $\LG{2}$s and $\LG{3}$s are independently satisfied or violated. We find significant LG violations persist even with significant smoothing of projectors.  In Section \ref{sec:higher}, we analyze the LG inequalities in the higher excited states of the QHO, and observe the expected classicalization, with the magnitude of LG violations rapidly decreasing as energy increases.

	In Section \ref{sec:mlvl}, we use a more general dichotomic observable, defined for arbitrary regions of space.  Using this more general variable we find LG violations even in the ground state.  
	
	In Section \ref{sec:morse}, to demonstrate the versatility of the techniques developed in Section \ref{sec:calc}, we calculate temporal correlators for the Morse potential, and perform a brief LG analysis, finding significant LG3 and LG4 violations.  We summarize and conclude in Section \ref{sec:conc}.
	
	The fine details of the complicated machinery required to calculate temporal correlators in the QHO, are largely relegated to a series of technical appendices.
	
	\section{Temporal correlators in bound systems}\label{sec:calc}
	
	\subsection{Conventions and Strategy}
	\label{subsec:cands}
	%We will consider variables constructed through the coarse-graining of position.  This converts the continuous variable into a large set of possible discrete variables, for which the LG inequalities are well suited to testing for an underlying MR description.  For simplicity of presentation, we first work with the simplest dichotomic variable,
	For most of this chapter we work with the simplest choice of dichotomic variable $ Q=\text{sgn}(x)$, 
	however the method we develop also gives the result for more general observables involving arbitrary regions of space.  This is investigated further in Section ~\ref{sec:mlvl}.
	We will work with a general bound one-dimensional system with Hamiltonian $\hat H$ and energy eigenstates $\ket{n}$ with corresponding eigenvalues $E_n$.  
	
	We now outline our strategy.  We make use of the two-time quasi-probability Eq.~(\ref{quasi}), with the projectors
%	
%	
%	introduced in Ref.  \cite{halliwell2016b}, given by,
%	%is a very convenient mathematical object to use in the study of macrorealism.  For a dichotomic variable, the quasi-probability has definition 
%	\beq
%	\label{eq:qp}
%	q(s_1, s_2)=\text{Re} \Tr (P_{s_2}(t_2)P_{s_1}(t_1)\rho),
%	\eeq
%	where $s=\pm1$, and 
	\beq
	\label{eq:proj}
	P_s=\frac{1}{2}(1+ s \hat Q)=\theta(s \hat x),
	\eeq
	where $\theta(x)$ is the Heaviside step function. Time dependence is then handled in the Heisenberg picture.
	
	We will use the quasi-probability in two ways. Firstly, it is proportional to the two-time LG inequalities, so negativity signifies an LG2 violation  \cite{halliwell2016b}. (The quasi-probability has a maximum negative value of $-\tfrac{1}{8}$).
	Secondly, it yields a simple way to extract the temporal correlators needed for the three-time LG inequalities, via its useful moment expansion Eq.~(\ref{mom}).
%	\beq
%	\label{momexp}
%	q(s_1, s_2)=\frac{1}{4}\left(1+s_1 \expval{\hat Q_1}+s_2 \expval{\hat Q_2}+s_1 s_2 C_{12}\right).
%	\eeq
	We will hence first calculate the quasi-probability, and then go on to extract the temporal correlators, which we can then use in the LG inequalities Eqs.~(\ref{LG3a})--(\ref{LG3d}).  
	In contrast to the computation of correlators in simple spin models, these calculations are quite non-trivial.
	
	More detailed properties of the quasi-probability may be found in Ref.~\cite{halliwell2016b}, with its relation to the two-time measurement probability detailed in Chapter~\ref{chap:mlev}, but we make a few brief comments here. As indicated, it is used here purely as a mathematical tool and has no particular conceptual role beyond the fact that it is proportional to the two-time LG inequalities. It is clearly non-negative for commuting pairs of projections, so any negativity is associated with incompatibility of the pairs of measurements. It can usefully be expressed in terms of the Wigner-Weyl representation~\cite{halliwell2019c}, from which one can see that it can be negative even when the Wigner function of the initial state is non-negative, as is the case for gaussian initial states. We shall see an example of this later in this paper, and see also  Refs.~\cite{halliwell2019c, halliwell2021}. 
	
%	The quasi-probability is also very closely related to the two-time sequential measurement formula
%	\begin{equation}
%	p(s_1,s_2) = \Re \Tr \left(  P_{s_2} (t_2) P_{s_1} (t_1) \rho P_{s_1} (t_1)   \right),   
%	\end{equation}
%	This is always non-negative and has a moment expansion similar in form to Eq.~(2.3), with the only difference that $ \langle \hat Q_2 \rangle$  is replaced with $ \langle \hat Q_2^{(1)} \rangle$,  
%	the average of $Q_2$ in the presence of an earlier measurement at $t_1$ which has been summed out. (See Ref.~\cite{halliwell2016b}). 
%	
	Note in particular that both the quasi-probability and sequential two-time measurement formula have the same correlator given by Eq.~(\ref{corr2}).  	Physically, this corresponds to the fact that the same correlator is obtained using a pair of measurements in which the first measurement is projective or weak \cite{halliwell2016b, fritz2010}, or more generally, any one of a family of ambiguous measurements intermediate between these two options (as can be seen in Ref.~\cite{halliwell2019c}). The relevant non-invasive measurement protocols for LG tests of these quantities are described in more detail in Refs.~\cite{halliwell2016b, halliwell2017,halliwell2019b, halliwell2019a}.
%		\begin{equation}
%	\label{eq:qcorr}
%	C_{12} =  \frac{1}{2} \langle \hat Q_1 \hat Q_2 + \hat Q_2 \hat Q_1 \rangle.
%	\end{equation}

	%It should be noted that this calculation is quite non-trivial, especially when contrasted with the simplicity of determining the temporal correlators in the simple spin case. 
	
	\subsection{Calculating the quasi-probability}
	Using the definition of the two-time quasi-probability Eq.~(\ref{quasi}), and Eq.~(\ref{eq:proj}), we have
	\beq
	\label{qp2}
	q(s_1, s_2)=\text{Re} \Tr (e^{\frac{iHt_2}{\hbar}}\theta(s_1 \hat x )e^{-\frac{i H(t_2-t_1)}{\hbar}}\theta (s_2\hat x )e^{\frac{-iHt_1}{\hbar}}\rho).
	\eeq
	We will often work with the difference between measurement times, and so introduce the variable $\tau = t_2-t_1$.
	
	It is sufficient to calculate one element of the quasi-probability $q(+,+)$, where the full quasi-probability $q(s_1, s_2)$ may be reconstructed through symmetry arguments.  Working within the energy eigenbasis, we write Eq.~(\ref{quasi}) as
	\beq
	q(+,+)=\Re\sum_{m=0,n=0}^{\infty}\!\!\!\braket{\psi}{m}\!\!\!\mel{m}{P_+(t_2)P_{+}(t_1)}{n}\!\!\braket{n}{\psi},
	\eeq
	which equals
	\beq
	q(+,+)=\Re \sum_{m=0,n=0}^{\infty}\braket{\psi}{m}\braket{n}{\psi}q_{mn},
	\eeq
	where
	\beq
	\label{eqn:qmn2}
	q_{mn}=
	\mel**{m}{e^{\frac{i H t_2}{\hbar}}\theta(\hat x)e^{-\frac{iH\tau}{\hbar}}\theta(\hat x)e^{-\frac{i H t_1}{\hbar}}}{n}.
	\eeq
	Using the position representation for the step functions $\theta(\hat x)=\int^\infty_0 \vert x \rangle \langle x \vert \mathop{dx}$, this may be written explicitly as
	\beq
	\label{eq:qmn}
	q_{mn}=e^{i\frac{E_m}{\hbar}t_2-i\frac{E_n}{\hbar}t_1}\int_{0}^{\infty}\int_{0}^{\infty}\mathop{dx}\mathop{dy}\braket{m}{x}\!\!\mel{x}{e^{-\frac{i H \tau}{\hbar}}}{y}\!\braket{y}{n},
	\eeq
	where from here the calculation forks, with two obvious ways to proceed. One approach is to insert a resolution of unity in the energy eigenbasis, and truncate the infinite sum to yield an approximation to the correlators, which we will proceed with first.  The second approach is to insert the expression for the propagator, if known, which will allow us to calculate exact expressions for the quasi-probability.
	
	The single time averages are given by 
	\beq
	\ev*{\hat{Q}}=\ev{\sgn(\hat x)}{n}.
	\eeq
	In the special case of symmetric potentials, since $\text{sgn}(\hat x)$ flips the parity of $\ket n$, this represents the overlap between an odd and an even state. Hence in this case, the single time averages in the moment expansion Eq.~(\ref{mom}) are identically zero.  This yields a simple relationship between the correlator and the quasi-probability,
	\beq
	\label{eq:qeig}
	q(+,+)=\frac{1}{4}(1+C_{12}).
	\eeq
	This is clearly non-negative, and so LG2 violations are not possible for single energy eigenstates, however this is not true for superposition states, where the relationship is not as simple, as we cover in Sec.~\ref{c4mel}. 
	
	We adopt the notation $C_{12}^{\ket{n}}$, to denote the temporal correlator between times $t_1$ and $t_2$, for a given energy eigenstate $\ket{n}$.
	
	\subsection{Quasi-probability for energy eigenstates}
	\label{subsec:inf}
	We use the approximation approach first, whereupon inserting the resolution of unity, we have
	\beq
	q_{mn}=e^{i\frac{E_m}{\hbar}t_2-i\frac{E_n}{\hbar}t_1}\sum_{k=0}^{\infty}e^{-i \frac{E_k}{\hbar}(t_2-t_1)}\int_{0}^{\infty}\int_{0}^{\infty}\mathop{dx}\mathop{dy}\braket{m}{x}\!\braket{x}{k}\!\braket{k}{y}\!\braket{y}{n}.
	\eeq
	The integration is now separable into two integrals, the partial overlap of energy eigenstates,
	\begin{equation}
	J_{k\ell}=\int_0^{\infty}\mathop{dx}\braket{k}{x}\braket{x}{\ell}.
	\end{equation}
	Surprisingly, as detailed in Appendix \ref{app:wronski}, these integrals can be completed for generic potentials over arbitrary boundaries $x_1$, $x_2$, with result
	\begin{equation}
	\label{eqn:wronski}
	J_{k\ell}(x_1, x_2)=\frac{1}{2(\varepsilon_\ell- \varepsilon_k)}\left[\psi_k'(x_2)\psi_\ell(x_2)-\psi_\ell'(x_2)\psi_k(x_2)-\psi_k'(x_1)\psi_\ell(x_1)+\psi_\ell'(x_1)\psi_k(x_1)\right],
	\end{equation}
	where $\psi_k(x)=\braket{x}{k}$.  Hence the matrix elements of the quasi-probability are
	\beq
	q_{mn}=e^{i\frac{E_m}{\hbar}t_2-i\frac{E_n}{\hbar}t_1}\sum_{k=0}^{\infty}e^{-i\frac{E_k}{\hbar}(t_2-t_1)}J_{mk}J_{nk}.
	\eeq
	The quasi-probability for a single energy eigenstate is the real part of the diagonal elements,
	\beq
	\label{eqn:qpapprox}
	q_n(+,+)=\Re e^{i \frac{E_n}{\hbar} \tau }\sum_{k=0}^{\infty}e^{-i \frac{E_k}{\hbar}\tau }J_{nk}^2.
	\eeq

	By truncating this sum, we are able to calculate the quasi-probability for any coarse-graining of space, and for any soluble bound system, purely in terms of the spectrum of its Hamiltonian.  We will primarily use this result with the QHO in this work, however in Sec.~\ref{sec:morse}, we consider its application to other systems.  Further details about implementation of this formula, including a truncation error estimate $\Delta_n(m)$ defined in Eq.~(\ref{eq:trunc}), are found in Appendix \ref{app:calcdeets}.
	
	We also note in passing that Eq.~(\ref{eqn:wronski}) unlocks the result of several indefinite integrals in the form of partial spatial overlaps of special functions, which we have not been able to find in the literature. These include; generalised Laguerre polynomials (Morse potential \cite{dahl1988}), Mathieu functions (quantum pendulum \cite{aldrovandi1979,baker2002}), Airy functions (triangular potential \cite{khonina2013}), and Hypergeometric functions (transformation technique \cite{morales2015}), with the exactly soluble potential leading to the result bracketed.

	\subsection{Application to the QHO}
	\label{subsec:qhoconv}
	The results so far have been general to any bound potential.  We now apply them to the systems defined exactly (or approximately) by the harmonic oscillator Hamiltonian,
	\beq
	\hat H=\frac{\hat p_{\rm{phys}}^2}{2m}+\frac12 m\omega^2 \hat x_{\rm{phys}}^2,
	\eeq
	with physical position and momentum $x_{\rm{phys}}$ and $p_{\rm{phys}}$.  We continue to denote energy eigenstates by $\ket n$, and use natural units, so we work with dimensionless position and momentum $x$ and $p$, defined by $\sqrt{\hbar/(m\omega)} x=x_{\rm{phys}}$ and $p \sqrt{(m\omega)/\hbar}=p_{\rm{phys}}$.  We will similarly write energy in terms of $\hbar \omega$, with $E_n=(n+\frac12)\hbar \omega=\varepsilon_n \hbar \omega$.  The energy eigenstates within the position basis are then given by
	%$\sqrt{\hbar/(m\omega)}$ and energy in units of $\hbar \omega$.  ,
	\beq
	\braket{n}{x}=\frac{1}{\sqrt{2^n n!}}\pi^{-\frac{1}{4}}\exp(-\frac{1}{2}x^2)H_n(x),
	\eeq
	where $H_n(x)$ are the Hermite polynomials.
	
	For the ground state, the infinite series needs many terms to reach good convergence, and so we instead turn to $q_{11}$, whereupon taking the real part, we have the quasi-probability $q(+,+)$, for the first excited state.  We calculate the first few $J_{nm}$ using Eq.~(\ref{eq:jijlim}). We find
	$J_{1,0}^2=\frac{1}{2\pi}$, $J_{1,1}^2=\frac14$, $J_{1,2}^2=\frac{1}{4\pi}$, $J_{1,3}^2=0$, and $J_{1,4}^2=\frac{1}{48\pi}$.  Using just the first three terms leads to a very good approximation to the correlator, with $\Delta_1(2)=0.011$.  The sum may be explicitly written as  $	q_{11}=\frac{1}{4}+\frac{1}{4\pi}e^{-i\omega\tau}+\frac{1}{2\pi}e^{i\omega \tau}$. Taking the real part, we find
	\beq
	\label{eq:approx}
	q(+,+)=\frac{1}{4}+\frac{3}{4\pi}\cos \omega \tau.
	\eeq
	By comparing Eq.~(\ref{eq:approx}) to the expression for the quasi-probability of a pure eigenstate Eq.~(\ref{eq:qeig}), we can identify the correlator as 
	\beq
	\label{eq:cos}
	C_{ij}^{\ket1}\approx\cos\omega\tau,
	\eeq
	where we have dropped the coefficient of $\tfrac{3}{\pi}\approx0.955$, by the requirement that $C_{ij}\to 1$ as $\tau \to 0$.
	
	This approximation is interesting since it says that for the first excited state, the QHO to a good approximation behaves just like the canonical simple spin-$\frac{1}{2}$ example used in much LG research.  We hence may borrow intuition from this simpler system, and we know to expect LG violations.  
	This similarity can be understood by explicitly calculating the temporal correlator using the three-term approximation to the post-measurement state.  The temporal correlator here is defined as
	\begin{equation}
	C_{12}^{\ket1}=\Re \expval{\hat Q_2 \hat Q_1}{1}.
	\end{equation}
	From Eq.~(\ref{eq:proj}), we have $\hat Q=2\theta(\hat x)-1$, so using the three-term approximation and taking $t_1=0$, we find
	\begin{equation}
	\label{1approx}
	\hat Q_1 \ket{1}\approx\frac{2}{\sqrt{2\pi}}\ket0+\frac{1}{\sqrt{\pi}}\ket2.
	\end{equation}  
	Similarly, taking $t_2=\tau$, we find
	\begin{equation}
	\bra{1}\hat Q_2 \approx e^{i\omega\tau}\left(\frac{2}{\sqrt{2\pi}}\bra0+e^{-2i \omega\tau}\frac{1}{\sqrt{\pi}}\bra2.\right).
	\end{equation}
	Hence like the spin-$\frac{1}{2}$ case, it boils down to just two states.
	Contracting these result, and taking the real part yields
	\begin{equation}
	\label{cosapprox}
	C_{12}^{\ket1}=\frac3\pi \cos(\omega \tau) \approx \cos(\omega \tau) 
	\end{equation}
	
	We note that taking the next order of approximation we have $\Delta_1(4)=0.005$, and clearly a much better approximation.  The expression for the correlator (again normalized to 1 as $\tau\to0$), is then
	\beq
	C_{ij}^{\ket1}\approx \frac{36}{37}\cos \omega \tau + \frac{1}{37}\cos 3\omega \tau.
	\eeq
	
	\subsection{Exact Correlators for the QHO}
	For the QHO, we are also able to calculate exact expressions for the correlators,  whereupon inserting the expression for the QHO propagator into Eq.~(\ref{eq:qmn}), we find
	\begin{multline}
	\label{eq:qmnfull}
	q_{mn}=\mathcal{N}_{mn}(\tau) e^{i\frac{E_m}{\hbar}t_2-i\frac{E_n}{\hbar}t_1}\times\\\int^{\infty}_0\int^{\infty}_0 \mathop{dr}\mathop{ds}H_m(r)H_n(s)e^{-\frac{r^2}{2}}e^{-\frac{s^2}{2}}\exp({i \frac{1}{2\tan(\omega \tau)}(r^2+s^2)- i\frac{1}{\sin(\omega \tau)}rs}),
	\end{multline}
	where $\mathcal{N}_{mn}(\tau)$ is a dimensionless normalisation factor, defined by
	\beq
	\mathcal{N}_{mn}(\tau)=\frac{1}{\pi}\frac{1}{\sqrt{2^n n!}}\frac{1}{\sqrt{2^m m!}}\frac{1}{\sqrt{2i \sin(\omega\tau)}}
	\eeq
	This integral can be completed, via a generating integral approach, which is detailed in Appendix~\ref{app:gen}.  Using this result, we are able to write the exact results for the temporal correlators.  To do so, it is helpful to introduce the function
	\beq
	f(\tau)=-ie^{-i\frac{\omega\tau}{2}}\sqrt{2i\sin\omega\tau}.
	\eeq
	The correlators for the groundstate and first excited state may then be written as
	\begin{align}
	\label{eq:arctan}
	C_{ij}^{\ket0}&=\frac{2}{\pi}\Re\arctan\left(\frac{1}{f(\tau)}\right),\\
	\label{cor:1s}
	C_{ij}^{\ket1}&=\frac{2}{\pi}\Re\left[\arctan\left(\frac{1}{f(\tau)}\right)+f(\tau)\right].
	\end{align}
	These two rather non-intuitive expressions are plotted in Fig. \ref{fig:corrs}, where they are easily seen to have the broadly expected physical behaviours
	The expressions for the first nine correlators are included in Appendix~\ref{app:corrs}. 
	\subsection{Matrix Elements for a two-state oscillator}
	\label{c4mel}
	It will turn out that very much can be studied in the QHO, with initial superposition states of just the ground state and first excited state.  That is working with states of the form 
	\beq
	\label{eq:state}
	\ket \psi = a \ket 0 + b \ket 1.
	\eeq
	To determine the off-diagonal elements of $q_{mn}$ defined in Eq.~(\ref{eqn:qmn2}), it is simplest to do so term-by-term in the moment expansion Eq.~(\ref{mom}).
	
	The quantum mechanical temporal correlator is given by Eq.~(\ref{corr2}), with matrix elements
	\beq
	\mel{m}{\hat C_{ij}}{n}=\frac12\mel*{m}{\hat Q_1 \hat Q_2 + \hat Q_2 \hat Q_1}{n}.
	\eeq
	We now note that  $\hat Q_1 \hat Q_2 + \hat Q_2 \hat Q_1$ is invariant under reflections. This means that $(\hat Q_1 \hat Q_2 + \hat Q_2 \hat Q_1)\ket{n}$ must have the same parity as $\ket{n}$.  Hence in cases where $m$ is odd and $n$ even, or vice-versa, the correlator represents the overlap of an even state with an odd state, and is therefore zero in these cases.  Hence for the general state Eq.~(\ref{eq:state}), the correlator is simply given as the mixture
	\beq
	\label{supcor}
	C_{ij}=\abs{a}^2 C_{ij}^{\ket0}+ \abs{b}^2 C_{ij}^{\ket1}.
	\eeq
	We note as well that for $Q=\textrm{sgn}(x)$, from the orthogonality of energy eigenstates that 
	
	We now determine the matrix elements of the quasi-probability.  Owing to the argument preceding Eq.~(\ref{eq:qeig}), the diagonal elements take the simple form 
	\beq
	q_{nn}(s_1, s_2)=\frac{1}{4}\left(1+s_1 s_2 C_{ij}^{\ket n}\right).
	\eeq
	Similarly, with the vanishing of the correlators on the off-diagonals, we see that
	\beq
	q_{01}(s_1, s_2)=\frac{1}{4}\mel{0}{s_1 \hat Q_1+s_2 \hat Q_2}{1}.
	\eeq
	As these averages involve only a single time each, the time dependence is trivial, and we just require the value of
	\beq
	\ev*{\hat{Q}}=\mel{0}{\text{sgn}(\hat x)}{1},
	\eeq
	which is readily calculated using Eq.~(\ref{eq:jijlim}) as $\ev*{\hat{Q}}=\sqrt{\frac{2}{\pi}}$.

	We hence find the two-time quasi-probability for the state Eq.~(\ref{eq:state}) to be
	\beq
	\label{eq:q2so}
	q(s_1, s_2)=\frac{1}{4}\left[1+s_1\left(2\sqrt{\frac{2}{\pi}}\Re a^*b e^{i\omega t_1}\right)+s_2\left(2\sqrt{\frac{2}{\pi}}\Re a^*b e^{i\omega t_2}\right)+s_1 s_2\left(\abs{a}^2C_{ij}^{\ket0}+\abs{b}^2C_{ij}^{\ket1}\right)\right].
	\eeq
	\subsection{The classical analogue}
	\begin{sloppypar}
	It is useful in many of these calculations to compare the correlators obtained with their classical analogue. 
	This is readily found from the classical analogue of the quasi-probability, namely $\langle \theta (x) \theta (x(\tau) \rangle $, with phase-space initial state $f(x^2 + p^2)$, which is normalised to ${2\pi\int_0^{\infty}\mathop{dr}r f(r)=1}$.  This choice covers fixed energy states, and mixtures thereof.  We then have,
	\end{sloppypar}
	%The quasi-probability has a simple classical analogue, where the expectation value in a classical phase-space is computed, with position given by the classical equations of motions.  That is,
	\beq
	\mathbbmsl{q}(0, \tau)=\int_{-\infty}^\infty\int_{-\infty}^\infty \mathop{dx}\mathop{dp}f(x^2 + p^2)\theta(x) \theta(x \cos \omega \tau+p\sin \omega \tau).
	\eeq
	The step-functions are easiest handled in polar coordinates as
	\beq
	\mathbbmsl{q}(0, \tau)=\int_{0}^{\abs{\pi-\omega \tau}}\mathop{d\theta}\int_0^\infty \mathop{dr}r f(r),
	\eeq
	where inserting the normalisation condition on $f(r)$ yields
	\beq
	\label{eq:class}
	\mathbbmsl{q}(0, \tau)=\frac{\lvert{\pi-\omega \tau}\rvert}{2\pi},
	\eeq
which has the corresponding classical correlator of $\mathbbmsl{C}_{12}=-1+\frac{2}{\pi}\lvert{\pi-\omega \tau}\rvert$.
	
	We now plot two of the QHO correlators, alongside the classical analogue, in Fig.~\ref{fig:corrs}.  We also plot the correlator for the first excited state, alongside the approximation Eq.~(\ref{eq:cos}), which also serves to compare the behaviour of the first excited state of the QHO with the canonical simple spin model typically used in LG research. 
	\begin{figure}
		\subfloat[]{{\includegraphics[height=5.1cm]{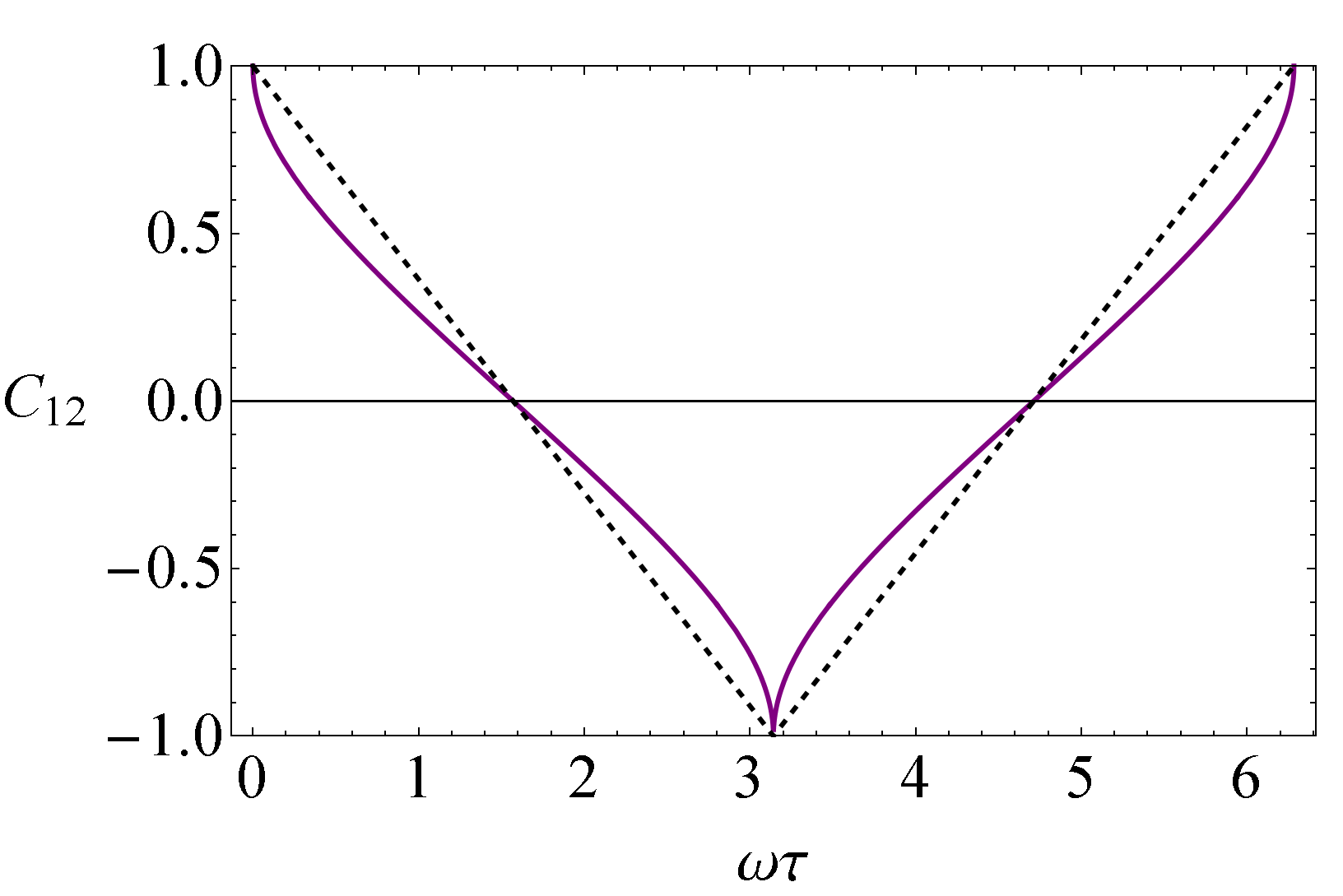}}}%
		\qquad
		\subfloat[]{{\includegraphics[height=5.1cm]{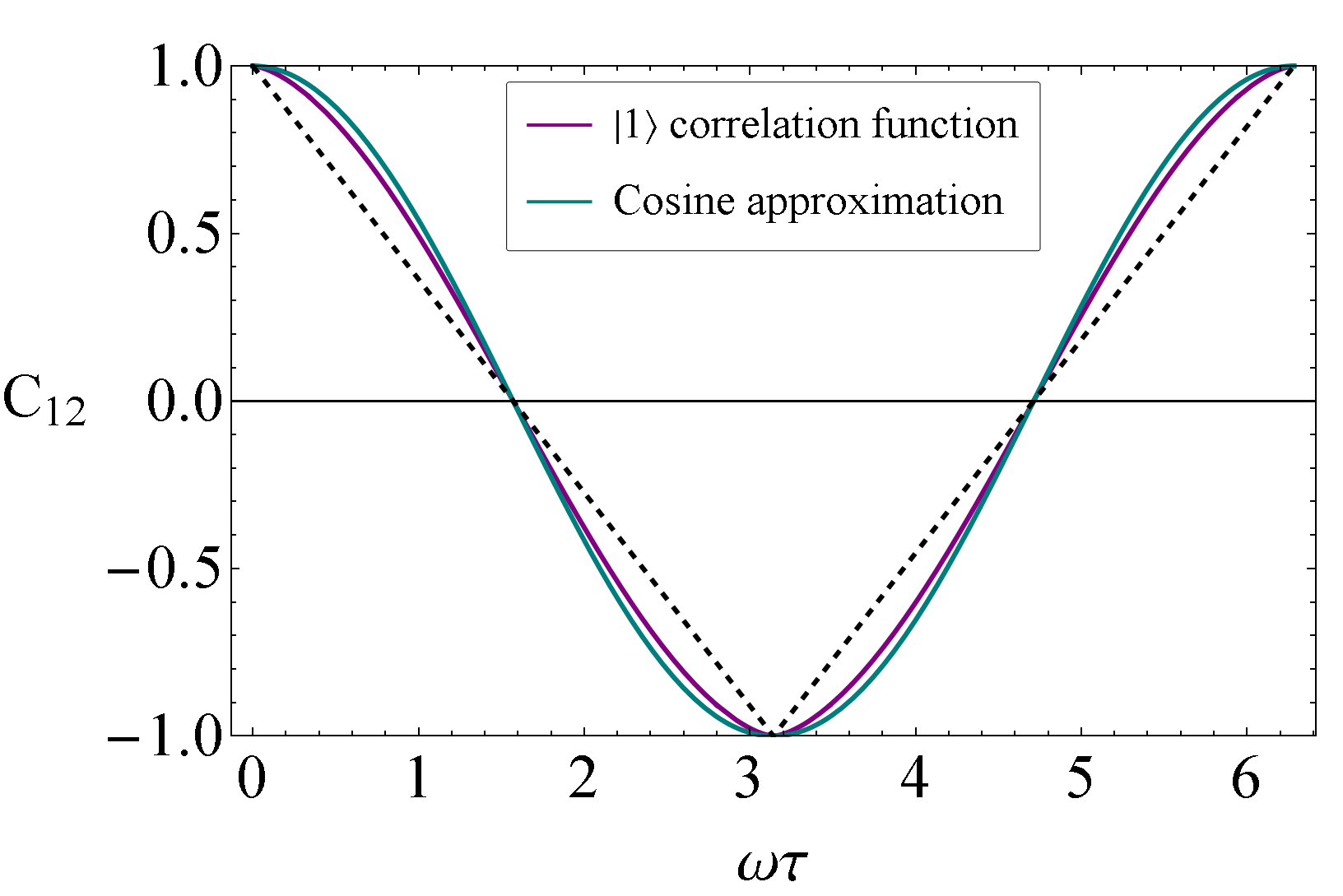}}}
		\caption[Correlators for QHO energy eigenstates, and comparison to spin-$\frac12$ case]{In (a), the temporal correlator for the ground state of the QHO Eq.~(\ref{eq:arctan}) is plotted, alongside the classical analogue Eq.~(\ref{eq:class}) (dashed).  In (b), the temporal correlator for the first excited state Eq.~(\ref{cor:1s}) is plotted, alongside the correlator for the simple spin case, showing the close similarity mentioned in the text.  The classical analogue is again shown (dashed).}%
		\label{fig:corrs}%
	\end{figure}

	\section{Leggett-Garg Violations in the two-state QHO}\label{sec:1and0}
	\subsection{Four Regimes}
	In this section, we study the LG inequalities in the scenario where we have access only to the ground state and the first excited state, the two-state oscillator.  
	This allows us to take full advantage of the simplicity of the expression for the correlator in this two-state superposition, Eq.~(\ref{supcor}).  We will find that even in this state space, the LG2s and LG3s can be independently violated or satisfied, giving access to each of four logical possibilities which are laid out in Table \ref{tbl:reg}.
	{
	\begin{table}[b]
	\centering
		\def\arraystretch{1.125}
		{\setlength{\tabcolsep}{0.5em}
			\begin{tabular}{ Sc | Sc | Sc  }
				\textbf{Regime} & \textbf{LG3s} & \textbf{LG2s} \\
				\hline
				I & \includegraphics[scale=0.05]{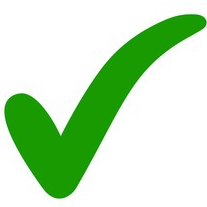} & \includegraphics[scale=0.05]{check} \\
				II & \includegraphics[scale=0.05]{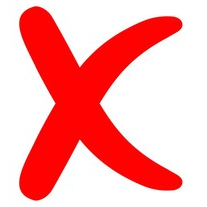} & \includegraphics[scale=0.05]{check}  \\
				III & \includegraphics[scale=0.05]{check} & \includegraphics[scale=0.05]{cross} \\
				IV & \includegraphics[scale=0.05]{cross} & \includegraphics[scale=0.05]{cross}
		\end{tabular}}
		\caption[The four regimes testable with the $\LG{2}$ and $\LG{3}$ inequalities]{The four regimes testable with the $\LG{2}$ and $\LG{3}$ inequalities.  A tick denotes that the complete set of inequalities is satisfied, whereas a cross indicates that one or more of the inequalities in that set is violated.}
		\label{tbl:reg}
	\end{table}
	}
	These four regimes take their importance from the fact that, for the data set consisting of the three $\langle Q_i \rangle$ and three correlators,
	the LG inequalities form necessary and sufficient conditions for MR only when all the $\LG{2}$ and $\LG{3}$ inequalities are satisfied, as discussed in Sections~\ref{sec:fine}, \ref{c2s2} and \ref{c3s2}.  The QHO limited to the superpositions of the $\ket0$ and $\ket1$ states makes a complete playground for the experimentalist, exhibiting all variants of MR tested by the LG inequalities.  We analyze and provide examples of each of the regimes in Table~\ref{tbl:reg}. In what follows, we survey the parameter-space of these two-state superpositions to determine where each of the four regimes lie.
	
	%	 This has a couple of interesting consequences in cases where the $\LG{3}$ inequalities are satisfied.  From the sufficiency aspect; although typically just $\LG{3}$s are experimentally tested, their being satisfied does not in fact indicate the system behaves macrorealistically, since the $\LG{2}$s must also be verified as satisfied.  This means using solely the $\LG{3}$s to test for MR leaves space open for false positives.  From the necessity aspect; since all conditions must be satisfied, it means the violation of one or more of the $\LG{2}$s is enough to indicate the failure of MR, which since it involves just two times, may be easier to implement experimentally.
%	These four regimes have previously been explored both experimentally and theoretically in simple spin models \cite{majidy2021a}.
	\subsection{LG inequalities for pure eigenstates}

	We first consider initial states which are energy eigenstates. The LG2s are trivially satisfied for these states since $\ev{Q_i}=0$, so we can only access regimes I and II in Table \ref{tbl:reg}. This is not the case for superpositions.
	
	Consider now the $\LG{3}$s, $L_{i}^{\ket n}$, where $i$ indexes one of the four $\LG{3}$ kernels Eqs.~(\ref{LG3a})--(\ref{LG3d}), with the initial state $\ket n $.  For example,
	\begin{equation}
	L_{1}^{\ket 0}=1+C_{12}^{\ket0}+C_{13}^{\ket0}+C_{23}^{\ket0}.
	\end{equation}
	
	These four inequalities are plotted in Fig.~\ref{fig:lgpure}, (although two of them coincide).  For the ground state, the $\LG{3}$ inequalities are satisfied always.  Since the $\LG{2}$ inequalities are trivially satisfied as well, this means that for the ground state lies in regime I of Table \ref{tbl:reg}.  For the first excited state, we see the $\LG{3}$ are violated always (except at points of measure zero). The magnitude of the violation is significant, reaching approximately 73\% of the L\"{u}ders bound of $-\frac12$.  This means that the first excited state corresponds to regime II.
		
	\begin{figure}
		\subfloat[]{{\includegraphics[height=5.1cm]{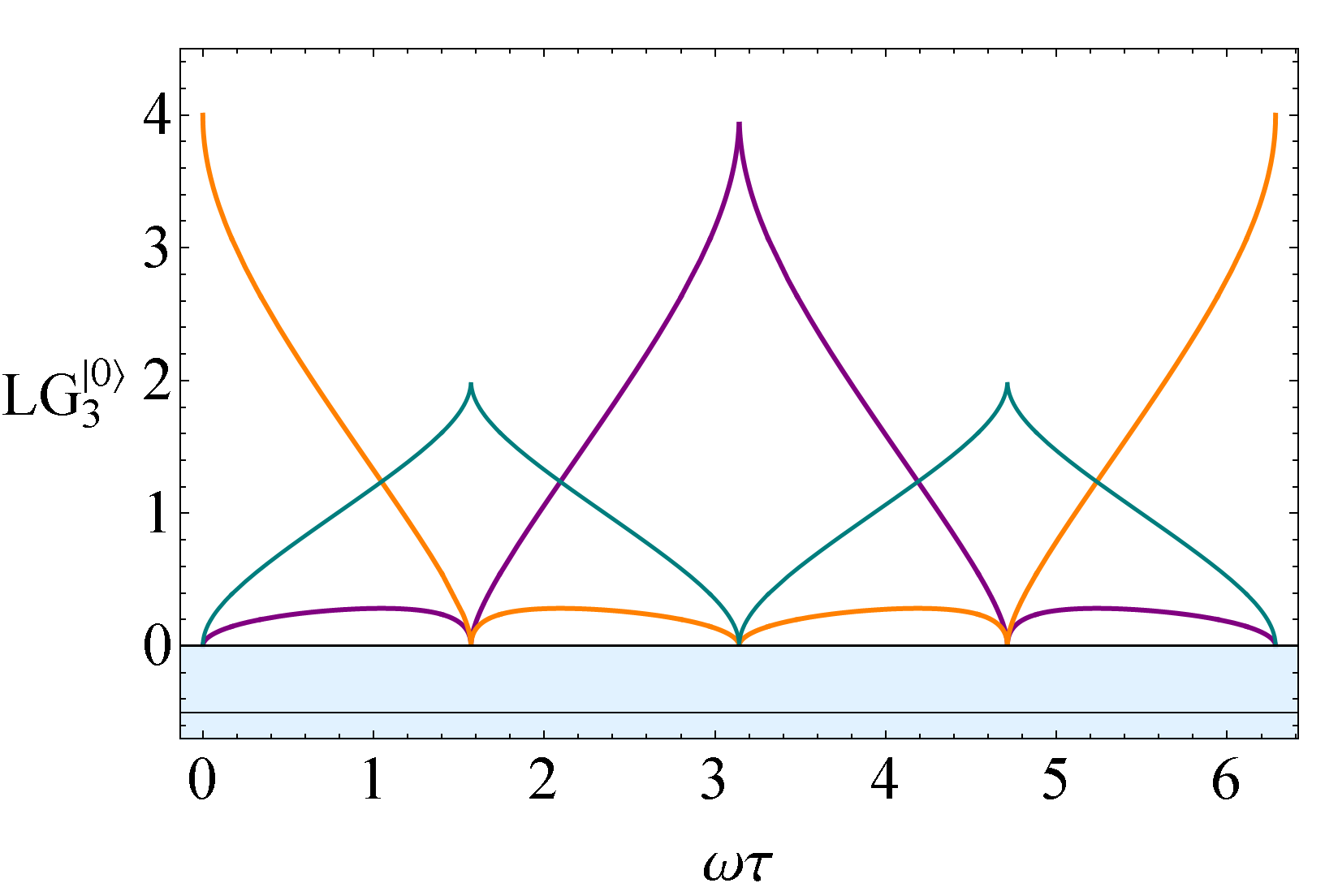}}}%
		\qquad
		\subfloat[]{{\includegraphics[height=5.1cm]{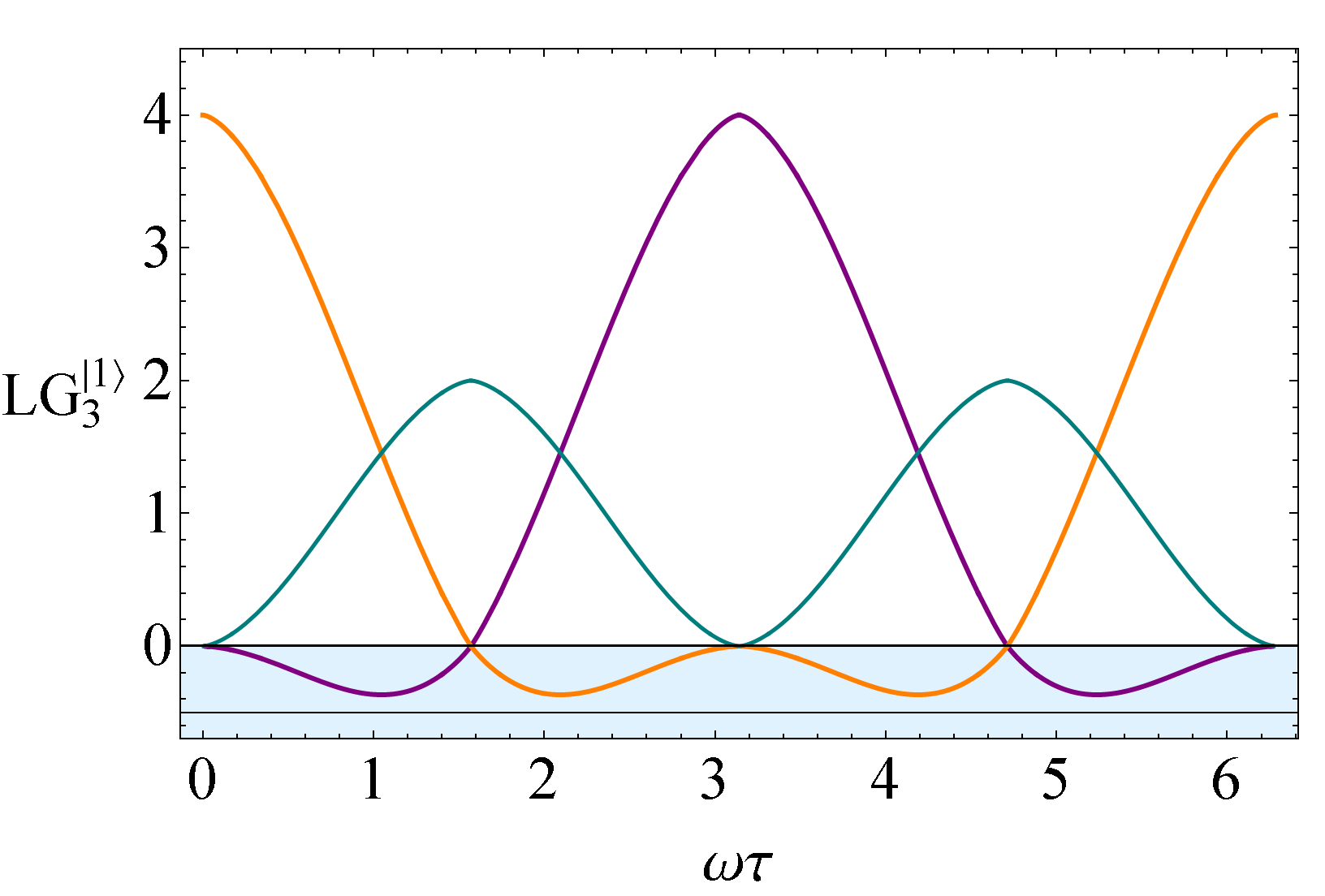}}}
		\caption[LG3 inequalities for $\ket0$ and $\ket1$ states]{In (a), the three-time LG inequalities are plotted for the ground state of the QHO, where they are satisfied at all times.  In (b), the same inequalities are shown for the first excited state, where they are violated everywhere, except for points of measure zero.  These violations come close to the maximal violation, the L\"{u}ders bound of $-\frac12$.}%
		\label{fig:lgpure}%
	\end{figure}
	\subsection{LG inequalities for superposition states}
	
	Since the matrix elements of the correlator are diagonal for
	superpositions of $\ket0$ and $\ket1$, the LG3s will simply be a convex sum of $L_i^{\ket0}$ and $L_i^{\ket1}$.  We can hence continuously transition between states with no $\LG{3}$ violation, and those with $\LG{3}$ violation everywhere.
	We begin by parametrising the two-state superpositions as
	\beq
	\label{eq:param}
	\ket{\psi}=\cos \tfrac{\theta}{2} \ket 0+e^{i\phi}\sin \tfrac{\theta}{2} \ket 1,
	\eeq
	where $0\leq \theta \leq \pi$, and $0\leq \phi \leq 2\pi$.

	Since the $\LG{3}$ inequalities are constructed purely from correlators, using Eq.~(\ref{supcor}), we find
	\beq
	L_i(\theta)=\cos^2\tfrac{\theta}{2} ~L_i^{\ket0}+\sin^2\tfrac{\theta}{2}~L_i^{\ket1},
	\eeq
	
\begin{figure}
	\begin{center}
		{\includegraphics[height=6.4cm]{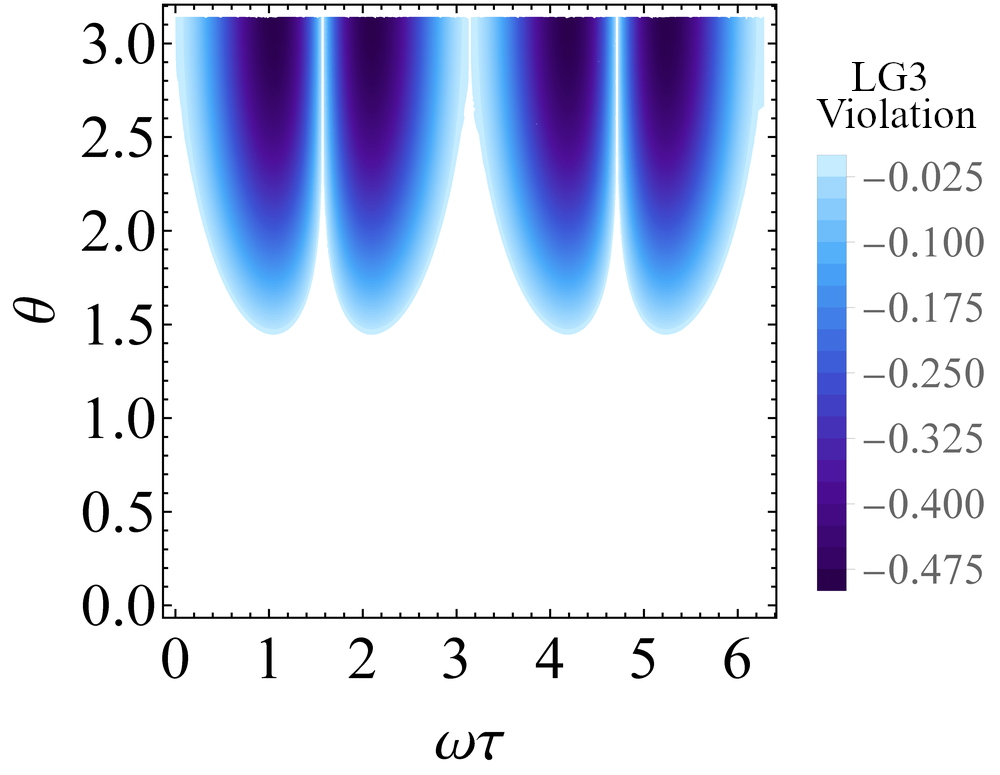}}
		\end{center}
		\caption[LG3 inequalities for superpositions $a\ket0+b\ket1$]{The regions where at least one of the $\LG{3}$ inequalities is violated are shown as a function of time between measurements, and superposition coefficients.  The shading corresponds to the magnitude of the $\LG{3}$ violation.}%
		\label{fig:lgsup}%
	\end{figure}
and so a two-dimensional space of $\theta$ and $\tau$ defines whether the $\LG{3}$ inequalities for the superposition are satisfied.  This space is plotted in Fig.~\ref{fig:lgsup}, where for a given $\theta$, there are three possible cases; the $\LG{3}$ are violated for all $\tau$, the $\LG{3}$ are satisfied for all $\tau$, or a non-trivial mixture of the two, where they are satisfied for some ranges of $\tau$, and violated 	for others.

	We now analyze the behaviour of the two-time LG inequalities.  Due to interference terms, the averages $\langle Q_i\rangle$ are non-zero for superposition states, which can produce $\LG{2}$ violations.  Using the parametrisation Eq.~(\ref{eq:param}) in the expression for the quasi-probability Eq.~(\ref{eq:q2so}), we find
	\begin{multline}
	q(s_1, s_2)=\frac14\Bigg[1+s_1\left(\sqrt{\frac{2}{\pi}}\sin \theta\cos (\phi+\omega t_1)\right)+s_2\left(\sqrt{\frac{2}{\pi}}\sin \theta\cos(\phi+\omega t_2)\right)\\+s_1 s_2\left(\cos ^2\tfrac{\theta}{2}~C_{ij}^{\ket0}+\sin^2\tfrac{\theta}{2}~C_{ij}^{\ket1}\right)\Bigg].
	\end{multline}
	Without loss of generality, we set $t_1=0$, and in Fig.~\ref{fig:lg2frac}, we present the full parameter-space for the two-state superpositions, showing where the $\LG{3}$ and $\LG{2}$ inequalities are violated.  Hence with superposition states, we can reach regimes III and IV from Table \ref{tbl:reg}.
	
	\begin{figure}
	
	\begin{center}
		\subfloat[]{{\includegraphics[height=7cm]{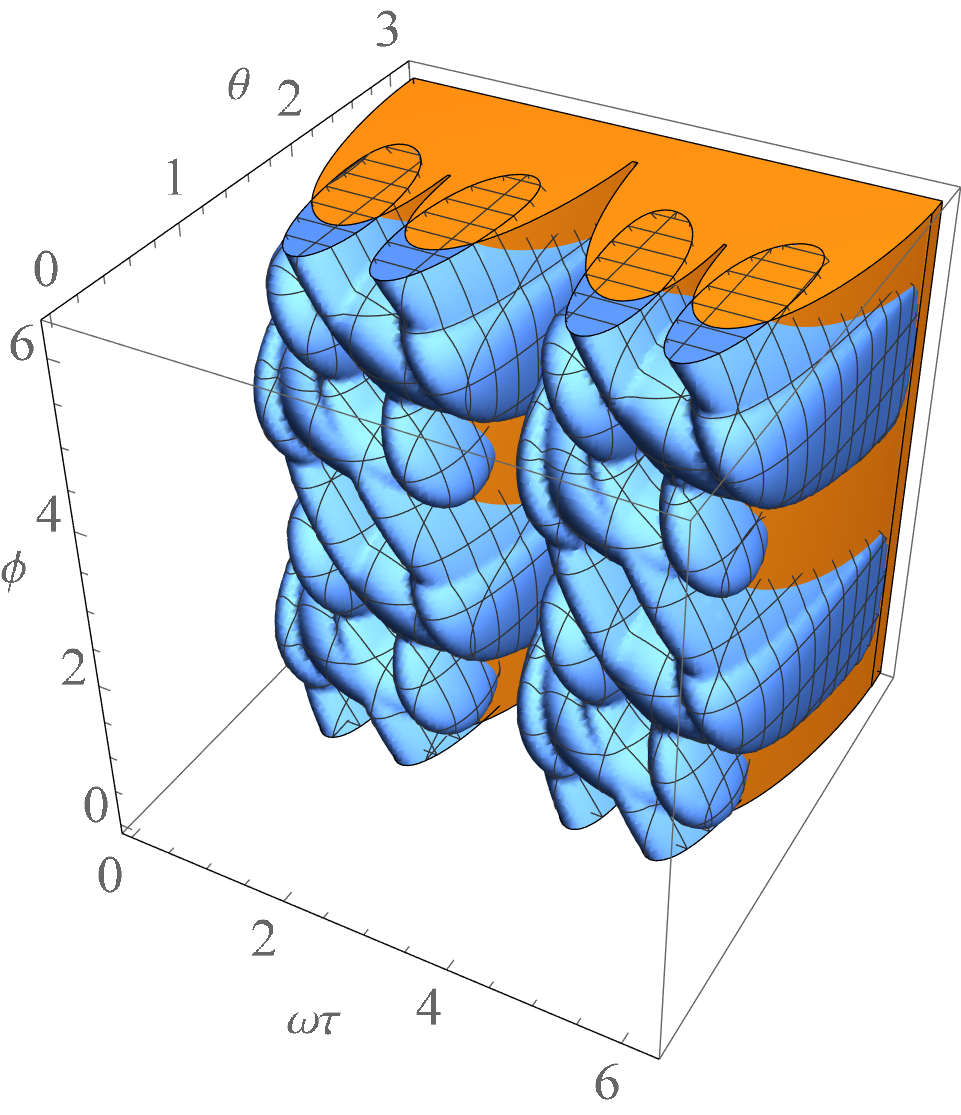}}}	%
		\qquad
		\subfloat[]{{\includegraphics[height=6.3cm]{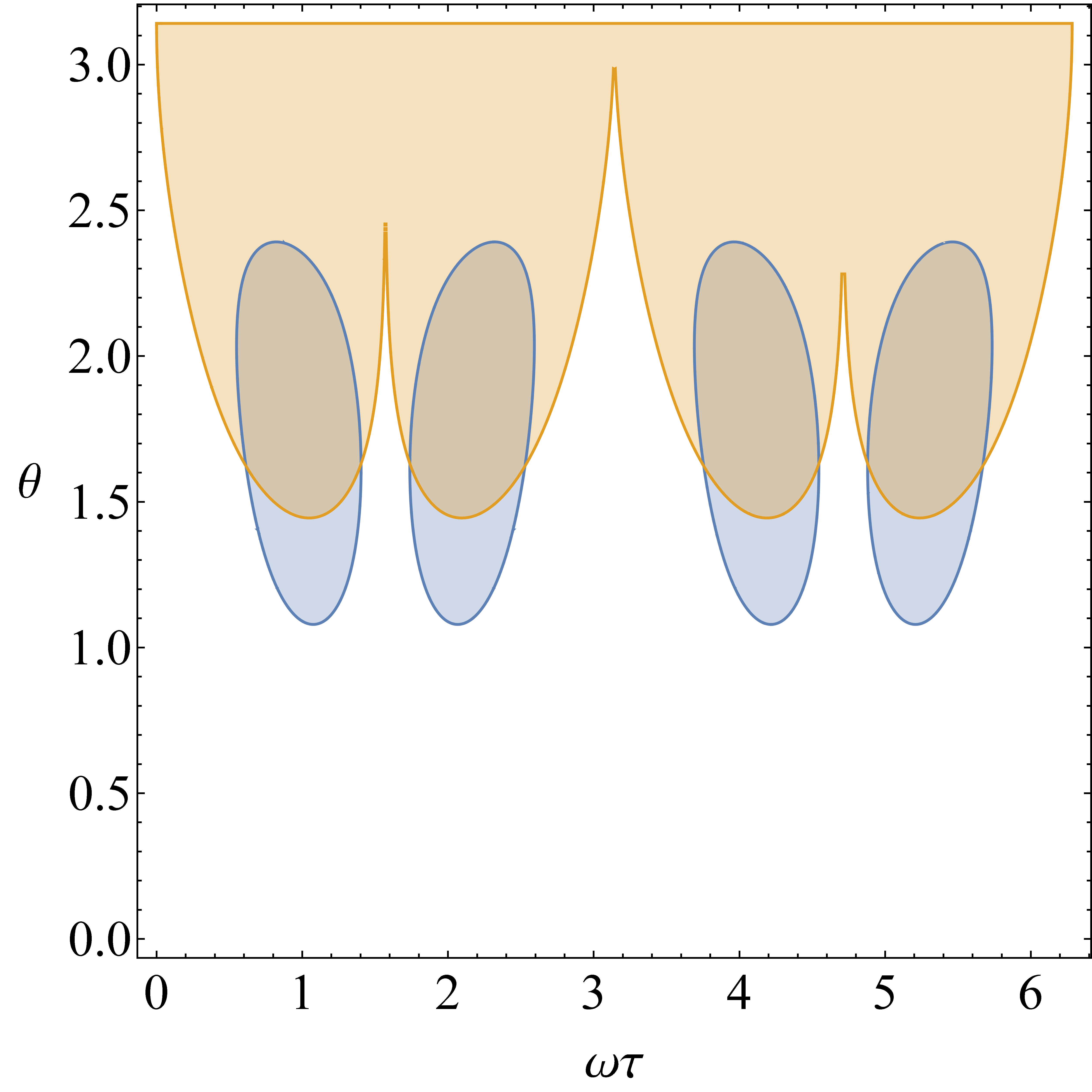}}}
		\end{center}
		\caption[Exploration of overlaps of LG2 and LG3 inequality violations for $a\ket0 +b\ket1$]{In (a) the behaviour of the $\LG{2}$ and $\LG{3}$ inequalities is plotted, for all possible superpositions of $\ket0$ and $\ket1$.  Within the orange (non-meshed) region, at least one of the four $\LG{3}$ inequalities are violated, and within the blue (meshed) region, at least one of the twelve $\LG{2}$ inequalities are violated.  In (b), a slice of this parameter space at $\phi=\pi$ is presented. }%
		\label{fig:lg2frac}%
	\end{figure}
	
	In Fig.~\ref{fig:2viol3sat}, the complete set of $\LG{2}$ and $\LG{3}$ inequalities are plotted for the state $\theta=0.7$, and $\phi=\pi$.  This state has been chosen to represent the important regime III, where the only violation occurs in the $\LG{2}$ inequalities between $t_2$ and $t_3$, despite the LG3s being satisfied.
	
	\begin{figure}
		\includegraphics[width=16cm]{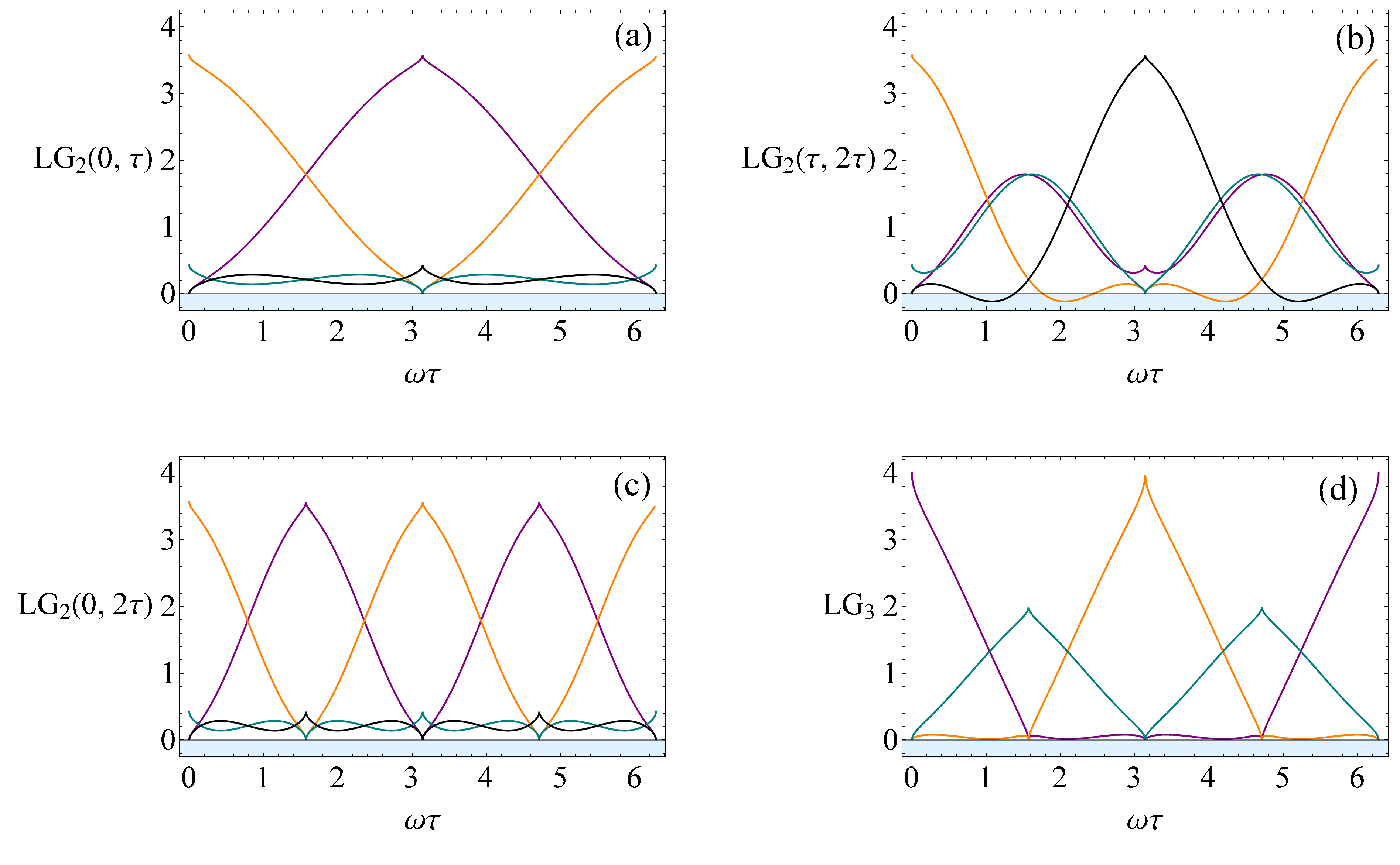}
		\caption[Set of LG2 and LG3 inequalities with only LG2 violation]{The complete set of two- and three-time LG inequalities are plotted for the state $\theta=1.4$, $\phi=\pi$. For this state, the $\LG{3}$ (d) inequalities are always satisfied, however there is still an LG violation present, at the level of two times, seen by the violation of the $\LG{2}$ between $\tau$ and $2\tau$ in (b).}
		\label{fig:2viol3sat}
	\end{figure}
	\subsection{LG4s and higher}
		\begin{figure}
		\subfloat[]{{\includegraphics[height=5.4cm]{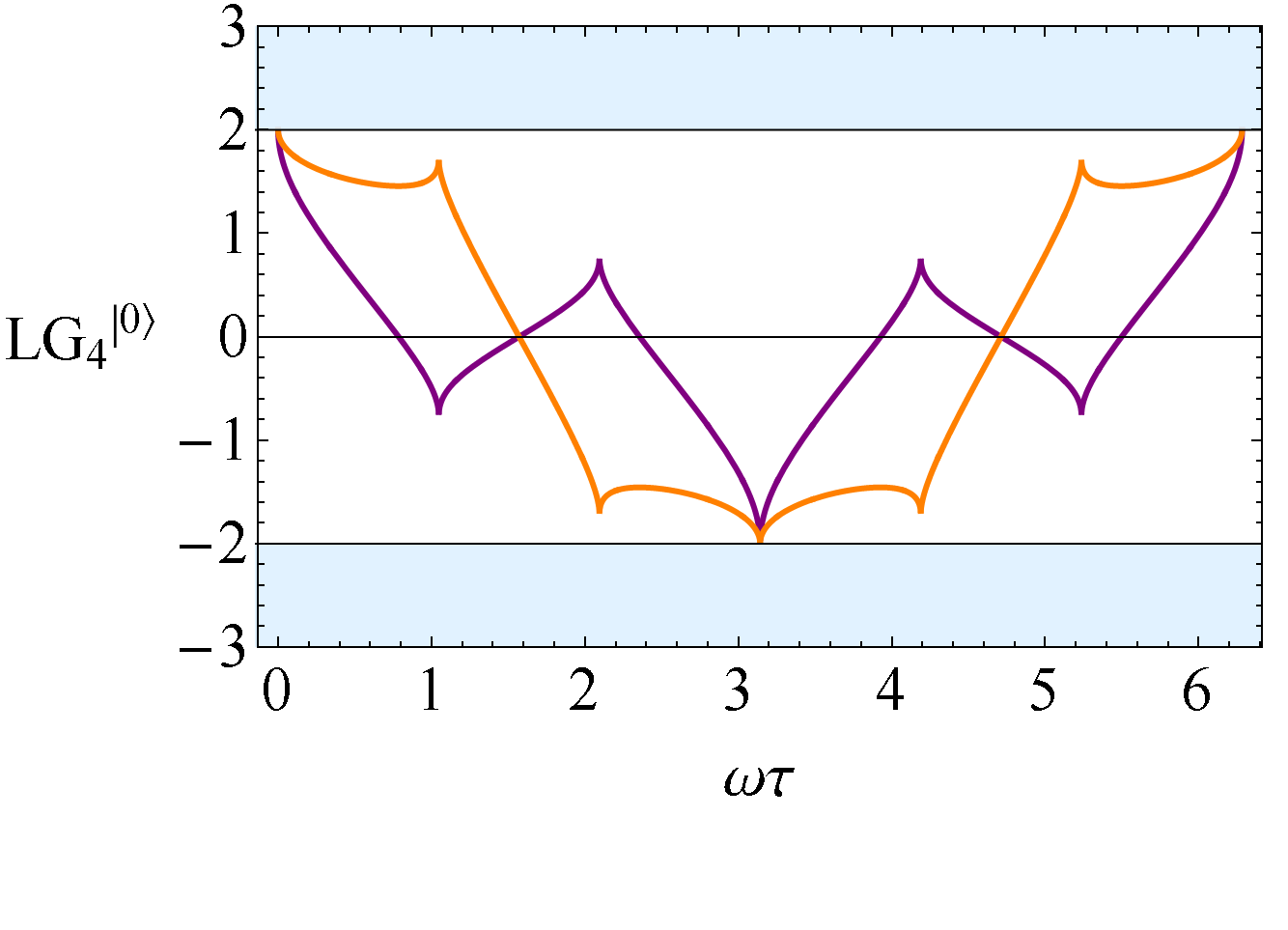}}}%
		\qquad
		\subfloat[]{{\includegraphics[height=5.4cm]{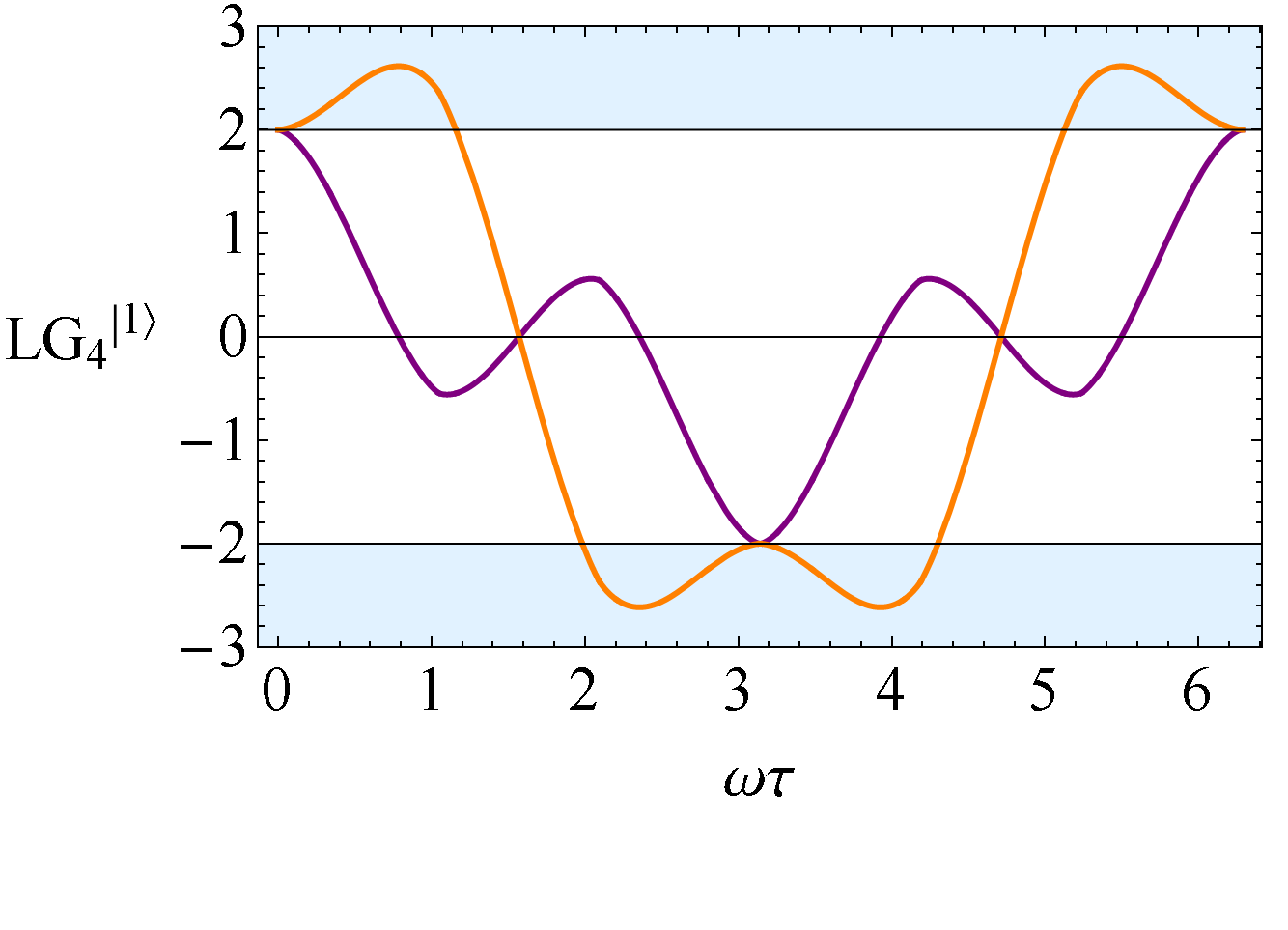}}}
		\caption[LG4 inequalities for $\ket0$ and $\ket1$ states]{In (a) the $\LG{4}$ inequalities are plotted for the ground state, where they are satisfied for all times.  In (b), the inequalities are plotted for the first excited state, showing some regions of violation.  The maximal violation is approximately $2.615$, which represents around $92\%$ of the maximal violation of $2\sqrt{2}$.}%
		\label{fig:lg4}%
	\end{figure}
	We now consider the behaviour of the LG inequalities when more measurements are made.  We consider the case of the $\LG{4}$ inequalities, and the LGn inequalities from Chapter~\ref{chap:ntime}.  Constructed purely from sums of correlators, it is apparent that the $\LG{n}$ for two-state superpositions  will again just be mixtures of the $\ket0 $ and $\ket 1$ cases, and for $n>2$, we have 	
	\beq
	L_i(n)=\cos^2\tfrac{\theta}{2}~ L_i^{\ket0}(n)+\sin^2\tfrac{\theta}{2}~ L_i^{\ket 1}(n).
	\eeq
%	The $\LG{4}$ inequalities take the form 
%	\begin{equation}
%	-2\leq C_{12}+C_{23}+C_{34}-C_{14}\leq 2,
%	\end{equation}
%	together with the six more inequalities obtained by moving the minus sign to the other three locations.  The necessary and sufficient conditions for MR at four times consist of these eight LG4 inequalities, together with the set of sixteen LG2s for the four time pairs \cite{halliwell2016b, halliwell2019, halliwell2017,halliwell2019b}.  
We plot the $\LG{4}$ inequalities Eqs.~(\ref{LG4a})--(\ref{LG4d}) in  Fig.~\ref{fig:lg4}.  We again have the property that the LG4 inequalities are always satisfied for the ground state, and are violated for the first excited state.  However, in contrast to the $\LG{3}$ inequalities, the $\LG{4}$ inequalities are not violated everywhere, and have large regions of parameter space where they are satisfied.  They allow access to interesting combinations of MR violations, whilst maintaining the experimental simplicity of working with a single energy eigenstate.
	
	For the LGn case, we found no violations for the groundstate. Since the excited state correlator Eq.~(\ref{eq:cos}) is very similar to the simple spin case, the asymptotic behaviour for large $n$ is the same as that of the example analyzed in our earlier paper \cite{halliwell2019}.
	
\subsection{Smoothed Projectors}
\label{subsec:smooth}
	A natural question that arises is whether the observed LG violations are an artefact of using sharp projectors, that may fade when physically realisable measurements are used.  To provide an indication that this is not the case, in Appendix \ref{sec:smooth} we repeat the calculation of $C_{12}^{\ket1}$, but with a spatially smoothed projector.  We find violations persist while the projector smoothing is less characteristic length-scale of the oscillator.
	
	\section{Higher states}\label{sec:higher}
		\begin{figure}
		\subfloat[]{{\includegraphics[height=4.6cm]{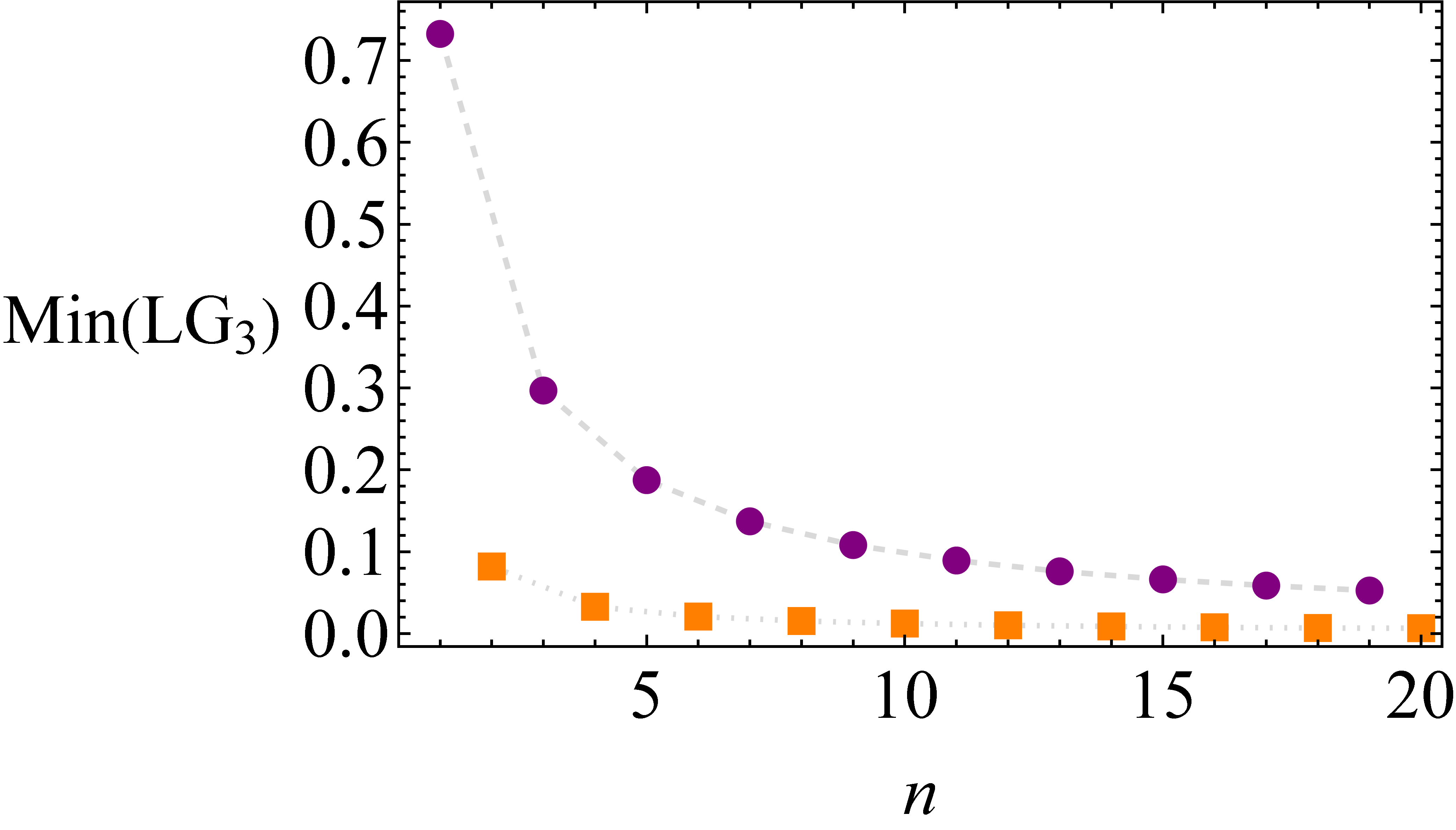}}}%
		\hspace{5mm}
		\subfloat[]{{\includegraphics[height=4.6cm]{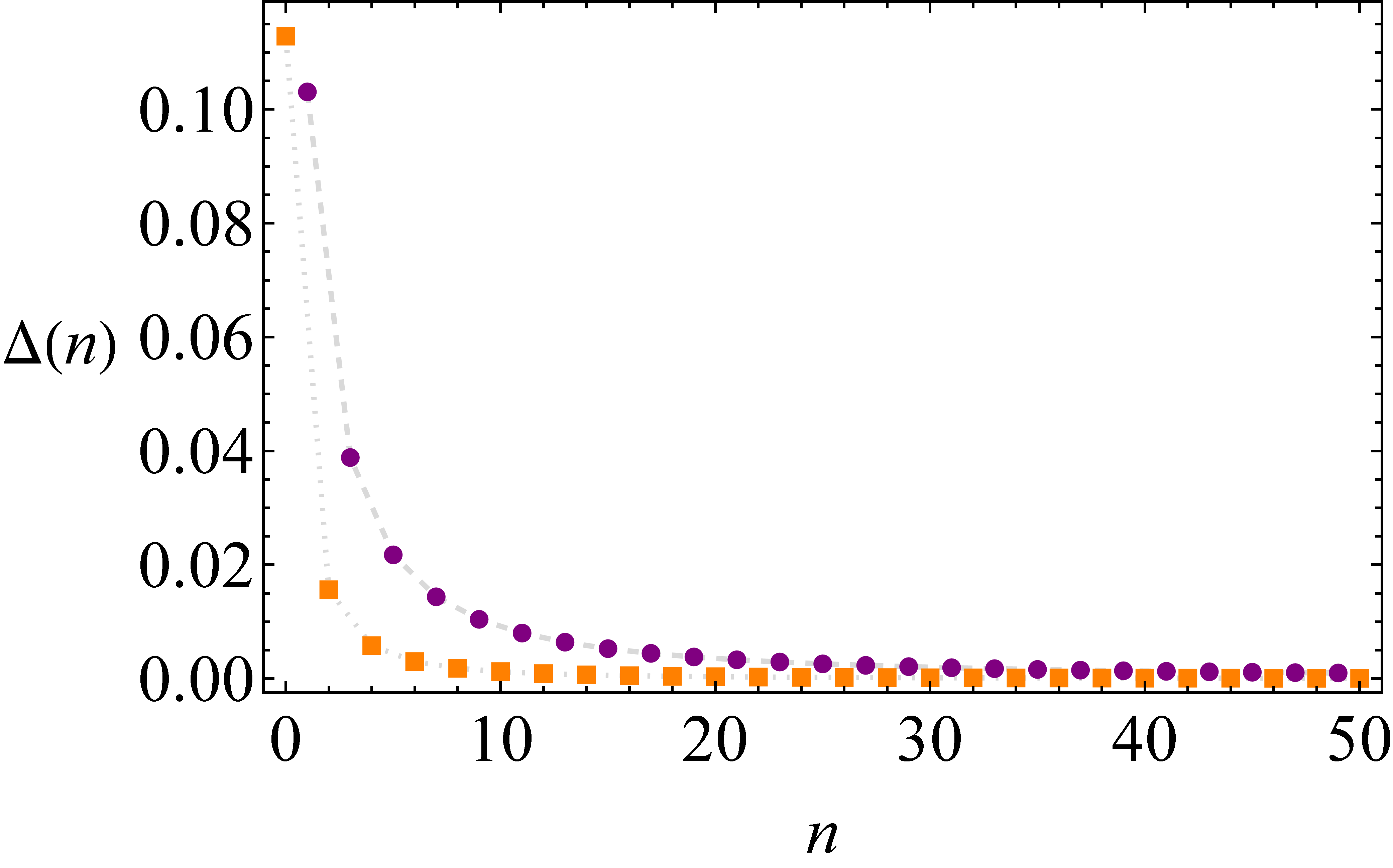}}}
		\caption[Classicalisation for higher energy eigenstates]{In (a), the maximal violation for a given energy eigenstate is plotted as a fraction of the L{\"u}ders bound.  The odd eigenstates are represented by the dashed line (circles), and always have significantly more violation than the even states represented by the dotted line (squares).  In (b), we plot the average magnitude of difference between the quantum correlators and the classical correlator, showing expected classicalization for large $n$. }%
		\label{fig:numopt}%
	\end{figure}
	Following the analysis in Section \ref{sec:1and0} of just the $\ket0$ and $\ket1$ states, it is natural to ask whether the LG inequalities behave significantly differently when including more of the energy eigenstate spectrum.  By two methods, we establish that for large $n$, the predictions of the QHO tend towards classical statistics.
	%\subsection{Maximal Violation}

	Firstly, by plotting the higher eigenstate correlators, we can visually see that they tend toward the classical correlator Eq.~(\ref{eq:class}).  To make this exact, we calculate the average distance from the classical correlator, as a function of $n$.  For this, we define
	\begin{equation}
	\Delta(n)=\frac{1}{2\pi}\int_0^{\frac{2\pi}{\omega}}\mathop{dt}\lvert \mathbbmsl{C}_{12} - C_{12}^{\ket n}\rvert,
	\end{equation}
	which we plot up to $n=50$ in Fig.~\ref{fig:numopt}, where we see that the quantum temporal correlators very rapidly match the classical correlator.
	
	Secondly, with details in Appendix \ref{app:classic}, we perform an asymptotic analysis on the infinite sum representation of the quasi-probability, and show that for large $n$, this series tends towards the Fourier series of a triangle wave, matching the classical result.

	\section{Different choices of Dichotomic Variables} \label{sec:mlvl}

	We have so far worked with the simplest choice of dichotomic variable, $Q=\sgn(x)$.  We now consider what new effects may be discovered using different choices. We first consider a $Q$ defined
	using coarse grainings over more general spatial regions. Secondly, we consider a more novel choice of $Q$, namely the parity operator, which does not necessarily have a macrorealistic description but reveals some interesting features.
	
	\subsection{Quasi-probability for arbitrary regions in the ground state}
	\label{subsec:finegrain}
	
	We now demonstrate how to calculate the quasi-probability where each measurement is taken over an arbitrary coarse graining of space.  In this work, we consider just a pair of dichotomic variables, however the techniques presented are readily applied to a full many-valued variable LG analysis as introduced in Chapter~\ref{chap:mlev}.
		
	We define projectors over arbitrary regions of space as 
	\beq
	\label{eq:genproj}
	E(\alpha)=\int_{\Delta(\alpha)}\mathop{dx}\dyad{x},
	\eeq
	where $\Delta(\alpha)$, where $\alpha=\pm$, is a set of intervals which partition the real line, which may be chosen to be different for each measurement time. The dichotomic variable $Q$ is then defined by $Q = 2 E(+) - 1 $, and discerns whether a particle is within a given region of space ($+$), or outside of it ($-$).  The two-time quasi-probability is then given by 
	\beq
	\label{eq:mlvl}
	q(+, +) =  \Re\ev{E_2(+) E_1(+)}
	\eeq
	and the LG2 inequalities are then simply $q(+,+) \ge 0 $. 
	
	Remarkably, by using these more general dichotomic variables we can find $\LG{2}$ violations in the ground state of the QHO.
	To demonstrate this, we write the quasi-probability using the QHO propagator Eq.~(\ref{eq:qmnfull}), with the projectors Eq.~(\ref{eq:genproj}) for the special case of the ground state, 
	\begin{equation}
	\label{eq:shortspace}
	q(+,+)=\Re\mathcal{N}_{00}(\tau)\int^{b}_a\int^{d}_c \mathop{dr}\mathop{ds}e^{-\frac{r^2}{2}}e^{-\frac{s^2}{2}}\exp({i \frac{1}{2\tan(\omega \tau)}(r^2+s^2)- i\frac{1}{\sin(\omega \tau)}rs}).
	\end{equation}
	Since the real part of the integrand oscillates around zero, negative values of $q(+,+)$ can clearly by achieved by suitable choice of the spatial intervals $[a,b]$ and $[c,d]$.
	
	In Fig.~\ref{fig:mlvl}(a), we plot the result of
	the integration, with the first interval  $\Delta(+)=[0,\infty)$, and then plot the remaining integrand of the second integral, in the regions where it is negative.  This indicates the region in which to make a second measurement, which will lead to negativity of the quasi-probability, and hence $\LG{2}$ violations.
	
	To calculate the quasi-probability with arbitrary coarse-grainings, it is most efficient to use the techniques in Sec.~\ref{subsec:inf} since this approach already encapsulates these more general projectors.  The details of this calculation are found in Appendix \ref{app:arbproj}.
	
	\begin{figure}
		\begin{center}
		\subfloat[]{{\includegraphics[height=5.8cm]{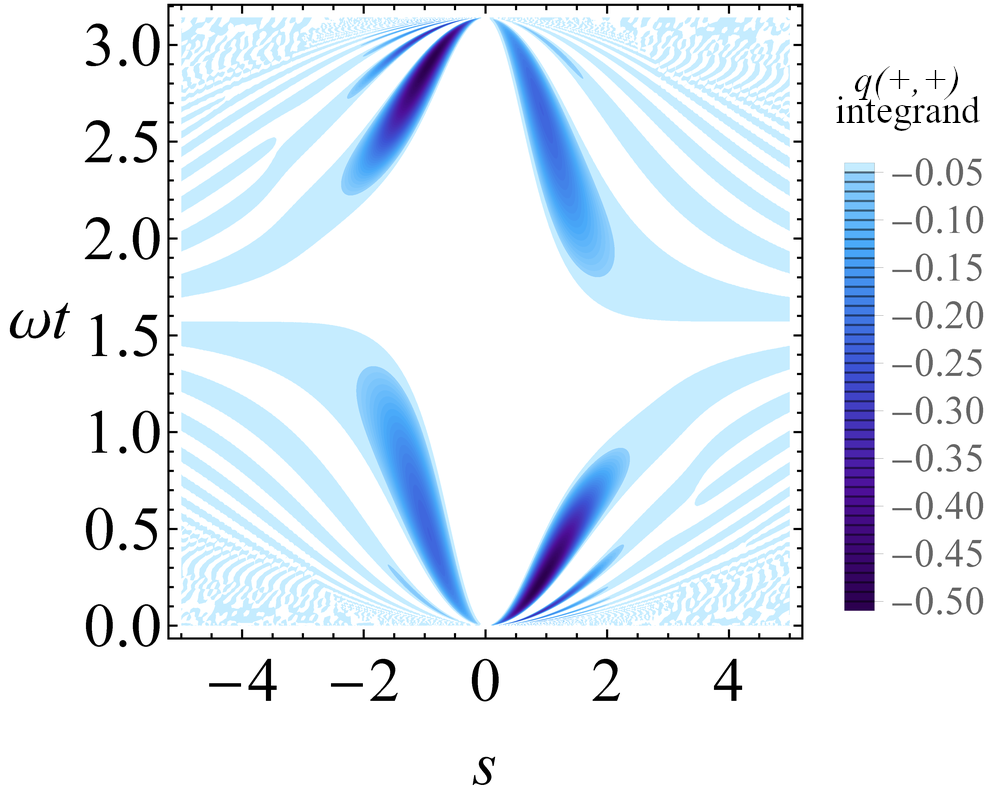}}}%
		%\qquad
		\subfloat[]{{\includegraphics[height=5.8cm]{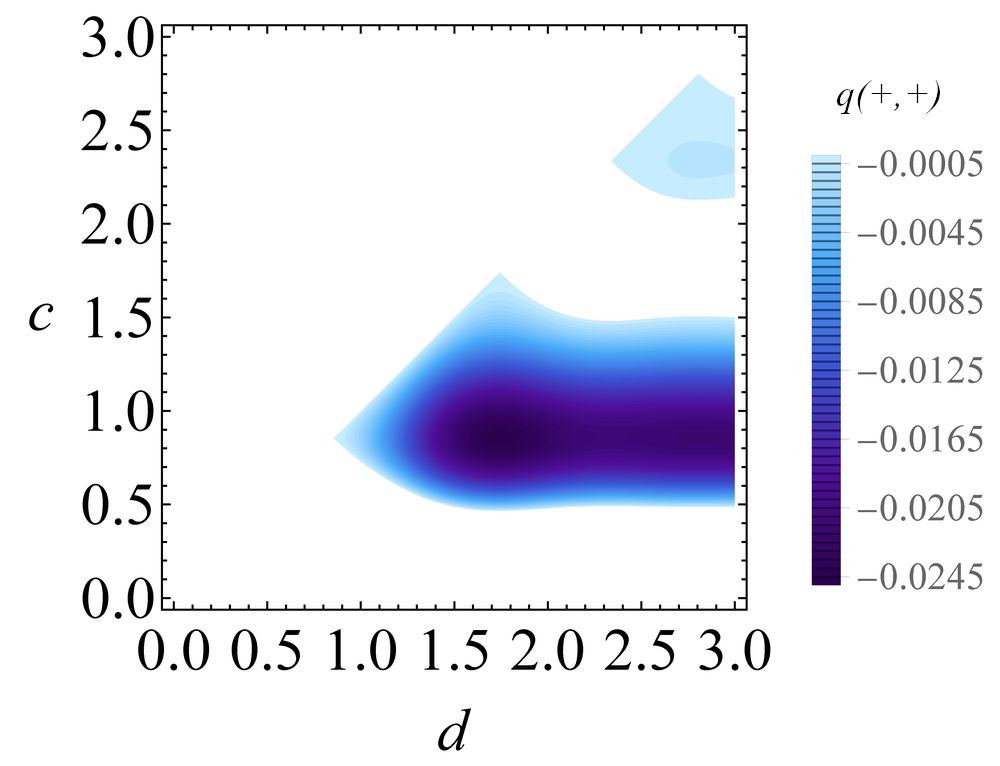}}}
		\end{center}
		\caption[LG violations in the ground state using finer grained variables]{In (a), we plot the result of the first integration in Eq.~(\ref{eq:shortspace}) over $[0,\infty]$, which serves as a rough map to choosing a second measurement which will lead to a negative quasi-probability.  In (b), we plot the quasi-probability, for $\omega t=2.77$, with the first measurement over $[0,\infty]$, and the second over the interval $[c,d]$.  The quasi-probability reaches around $20\%$ of the maximal violation of $-0.125$.}
		\label{fig:mlvl}
	\end{figure}
	
	In Fig.~\ref{fig:mlvl}(b), we plot the quasi-probability for $\omega \tau=2.77$, over a range of possible second measurement windows.  From this, we observe that there are a wide range of choices for the second measurement which lead to an LG2 violation, which can reach $20 \%$ of the maximum possible violation.
	We also observe that these measurements do not not need to be over particularly small regions of space, and that simply shifting the axis of the second measurement from $[0,\infty)$ to $[c,\infty)$, compared to the usual choice of dichotomic variable, is sufficient for a good LG2 violation in the ground state.	
	
	The ground state wave function has positive Wigner function and from a macrorealistic point of view, is sometimes thought of as a ``classical'' state describing a particle localized at a phase space point, yet we see here that a LG2 violation is possible. A similar phenomenon was noted in Ref.~\cite{halliwell2021}, which used a coherent state with $\langle x \rangle =0$. The origin of this effect is explored in Chapter~\ref{chap:QHO2}. We have also looked for LG3 and LG4 violations in the ground state but an extensive parameter search yields no result.

	%With the most natural choice of dichotomic variables used in Section~\ref{sec:1and0}, the groundstate of the QHO was ostensibly classical.  However with the different variable choice employed in this section, a failure of MR is detected.  Although $\LG{2}$ violations are possible with the simpler dichotomic variable from earlier, they required tuned superposition states, whereas with a different choice of variable, we detect $\LG{2}$ violations even in pure eigenstates.
	
	%While interesting in itself, this result indicates directly, that using a broader choice of variables can lead to much richer and more powerful tests of quantumness.  For conciseness, the analysis in this section has been limited to the case of the $\ket0$ state, and the case where the first measurement is over the half plane.  However the results presented are valid for higher energy states, and lend themselves to a full many-valued LG analysis, as introduced in Ref~\cite{halliwell2020}.
	
	\subsection{LG violations using the parity operator}
	
	Our second choice of different dichotomic operator arises from the observation that the variables $Q_1$, $Q_2$ etc used in LG inequalities do not necessarily have to be the time evolution of a given dichotomic variable $Q$, but could be a set of {\it any} dichotomic variables. The derivation of the LG inequalities still holds so an LG test is still possible as long as the variables can be measured.
	Such an approach has been used in some experiments \cite{goggin2011}. Note at this point that we are now departing from the quantum harmonic oscillator, at least in terms of time evolution, as the physical system of interest (although the QHO is still the natural arena in which to create the gaussian state used below).
	
	Two interesting variables to use in this context are the parity operator,
	\begin{equation}
	\label{parityop}
	\Pi = \int_{-\infty}^\infty dx \ | x \rangle \langle- x |,
	\end{equation}
	and the parity inversion operator,
	\begin{equation}
	R = i\int_{-\infty}^\infty dx \ {\rm sgn} (x) \ | x \rangle \langle -x |.
	\end{equation}
	These two operators, taken together with the usual choice $Q = {\rm sgn} ( x) $ have the interesting property that they obey the same algebra as the Pauli matrices
	$ [Q, R ] = 2 i  \Pi $ etc and they also all anticommute with each other. By comparison with the spin-$\frac{1}{2}$ model, this is a clear suggestion that LG violations are readily possible with such variables. These variables were previously used in 
	Ref. \cite{praxmeyer2005} to demonstrate a Bell inequality violation for a gaussian two-particle entangled state of the Einstein-Podolsky-Rosen type.
	
	We consider the LG2
	\begin{equation}
	1 - \langle Q_1 \rangle - \langle Q_2 \rangle + C_{12} \ge 0,
	\end{equation}
	and we take $Q_1 = Q = {\rm sgn} (\hat x) $ and $Q_2 = \Pi$. Since these operators anticommute the correlator is zero and the LG2 is then
	\begin{equation}
	1 - \langle Q \rangle - \langle \Pi \rangle \ge 0.
	\end{equation}
	We take a gaussian state of width $\sigma$, $\langle x \rangle = q $ and $\langle p \rangle = 0$ and readily find that the LG2 inequality is
	\begin{equation}
	1 - {\rm erf} \left( \frac{q}{\sqrt{2} \sigma } \right)  - \exp( -\frac{q^2}{2 \sigma^2})  \ge 0.
	\end{equation}
	The left-hand side is readily shown to take its minimum value of approximately $ -0.3024 $ at $ q/\sigma = \sqrt{2/\pi}$, hence there is a clear LG2 violation.
	
	Although the parity operator is measurable for some systems and therefore the above inequality can be tested experimentally, it has no macrorealistic counterpart (unlike Q = $\textrm{sgn}(x)$). Still, the fact that LG2 violations are obtained so easily with these variables could give interesting clues to understand violations in more physical cases
	
	\section{Leggett-Garg violations in the Morse potential}
	\label{sec:morse}	
	In Sec. \ref{subsec:inf}, we laid out a general technique to calculate temporal correlators, when the energy eigenspectrum is known.  As a demonstration, we now apply this result to a different exactly soluble potential, the Morse potential \cite{morse1929}, an asymmetric potential with minimum at $r_c$, which combines a short-range repulsion, with a long-range attraction.  The potential is defined by
	\begin{equation}
	V(r) = D_e\left(e^{-2a(r-r_e)}-2e^{-a(r-r_e)}\right),
	\end{equation}
	where $D_e$ corresponds to the well depth, and $a$ to its width.
	
	The details and results of applying our method to the Morse potential may be found in Appendix \ref{app:morse}, where we find qualitatively similar behaviour to the QHO, and observe significant $\LG{3}$ and $\LG{4}$ violations in the $\ket1$ state.  There are several other systems for which the Schr\"{o}dinger equation may be solved exactly, including the Hydrogen-like atom \cite{landau1977}, the linear potential \cite{khonina2013}, and the quantum pendulum \cite{aldrovandi1979,baker2002}.  The result presented in Sec.~\ref{subsec:inf} hence allows future exploration of the behaviour of the LG inequalities in several bound systems.
	
	\begin{figure}
		\subfloat[]{{\includegraphics[height=5.2cm]{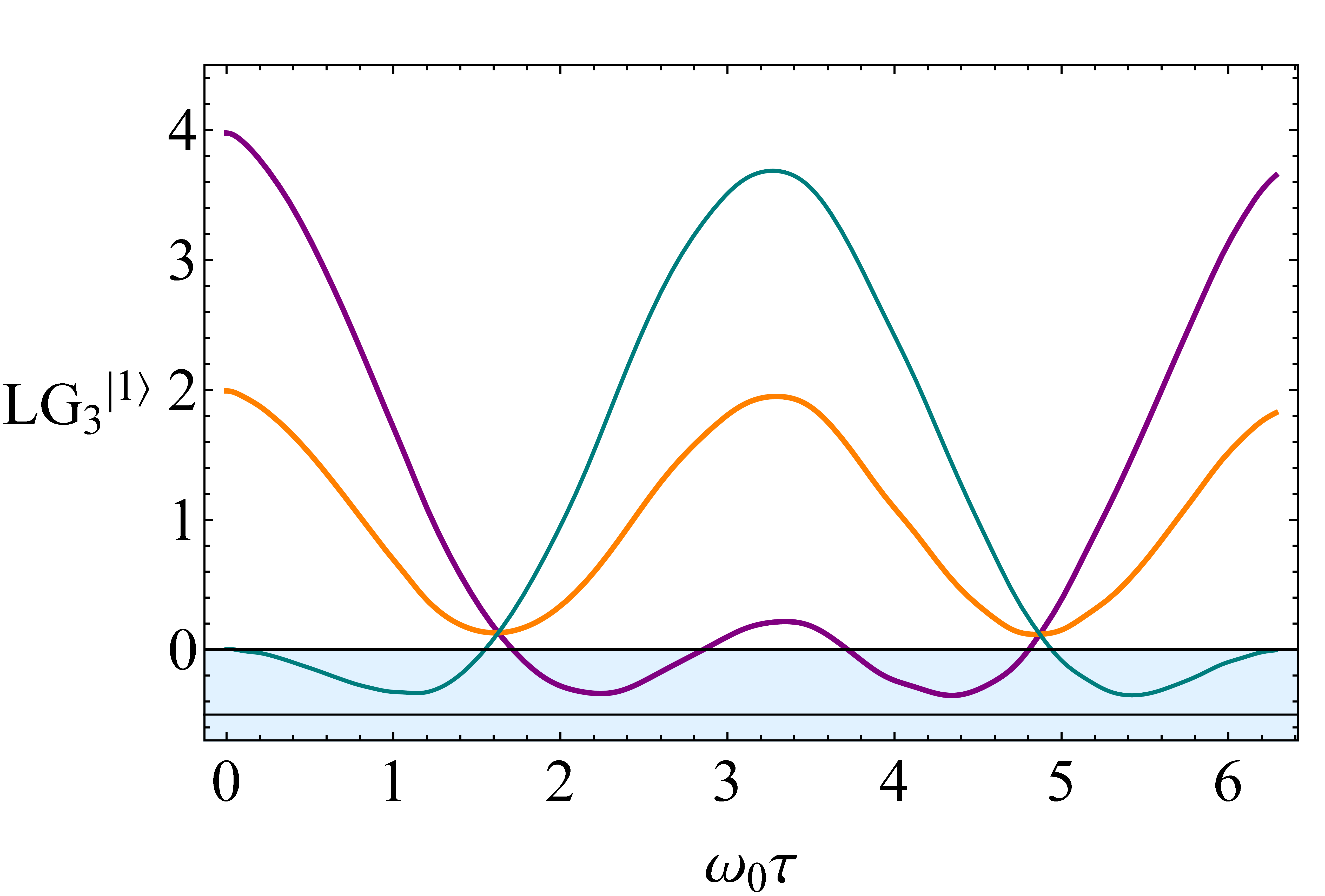}}}%
		\qquad
		\subfloat[]{{\includegraphics[height=5.2cm]{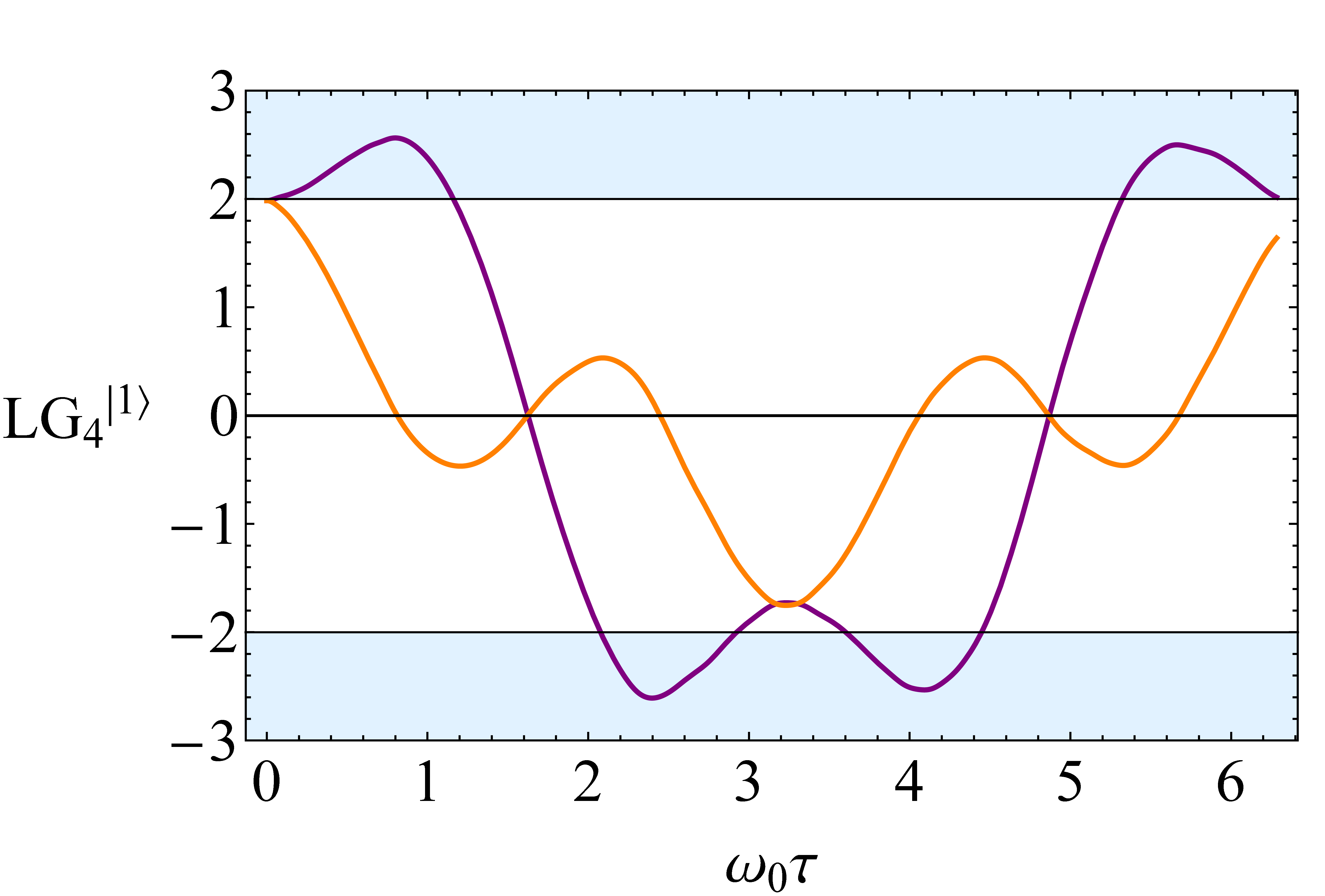}}}
		\caption[LG3 and LG4 Violations in the Morse potential]{In (a), the LG3 inequalities for the $\ket1$ state of the Morse potential with $\lambda=50$ is plotted.  In (b), the LG4 inequalities are plotted.  There are significant violations for both, reaching $70\%$ and $92\%$ of their maximal violations respectively.}%
		\label{fig:morse2}%
	\end{figure}

	\section{Summary and Conclusion}
	\label{sec:conc}
	We have presented an analytic investigation of the LG approach to macrorealism in bound systems with an in-depth study of the QHO.  
	For a dichotomic variable given by the sign of the oscillator position,
	we developed a method to calculate the temporal correlators for energy eigenstates of any bound system with a known energy eigenspectrum.
	
	We found that all the interesting variants of MR conditions tested for by combinations of the \LG{2}, \LG{3} and \LG{4} inequalities can be observed in states of the QHO involving superpositions of only the $\ket0$ and $\ket1$ states.  
	Intriguingly, the temporal correlator of the $\ket1$ state was shown to be, to a very good approximation, a cosine, like the temporal correlators in the canonical simple spin systems used in much LG research.  	We also found that the higher energy eigenstates rapidly begin to exhibit statistics in line with a classical model.
	
	We found that LG violations persist even with a significant amount of smoothing of the projectors.  This is important, since it means the LG violations are not just an artefact of non-physical, sharp projectors, and therefore the LG violations described here should be directly observable in experiment.
	
	Although the bulk of the work in this paper has been conducted using the simplest dichotomic variable $ Q =\text{sgn}(x)$, all the results presented are general enough to also allow for dichotomic variables defined on arbitrary regions of space.  We briefly investigated the effect of using a different variable, and found it provides a richer and more powerful test of MR, with two-time LG violations even in the $\ket0$ state, despite being a state with positive Wigner function. (See also Ref. \cite{halliwell2021}).  In addition, we found LG2 violations if one of the dichotomic variables at one of times is taken to be the parity operator.
	
	To demonstrate the method, we calculated the temporal correlators for another exactly soluble system, the Morse potential, where we found significant LG3 and LG4 violations for the first excited state.
	%In conclusion, we have shown that there is great 	 to extend the LG experimental program towards macroscopic systems, via the route of coarse-grained position measurements on bound systems.  
	
	In conclusion, we have derived a number of results on LG inequalities for coarse-grained position measurements in bound systems. These results provide indications as to what sort of states to 
	create in experiments on macroscopic systems in order to find evidence of macroscopic coherence.
%	
%	Future work will explore in more depth the phenomenon of LG violations in these systems for initial states with positive Wigner function. We also propose to develop non-invasive measurement protocols suitable for continuous variable systems by adapting the continuous in time velocity measurement scheme proposed and utilized in Refs. \cite{halliwell2016a,majidy2019a}.

	%	\blue{Use this refs for the Wigner function: V. I. Tatarskii, Sov. Phys. Usp. 26, 311 (1983); M. Hillery,
	%		R. O’Connell, M. Scully, and E. Wigner, Phys. Rep. 106, 121
	%		(1984).
	%		Add  Phys. Rev. A 96, 012121 (2017) (my paper) to Ref.[5].
	%		Remove all the ISSN numbers etc from the refs.
	%	}
	%	
	
	\begin{subappendices}

	\section{Quasi-probability for generic bound potentials}
	\label{app:general}
	\subsection{Partial overlap integrals}
	\label{app:wronski}
		To calculate the quasi-probability most generally, we need the integrals
	\begin{equation}
	J_{k\ell}(x_1, x_2)=\int_{x_1}^{x_2}\mathop{dx}\braket{k}{x}\braket{x}{\ell},
	\end{equation}
	where we adopt the notation that when $J_{k\ell}$ appears without an argument, $J_{k\ell}=J_{k\ell}(0,\infty)$. It is possible to compute these integrals, by expanding a result found in Ref. \cite{moriconi2007a}, which bears resemblance to Abel's identity.  We first denote energy eigenstates in position space as
	\begin{equation}
	\psi_n(x) = \braket{x}{n}.
	\end{equation}
	We then construct the Wronskian between the two states $k$ and $\ell$,
	\begin{equation}
	W(x)=\psi_k'\psi_\ell - \psi_\ell'\psi_k.
	\end{equation}
	Differentiating the Wronskian, we find
	\begin{equation}
	\label{eq:wdiff}
	W'(x)=\psi_k''\psi_\ell - \psi_\ell'' \psi_k.
	\end{equation}
	Since we are working with energy eigenstates, from the Schr{\"o}dinger equation we have
	\begin{equation}
	\label{eq:schrono}
	\psi_n'' =2(V-\varepsilon_n)\psi_n,
	\end{equation}
	where $\varepsilon_n$ are the dimensionless energy eigenvalues. Substituting this into Eq.~(\ref{eq:wdiff}), we find
	\begin{equation}
	W'(x)=2(\varepsilon_\ell-\varepsilon_k)\psi_k\psi_\ell,
	\end{equation}
	where any dependence on the form of the potential has disappeared from this equation, being contained implicitly in the spectrum of the Hamiltonian.  We are now free to integrate both sides over any given region of space, which yields
	\begin{equation}
	\frac{1}{2(\varepsilon_\ell-\varepsilon_k)}\eval{W(x)}_{x_1}^{x_2}=\int_{x_1}^{x_2}\mathop{dx}\psi_k\psi_\ell.
	\end{equation}
	The right-hand side of the equation is exactly the matrix elements $J_{k\ell}$, and so completing the integration, we find
	\begin{equation}
	\label{eqn:wronskiapp}
	J_{k\ell}(x_1, x_2)=\frac{1}{2(\varepsilon_\ell- \varepsilon_k)}\left[\psi_k'(x_2)\psi_\ell(x_2)-\psi_\ell'(x_2)\psi_k(x_2)-\psi_k'(x_1)\psi_\ell(x_1)+\psi_\ell'(x_1)\psi_k(x_1)\right].
	\end{equation}
	\subsection{Another derivation of the $J_{mn}$ matrices}
	The generality of the result for the $J_{mn}$ matrices is surprising and useful.  Despite being elementary, we have not found it elsewhere in the literature, and so it is worth giving another derivation for.  Using the time evolution of energy eigenstates, we introduce some time dependence into the definition of $J_{mn}$, as
\begin{equation}
J_{mn}=e^{-i(E_m-E_n)\tau}\mel{m}{\theta(\hat x(\tau))}{n}.
\end{equation}
Now by writing the theta-function as the time integral of the current, we find
\begin{equation}
J_{mn}=e^{-i(E_m-E_n)\tau}\int_0^{\tau}\mathop{dt} \mel{m}{J(t)}{n},
\end{equation}
where \begin{equation}
J(t)=\frac{1}{2}e^{iHt}(\hat p \delta(\hat x)+\delta(\hat x)\hat p)	e^{-iHt}.
 \end{equation}
Using this, we find
\begin{equation}
J_{mn}=e^{-i(E_m-E_n)\tau}\frac{1}{2}\left(\mel{m}{p}{0}\braket{0}{n} + \braket{m}{0}\mel{0}{p}{n}\right)\int_0^\tau \mathop{dt} e^{i(E_m-E_n)t}
\end{equation}
Completing the integration, and using $p=-i\frac{d}{dx}$, we recover the result
\begin{equation}
J_{mn}=\frac{1}{2(E_m-E_n)} \left(\psi_m^{'}(0)\psi_n(0)-\psi_m(0)\psi^{'}_n(0)\right)
\end{equation}
	\subsection{Calculational Details}
	\label{app:calcdeets}
	
	We continue the analysis coarse-graining onto the left and right hand side of the axis, in the important special case of symmetric potentials.  In this case, we have
	\begin{equation}
	\label{eq:jijlim}
	J_{ij}=\int_{0}^{\infty}\mathop{dx}\psi_i\psi_j=\frac{1}{2(\varepsilon_j-\varepsilon_i)}\left[\psi_j'(0)\psi_i(0)-\psi'_i(0)\psi_j(0)\right],
	\end{equation}
	since wavefunctions must vanish at infinity.  With a symmetric potential, we have for $n$ odd, $\psi_n(0)=0$, and for $n$ even, we have $\psi_n'(0)=0$.  Eq.~(\ref{eq:wdiff}) is not defined for $i=j$, however for a symmetric potential, we know $J_{ii} = \frac{1}{2}$, as so the $k=n$ term contributes $\frac14$ is to both results.
	
	Hence, if $n$ is odd, we have
	\begin{equation}
	\label{eq:qgen}
	q(n)=\frac14+\Re e^{i \frac{E_n}{\hbar} \tau}\psi_n'(0)^2\sum_{k=0, k\neq n}^{\infty}e^{-i \frac{E_k}{\hbar} \tau}\frac{1}{4(\varepsilon_k-\varepsilon_n)^2}\psi_k(0)\psi_k(0).
	\end{equation}
	If $n$ is even, we have
	\begin{equation}
	q(n)=\frac14+\Re e^{i\frac{E_n}{\hbar} \tau}\psi_n(0)^2\sum_{k=0,k\neq n}^{\infty}e^{-i \frac{E_k}{\hbar} \tau}\frac{1}{4(\varepsilon_k-\varepsilon_n)^2}\psi_k'(0)\psi_k'(0).
	\end{equation}
	
	To be computationally feasible, we must truncate this sum for some $m$.  To estimate the error involved with a given truncation, we note that $J_{k\ell}$ form the coefficients of the expansion of $\theta(\hat x)\ket{k}$. This represents the probability of finding the particle on either side of the axis.  In the case of symmetric potentials, these are  (anti-)symmetric states, and so have norm $\frac12$.  Hence this gives us the truncation 
	\beq
	\label{eq:trunc}
	\Delta_n(m)=\frac12-
	\sum_{k=0}^{m}J_{nk}^2.
	\eeq
	\section{Exact Quasi-probability in the QHO}
	\label{app:gen}
	\subsection{Strategy}
	\label{app:strat}
	To calculate the quasi-probability exactly, we begin by writing Eq.~(\ref{eq:qmnfull}) as 
	\beq
	q_{mn}=\mathcal{N}e^{i\frac{E_m}{\hbar}t_2-i\frac{E_n}{\hbar}t_1}\int^{\infty}_0\int^{\infty}_0 \mathop{dr}\mathop{ds}H_m(r)H_n(s)\exp(\alpha r^2 + \beta s^2+ i \gamma r s),
	\eeq
	where we have introduced the variables
	\begin{align}
	\alpha = \beta &= \frac{i}{2 \tan(\omega \tau)}-\frac{1}{2},\\
	\gamma &= -\frac{1}{\sin(\omega \tau)}.
	\end{align}
	We now note that as a product of Hermite polynomials, the term $H_m(r)H_n(s)$ will itself just be a polynomial involving powers of $r$ and $s$. Hence $q_{mn}$ may be broken up into a sum of terms of the form
	\beq
	X_{k\ell}=\int_0^{\infty}\int_0^{\infty}\mathop{dr}\mathop{ds}r^k s^\ell \exp(\alpha r^2 +\beta s^2 + i\gamma r s).
	\eeq
	To calculate these terms, we consider the generating integrals
	\begin{align}
	\label{genI}
	I(\alpha, \beta,\gamma)&=\int^{\infty}_0\int^{\infty}_0\mathop{dr}\mathop{ds}\exp(\alpha r^2+\beta s^2 + i\gamma r s),\\
	\label{genJ}
	J(\alpha, \beta,\gamma)&=\int^{\infty}_0\int^{\infty}_0\mathop{dr}\mathop{ds}r\exp(\alpha r^2+\beta s^2 + i\gamma r s).
	\end{align}
	We may then calculate any of the $X_{k\ell}$ integrals through repeated use of partial differentiation.  In particular, if $k$ and $\ell$ are both even, we have
	\beq
	X_{k\ell}=\pdv{\alpha^{\frac{k}{2}}}\pdv{\beta^{\frac{\ell}{2}}}I(\alpha,\beta,\gamma).
	\eeq
	Similarly, if $k$ and $\ell$ are both odd, we have 
	\beq
	X_{k\ell}=-i\pdv{\gamma}\pdv{\alpha^{\frac{k-1}{2}}}\pdv{\beta^{\frac{\ell-1}{2}}}I(\alpha,\beta,\gamma).
	\eeq
	Finally, for the case where say $k$ is odd, and $\ell$ is even, we have
	\beq
	X_{k\ell}=\pdv{\alpha^{\frac{k-1}{2}}}\pdv{\beta^{\frac{\ell}{2}}}J(\alpha, \beta, \gamma),
	\eeq
	and vice-versa for $\ell$ odd, $k$ even.  This approach although a handful on paper, is simple to implement in computer algebra software.
	\subsection{Generating Integrals}
	We now proceed with calculating $I(\alpha,\beta,\gamma)$ Eq.~(\ref{genI}), by completing the square in the exponential function.  This yields
	\begin{equation}
	I(\alpha,\beta,\gamma)=\int_0^\infty \int_0^\infty \mathop{dr}\mathop{ds}\exp\left(\frac{r^2(4\alpha \beta + \gamma^2)}{4\beta}\right)\exp\left(\beta s+i \frac{\gamma r}{2\sqrt{\beta}}\right).
	\end{equation}
	We introduce the short-hand $\delta=\frac{4\alpha \beta + \gamma^2}{4\beta}$.  The $s$ integral may now be completed in terms of the error function as
	\begin{equation}
	\label{eq:iint}
	I(\alpha,\beta,\gamma)=\frac{\sqrt{\pi}}{2\sqrt{-\beta}}\int_0^\infty \mathop{dr}\exp\left(\delta r^2\right)\left(1+i\erfi\left(\frac{\gamma r}{2\sqrt{-\beta}}\right)\right).
	\end{equation}
	For simplicity of presentation, we separate the integral into two parts,
	\begin{align}
	I_1(\alpha, \beta,\gamma)&=\frac{\sqrt{\pi}}{2\sqrt{-\beta}}\int_0^\infty \mathop{dr}\exp\left(\delta r^2\right),\\
	I_2(\alpha, \beta,\gamma)&=\frac{i\sqrt{\pi}}{2\sqrt{-\beta}}\int_0^\infty \mathop{dr}\exp\left(\delta r^2\right)\erfi\left(\frac{\gamma r}{2\sqrt{-\beta}}\right).
	\end{align}
	The first integral is simply the Gaussian integral on the half-plane, which has the result
	\begin{equation}
	I_1(\alpha, \beta,\gamma)=\frac{\pi}{4\sqrt{-\beta}}\sqrt{-\frac{1}{\delta}}.
	\end{equation}
	To proceed with $I_2(\alpha, \beta,\gamma)$, we rescale the variables to $u=\frac{\gamma r}{2\sqrt{-\beta}}$, leading to
	\begin{equation}
	I_2(\alpha,\beta,\gamma)=\frac{i\sqrt{\pi}}{\gamma}\int_0^{\infty}\mathop{du}\exp(-u^2\frac{4\alpha\beta+\gamma^2}{\gamma^2})\erf(u).
	\end{equation}
	This integral takes the form
	\begin{equation}
	\int_0^\infty \mathop{du}e^{-c u ^2}\erf{u}=\frac{1}{\sqrt{\pi}\sqrt{c}}\arctan\left(\frac{1}{\sqrt{c}}\right),
	\end{equation}
	which is convergent for $\Re(c)>0$, which may be confirmed for the case with $c=\frac{4\alpha \beta +\gamma^2}{\gamma^2}$.  Overall, this gives the result 
	\begin{equation}
	I_2(\alpha,\beta,\gamma)=\frac{1}{\sqrt{4\alpha \beta + \gamma^2}}\arctan\left(\frac{i\gamma}{\sqrt{4\alpha \beta + \gamma^2}}\right).
	\end{equation}
	Hence the complete result is
	\begin{equation}
	I(\alpha,\beta,\gamma)=\frac{\pi}{4\sqrt{-\beta}}\sqrt{-\frac{1}{\delta}}\left(1+\frac{2}{\pi}\arctan\left(\frac{i\gamma}{\sqrt{4\alpha \beta + \gamma^2}}\right)\right).
	\end{equation}
	
	To calculate $J(\alpha, \beta,\gamma)$, we pick up the calculation from Eq.~(\ref{eq:iint}), with
	\beq
	J(\alpha,\beta,\gamma)=\frac{\sqrt{\pi}}{2\sqrt{-\beta}}\int_0^\infty \mathop{dr}r\exp\left(\delta r^2\right)\left(1+i\erfi\left(\frac{\gamma r}{2\sqrt{-\beta}}\right)\right),
	\eeq
	which for clarity, we again separate into two parts, 
	\begin{align}
	J_1(\alpha, \beta,\gamma)&=\frac{\sqrt{\pi}}{2\sqrt{-\beta}}\int_0^\infty \mathop{dr}r\exp\left(\delta r^2\right),\\
	J_2(\alpha, \beta,\gamma)&=\frac{i\sqrt{\pi}}{2\sqrt{-\beta}}\int_0^\infty \mathop{dr}r\exp\left(\delta r^2\right)\erfi\left(\frac{\gamma r}{2\sqrt{-\beta}}\right).
	\end{align}
	These integrals are all simpler to compute, owing to the presence of the factor of $r$.  The first integral is easily calculated using $\pdv{r}e^{ar^2}=2a r e^{ar^2}$, and is
	\beq
	J_1(\alpha,\beta,\gamma)=-\frac{\sqrt{\pi}}{4\sqrt{-\beta}}\frac{1}{\delta}.
	\eeq
	The second integral is amiable to integration by parts, where we find the result
	\beq
	J_2(\alpha,\beta,\gamma)=-\frac{i \sqrt{\pi } \gamma }{2 \beta  \sqrt{\frac{\alpha  \beta }{4 \alpha  \beta +\gamma ^2}} \left(-\frac{4 \alpha  \beta +\gamma ^2}{\beta }\right)^{3/2}}.
	\eeq
	In total, this gives the result
	\begin{equation}
	J(\alpha,\beta,\gamma)=\frac{\sqrt{\pi}}{2\sqrt{-\beta}}\left(-\frac{2\beta}{4\alpha \beta + \gamma^2}+i\frac{\gamma}{\sqrt{-\beta}\sqrt{\frac{\alpha \beta}{4\alpha \beta + \gamma^2}}\left(-\frac{4\alpha \beta + \gamma^2}{\beta}\right)^{\frac{3}{2}}}\right).
	\end{equation}
	\subsection{Example Calculations}
	With the generating integrals completed, we can find the expressions for correlators using the approach in Appendix~\ref{app:strat}.
	
	The final results are significantly simplified by the identity
	\beq
	\label{eqn:id}
	\frac{\beta}{4\alpha \beta + \gamma^2}=-\frac{1}{4},
	\eeq
	which is simple to prove, by substituting in the definitions for $\alpha,\beta $ and $\gamma$ as
	\begin{equation}
	\frac{\beta}{4\alpha \beta + \gamma^2}=\frac12 \frac{-1+i \cot \omega \tau}{1-2i \cot \omega \tau -\cot^2 \omega \tau+\csc^2\omega \tau},
	\end{equation}
	where using the identity $\csc^2x-\cot^2x=1$ yields the required result.  
	
	It is important that this identity only be applied after completing all the differentiation steps of the algorithm in Appendix~\ref{app:strat}.  
	
	We also note the use of the factor of $\frac{1}{\sqrt{-\beta}}$
	%Making the back substitutions, we find that the overall i now, where writing by $\beta$ explicitly, we find the relationship
	\beq
	\frac{1}{\sqrt{2i\sin\omega \tau}}=e^{-i\frac{\omega \tau}{2}}\sqrt{-\beta},
	\eeq
	which allows us to take care of the time-dependence of the propagator prefactor.
	
	To for example calculate the correlator for the ground state, we find
	\begin{equation}
	q_{00} = \frac{1}{\pi} e^{-i\frac{\omega \tau}{2}}\sqrt{-\beta}e^{i\frac{\omega t_2}{2}-i\frac{\omega t_1}{2}}I(\alpha,\beta,\gamma).
	\end{equation}
	Making the appropriate cancellations and substitutions we find
	\begin{equation}
	q_{00}=\left(\sqrt{-\frac{\beta}{4\alpha \beta + \gamma^2}}+\frac{2}{\pi}\sqrt{-\frac{\beta}{4\alpha \beta + \gamma^2}}\arctan\left(\frac{i\gamma}{\sqrt{4\alpha\beta + \gamma^2}}\right)\right).
	\end{equation}
	Making use of the identity Eq.~(\ref{eqn:id}), we find
	\begin{align}
	\label{q00}
	q_{00}&=\frac{1}{4}\left(1+\frac{2}{\pi} \arctan(\frac{\gamma}{\sqrt{-4 \alpha \beta - \gamma^2}})\right),
	\end{align}
	whereupon back-substitution of the $\alpha,\beta$ and $\gamma$, and taking the real part, we find
	\begin{align}
	q^{\ket0}(+,+)=\frac14 \left(1+\frac{2}{\pi}\Re \arctan(\frac{ie^{i\frac{\omega\tau}{2}}}{\sqrt{2i\sin\omega\tau}})\right).
	\end{align}
	Care must be taken with the branch-cut of the square-root function, and the results presented here are consistent with the choice of taking the branch cut along the negative real axis.
	\subsection{Temporal Correlators}
	\label{app:corrs}
	We tabulate the exact correlators for the first nine energy eigenstates of the QHO.  We again rely on the function $f(\tau)=-i e^{-i\frac{\omega \tau}{2}}\sqrt{2i\sin\omega \tau}$.
	
	\begin{align}
	C_{ij}^{\ket0}&=\frac{2}{\pi}\Re\left[\arctan\left(\frac{1}{f(\tau)}\right)\right],\\
	C_{ij}^{\ket1}&=\frac{2}{\pi}\Re\left[\arctan\left(\frac{1}{f(\tau)}\right)+f(\tau)\right],\\
	C_{ij}^{\ket2}&=\frac{2}{\pi}\Re\left[\arctan\left(\frac{1}{f(\tau)}\right)+\frac{1}{2}f(\tau)\right],\\
	C_{ij}^{\ket3}&=\frac{2}{\pi}\Re\left[\arctan\left(\frac{1}{f(\tau)}\right)+\frac{5+e^{-2i\omega\tau}}{6}f(\tau)\right],\\
	C_{ij}^{\ket4}&=\frac{2}{\pi}\Re\left[\arctan\left(\frac{1}{f(\tau)}\right)+\frac{14+e^{-2i\omega\tau}}{24}f(\tau)\right],\\
	C_{ij}^{\ket5}&=\frac{2}{\pi}\Re\left[\arctan\left(\frac{1}{f(\tau)}\right)+\frac{94+17e^{-2i\omega\tau}+9e^{-4i\omega\tau}}{120}f(\tau)\right],\\
	C_{ij}^{\ket6}&=\frac{2}{\pi}\Re\left[\arctan\left(\frac{1}{f(\tau)}\right)+\frac{148+14e^{-2i\omega\tau}+3e^{-4i\omega\tau}}{240}f(\tau)\right],\\
	C_{ij}^{\ket7}&=\frac{2}{\pi}\Re\left[\arctan\left(\frac{1}{f(\tau)}\right)+\frac{1276+218^{-2i\omega\tau}+111e^{-4i\omega\tau}+75e^{-6i\omega \tau}}{1680}f(\tau)\right],\\
	C_{ij}^{\ket8}&=\frac{2}{\pi}\Re\left[\arctan\left(\frac{1}{f(\tau)}\right)+\frac{8528+904e^{-2i\omega\tau}+258e^{-4i\omega\tau}+75e^{-6i\omega \tau}}{13440}f(\tau)\right].
	\end{align}
	\section{QHO Results}
	\subsection{Smoothed Projectors}
	\label{sec:smooth}
	It is simplest to calculate the effect of smoothed projectors, working with the quasi-probability expressed as a truncated infinite sum.  To do this, we switch from using $\theta(\hat x)$ as our projector, to the continuous $\frac12(1+ \erf(\sqrt{\frac{m\omega}{\hbar}}\frac{\hat x}{a}))$. Where $a$ is a dimensionless parameter characterising the degree of smoothing.  For small $a$, we expect to recover the sharp projector result, and $a=1$ corresponds to a smoothing on the characteristic length scale of the QHO, $\sqrt{\frac{\hbar}{m\omega}}$.	This adjusts the matrix elements to be
	
	\beq
	\label{Jil}
	J_{k\ell}=\frac{1}{\sqrt{\pi2^{k+\ell} k!\ell!}}\int_{-\infty}^{\infty}\mathop{dr}H_{k}(r)H_{\ell}(r)e^{-r^2}\frac{1+ \erf(\frac{r}{a})}{2}.
	\eeq
	
	We compute these integrals numerically, and then investigate the LG violations possible with different values of $a$.  In Fig.~\ref{fig:smoothp}(a), we plot one of the $\LG{3}$ inequalities, varying the value of $a$.  The smoothed projectors result in a similar (although not identical) shape, but with a reduced amplitude.  In Fig.~\ref{fig:smoothp}(b), we plot the minimal value taken by the $\LG{3}$ inequalities for each value of $a$, where we can see that once the smoothing reaches the characteristic length-scale of the QHO, LG violations vanish.
	\begin{figure}
		\begin{center}
		\subfloat[]{{\includegraphics[height=4.6cm]{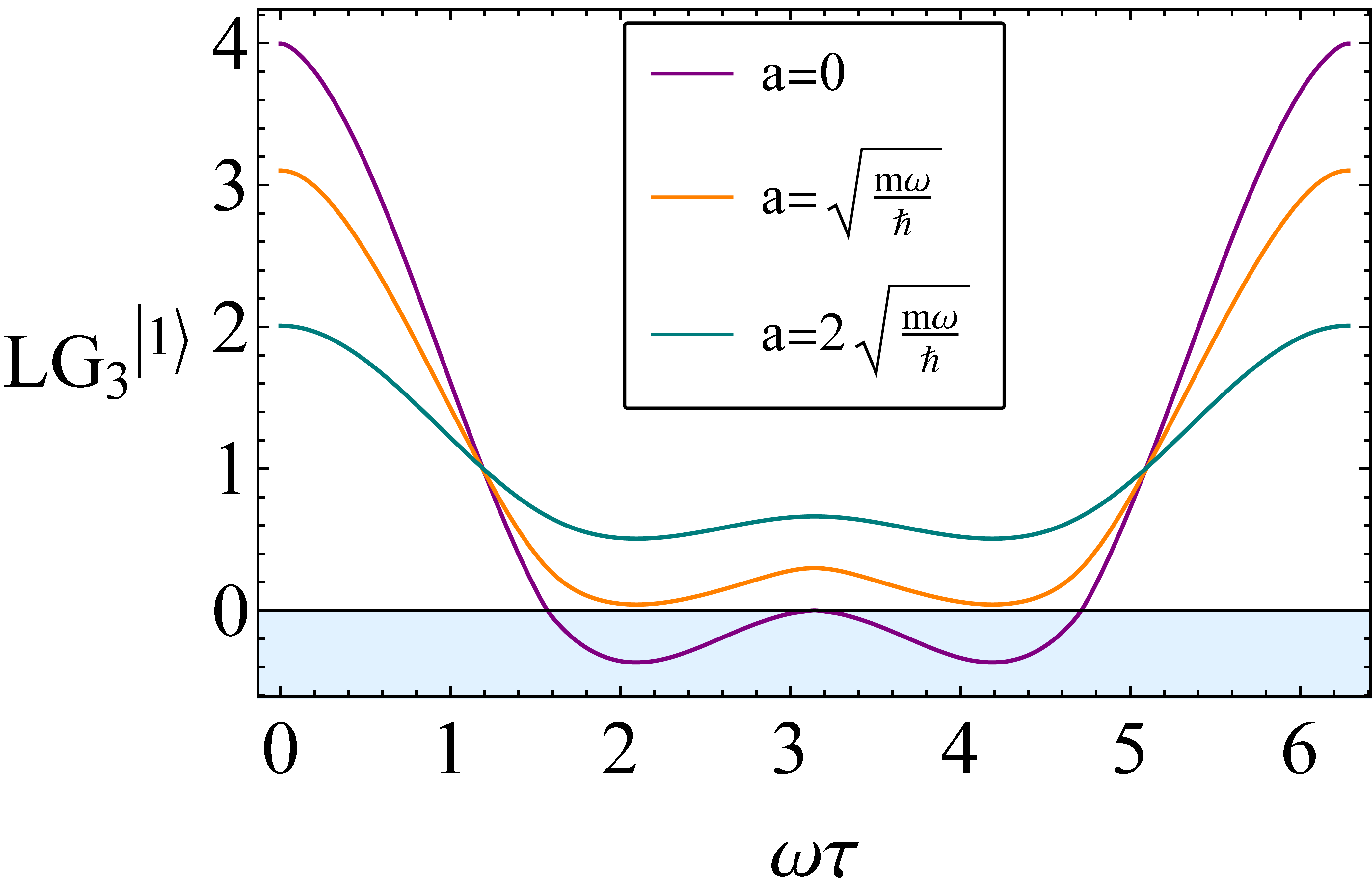}}}%
		%\qquad
		\hspace{5mm}
		\subfloat[]{{\includegraphics[height=4.6cm]{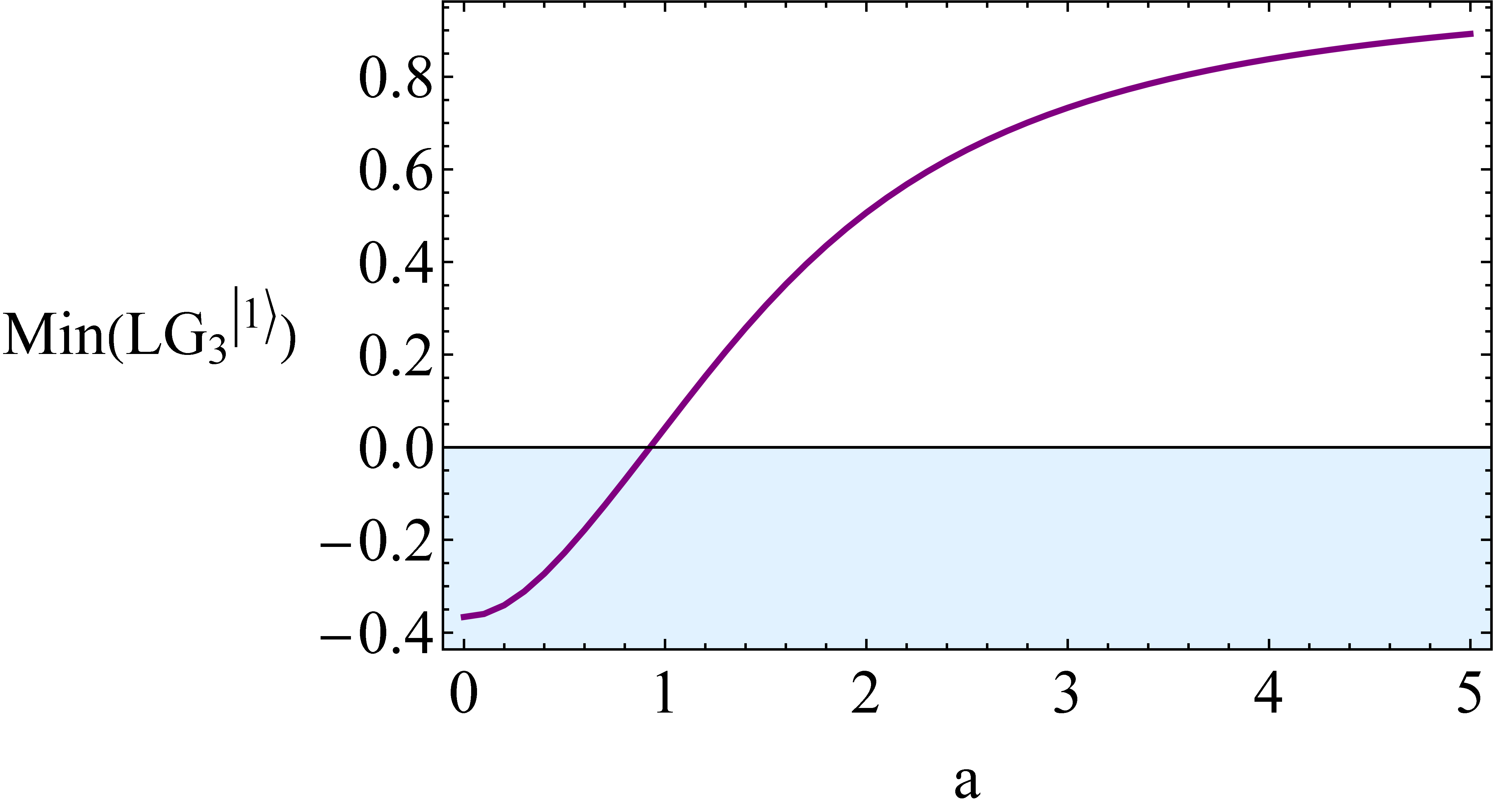}}}
		\end{center}
		\caption[Persistence of LG violations with smoothed projectors]{In (a), we plot one of the $\LG{3}$s, for different values of $a$, showing the qualitative effect of projector smoothing.  In (b), we plot the minimum value taken by the $\textrm{LG}_3$ inequalities, for a given value of $a$.}%
		\label{fig:smoothp}%
	\end{figure}
	\subsection{Classicalization}
	\label{app:classic}
	To understand the classicalization of the QHO, we look at the large $n$ asymptotic behaviour of the quasi-probability, using the original sharp projectors. Using Eq.~(\ref{eq:qgen}) for the QHO, we find
	\begin{equation}
	q(n)=\frac14+\Re \psi_n'(0)^2 \sum_{k=0}^{\infty}e^{-i(k-n)\tau}\frac{1}{4(k-n)^2}\psi_k(0)^2,
	\end{equation}
	where the sum is over even $k$.  We first relabel the sum to simplify the denominator,
	\begin{equation}
	q(n)=\frac14+\Re \psi_n'(0)^2 \sum_{k=-n}^{\infty}e^{-ik\tau}\frac{1}{4k^2}\psi_{k+n}(0)^2,
	\end{equation}
	which also shifts it to a sum over only odd $k$.  We now make the first of two observations relevant in the large $n$ regime.  Since the Hermite functions evaluated at $0$ remain finite, and of order of magnitude $1$, the magnitude of the terms of this sum will fit into the envelope given by $\frac{1}{k^2}$, and so, for large $n$, we can safely extend the lower limit of the sum to $-\infty$, yielding
	\begin{equation}
	q(n)=\frac14+\Re \psi_n'(0)^2 \sum_{k=-\infty}^{\infty}e^{-ik\tau}\frac{1}{4k^2}\psi_{k+n}(0)^2.
	\end{equation}
	Using the standard recurrence relations of the Hermite functions, we find that at the origin we have
	\begin{equation}
	\frac{\psi_{n+1}(0)}{\psi_{n-1}(0)}=\sqrt{\frac{n}{n+1}}.
	\end{equation}
	This leads to the second observation, that in the large $n$ regime, the rate at which $\psi_{n+k}(0)$ changes is negligible compared to the rate of change of $\frac{1}{k^2}$, and hence may be considered approximately constant in the sum, and so we have
	\begin{equation}
	\label{eq:progress}
	q(n)\approx\frac14+\Re \psi_n'(0)^2 \psi_{n+1}(0)^2\sum_{k=-\infty}^{\infty}e^{-ik\tau}\frac{1}{4k^2}.
	\end{equation}
	To proceed, we look at the generating function of the Hermite polynomials, 
	\begin{equation}
	\label{eqn:gen}
	e^{2xt-t^2}=\sum_{n=0}^{\infty}H_n(x)\frac{t^n}{n!}
	\end{equation}
	Evaluating at $x=0$, and using the Taylor series for the Gaussian, we find
	\begin{equation}
	\sum_{k=0}^\infty (-1)^k\frac{t^{2k}}{k!}=\sum_{n=0}^{\infty}H_n(0)\frac{t^n}{n!},
	\end{equation}
	whereby comparing powers of $t$, we find for even $n$, $\abs{H_n(0)}=\frac{n!}{(\frac{n}{2})!}$.  Performing a similar analysis for the derivative term, we find that for odd $n$, $\abs{H'_n(0)}=\frac{(n+1)!}{(\frac{n+1}{2})!}$.  The product term, with normalisation re-included is then
	\begin{equation}
	\psi_n'(0)\psi_{n+1}(0)=\frac{(n!)^2}{2^{n-1} \sqrt{2 n} (n-1)!\left(\frac{n}{2}!\right)^2},
	\end{equation}
	which through computer algebra software is found to have the limit
	\begin{equation}
	\lim\limits_{n\to \infty}\psi_n'(0)\psi_{n+1}(0) = \frac{2}{\pi}.
	\end{equation}
	Using this in Eq.~(\ref{eq:progress}), we find
	\begin{equation}
	q(n)\approx\frac14+\Re\sum_{k=-\infty}^{\infty}e^{-ik\tau}\frac{1}{\pi^2 k^2}.
	\end{equation}
	We can now identify the sum as the exponential Fourier series for a symmetric triangle wave, with amplitude $\frac14$, which is exactly the classical correlator.  This represents the classicalization of the QHO, as $n$ becomes large. 
	\subsection{QHO quasi-probability with arbitrary coarse-grainings}
	\label{app:arbproj} 
    To calculate the quasi-probability in the QHO with arbitrary coarse-grainings, we apply the method in Appendix \ref{app:general}, using the integrals $J_{k\ell}(a,b)$.
    
	We first note Eq.~(\ref{eqn:wronski}) is undefined for $k=l$, and must be calculated by hand as 
	\begin{equation}
	J_{kk}(a,b)=\int_{a}^{b}\mathop{dx}\psi_k(x)\psi_k(x).
	\end{equation}
	The quasi-probability is then calculated through Eq.~(\ref{eqn:qpapprox}) as
	\begin{equation}
	q_{nn}(+,+)=e^{i\frac{E_n \tau}{\hbar}}\sum_{k=0}^{\infty}e^{-i\frac{E_k \tau}{\hbar}}J_{nk}(a,b)J_{nk}(c,d).
	\end{equation}
	We note again, that these $\psi_k(x)$ and $E_k$ are not yet the QHO eigenstates and energies, and that this result is generic. 
	
	%To avoid extraneous complexity, we will consider just the $\ket0$ state of the QHO, and we will consider that the first measurement is taken over the range $[0,\infty]$.  Altogether, this leads to a quasi-probability that is dependent on time, and the interval of the second measurement, $[c,d]$,
	To make things concrete, we will now calculate this for the ground state of the QHO. We take the first pair of intervals $\Delta(\alpha_1)$ to be $\Delta(+)=[0,\infty)$ and $\Delta(-)=(-\infty, 0]$.  The second pair of intervals $\Delta(\alpha_2)$ are taken to be $\Delta(+)=[c,d]$, and $\Delta(-)$ its complement. The quasi-probability for the ground state is,
	\begin{equation}
	\label{eq:qmult}
	q(+,+)=\frac12 J_{00}(c,d)+\Re\sum_{k=0}^{\infty}e^{-i \omega \tau k}J_{0k}(0,\infty)J_{0k}(c,d).
	\end{equation}  
	Again, due to it being a symmetric potential, we have $J_{00}(0,\infty)=\frac12$ as before.  Doing the integral manually for the ground state, we find
	\begin{equation}
	J_{00}(c,d)=\frac12(\erf(d)-\erf(c)).
	\end{equation}
	Using this result and the QHO eigenspectrum in Eq.~(\ref{eqn:wronskiapp}), we are able to calculate the quasi-probability using Eq.~(\ref{eq:qmult}).
	\section{Morse Potential}
	\label{app:morse}
		\begin{figure}
		\begin{center}
		\subfloat[]{{\includegraphics[height=4.9cm]{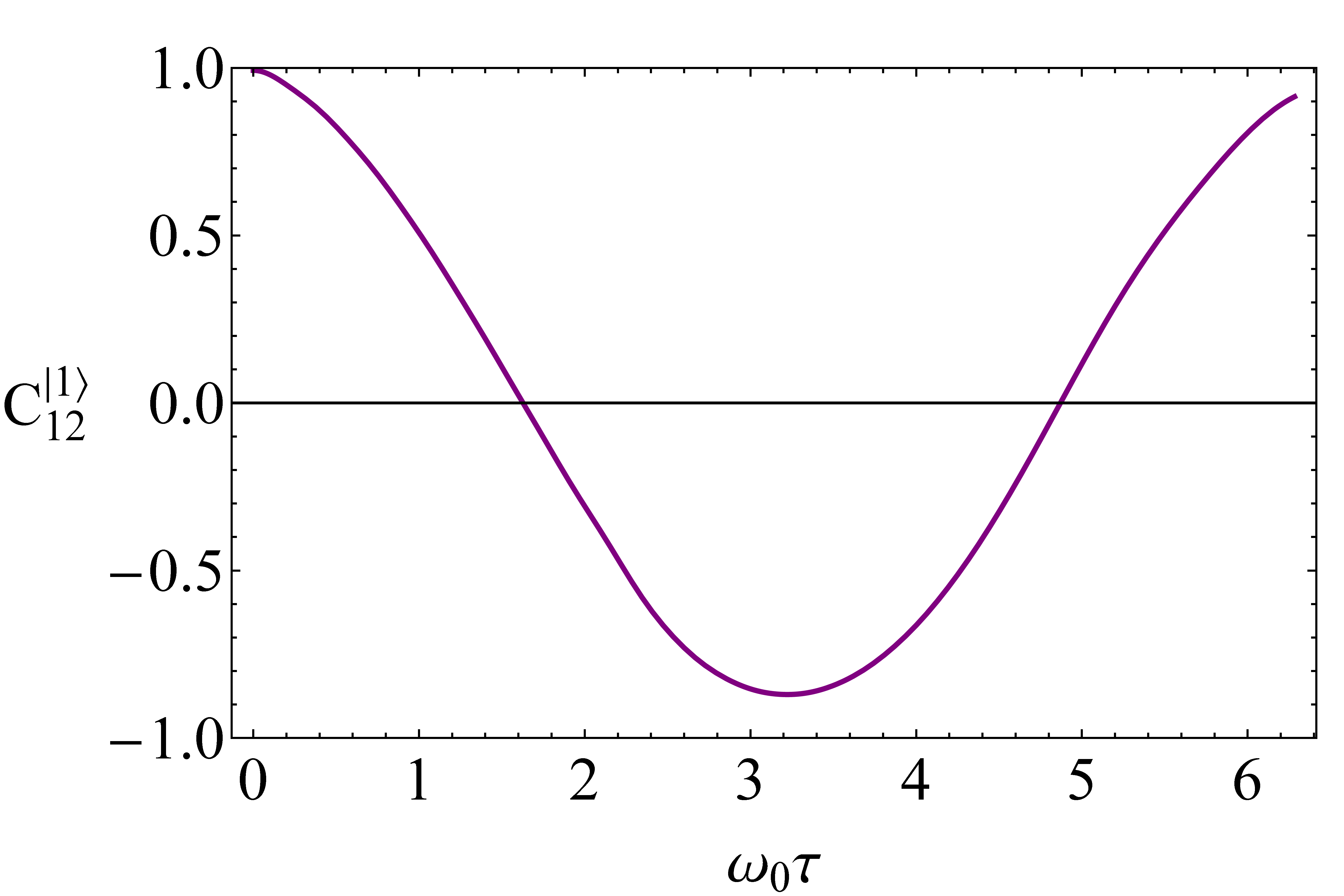}}}%
		\qquad
		\subfloat[]{{\includegraphics[height=4.9cm]{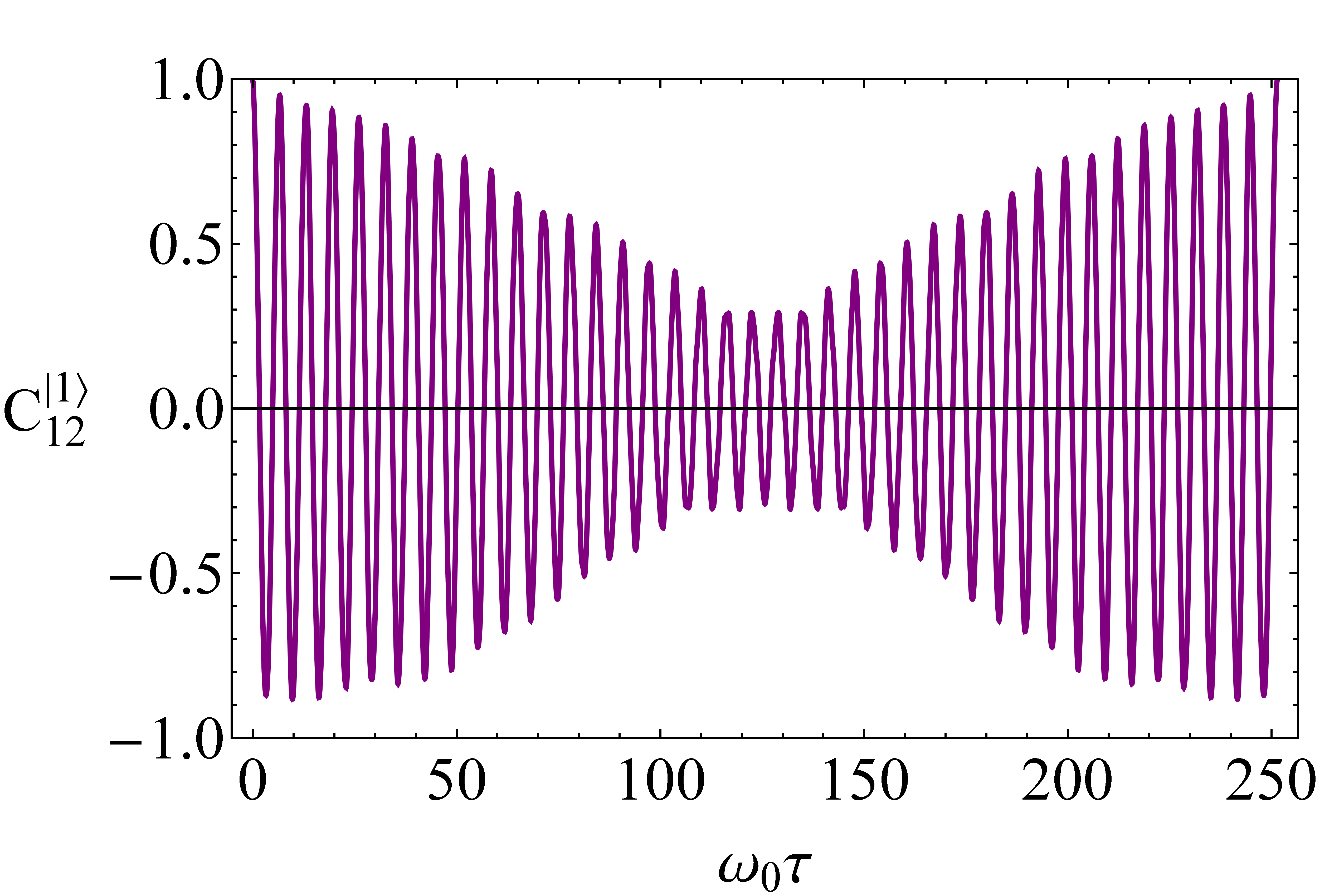}}}
		\end{center}
		\caption[Temporal correlators in the Morse potential]{
			The temporal correlator in the $\ket1$ state for the Morse potential over time interval $\omega_0 \tau = 2\pi$ is shown in (a). It is not quite periodic over this time interval but becomes exactly periodic over a much longer period of time, as shown in (b).
		}%
		\label{fig:morse}%
	\end{figure}
	The Morse potential is an asymmetric potential with minimum at $r_c$, which combines a short-range repulsion, with a long-range attraction.  The potential is defined by
	\begin{equation}
	V(r) = D_e\left(e^{-2a(r-r_e)}-2e^{-a(r-r_e)}\right),
	\end{equation}
	where $D_e$ corresponds to the well depth, and $a$ to its width.  The Morse potential supports up to $\lfloor \lambda-\frac12 \rfloor$ bound states \cite{dahl1988}, where $\lambda=\frac{\sqrt{2mD_e}}{a\hbar}$ and $\lfloor x \rfloor$ is the floor function.  The eigenstates and energies are
	\begin{align}
	\varepsilon_n&=-\frac12\left(\lambda-n-\frac12\right)^2,\\
	\psi_n(z)&=N_n z^{\lambda-n-\frac12}e^{-\frac12 z}L_n^{2\lambda-n-1}(z),
	\end{align}
	where $L_n^{(\alpha)}(z)$ are the generalized Laguerre polynomials, and $z$ is a scaled spatial coordinate defined as $z=2\lambda e^{-a(r-r_c)}$.
	In the standard non-dimensionalization, physical energy eigenvalues relate to dimensionless as $E_n=\frac{2\hbar \omega_0}{\lambda}\varepsilon_n$, 	where $\omega_0=a \sqrt{\frac{2D_e}{m}}$ is the frequency of small oscillations in the potential.  Normalisation is given by
	\begin{equation}
	N_n=\left(\frac{n!(2\lambda-2n-1)}{\Gamma(2\lambda-n)}\right).
	\end{equation}
	In the Morse potential, states are constrained to $r>0$, however we can still consider coarse-graining onto the left and right halves of the well.  The energy eigenstates similarly still vanish at $r=0$ and $r\to\infty$, so many of the terms in Eq.~(\ref{eqn:wronski}) still vanish.  Filling out Eq.~(\ref{eqn:qpapprox}), we find
	\begin{equation}
	\label{eqn:morseq}
	q_{nn}=J_{nn}^2+e^{i\frac{\varepsilon_n}{\lambda}\omega_0\tau}\sum_{k=0,k\neq n}^{\lfloor \lambda-\frac12 \rfloor}e^{-i \frac{\varepsilon_k}{\lambda}\omega_0\tau}\frac{1}{(\varepsilon_n-\varepsilon_k)^2}\left(\psi_n^\prime(x_c)\psi_k(x_c)-\psi_k^\prime(x_c)\psi_n(x_c)\right)^2,
	\end{equation}	
	Although this sum is finite, by choosing a large enough $\lambda$, and just looking at low energy states we can approximate it well.  Physically, this corresponds to measurements in shallower wells being likely to eject the particle from the well, where we would then have to consider the continuous positive energy solutions as well.  We can again estimate the error for a particular $n$ and $\lambda$ as
	\begin{equation}
	\Delta(n,\lambda)=J_{nn}-\sum_{k=0}^{\lfloor \lambda-\frac12 \rfloor} J_{nk}^2.
	\end{equation}
	
	Owing to the fact the Morse potential is non-symmetric, we must calculate the terms $J_{nn}$ by hand, which for the low $n$ is a simple integration.  The truncation error for the first excited state reaches $0.01$ for $\lambda=15$, and similar to the QHO, is remains higher for the ground state.  For accuracy, we hence calculate just the $\ket1$ state, and choose $\lambda=50$, yielding $\Delta(n,\lambda)=0.001$.
	
	We extract temporal correlators from Eq.~(\ref{eqn:morseq}) using the moment expansion Eq.~(\ref{mom}), which we plot in Fig.~\ref{fig:morse}(a).  At first glance, this correlator is qualitatively very similar to the $\ket1$ correlator for the QHO, and hence similar to the simple cosine correlator for spin-$\frac12$ models.  However owing to the asymmetry of the Morse potential, the correlator never reaches the value $-1$,  and due to the anharmonicity, the correlator is periodic over a much longer time scale, which is shown in Fig.~\ref{fig:morse}(b).
	
	In Fig.~\ref{fig:morse2}(a), we plot the LG3 inequalities for the first excited state of the Morse potential, where there are significant violations, reaching $70\%$ of the L\"{u}ders bound.  The LG4 inequalities are plotted in Fig.~\ref{fig:morse2}(b), with significant violations reaching $92\%$ of the L\"{u}ders bound.  Violations diminish in magnitude for the large-time behaviour, but remain present for nearly all intervals of $\omega_0 \tau$.

	\end{subappendices}
%\end{document}

%\renewcommand{\headrulewidth}{0pt}
%\fancyfoot{\makebox[\textwidth][c]{\hyperref[link:5]\thepage}}

%\fancypagestyle{plain}{%redefine \pagestyle{plain} to add the link <<<
%    \fancyhf{} 
%\fancyfoot{\makebox[\textwidth][c]{\hyperref[link:5]\thepage}}
 %   \renewcommand{\headrulewidth}{0pt}
 %   \renewcommand{\footrulewidth}{0pt}
%}%

\fancyhf{}
\renewcommand{\headrulewidth}{0pt}
\fancyfoot{\makebox[\textwidth][c]{\hyperref[link:6]\thepage}}

\fancyhfoffset[LE,RO, RE, LO]{0cm}
\renewcommand{\chaptermark}[1]{ \markboth{#1}{} }
\renewcommand{\sectionmark}[1]{ \markright{#1}{} }
\fancyhf{}
\fancyhead[L]{\textsl{\thesection~~ \rightmark}}
\fancyhead[R]{\hyperref[link:6]\thepage}
\renewcommand{\headrulewidth}{1pt}
\chapter{Leggett-Garg Violations for Continuous-Variable Systems with Gaussian States}
\label{chap:QHO2}
\bookepigraph{3.5in}{The day you teach the child the name of the bird, \\the child will never see that bird again.}{Jiddu Krishnamurti}{}{1}
\vspace{-2em}
\section{Introduction}

In this chapter, we further develop the programme of LG tests in the QHO, which follows the experimental procedure conceived in Refs.~\cite{bose2018,das2022a}. Building on the results of Chapter~\ref{chap:QHO1}, we are able to analyse the situation where the initial state is a coherent state of the QHO.  The first main aim of this chapter is to undertake a thorough analysis of LG2, LG3 and LG4 inequalities for an initial coherent state. This is carried out in Section \ref{sec:corecalc}, where we set out the formalism, and calculate temporal correlators, drawing results from Chapter~\ref{chap:QHO1}.  We carry out a detailed parameter search and find the largest LG violations possible for a coherent state.

A second aim is to explore the physical origins of the LG violations. So, in Section \ref{sec:physmec} we examine the difference between QM currents and their classical counterparts for initial coherent states projected onto the positive or negative $x$-axis.  In this section we also provide a second approach to calculating correlators in the small-time limit, which is in fact valid for general states.  We also calculate the Bohmian trajectories, to give further physical portrait of what underlies the observed LG violations.  Finally, we examine the measurement process in the Wigner representation, noting that the initially positive Wigner function of the coherent state acquires negativity as a result of the projective measurement process. 

Modifications to the above framework are considered in Section \ref{sec:mf}. We investigate what violations are possible using the Wigner LG inequalities \cite{saha2015, naikoo2020}.  We briefly discuss other types of measurements beyond the simple projective position measurements used so far and also consider LG2 violations with projections onto coherent states. We also determine how the LG violations may be modified for squeezed states or thermal states. We summarize in Section \ref{sec:sum}.  We relegate to a series of appendices the grisly details of the calculations involved in this analysis.

\section{Calculation of Correlators and LG Violations}
\label{sec:corecalc}
\subsection{Conventions and Strategy}
For most of this chapter, we will work with coherent states of the harmonic potential,  which can of course be thought of as the ground state of the QHO, shifted in phase-space.  We continue to use the Hamiltonian and naming conventions as defined in Sec.~\ref{subsec:qhoconv} in the previous chapter.  The intricacy of calculating temporal correlators within QM stems from the complexity in the time evolution of a post-measurement state.  By considering a co-moving frame for the post-measurement state, we develop a time evolution result which explicitly separates the quantum behaviour from the classical trajectories.
%We will work with systems defined exactly (or approximately) by the harmonic oscillator Hamiltonian,
%\begin{equation}
%\hat H = \frac{\hat p_{\text{phys}}^2}{2m}+\frac12 m\omega^2 \hat x_{\text{phys}}^2,
%\end{equation}
%with physical position and momenta $x_{\text{phys}}$ and $p_{\text{phys}}$.  In calculations we use the standard dimensionless variables $x\sqrt{\hbar/(m\omega)}=x_{\text{phys}}$ and $p\sqrt{\hbar m\omega}=p_{\text{phys}}$. We denote energy eigenstates $\ket n$, writing $\psi_n(x)$ in the position basis, with corresponding energies $E_n=\hbar \omega(n+\frac12)=\varepsilon_n \hbar \omega$.

We write coherent states as $\ket \alpha$, where its eigenvalue $\alpha$ relates to rescaled variables as
\begin{align}
\label{eq:alphatox}
\ev{\hat x(t)}&=\sqrt2 \Re \alpha(t),\\
\ev{\hat p(t)}&=\sqrt2 \Im \alpha(t).
\end{align}
These are the classical paths underlying the motion of coherent states, and we adopt the short-hand $x_1=\ev{\hat x(t_1)}$ and likewise $p_1=\ev{\hat p(t_1)}$.  A coherent state may be represented in terms of $\alpha$ and an initial phase, however in this work we will largely represent them in terms of the initial averages $x_0$ and $p_0$, for clarity of physical understanding.  We construct coherent states with the unitary displacement operator $D(\alpha)=\exp(\alpha a^\dagger-\alpha^* \hat a)$ operating on the ground state,
\begin{equation}
\ket \alpha = D(\alpha)\ket 0,
\end{equation}
which results in the wave-function
\begin{equation}
\psi^{\alpha}(x,t)=\frac{1}{\pi^\frac14}\exp(-\frac12(x-x_t)^2 + i \frac{p_t}{\hbar} x+i\gamma(t)).
\end{equation}
We proceed following the same strategy as in Chapter~\ref{chap:QHO1}, i.e. calculating the two-time quasi-probability Eq.~(\ref{quasi}), then extracting the quantities $\ev{Q_i}$, $C_{ij}$ by comparison with the moment expansion Eq.~(\ref{mom}).

We have not been able to find the maximally violating state but Eq. (\ref{eq:maxQP}) suggests it is probably discontinuous at $x=0$. This in turn suggests that we will not able to get close to maximal violations with the simple gaussian states explored here.
\subsection{Calculation of the Correlators}
We now calculate the temporal correlators for the case $\hat Q = \sgn(\hat x)$, as in the previous chapter Eq.~(\ref{eq:proj}).  The QP Eq.~(\ref{quasi}) is given by
%The quasi-probability for coherent states with the left-right dichotomic variable, $P_s=\theta(s\hat x)$, is given by

\begin{equation}
q(s_1,s_2)=\Re\ev{e^{i H t_2}\theta(s_2\hat x)e^{-i H \tau}\theta(s_1\hat x)e^{-i H t_1}}{\alpha}.
\end{equation}
By considering the displacement operator as acting on the measurements instead of the state, the quasi-probability is shown in Appendix \ref{app:timeev} to be
\begin{equation}
\label{eqn:quasgen}
q(s_1,s_2)=\Re e^{\frac{i\omega \tau}{2}} \ev{\theta(s_2(\hat x + x_2))e^{-iH\tau}\theta(s_1(\hat x + x_1))}{0},
\end{equation}
which reveals that the quasi-probability for coherent states can be understood as the quasi-probability for the pure ground state, with measurement profiles translated according to the classical paths.  This shows that any LG test on any coherent state may be directly mapped to an LG test on the ground-state of the QHO with translated measurements, as calculated in Section~\ref{subsec:finegrain}.

Using the surprising result that $\int_a^b \psi_n(x)\psi_m(x) dx$ has an exact and general solution as shown in Appendix~\ref{app:wronski},  we calculate the temporal correlators as an infinite sum
\begin{equation}
\label{eqn:corr}
C_{12}=\erf(x_1)\erf(x_2)+ 4\sum_{n=1}^\infty \cos (n\omega \tau) J_{0n}(x_1,\infty)J_{0n}(x_2,\infty).
\end{equation}

The details of this calculation are given in Appendix ~\ref{app:timeev}.  The infinite sum may be evaluated approximately using numerical methods, by summing up to a finite $n$.  This calculation matches the analytically calculated special case of $x_0=0$ given in Ref. \cite{halliwell2021}.
%We now observe that the $J_{0n}$ terms in the infinite sum will be suppressed if either $\lvert x_1 \rvert$ or $\lvert x_2 \rvert$ become large, since each term involves a factor of either $\psi_0(x)$ or $\psi'_0(x)$, resulting in double-exponential suppression.  We can place a limit on the magnitude of the sum in Eq.(\ref{eqn:cohquas}), by considering it as the overlap between $\theta(x-x_1)\psi_0(x)$ and $\theta(x-x_2)\psi_0(x)$, less $J_{00}(x_1, \infty)J_{00}(x_2, \infty)$.  This yields the limit
%\begin{equation}
%\abs{\sum_{n=1}^\infty e^{-in\omega \tau}J_{0n}(x_1,\infty)J_{0n}(x_2,\infty)}\leq\frac{1}{2} \theta (x_1-x_2) (\text{erf}(x_2)+\text{erfc}(x_1)-1)-\frac{1}{4} (\text{erfc}(x_1)-2) \text{erfc}(x_2).
%\end{equation}
%A quick evaluation shows that with either $|x_1|$ or $|x_2|$ greater than $3$, the magnitude of the sum is limited to the order of $10^{-6}$.  This means that if the coherent state is a significant distance from the axis at either of the measurement times, the sum containing the quantum effects will vanish, and the quasi-probability will be non-negative.
%
%The same applies to the correlators Eq.~(\ref{eqn:corr}) used in the LG inequalities, where any $x_i>3$ renders the correlators involving that measurement time classical.  This does not rule out the possibility of LG3/4 violations with $x_i>3$, but it does mean at least two of the correlators will be forced to be classical, which will clearly reduce in magnitude any possible violations.
\subsection{LG Violations}
\label{sec:LGresults}
\begin{figure}
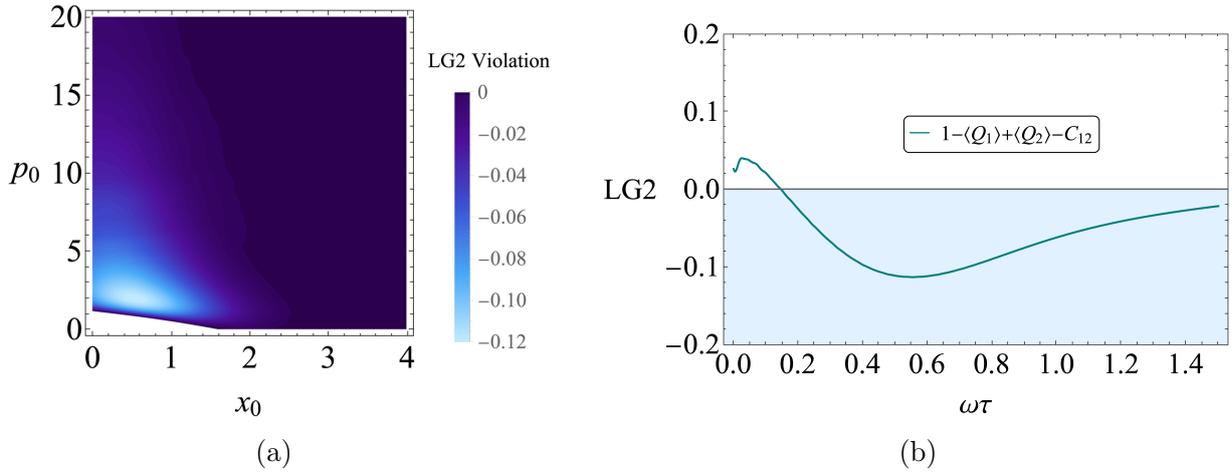

	\subfloat[]{{\includegraphics[height=5.6cm]{fig1a.png}}}%
	%\qquad
	\hspace{5mm}
	\subfloat[]{{\includegraphics[height=5.6cm]{fig1b.png}}}
	\caption[Parameter space exploration for LG2 violations, and temporal behaviour]{Plot (a) is a parameter space exploration, showing the largest LG2 violation for a coherent state with $x_0=0.55$, $p_0=-1.93$. Plot (b) shows the temporal behaviour of the LG2s for a state leading to the largest violation of $-0.113$.}%
	\label{fig:lg2}%
\end{figure}
The freely chooseable parameters are the initial parameters of the coherent state $x_0$, $p_0$, and the time interval between measurements, $\tau$. 
The parameter space explorations are symmetric in both $x_0$ and $p_0$, since any difference in phase-space quadrant can be absorbed by changing $s_1$, $s_2$ in Eq.~(\ref{eqn:quasgen}), and using the property that a half-cycle evolution leads to a reflection, with similar holding true for LG3s and LG4s.  Where there are more than two measurements, we use equal time-spacing. In Fig.~\ref{fig:lg2} (a), we plot a parameter space exploration of violations for the LG2 inequalities, where for a given $x_0$, $p_0$, we have numerically searched for the largest violation for that state. The LG3 and LG4 inequalities have a similar distribution, but with progressive broadening. Figures for the LG3 and LG4 inequalities are included in Appendix~\ref{app:LGresults}.  
 \begin{table}
\centering
\setlength{\tabcolsep}{0.5em}
\begin{tabular}{|c|c|c|c|}
\hline
Inequality         & Largest Violation & Percent of L{\"u}ders Bound & Location ($x_0, p_0$) \\ \hline
LG2 & -0.113            & 22\%                    & ($\pm 0.550$, $\pm 1.925$)        \\
LG3                        & -0.141            & 28\%                    & ($\pm 0.859$, $\pm 3.317$)           \\
LG4                        & 2.216              & 26\%                    & ($\pm 0.929$, $\pm 3.666$)          
\\
\hline
\end{tabular}
\caption[Largest violations for LG2, LG3 and LG4 inequalities for coherent states]{\label{tab:1}Tabulation of parameter space results.}
\end{table}

The parameters leading to the largest violation, and the magnitude of those violations are reported in Table~\ref{tab:1}.  In Fig.~\ref{fig:lg2}(b) and Fig.~\ref{fig:lg34}, we plot the temporal behaviour of the LG2, LG3 and LG4 violations for the states in Table~\ref{tab:1} leading to the largest violations.  

\begin{figure}
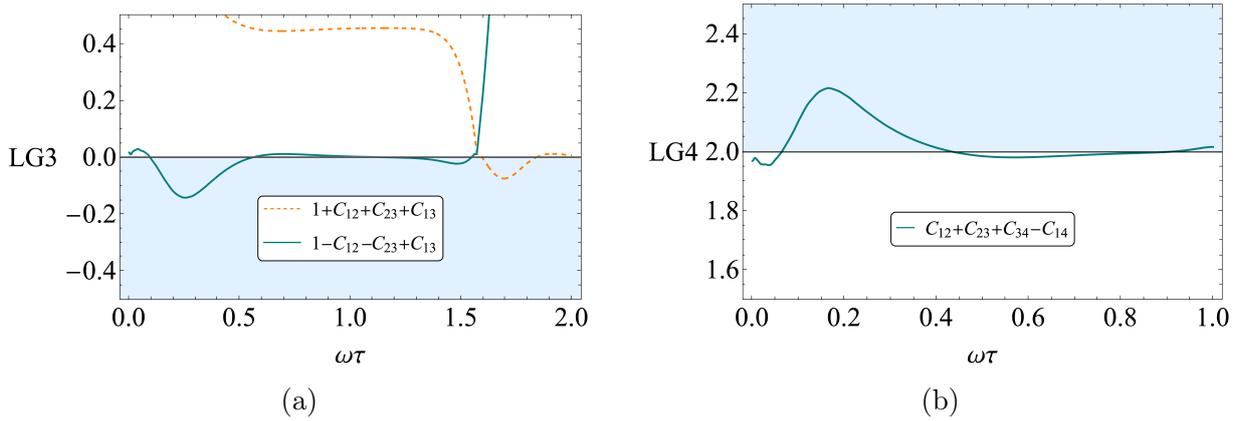

	\subfloat[]{{\includegraphics[height=5.2cm]{fig2a.png}}}%
	%\qquad
	\hspace{5mm}
	\subfloat[]{{\includegraphics[height=5.2cm]{fig2b.png}}}
	\caption[LG3 and LG4 violations for a coherent state as a function of time]{Plot (a) shows temporal behaviour of the LG3s for a coherent state with $x_0=0.85$ and $p_0=3.30$, yielding the largest violation $-0.141$. Plot (b) shows the same for the LG4s with a coherent state with $x_0=0.93$ and $p_0=3.67$, leading to the largest violation of $2.216$.}%
	\label{fig:lg34}%
\end{figure}

\section{Physical Mechanisms of Violation}
\label{sec:physmec}
Given the LG violations exhibited in the previous section a natural question to ask concerns the underlying physical effects producing the non-classical behaviour  responsible for the violations. Since $ Q(t) = \sgn(x(t) )$, a classical picture of the system would involve a set of trajectories $x(t)$ and probabilities for those trajectories. It is then natural to look at the parallel structures in quantum theory and compare with the classical analogues. We therefore look at the quantum-mechanical currents associated with the LG inequalities, which correspond to the time evolution of certain probabilities, and also to the Bohm trajectories associated with those currents, in terms of which the probability flow in space-time is easily seen.
What we will see is that the departures from classicality are essentially the ``diffraction in time’’ effect first investigated by Moshinsky \cite{moshinsky1952, moshinsky1976}. This effect was discovered considering the time evolution of an initial plane wave in one dimension initially restricted to $x<0$, where upon release, the current across the origin displays an oscillatory behaviour.  This behaviour is considered non-classical, as it implies some motion in the opposite direction of the initial momentum $p$.  The key mathematical object is the Moshinsky function 
\begin{equation}
\label{eq:moshfunc}
M(x,p,t)=\langle x\lvert e^{-iHt}\theta(\hat x)\rvert p \rangle 
\end{equation}
for an initial momentum state $\lvert p\rangle$, which is directly related to this work since represents the momentum basis matrix element for the evolution following a measurement. We note there have been two experimental studies of the diffraction in time effect \cite{szriftgiser1996,tirole2023}.

\subsection{Analysis with Currents}
\label{subsec:currents}
As we saw in Section.~\ref{sec:LGresults}, the quasi-probability component $q(-,+)$ exhibits a healthy degree of negativity.  We start by writing it as
\begin{equation}
\label{eq:dqdt}
q(-,+)=\int_{0}^{\tau}\mathop{dt}\frac{dq}{dt}.
\end{equation}
 It is then simple to relate $\frac{dq}{dt}$ to a set of standard quantum mechanical currents, which can be calculated analytically.  Overall negativity of the QP can then be spotted by the non-classicality or negativity of certain combinations of currents.  With details in Appendix \ref{app:curr}, we are able to write the quasi-probability as the following combinations of currents at the origin,
\begin{equation}
\label{eq:qpdiffs}
q(-,+)=\int_{t_1}^{t_2}\mathop{dt}J_-(t)+\frac12 \int_{t_1}^{t_2}\mathop{dt}\left(\mathbbmsl{J}_-(t)-J_-(t) +\mathbbmsl{J}_+(t)-J_+(t)\right),
\end{equation}
where $J_\pm(t)$ which we refer to as the \textsl{chopped current}, is the current following a measurement of $\hat Q$ at $t_1$, with result $s_1 = \pm$, with $\mathbbmsl{J}_\pm(t)$ the classical analogues.  The chopped currents contains the complexity of the influence of the earlier measurement, and are hence quite complicated and given in Appendix~\ref{app:chopcurr}.
 
Note that the first time integral is simply the sequential measurement probability $p_{12}(-,+)$, which is non-negative.  The negativity of the quasi-probability therefore arises as a result of the difference between the classical and quantum chopped currents. 

The classical and quantum chopped currents and the current combination appearing in Eq.~(\ref{eq:qpJ}) are all plotted in Figure~\ref{fig:currs} for the initial state giving the LG2 violation described in Section \ref{sec:LGresults}. The departures from classicality are clearly seen and are consistent with a broadening of the momentum distribution produced by the measurement. The quantum chopped currents diverge initially due to the sharpness of the measurement, however as shown in Sec.~\ref{sec:smooth}, LG violations do persist when the discontinuity in the step-function is smoothed over.

Most importantly, we see that the combination of currents appearing in the quasi-probability Eq.~(\ref{eq:qpJ}) will clearly produce an overall negativity when integrated over time, thereby confirming the LG2 violation shown in Fig.~\ref{fig:lg2}. Furthermore, we have integrated Eq.~(\ref{eq:qpJ}) numerically and find an exact agreement with the calculation of Section \ref{sec:corecalc}, Eq.~(\ref{eqn:cohquas}) thereby providing an independent check of this result. 
\begin{figure}
	\subfloat[]{{\includegraphics[height=5.2cm]{fig3a.png}}}%
	%\qquad
	\hspace{5mm}
	\subfloat[]{{\includegraphics[height=5.2cm]{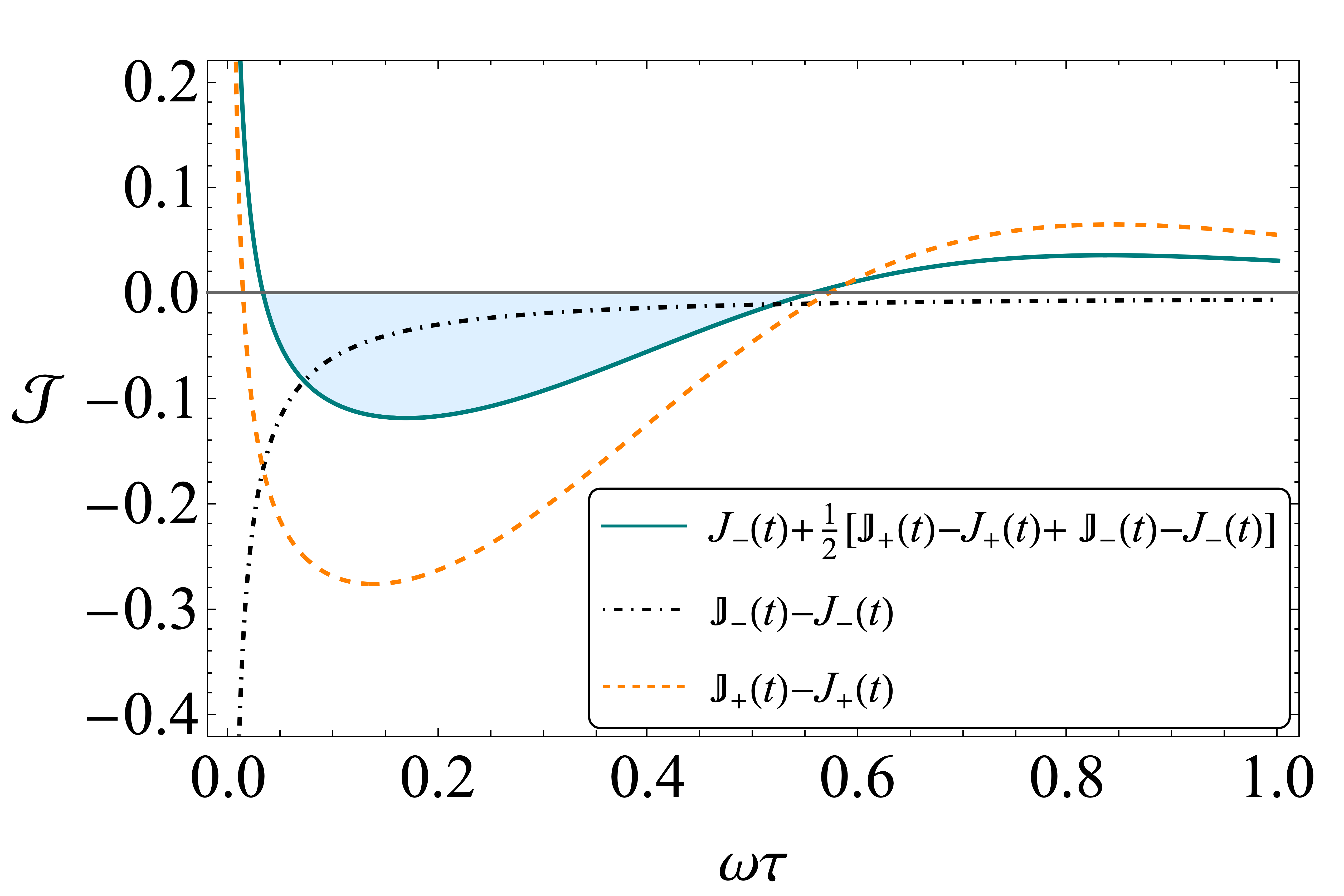}}}
	\caption[Analysis of LG2 violations using probability currents]{Plot (a) shows the post-measurement quantum currents $J_\pm(t)$, and their classical analogues $\mathbbmsl{J}_\pm(t)$, normalised to $\mathcal{J}=\tfrac{J}{\omega}$, for the coherent state with $x_0=0.55$, $p_0=-1.925$. Plot (b) shows their combination  appearing in the time derivative of $q(-,+)$, Eq.~(\ref{eq:qpdiffs}).}%
	\label{fig:currs}%
\end{figure}

It is also convenient to explore the currents in the small time limit. This is done in App \ref{app:tJ} and gives a clear analytic picture of the departures from classicality. Since these expressions are valid for any initial state they could provide a useful starting point in the search for other initial states giving  LG2 violations larger than the somewhat modest violations found here.

\subsection{Bohm trajectories}
\label{subsec:bohm}
\begin{figure}
	\begin{center}
	\includegraphics[height=9.4cm]{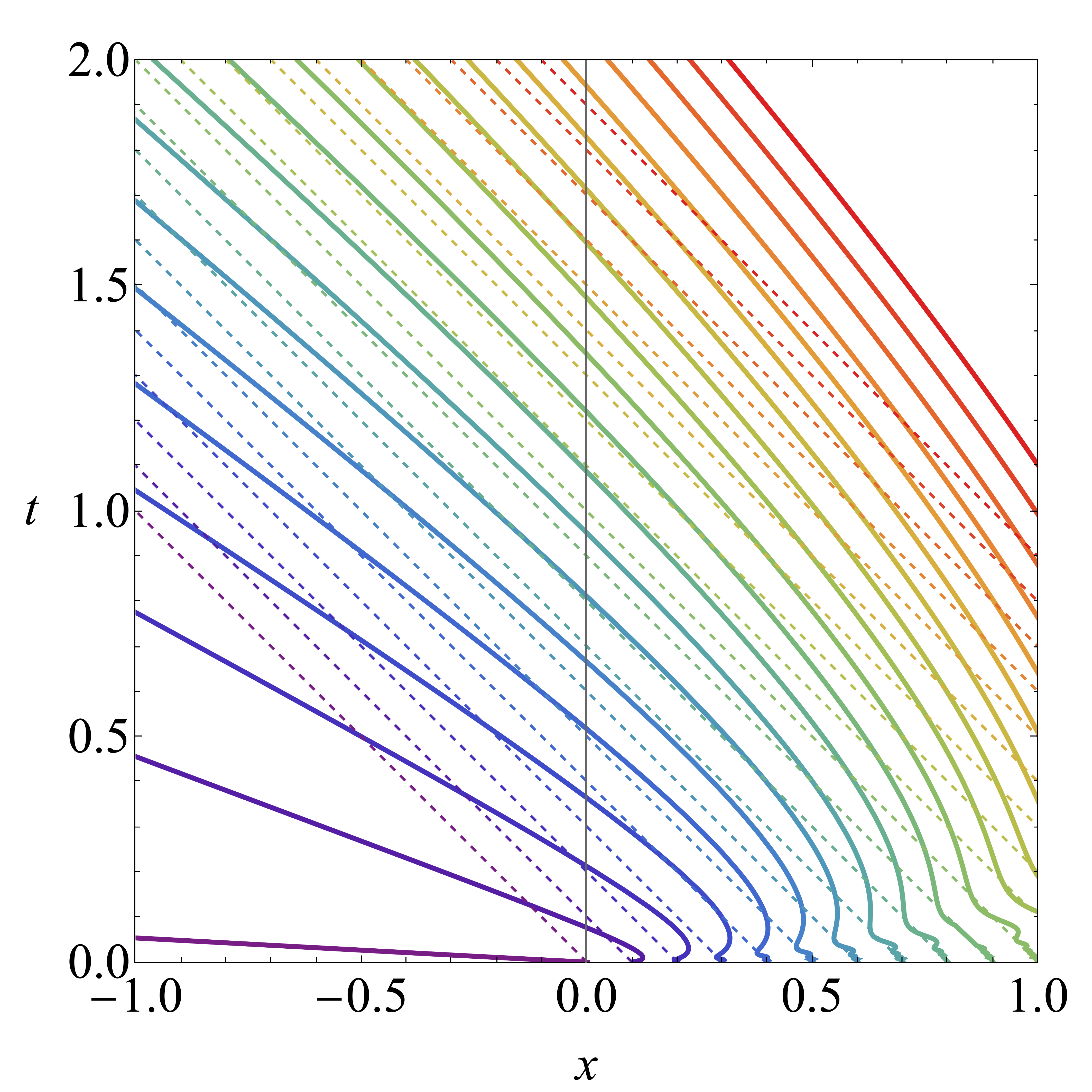}
	\end{center}
	\caption[Bohm trajectories showing diffraction in time effects]{The Bohm trajectories associated with the Moshinsky function, $\langle x|e^{-iHt}\theta(\hat x)|p\rangle$ with $p = -1$.  The equivalent classical paths are shown dotted.}%%
	\label{fig:bohmMosh}
\end{figure}
To give a visual demonstration of how the measurements influence motion in a way to lead to LG2 violations, we now calculate and plot the de Broglie-Bohm trajectories~\cite{bohm1952,bohm1952a,debroglie1927}.

As noted earlier, the Moshinsky function underlies the behaviour of the quasi-probability for these measurements, so we initially examine the Bohm trajectories for this scenario.  Using Moshinsky's calculation (free particle dynamics), we calculate the quantum-mechanical current $J_M(x,t)$, which we then use in the guidance equation for Bohm trajectories,
\begin{equation}
\dot{x}(t)=\frac{J_M(x,t)}{\lvert M(x,p,t)\rvert^2},
\end{equation}
which we proceed to solve numerically.

In Fig.~\ref{fig:bohmMosh}, we plot the trajectories for a state initially constrained to the right-hand side of the axis, with a left-ward momentum, with classical trajectories shown dotted.  

From this we see two distinct phases of deviation from the classical result.  Initially the trajectories rapidly exit the right hand side, with a negative momentum larger than in the classical case, an anti-Zeno effect \cite{kaulakys1997}.  After a short while, a Zeno effect \cite{turing2001, misra1977} happens, and the trajectories bend back relative to the classical trajectories, staying in the right hand side longer than in the classical case.  We will see both of these behaviours at play in the case studied in this chapter.

Using the expressions for the chopped current Eq.~(\ref{eq:chopJ}) and chopped wave-function Eq.~(\ref{eq:chopwav}), we can write the guidance equation for the harmonic oscillator case,
\begin{equation}
	\dot x (t) = \frac{J_{\pm}(x,t)}{\lvert \phi_\alpha^{\pm}(x,t)\rvert^2},
\end{equation}
which we again solve numerically.

In Fig.~\ref{fig:bts}, we show the Bohm trajectories for the state with $x_0=0.55$, $p_0=-1.925$, initially found on the right hand side of the well.  This corresponds to the behaviour of the current $J_+(x,t)$ from the previous section.  Looking at the zoom of the trajectories in Fig.~\ref{fig:bts}(b), we can observe the same behaviour that is seen in the Moshinsky case -- initially an anti-Zeno effect, during which the trajectories exit faster than they would classically, followed by a Zeno effect a short while later, where trajectories exit more slowly than in the classical case.

This lines up with the behaviour of $\mathbbmsl{J}_+(t)-J_+(t)$ displayed in Fig.~\ref{fig:currs}(b), and is hence a representation on the trajectory level of the source of the LG2 violations.

\begin{figure}
	\begin{center}
	\subfloat[]{{\includegraphics[height=7.2cm]{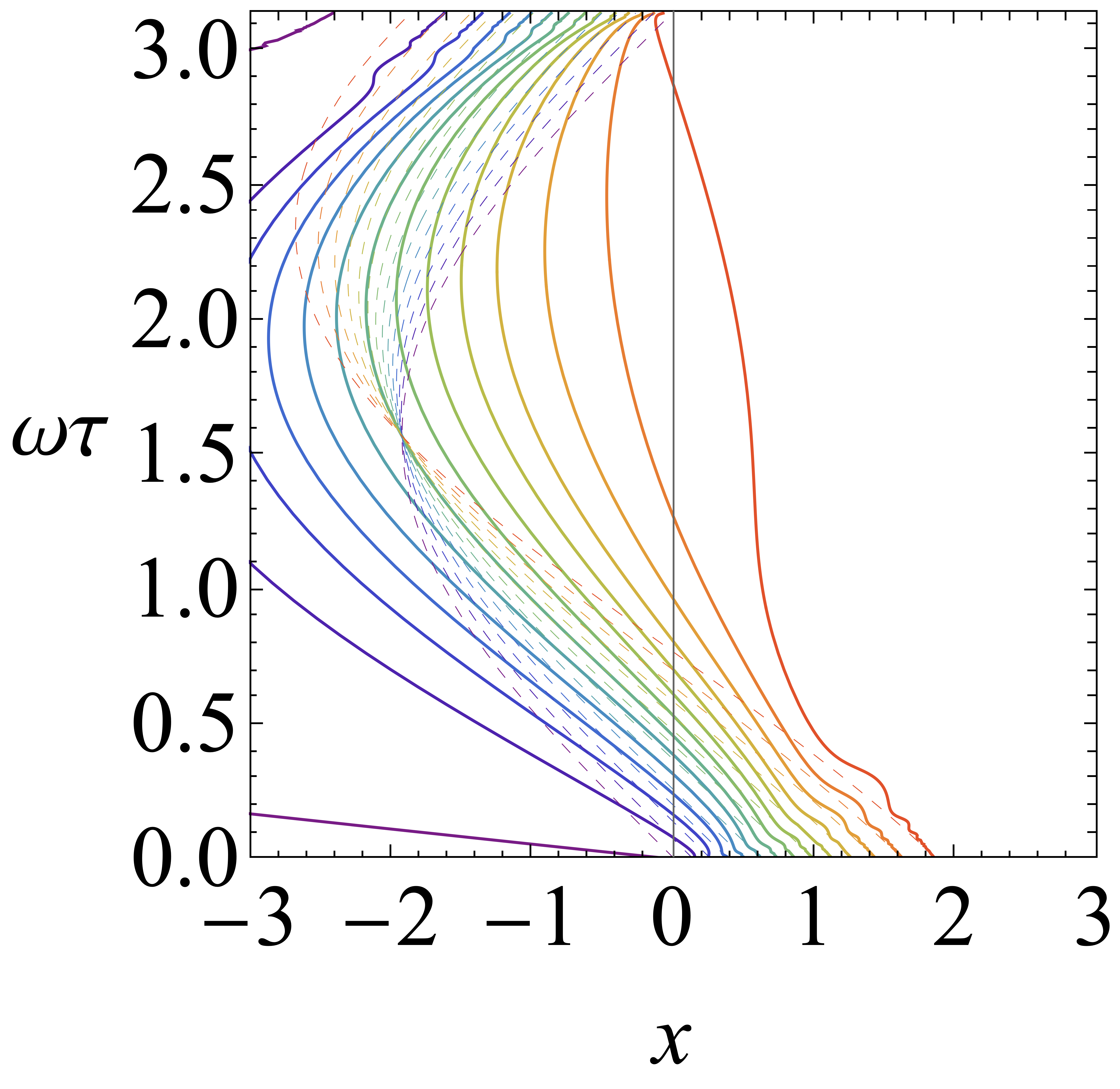}}}%
	%\qquad
	\hspace{5mm}
	\subfloat[]{{\includegraphics[height=7.2cm]{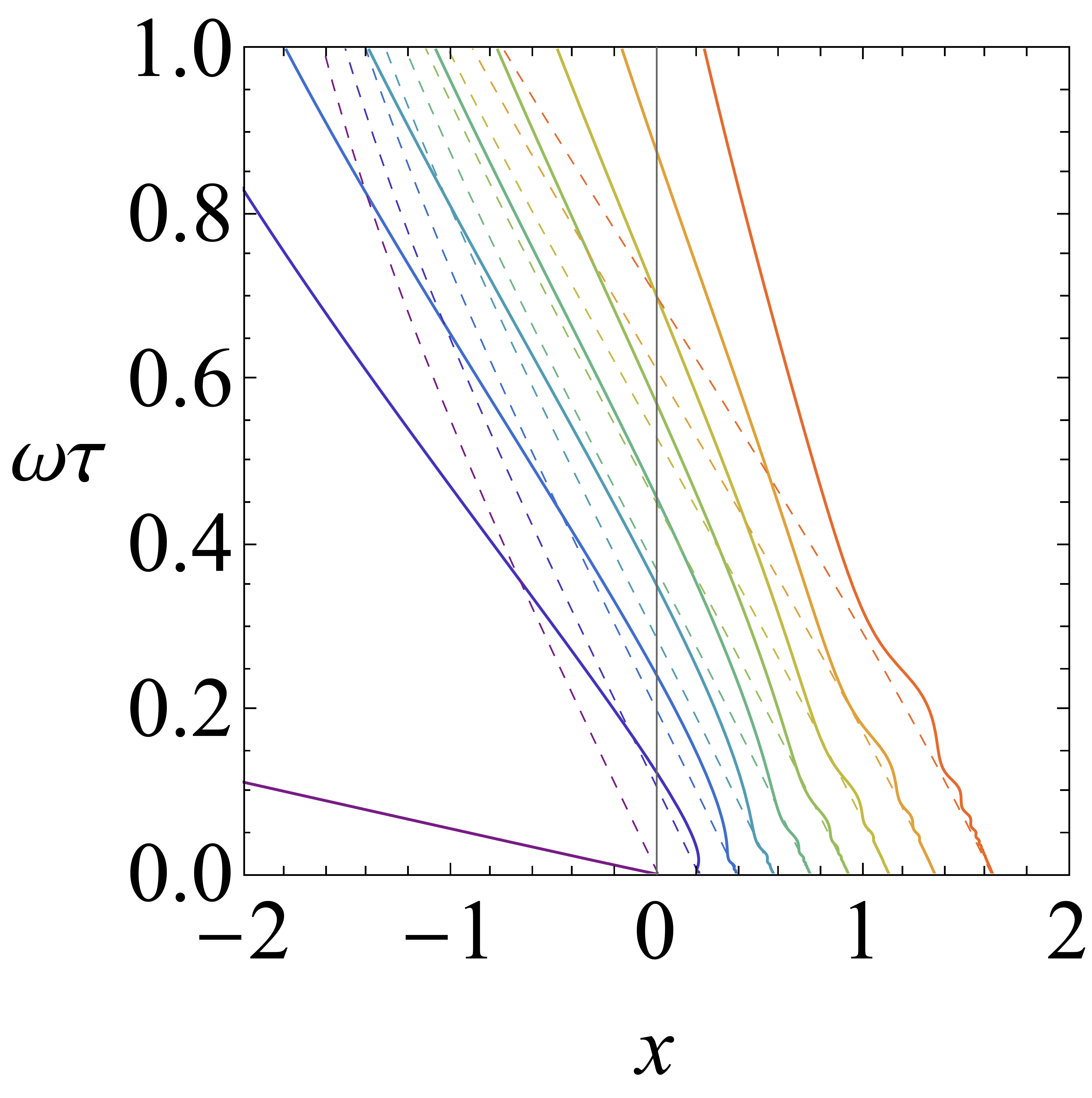}}}
	\end{center}
	\caption[The Bohm trajectory behaviour associated with LG2 violations]{The Bohm trajectories for case of the particle being initially found on the right hand side of the axis.  In (a), trajectories are separated such that two adjacent lines bound the evolution of $6.67\%$ of the probability density, and in (b) they bound $10\%$ of the probability density.}%
	\label{fig:bts}%
\end{figure}

\subsection{Wigner Function Approach}
Another way to understand the LG2 violations, is within the Wigner representation \cite{wigner1932a, hillery1984a, tatarskii1983a, case2008, halliwell1993a}.  Since coherent states have non-negative Wigner functions, the source of MR violation lies in the non-gaussianity of the Wigner transform of the operators describing the measurement procedure.  We calculate these transformations in Appendix ~\ref{app:wig}, which 
allows us to write the quasi-probability as a phase-space integral,
\begin{equation}
	q(s_1,s_2)=\int_{-\infty}^{\infty}\mathop{dX}\int_{-\infty}^{\infty}\mathop{dp}f_{s_1, s_2}(X,p)
\end{equation}
with the phase-space density $f_{s_1, s_2}(X,p)$ given by
\begin{equation}
\label{eq:wigfin}
f_{s_1, s_2}(X,p)=\frac12 W_{\rho}(X,p)\left(1+\Re\erf\left(i(p-p_0)+s_1 X\right)\right)\theta(s_2 X_{-\tau}),
\end{equation}
where $W_\rho(X,p)$ is the Wigner function of the initial state. In Fig.~\ref{fig:wig}, we plot this phase-space density $f_{-+}(X,p)$ for the state $x_0=0.55$, $p_0=-1.925$ and $\omega\tau=0.55$.  To make it clear that this integrates to a negative number, we numerically determine the marginals $f_{-+}(p) = \int_{-\infty}^{\infty}\mathop{dX}f_{-,+}(X,p)$ and likewise $f_{-+}(X)=\int_{-\infty}^{\infty}\mathop{dp}f_{-,+}(X,p)$, plotting them as insets in Fig.~\ref{fig:wig}.  It is clear from a simple inspection of these marginals that they will integrate to a negative number.

In Appendix~\ref{app:wig}, we plot the intermediate result Eq.~(\ref{eq:wigint}), where it is apparent how the choice of second measurement hones in on the negativity introduced by the initial measurement, ultimately leading to the LG2 violations in Section~\ref{sec:LGresults}.

\begin{figure}
	\begin{center}
	\includegraphics[height=9.4cm]{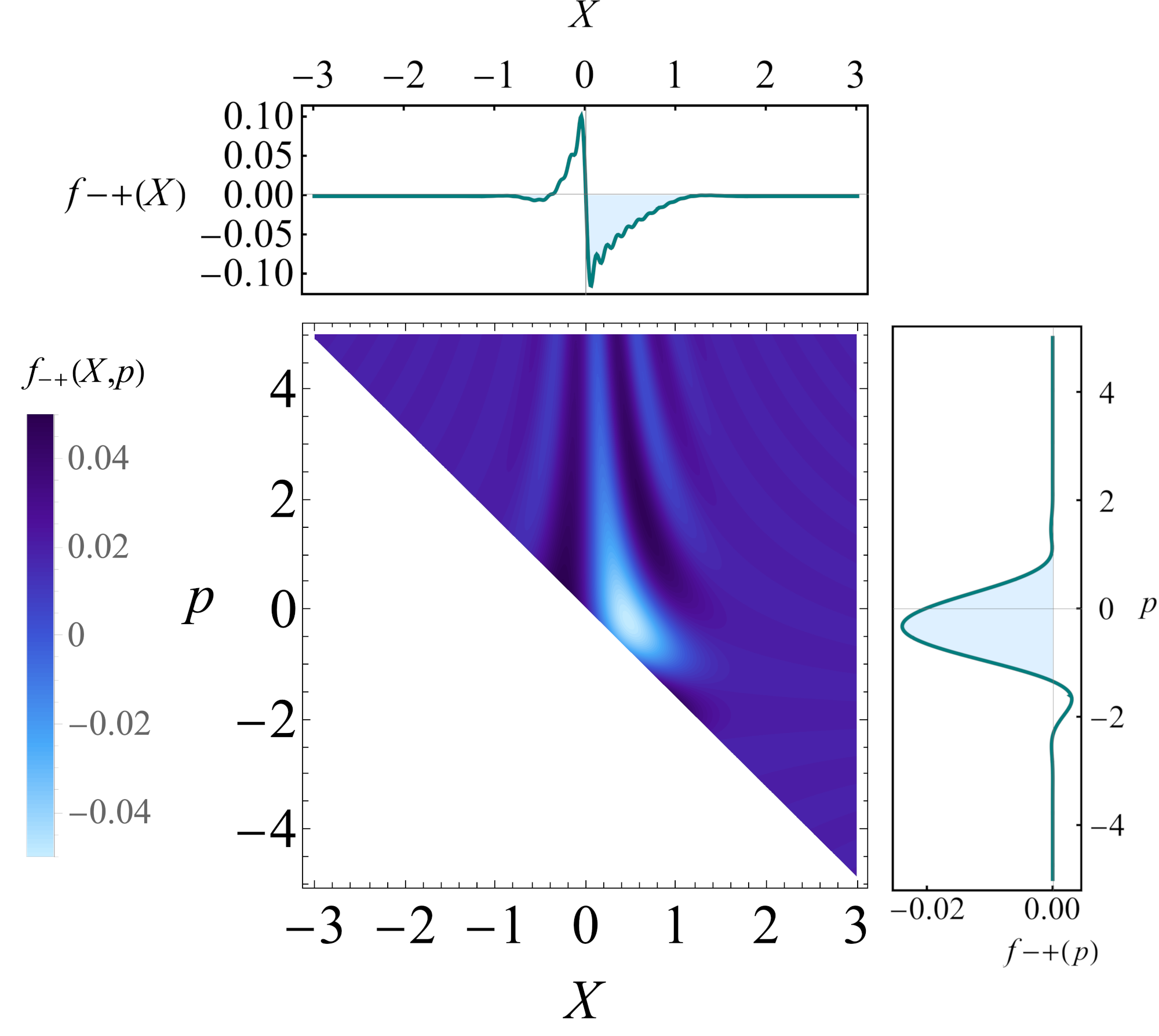}
	\end{center}
	\caption[Wigner function analysis of LG2 violations in phase-space]{We plot the phase-space density $f_{-,+}(X,p)$, in the case corresponding to an LG2 violation of $-0.113$. Inset are the $X$ and $p$ marginals, shaded over regions of negativity.}%
	\label{fig:wig}%
\end{figure}
\section{Modified Frameworks}
\label{sec:mf}
In this section, we broadly generalise our analysis, finding larger violations using the Wigner variant of the LG2s, and by analysing different measurements, and extending the analysis to squeezed and thermal coherent states.
\subsection{Achieving larger violations using the Wigner LG2 inequalities}
\label{subsec:wiglg}
Given the modest size of the LG violations obtained for a single gaussian, compared to the L{\"u}ders bound, it is natural to ask if there are modified situations in which larger violation can be obtained. One way of doing this is to examine slightly different types of inequalities known as the Wigner-Leggett-Garg (Wigner LG) inequalities \cite{saha2015, naikoo2020}. For the two-time case these arise as follows. The quasi-probability is readily rewritten as
\bea
q(s_1,s_2) &=& {\rm Re} \langle ( 1 - P_{-s_2} (t_2) ) ( 1 - P_{-s_1} (t_1) ) \rangle
\nonumber \\
&=& 1 - \langle P_{-s_1} (t_1)  \rangle -  \langle P_{-s_2} (t_2)  \rangle  + q(-s_1, -s_2).
\eea
However, from a macrorealistic perspective, there is nothing against considering the similar quasi-probability
\beq
\label{eqn:qw}
q^W (s_1,s_2) = 1 - \langle P_{-s_1} (t_1)  \rangle -  \langle P_{-s_2} (t_2)  \rangle 
+ p_{12} (-s_1,-s_2),
\eeq
(where recall $p_{12}$ is the sequential measurement probability Eq.~(\ref{eq:seqprob}))
since the two are the same classically. The relation $q^W (s_1,s_2) \ge 0 $ is  a set of Wigner LG2 inequalities (recalling the factor of $\tfrac{1}{4}$ difference between an LG2 and a QP). It differs from the usual LG2 inequalities by the presence of interference terms, which can be positive or negative, which indicates that violations larger the usual L\"uders bound (on the QP) of $- \frac{1}{8}$ might be obtained. The difference between them from an experimental point of view is that the original quasi-probability is measured from three different experiments (determining $\langle Q_1 \rangle$, $ \langle Q_2 \rangle $ and $C_{12}$) but the sequential measurement formula appearing in the Wigner version is measured in a single experiment.

To get a sense of how much larger the maximum violation might be, we take the simple case of one-dimensional projectors $P_{-s_1} (t_1) = | A \rangle \langle A |$ and 
$P_{-s_2}(t_2) = | B \rangle \langle B | $ and we find
 \beq
q^W (A,B) = 1 - | \langle \psi | A \rangle |^2  - | \langle \psi | B \rangle |^2 + 
| \langle \psi | A \rangle |^2  | \langle A | B \rangle |^2.
\label{qWAB}
\eeq
Simple algebra reveals the lower bound as  $ - \frac{1}{3}$, which is achieved with $ \langle A | B \rangle = 1 / \sqrt{3} $ and $ | \psi \rangle = (1 / \sqrt{6} ) ( | A \rangle + \sqrt{3} |B \rangle $). (It seems like that this is the most negative lower bound for all possible choices of projection but we have not proved this.)
This bound is significantly larger than the 
usual L\"uders bound on the QP of $ - \tfrac{1}{8}$.

In Section 3 we found that the quasi-probability $q(-,+)$ gives the greatest negativity so we compare with the corresponding Wigner expression,
\bea
q^W (-,+) &=&1 - \langle P_{+} (t_1)  \rangle -  \langle P_{-} (t_2)  \rangle 
+ p_{12} (+,-)
\nonumber \\
&=& q(-,+) + \left[ p_{12}(+,-) - q(+,-) \right],
\eea
where we have made use of Eq.~(\ref{eqn:qw}) for $q(-,+)$
We first note that Eq.~(\ref{eq:qpdiffs}) may be written $q(-,+) = p_{12} (-,+) + I$, where $I$ denotes the interference terms (i.e. the difference between the classical and quantum currents, the second term in Eq.~\ref{eq:qpdiffs}).
The analogous relations for $q(+,-)$ is readily derived and we find $ q(+,-) = p_{12}(+,-) - I $. (The difference in sign is expected on general grounds \cite{halliwell2016b}). We thus find
\beq
q^W (-,+) = p_{12} (-,+) + 2 I,
\eeq
so the interference term producing the violations is twice as large as the one in $q(-,+)$.

The computation of $p_{12} (-,+)$ can be carried out by integrating the chopped current $J_-(t)$, Eq.~(\ref{eq:chopcurr}) in Section 4.  Using the maximally violating state found in Section~\ref{sec:LGresults}, we find a largest violation of $-0.0881$, approximately three times larger than the standard LG2 violation, as well as a larger fraction of the conjectured Wigner LG2 L{\"u}ders bound of $-\frac{1}{3}$.  This is plotted alongside the standard LG2 in Fig.~\ref{fig:wigqp}, where the violation is both larger in magnitude, and present for a larger range of measurement intervals.

\begin{figure}
	\begin{center}
	\includegraphics[height=5.4cm]{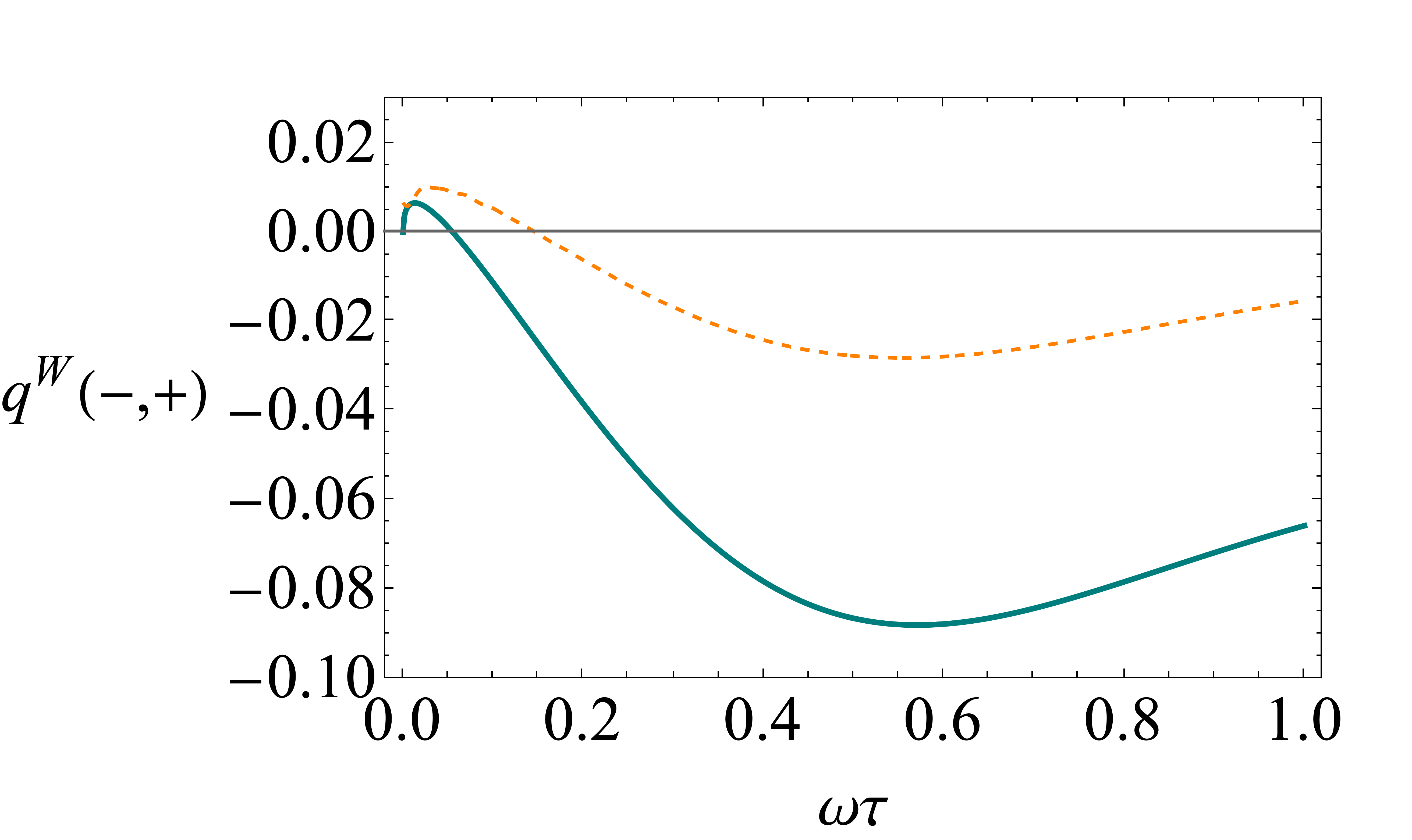}
	\end{center}
	\caption[Larger violations achieved using the Wigner LG2 inequalities]{Plot of $q^W(-,+)$ for the coherent state $x_0=-0.55$, $p_0=-1.925$,  as well as the standard quasi-probability (dashed), i.e. the LG2 in Fig.~\ref{fig:lg2}(b) times a factor of $\frac14$.}%
	\label{fig:wigqp}%
\end{figure}

%In fact, from the plot of the currents in Fig.~(\ref{fig:currs}) it is readily seen that this current is essentially zero for most of the range of interest which means that $p_{12} (-,+)$ will be approximately zero. It follows that we expect a Wigner LG2 violation approximately twice as large as the usual LG2 violation. 
%It will however be a  smaller fraction of the maximal Wigner LG2 violation, but it might be argued that what is most important here in the search for non-classical effects is the degree of departure for a genuine probability lying between $0$ and $1$ and the Wigner expression gives the larger departure.

As an aside, we note an interesting aspect of Eq.~(\ref{qWAB}), which is that the last term, corresponding to the sequential measurement probability, factors in two parts. This factoring will also hold for for more general projections at $t_2$ as long as the projection at $t_1$ is one-dimensional. This may have some advantages in terms of meeting the non-invasiveness requirement on the measurements. 
It seems plausible that one could find macrorealistic arguments implying that the sequential measurement probability factors. Then the first factor is the probability of finding $ |A \rangle$ in an initial state $| \psi \rangle $ and the second factor is the probability of finding $|B \rangle$ when the systems is prepared in state $|A\rangle$. These quantities could therefore be obtained in two different experiments with two different preparations with just a single measurement in each, for which there is no issue with invasiveness.

The other obvious way of getting larger violations is to consider von Neumann measurements, which involves making finer-grained measurements than the simple dichotomic ones used here and then coarse graining the probability to compute the correlators \cite{pan2018,dakic2014, budroni2014,wang2017,kumari2018}.  For example, one could make measurements onto three regions of the $x$-axis, $ x<0$, $ 0 \le  x \le L$ and $ x> L$ at the first time and then coarse-grain the two-time probabilities into probabilities for the usual coarse graining $x<0$ and $x>0$.
This produces extra interference terms which can enhance the violations. For the LG3 inequalities, von Neumann measurements can produce violations up to the algebraic maximum of $-2$. For the LG2 inequalities, the enhancement is smaller since there is only one correlator and the LG2 violations can be no more than $-1$. This corresponds to $-\frac{1}{4}$ in the quasi-probability, which we see is not as big as the violation of $ - \frac{1}{3}$  that can be produced produced by the Wigner LG2.

\subsection{Coherent state projectors}
\label{subsec:cohproj}
It is useful to know what else may be possible beyond using $\theta(\hat x)$ projectors.  We note investigations \cite{bose2018} into smoothed $\theta(\hat x)$ measurements, showing LG violations persist under smoothing of measurements up to the characteristic length-scale of the oscillator \cite{mawby2022}.  Modular variables such as $\cos(k\hat x)$ have also been investigated, and readily produce significant LG violations \cite{asadian2014}. These examples show that the LG violations are not due to the sharpness of projective measurements with $\theta(\hat x)$.  

In this section we will look at tests of macrorealism using coherent state projectors, which are interesting since they leave the post-measurement state gaussian. Projectors may be realised through heterodyne measurements \cite{yuen1978,yuen1983, dariano2003}.  Even though coherent states are not orthogonal, through considering a coherent state projector and its complement, it may be possible to implement an INRM protocol. This would lead to the two projectors $P_+=\ketbra{\beta}$ and $P_-=\id -\ketbra{\beta}$, defining a dichotomic variable in the usual way through Eq.~(\ref{eq:proj}). The quasi-probability is given by 
\begin{equation}
q(+,+)=	\Re \ev{e^{iHt_2}\ketbra{\beta_2}e^{-iH\tau}\ketbra{\beta_1}e^{-iHt_1}}{\alpha},
\end{equation}
where we make two simplifying observations.  Firstly,  all the time evolution may be absorbed into the measurement projectors.  Secondly, without loss of generality, we work with $\alpha=0$, where the change in phase-space location may be absorbed into $\beta_1$ and $\beta_2$.  It is hence entirely equivalent to analyze
\begin{equation}
q(+,+)=\Re \braket{0}{\gamma_1}\!\!\braket{\gamma_1}{\gamma_2}\!\!\braket{\gamma_2}{0},
\end{equation}
with the relation $\gamma_i=e^{-i\omega t_i}\beta_i-\alpha$.
The overlap between two coherent states is given by
\begin{equation}
\braket{\beta}{\alpha}=e^{-\frac12(\abs{\alpha}^2+\abs{\beta}^2-2\alpha \beta^*)}, 
\end{equation}
and we readily find
\begin{align}
q(+,+)&=\exp(-\lvert\gamma_1\rvert^2 -\lvert \gamma_2\rvert^2)\Re \exp(\gamma_1 \gamma_2^*),\\
q(+,-)&=\exp(-\lvert\gamma_1^2\rvert)\left(1-\Re\exp(\gamma_1 \gamma_2^* -\lvert\gamma_2\rvert^2)\right),
\end{align}
where $q(-,+)$ is found by a relabelling, and $q(-,-)$ does not lead to any violations.
To determine the largest violations, it is useful to note these quasi-probabilities depend only on the magnitude of $\gamma_1$ and $\gamma_2$, and the phase difference between them.  

In $q(+,+)$, $\lvert \gamma_1\rvert$ and $\lvert \gamma_2 \rvert$ appear in the same way, so we set them to be equal, and find a largest violation of $-0.0133$ at $\gamma_1=1.55$, $\gamma_2=1.55 e^{-1.047 i}$, which is about $10\%$ of the maximal violation.

For $q(-,+)$, since the violation is aided by the negative sign on $\Re \exp(\gamma_1 \gamma_2^* -\lvert\gamma_2\rvert^2)$, it is easy to see the largest violation will occur when both $\gamma_1$ and $\gamma_2$ are purely real.  We readily find that the largest violation is approximately $-0.1054$ with $\gamma_2=\frac12\gamma_1=0.536$, which is about $84\%$ of the maximum.

Violations meeting the L{\"u}ders bound may be achieved if a superposition state is chosen which satisfies Eq.~(\ref{eq:maxQP}.  The superposition state $\ket\psi = -\ket{\beta_1} -\ket{\beta_2}$ is properly normalized and gives a maximal violation for $q(+,+)$ if the coherent states are chosen so that $\braket{\beta_1}{\beta_2} = -\frac12$.  Similarly for $q(+,-)$, the state $\ket\psi = \ket{\beta_1} - \sqrt{3}\ket{\beta_2}$ leads to a maximal violation if we choose $\braket{\beta_1}{\beta_2}=\frac{\sqrt{3}}{2}$.
\subsection{Squeezed States}
The squeezed coherent state may be written~\cite{schleich2001},
\begin{equation}
\ket{\alpha,\zeta}=D(\alpha)S(\zeta)\ket0,
\end{equation}
where the squeezing operator is given by $S(\zeta)=\exp(\frac{1}{2} (\zeta^* \hat a^2-\zeta \hat a^{\dagger 2}))$.
While $S(\zeta)$ does not commute with the displacement operator $D(\alpha)$, there is a simple braiding relation, allowing us to write
\begin{equation}
	\ket{\alpha,\zeta}=S(\zeta)D(\beta)\ket0=S(\zeta)\ket\beta,
\end{equation}
with $\beta$ depending on both $\alpha$ and $\zeta$.  With the quasi-probability given by
\begin{equation}
q(+,+)=\Re \ev{\theta(\hat x)\theta(\hat x(t))}{\psi},	
\end{equation}
for $\ket\psi$ given by a squeezed coherent state. We can consider moving the $S(\zeta)$ in $\ket\psi$ onto each $\theta(\hat x)$ function, resulting in $S^\dagger(\zeta)\theta(\hat x)S(\zeta)$ twice.  Since the squeezing operator acts as a canonical transform, taking $\hat x$ and $\hat p$ into a linear combination of themselves, we have
\begin{equation}
S^\dagger(\zeta)\theta(\hat x)\theta(\hat x (t))S(\zeta)=\theta(a\hat x + b\hat p)\theta(c\hat x + d\hat p ),
\end{equation}
for some $a,b,c,d$ that may depend on $t$. 
We now note that $a\hat x + b\hat p$ may be written as $\lambda(\hat x \cos(t')+\hat p \sin(t'))$ for some $\lambda> 0$ and some $t'$, and since the theta-function is invariant under scaling, we see that
\begin{equation}
	S^\dagger(\zeta)\theta(\hat x)\theta(\hat x (t))S(\zeta)=\theta(\hat x(t_1'))\theta(\hat x(t_2')).
\end{equation}
This means the QP for a squeezed coherent state is equal to the QP for some other coherent state $\beta$, with different measurement times $t_1'$, $t_2'$.  Hence the operation of squeezing will not increase the largest possible violation reported in Section~\ref{sec:LGresults}, although for certain states with sub-optimal violation, squeezing can increase the amount of violation. 
\subsection{Thermal Coherent States}
\begin{figure}
	\begin{center}
\includegraphics[height=5.3cm]{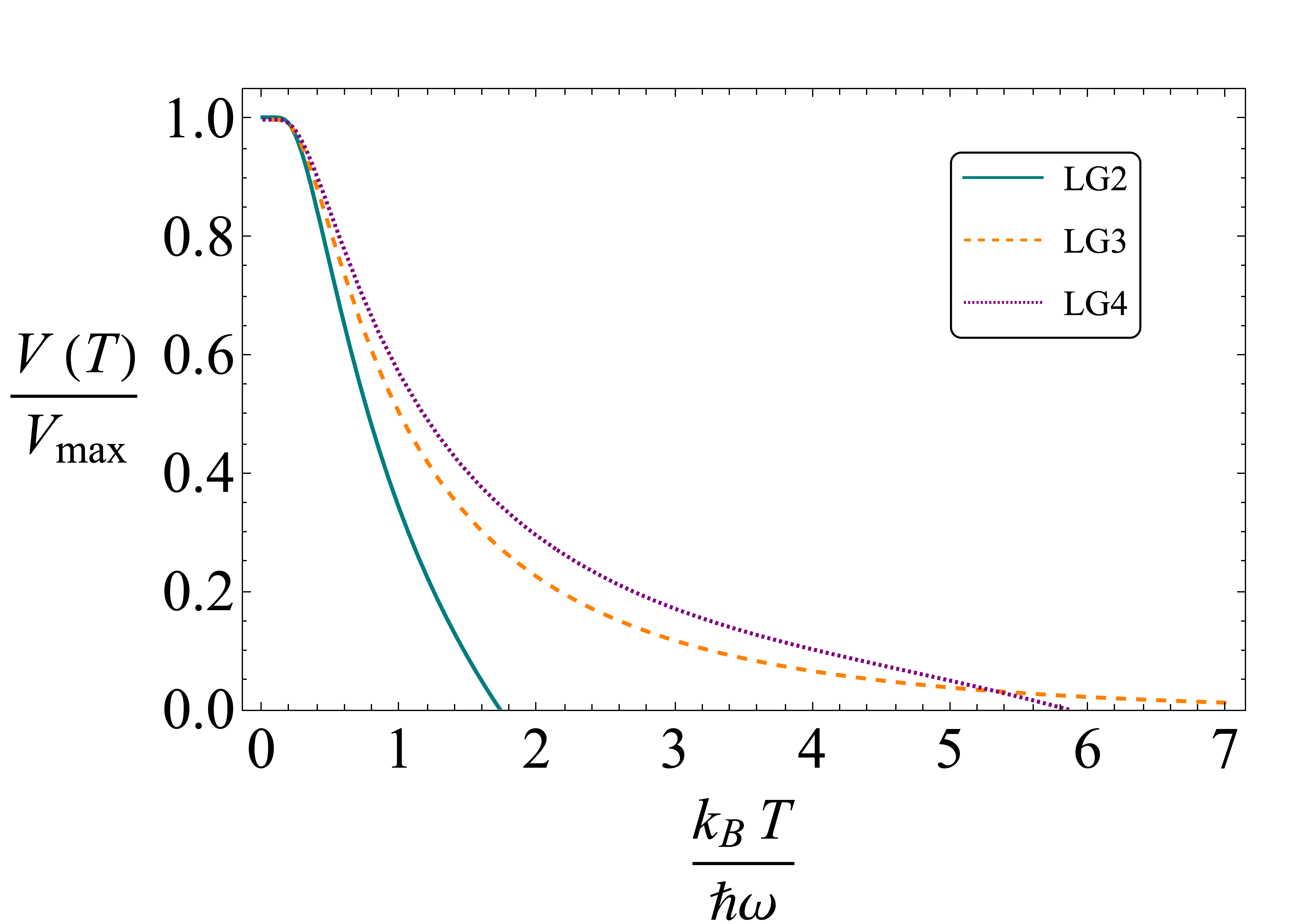}
\end{center}
	\caption[Classicalisation of LG violations through thermalisation]{The LG2, LG3 and LG4 violations are plotted as a fraction of their largest violation, for a state and time which realises $V_{\text{max}}$, with varying temperature.}%%
	\label{fig:thermLG}
\end{figure}
The thermal coherent state is produced by displacing a thermal mixed state, and at a temperature $T$ is given by
\begin{equation}
	\rho_{\text{th}}(\alpha, T)=\frac{1}{Z}\sum_{n=0}^{\infty}e^{-\frac{n\hbar\omega}{k_B T}} \ketbra{n,\alpha},
\end{equation}
where $k_B$ is the Boltzmann constant, $\ket{n,\alpha}$ are energy eigenstates displaced by $\alpha$ in phase-space~\cite{oz-vogt1991}. The partition function $Z$ is given by
\begin{equation}
Z=\frac{1}{1-e^{-\frac{\hbar \omega}{k_B T}}}.
\end{equation}
Since this state is a mixture, it is simple to update the calculation Eq.~(\ref{eqn:quassum}) to using this state, leading to

\begin{equation}
q(+,-)=-\frac{1}{Z}\Re\sum_{\ell=0}^{\infty}e^{-\frac{\ell\hbar\omega}{k_B T}} \sum_{n=0}^\infty e^{-i(n-\ell)\omega\tau}J_{\ell n}(x_1,\infty)J_{\ell n}(x_2,\infty)
\end{equation}
where the $J_{n\ell}$ matrices are given by Eq.~(\ref{eqn:wronskiapp}), except for the cases $n=\ell$, where they must be calculated explicitly.  A similar result may be calculated for the correlators using Eq.~(\ref{eqn:corr}), allowing the analysis of LG3 and LG4 inequalities.

In Fig.~\ref{fig:thermLG} using the states found in Section~\ref{sec:LGresults}, we plot the behaviour of the largest violation, as temperature is increased.  We see the violation persists up to temperatures $k_B T\approx \hbar \omega$, with some preliminary indication that LG3 and LG4 violations may be more robust against thermal fluctuations in the initial state.  After this, the violations sharply drop off, which is due to the thermal mixture $\rho_{\textrm{th}}(\alpha, T)$ containing more of the higher energy eigenstates, which we showed in Section~\ref{sec:higher} rapidly display only classical statistics.

\section{Summary}
\label{sec:sum}
We have undertaken a study of LG violations in the quantum harmonic oscillator for a dichotomic variable $Q = \text{sgn}(x)$ and for an initial state given by a coherent state and closely related states. In Section \ref{sec:corecalc}, building on our earlier work with energy eigenstates of the QHO in Chapter~\ref{chap:QHO1}, we showed how the quasi-probability, and hence the temporal correlators, may be expressed as a discrete infinite sum which is amenable to numerical analysis. We applied this analysis to the LG2, LG3 and LG4 inequalities and carried out parameter space searches. We found LG violations of magnitude 22\%, 28\% and 26\% of the maximum possible for the LG2, LG3 and LG4 inequalities, respectively, and gave the specific parameters for which these violations are achieved.  These violations appear to be robust under small parameter adjustments.  The LG2 violation in the case $x_0=0$ agrees with that reported in Ref.~\cite{halliwell2021}.

Our results are in significant disagreement with those presented in Ref.~\cite{bose2018}.  Their paper reports a maximal LG4 violation, for states with momenta up to three orders of magnitude larger than the states we found which yield a more modest LG violation.  We believe this is due to an error in the numerical study in their original paper.\footnote{Added note after acceptance of the thesis: the authors of Ref.~\cite{bose2018} have confirmed there was an error in their numerical code, which explains the discrepancy (private communication with S. Bose).}  Since this may well change the validity of the suggested experimental regimes, it is a significant discrepancy, however a later proposal by the authors is able to sidestep these issues \cite{das2022a}.  The numerical results in that paper are amenable to comparison with the analytic results derived in this work.

In Section \ref{sec:physmec} we sought a physical understanding of the mechanism producing the violations. We showed how to relate the quasi-probability (LG2) to a set of currents for projected initial states. We calculated and plotted these currents and also plotted their associated Bohm trajectories along with their classical counterparts. The plots showed the clear departures from classicality and give both a visual understanding and independent check of the LG2 violations described in Section \ref{sec:corecalc}.  We also provide a small-time expansion for the LG2s, which is valid for general states. We noted that the quantum effect producing the violations is essentially the diffraction in time effect first noted by Moshinsky~\cite{moshinsky1952,moshinsky1976}.

We explored the same issues from a different angle using the Wigner representation. The Wigner function of the initial coherent state is everywhere non-negative. We determined and plotted the Wigner function of the chopped initial state appearing in the quasi-probability. It has significant regions of negativity which are clearly the source of the LG2 violation.

In Section \ref{subsec:wiglg}, we extended our results to the slightly different Wigner LG inequalities, which are phrased in terms of the sequential measurement probability, and allow for larger violations, where we found a two-time violation three times greater in magnitude than the standard LG2s.  We also noted it is likely possible to increase the LG violations through the use of von Neumann measurements.    Due to in some cases the sequential measurement probability factorising into terms which may be measured in separate single-measurement experiments, we note a possible advantage of the Wigner LG2 inequalities in terms of meeting the non-invasiveness requirement.

We briefly noted in Section \ref{subsec:cohproj} that our work is readily generalized from pure projective measurements to smoothed step function projectors, gaussian projectors and modular variables such as $\cos ( \hat x) $. We also examined the LG2 inequality for the case in which both projectors are taken to be projections onto coherent states. We showed that decent violations are possible for an initial coherent state and that a maximal violation arises when the initial state is a superposition of two coherent states.

Finally we finished Section \ref{sec:mf} briefly discussing how the LG violations may be modified using families of states similar to a coherent state. We showed that the QP for any squeezed state is equal to the QP for some other coherent state, hence squeezing will not increase the largest violation found, however for a state with sub-optimal violation it may improve the violation.  We also considered a thermal initial state and estimated the degree to which thermal fluctuations may affect the degree of violation.

\begin{subappendices}
\section{Calculation of correlators}
\label{app:timeev}
The two-time quasi-probability for a coherent state with a generic position basis measurement $m(\hat x)$ is defined,
\begin{equation}
q(+,+)=\Re\ev{e^{\frac{i H t_2}{\hbar}}m(\hat x)e^{-\frac{i H \tau}{\hbar}}m(\hat x)e^{-\frac{i H t_1}{\hbar}}}{\alpha}.
\end{equation}
We are primarily interested in the case $m(\hat x) = \theta(\hat x)$, but what follows holds for more general $m(\hat x)$, e.g gaussian measurements.  Writing this in terms of the displacement operator, we have
\begin{equation}
q(+,+)=\Re\ev{D^{\dagger}(\alpha)e^{\frac{i H t_2}{\hbar}}m(\hat x)e^{-\frac{i H \tau}{\hbar}}m(\hat x)e^{-\frac{i H t_1}{\hbar}}D(\alpha)}{0}.
\end{equation}

Since the displacement operator is unitary, we have $D^\dagger(\alpha) D(\alpha)=\mathds{1}$. Hence if we may commute the two displacement operators to be neighbours, we will clearly reach a vast simplification of the calculation.

To make the exposition clearer, we consider splitting this expression into two states
\begin{align}
\ket{M(\alpha, t_1, \tau)}&=e^{-i\hat H \tau}m(\hat x)e^{-\frac{iHt_1}{\hbar}}D(\alpha) \ket0,\\
\bra{M(\alpha, t_2, 0)}&=\bra0 D^\dagger(\alpha)e^{\frac{i H t_2}{\hbar}}m(\hat x),
\end{align}
where we then have $q(+,+)=\Re \braket{M(\alpha, t_2, 0)}{M(\alpha, t_1, \tau)}$, where we have introduced the notation $\ket{M(\alpha, t, s)}$ to represent the coherent state measured with $m(\hat x)$ at time $t$, then evolved by time $s$.

Considering now the displacement operator acting to the left, we write $m(\hat x)D(\alpha)=D(\alpha)D^\dagger(\alpha)m(\hat x)D(\alpha)=D(\alpha)m(\hat x + x_\alpha)$, with $x_\alpha=\sqrt2 \Re \alpha$. We then have 

\begin{equation}
\ket{M(\alpha, t_1, \tau)} = e^{-\frac{i\omega t_1}{2}}e^{-i \hat H \tau}D(\alpha(t_1))m(\hat x + x_1)\ket 0.
\end{equation}
Using the standard result that $e^{-i H t}D(\alpha)e^{i H t}=D(\alpha(t))$, we can rewrite this as
\begin{equation}
\ket{M(\alpha, t_1, \tau)} =e^{-\frac{i\omega t_1}{2}}D(\alpha(t_2))e^{-i H\tau}m(\hat x + x_1)\ket0.
\end{equation}
This says that the post-measurement state is the evolution of the regular ground-state undergone a translated measurement, translated by the displacement operator, to a classical trajectory.  Proceeding similarly with the other term, we find
\begin{equation}
\bra{M(\alpha, t_2, 0)}=e^{\frac{i\omega t_2}{2}}\bra{0} m(\hat x + x_2)D^\dagger(\alpha(t_2)).
\end{equation}
Finally, contracting the two terms we are able to exploit the unitarity of $D(\alpha)$ to find
\begin{equation}
q(+,+)=\Re e^{\frac{i\omega \tau}{2}} \ev{m(\hat x + x_2)e^{-iH\tau}m(\hat x + x_1)}{0},
\end{equation}
\noindent
A calculation similar to that in Chapter~\ref{chap:QHO1} shows the quasi-probability is
\begin{equation}
q(+,+)=\Re e^{\frac{i\omega \tau}{2}}\sum_{n=0}^{\infty}e^{-i\omega \tau(n+\frac12)}\mel{0}{\theta(\hat x + x_2)}{n}\!\!\mel{n}{\theta(\hat x + x_1)}{0},
\end{equation}
and similarly for the other three components.  The matrix elements here are given by the $J_{mn}(x_1, \infty)=\mel{m}{\theta(\hat x + x_1)}{n}$ matrices from Appendix~\ref{app:wronski}.  The quasi-probability is then
\begin{equation}
\label{eqn:quassum}
q(+,+)=\Re \sum_{n=0}^\infty e^{-in\omega\tau}J_{0n}(x_1,\infty)J_{0n}(x_2,\infty)
\end{equation}
For $m\neq n$, the $J$ matrices are given by Eq.~(\ref{eqn:wronskiapp}). For the $n=m=0$ case, the integration is completed manually yielding
\begin{equation}
J_{00}(x, \infty)=\frac12(1-\erf(x)).
\end{equation}
Hence writing out the quasi-probability with $n=0$ case of the sum handled, we have
\begin{multline}
\label{eqn:cohquas}
q(s_1, s_2)= \frac{1}{4}\Bigg[1+ s_1 \erf(x_1) +s_2 \erf(x_2)+\\ s_1 s_2\Bigg( \erf(x_1)\erf(x_2)+ 4\sum_{n=1}^\infty\cos (n\omega \tau)J_{0n}(x_1,\infty)J_{0n}(x_2,\infty)\Bigg) \Bigg].
\end{multline}
Comparing to the moment expansion of the quasi-probability, we obtain the correlators
\begin{equation}
\label{eqn:corr}
C_{12}=\erf(x_1)\erf(x_2)+ 4\sum_{n=1}^\infty \cos (n\omega \tau) J_{0n}(x_1,\infty)J_{0n}(x_2,\infty).
\end{equation}
The infinite sum may be evaluated approximately using numerical methods, by summing up to a finite $n$.  This calculation matches the analytically calculated special case of $x_0=0$ given in Ref. \cite{halliwell2021}, and while it is possible to make an analytic calculation for the more general case, it turned out not to be as useful as the numerical evaluation.  The exact result is found in terms of Owen-T functions, but for complex arguments, which rendered the behaviour chaotic when computed \cite{owen1965}.

The only source of non-classicality here lies in the infinite sum, and  with the $J_{0n}$ matrices expressed in terms of the oscillator eigenstates, this means there is a double exponential suppression $e^{-x_1^2 -x_2^2}$ of this non-classical term. This corresponds to the requirement that at least two measurements must make a significant chop of the state, which fits the intuition that without significant chopping, there is no mystery attached to which side of the axis the particle may be found on.

\section{Determination of LG violations}
\label{app:LGresults}

\begin{figure}
	\begin{center}
	\subfloat[]{{\includegraphics[height=5.8cm]{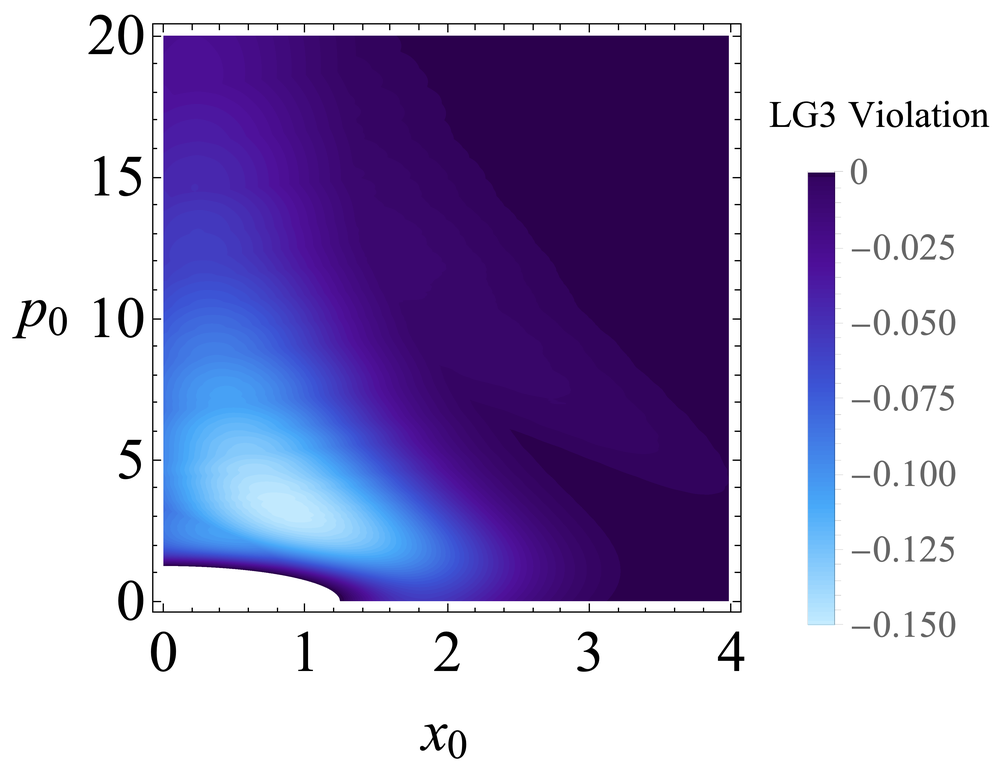}}}%
	%\qquad
	\hspace{5mm}
	\subfloat[]{{\includegraphics[height=5.8cm]{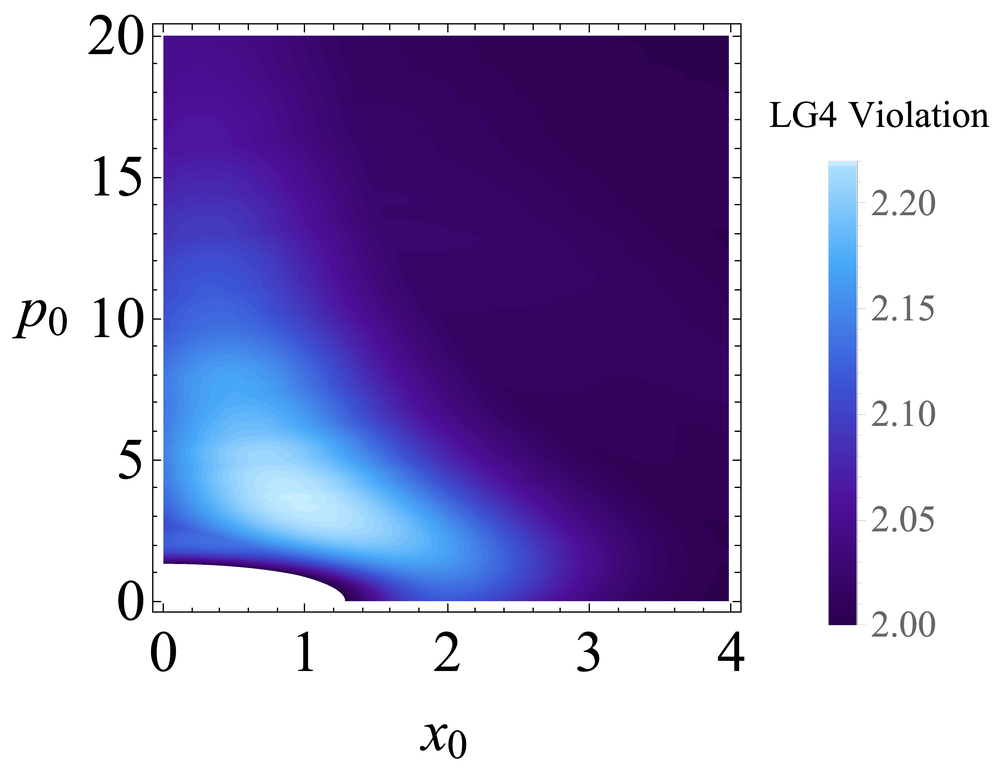}}}
	\end{center}
	\caption[Largest LG3 and LG4 violations across coherent state parameter space]{Plot (a) shows the greatest possible violations for the LG3 inequalities, plot (b) shows the same for the LG4 inequalities.}%
	\label{fig:param23}%
\end{figure}

In this appendix we fill in the details of the LG violations reported in Section \ref{sec:LGresults}.  Recall the variable parameters of the problem are $x_0$ and $p_0$, and the equal time spacing parameter~$\tau$.  We also note, that it is sufficient to explore a single quadrant of the $x_0, p_0$ parameter space, which we take to be the positive quadrant.  For states with $x_0<0$, the quasi-probability may be recovered by inverting the sign of $s_1$.  Likewise for states with $p_0<0$, by allowing the interval between measurements to take values $0<\tau\leq 2\pi$, their behaviour is included in the positive quadrant.  This same argument applies to the LG inequalities in general, where their different permutations correspond to flips of measurement signs.

To represent the three-dimensional parameter space, for each $x_0$, $p_0$, we use numerical minimisation over $0<\tau\leq 2\pi$ to find the largest possible violation for that coherent state.  In this numerical procedure, we take the largest possible violation from all of the inequalities involved.

The results of this parameter space search for the LG3 and LG4 inequalities is shown in Fig.~\ref{fig:param23}, which shows similar behaviour to the LG2 inequality parameter space behaviour in Fig.~\ref{fig:lg2}.  As more measurement intervals are included in the LG tests, a broader range of states lead to violation.

LG tests on QHO coherent states are mathematically equivalent to LG tests on the pure ground state $\ket0$, which are we found to have violations only when at least one of the $\theta(\hat x)$ measurements involved is displaced from the axis by order of magnitude $1$.  Hence at the centre of each of these parameter space plots is the region where the coherent state is too similar to the ground state to have any LG violation.
\begin{figure}
	\begin{center}
	\subfloat[]{{\includegraphics[height=5.2cm]{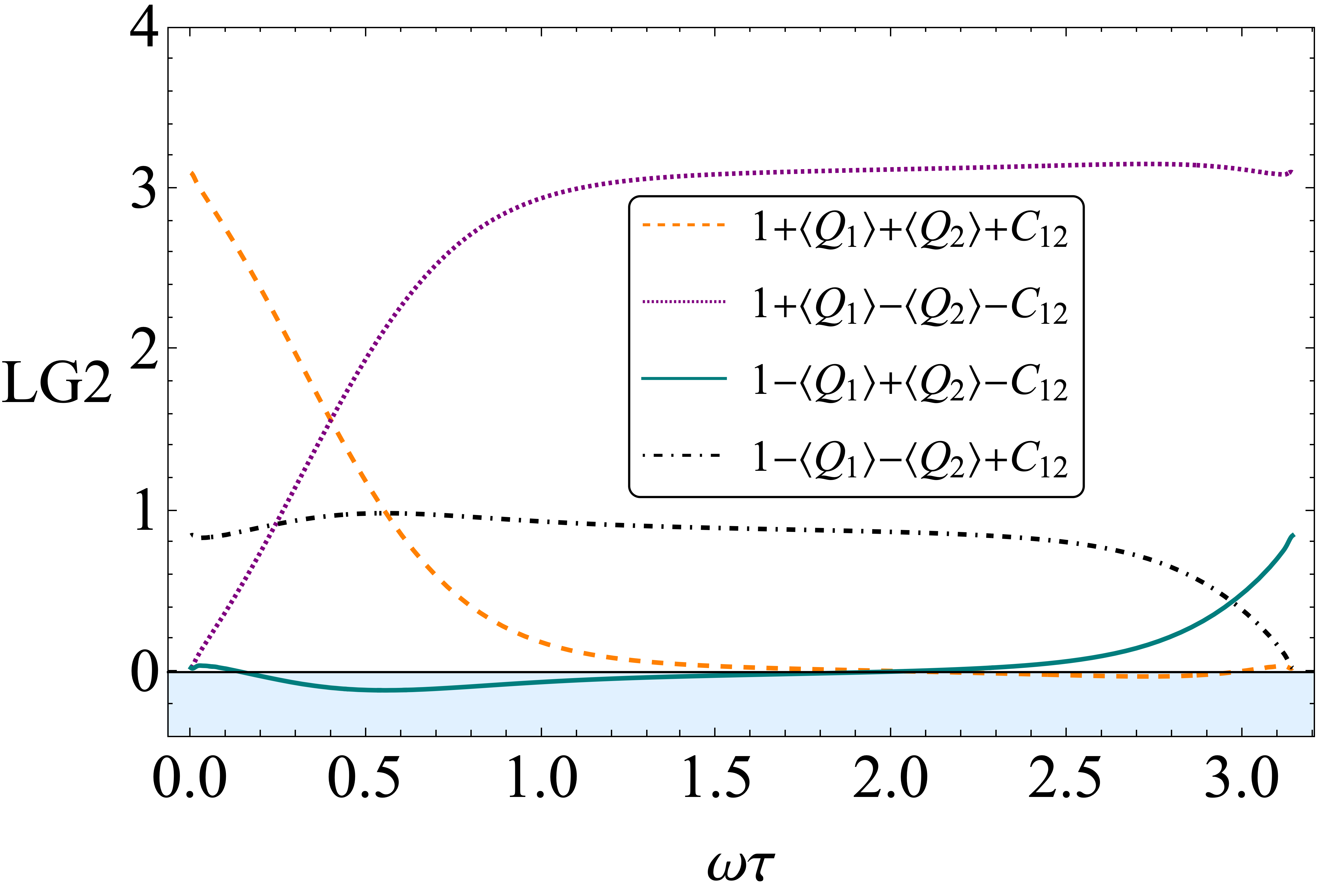}}}%
	%\qquad
	\hspace{5mm}
	\subfloat[]{{\includegraphics[height=5.2cm]{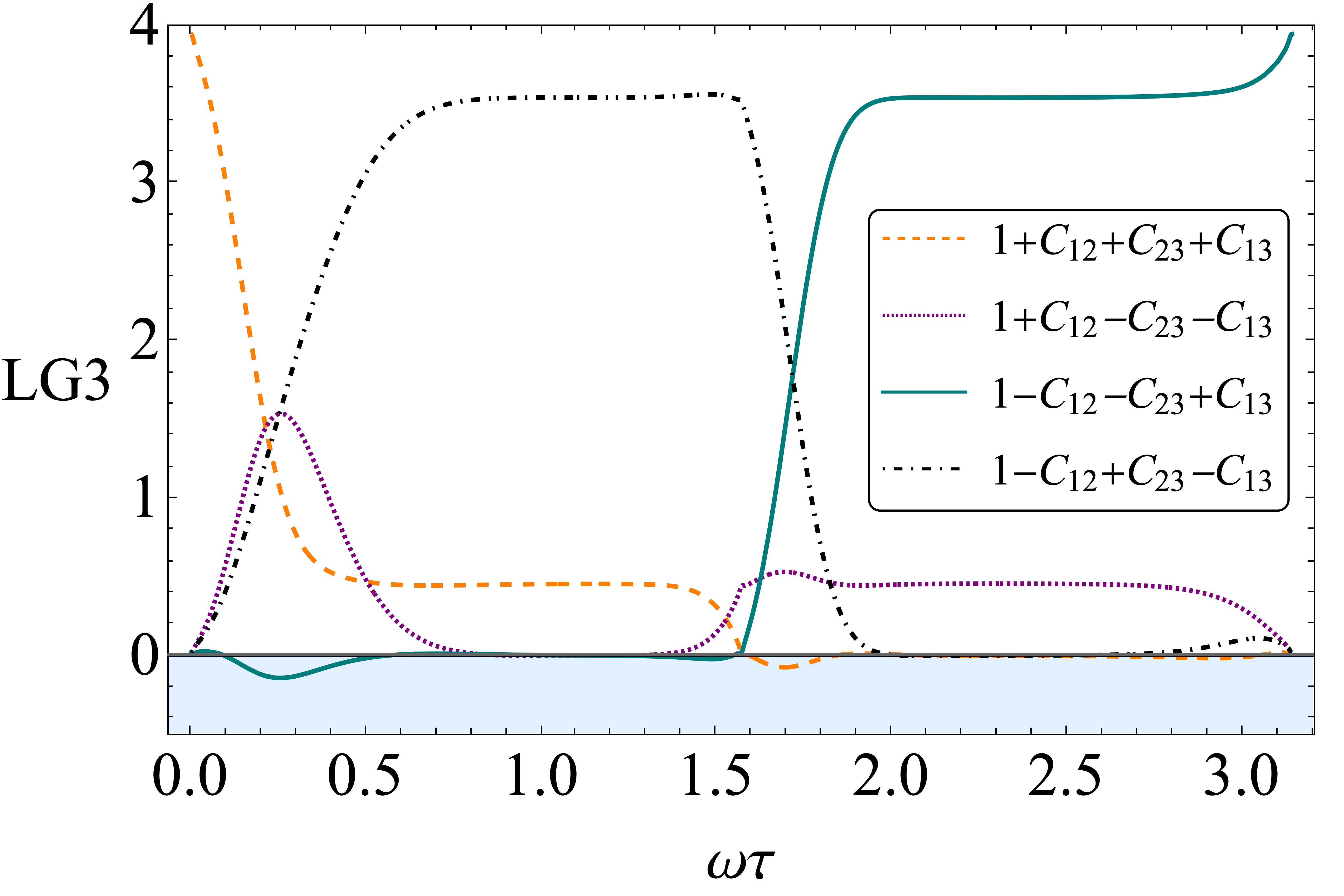}}}
	\end{center}
	\caption[Behaviour of all four LG2 and LG3 inequalities for their largest violations]{Plot (a) shows the four LG2 inequalities for the state with $x_0=0.55$, $p_0=-1.925$, which reaches a largest violation of $-0.113$ at  $\omega \tau = 0.555$.  Plot (b) shows the four LG3 inequalities for the state with $x_0=0.859$, $p_0=-3.317$, reaching a largest violation of $-0.141$ at $\omega\tau=0.254$. }%
	\label{fig:qpt}%
\end{figure}
\begin{figure}
	\begin{center}
	\subfloat[]{{\includegraphics[height=5.3cm]{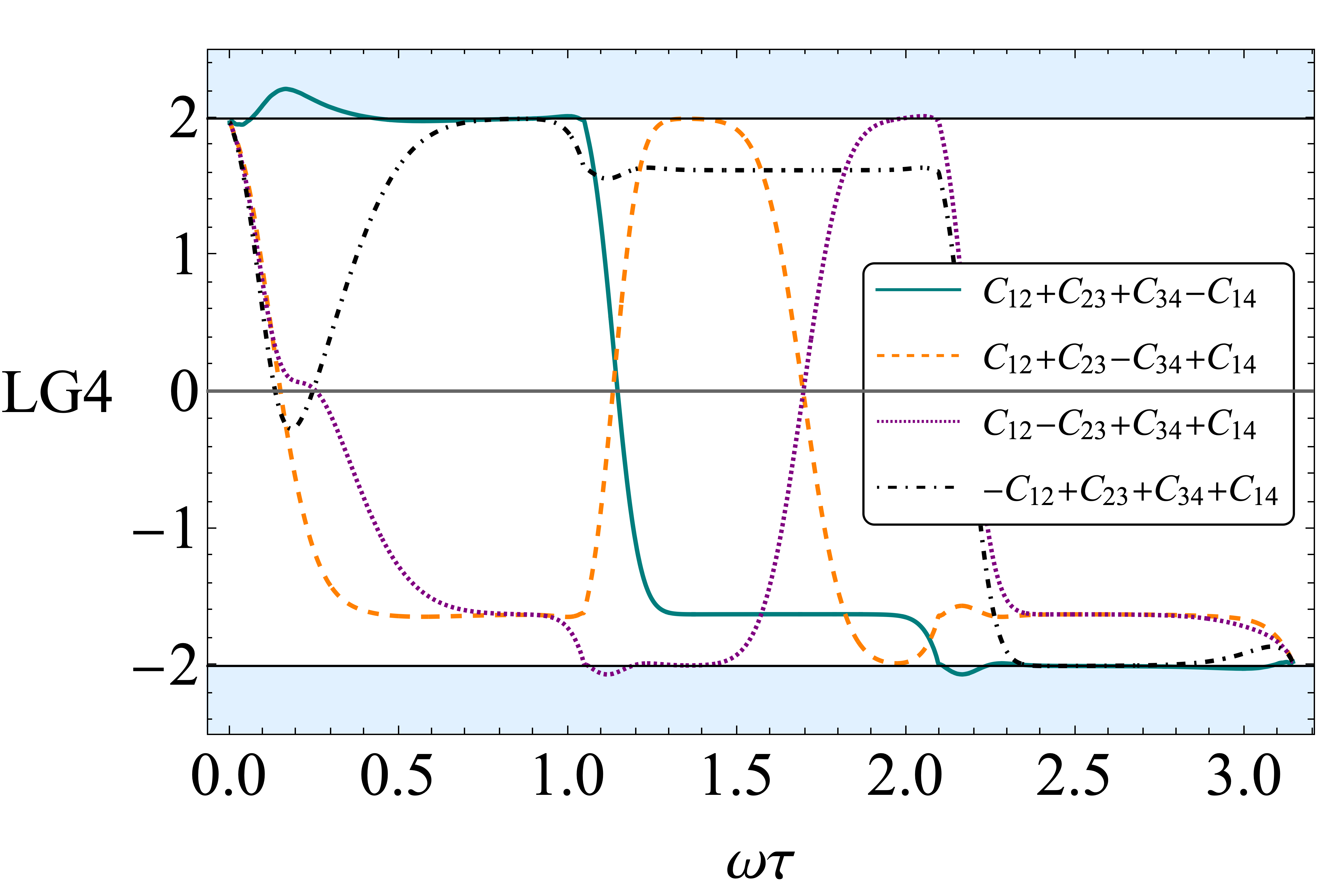}}}%
	\end{center}
	%\qquad
	\caption[Behaviour of all LG4 inequalities for their largest violation]{Plot of the four LG4 inequalities, for state $x_0=0.929$, $p_0=-3.666$, reaching a largest violation of $2.216$ at $\omega\tau=0.166$.}%
	\label{fig:LG4}%
\end{figure}

All the violations we have found are in states with initial position and momenta approximately on the length-scale of the width of the coherent state, $\sqrt{\hbar/(m\omega)}$.  However by appealing to Eq.~(\ref{eqn:corr}), we note that if one were to consider translating the measurement to $\theta(\hat x - x_i)$, the classical motion could be subtracted, and at least theoretically, the same magnitude of violations would exist for arbitrarily high $x_0$ and $p_0$.

In Figures \ref{fig:qpt} and \ref{fig:LG4}, we plot the temporal behaviour of the LG2s, LG3s and LG4s respectively, for the case in which the parameters are chosen to give the largest violation.

\section{Currents Analysis}
\subsection{Classical analogues}
\label{app:classan}
To understand the connection between the negativity of the quasi-probability Eq.~(\ref{eq:qpJ}) and the behaviour of the currents, it is very convenient to consider the analogous classical currents which are in general defined by
\begin{equation}
	\mathbbmsl{J}(t)=\int_{-\infty}^{\infty}\mathop{dp}\int_{-\infty}^{\infty}\mathop{dx} p(t) \delta(x(t))w(x,p),
\end{equation}
for a suitably chosen initial phase-space distribution $w(x,p)$.  For the un-chopped current $\mathbbmsl{J}(t)$ this is taken to be the Wigner function of the coherent state, $W(x,p,x_0, p_0)$ Eq.(\ref{eq:classWig}), which conveniently, is non-negative.  For the chopped curents it is taken to be $\theta(\pm x) W(x,p,x_0,p_0)$.  We then easily see that 
\begin{equation}
\label{eq:class}
	J(t)=\mathbbmsl{J}_-(t)+\mathbbmsl{J}_{+}(t),
\end{equation}
where $\mathbbmsl{J}_{\pm}(t)$ are the classical analogues to the post-measurement currents, where since we have used the coherent state Wigner function, we have $\mathbbmsl{J}(t)=J(t)$.

We begin by writing the classical phase-space density for the Gaussian state 
\begin{equation}
\label{eq:classWig}
\mathbbmsl{W}(X,p, x_0, p_0)=\frac{1}{\pi}\exp(-(X -x_0)^2 -(p-p_0)^2),
\end{equation}
where harmonic time-evolution leads to rigid rotation in phase-space,
\begin{equation}
	\mathbbmsl{W}(X,p,x_0,p_0,t)=\mathbbmsl{W}(X \cos\omega t -p \sin\omega t,p \cos \omega t +X \sin \omega t, x_0, p_0).
\end{equation}
We have a similar result for the measured classical state, with
\begin{equation}
\mathbbmsl{W}_{\pm}(X,p, x_0, p_0)=\frac{1}{\pi}\theta(\pm X)\exp(-(X -x_0)^2 -(p-p_0)^2),
\end{equation}
and
\begin{equation}
	\mathbbmsl{W}_{\pm}(X,p,x_0,p_0,t)=\mathbbmsl{W}_{\pm}(X \cos\omega t -p \sin\omega t,p \cos \omega t +X \sin \omega t, x_0, p_0).
\end{equation}
The chopped classical current is given by
\begin{equation}
\mathbbmsl{J}_{\pm}	(x,t)=\int_{-\infty}^{\infty}\mathop{dp}\int_{-\infty}^{\infty}\mathop{dX}p\delta(X-x)\mathbbmsl{W}_{\pm}(X,p,x_0, p_0,t)
\end{equation}
Completing the $X$ integral trivially, we have
\begin{equation}
\mathbbmsl{J}_{\pm}(x,t)=\int_{-\infty}^{\infty}\mathop{dp} p\mathbbmsl{W}_{\pm}(x,p,x_0, p_0,t).
\end{equation}
We are interested in the case of $x=0$, which we shorthand $\mathbbmsl{J}_\pm(0,t)=\mathbbmsl{J}_\pm(t)$, and is given by
\begin{equation}
	\mathbbmsl{J}_\pm(t)=\frac{1}{\pi}\int_{-\infty}^{\infty}\mathop{dp} p\,\theta (\mp p \sin\omega t) \exp \left(-(p \cos\omega t-p_0)^2-(-p \sin \omega t-x_0)^2\right).
\end{equation}
The step-function here just flips the integral between the positive or negative half-plane, dependent on $\sgn(\mp \sin \omega t)=1$ and $\sgn(\mp \sin \omega t)=-1$ respectively.  Computing the integral, this yields the result
\begin{equation}
\mathbbmsl{J}_{\pm}(t)=	\frac{1}{2\pi}e^{-p_0^2-x_0^2} \left(\mp\text{sgn}(\sin\omega t)+\sqrt{\pi } e^{g(x_0, p_0, t)^2} g(x_0, p_0, t) (1\mp\text{sgn}(\sin\omega t)\erf (g(x_0, p_0, t)))\right),\end{equation}
with $g(x_0, p_0, t)=p_0 \cos\omega t -x_0\sin\omega t$.
\subsection{Time derivative of the quasi-probability}
\label{app:curr}
\noindent
To calculate the time-derivative of the QP as appears in Eq.~(\ref{eq:dqdt}), we begin by writing the simple projector identity
\begin{equation}
\label{eq:id}
P\rho + \rho P = P\rho P - \bar P \rho \bar P +\rho,	
\end{equation}
where $\bar P = 1-P$.  Hence the quasi-probability, Eq.~(\ref{quasi}) is given by
\begin{equation}
q(-,+)=\frac12\Tr \left(P_{+}(t_2)\left(P_-(t_1) \rho P_-(t_1) - P_+(t_1) \rho P_+(t_1) +\rho\right) \right).
\end{equation}
This may be written in terms of the non-negative sequential measurement probabilities
\begin{equation}
\label{eq:seqprob}
p_{12}(s_1, s_2)=\Tr\left(P_{s_2}(t_2)P_{s_1}(t_1)\rho P_{s_1}(t_1)\right),
\end{equation}
in the form
\begin{equation}
q(-,+)=\frac12\left( p_{12}(-,+)-p_{12}(+,+)+\ev{P_+(t_2)}\right).
\end{equation}
It is now simple to take the derivative with respect to $t_2$, noting that 
\begin{equation}
\frac{d}{dt}\theta(\hat x(t))=\frac{1}{2m}\left(\hat p(t) \delta(\hat x(t)) + \delta(\hat x(t))\hat p(t)\right)=\hat J(t),
\end{equation}
yielding
\begin{equation}
\frac{d q(-,+)}{dt}=\frac12\Tr\left(\hat J(t)\left(P_-(t_1) \rho P_-(t_1) - P_+(t_1) \rho P_+(t_1) +\rho\right) \right).
\end{equation}
We can hence rewrite $q(-,+)$ as
\begin{equation}
\label{eq:qpJ}
q(-,+)=\frac12 \int_{t_1}^{t_2}\mathop{dt} \left(J_-(t)-J_+(t)+J(t)\right),
\end{equation}
where we have introduced the `chopped current' 
\begin{equation}
\label{eq:chopcurr}
J_{\pm}(t)=\ev{\theta(\pm \hat x) \hat J(t) \theta(\pm \hat x)}{\psi},
\end{equation}
which corresponds to the current at the origin, after the initial measurement.  The chopped currents are therefore the currents of the wave-functions $\langle x\lvert e^{-iHt}\theta(\pm\hat x)\rvert \psi\rangle$, and note the connection to the Moshinsky function when $|\psi\rangle$ is expanded in the momentum basis.

Using Eq.~(\ref{eq:class}), we may rewrite Eq.~(\ref{eq:qpJ}) in terms of the difference between quantum and classical post measurement currents, as 
\begin{equation}
q(-,+)=\int_{t_1}^{t_2}\mathop{dt}J_-(t)+\frac12 \int_{t_1}^{t_2}\mathop{dt}\left(\mathbbmsl{J}_-(t)-J_-(t) +\mathbbmsl{J}_+(t)-J_+(t)\right).
\end{equation}

\subsection{Chopped currents calculation}
\label{app:chopcurr}
We calculate the `chopped current' first, that is
\begin{equation}
J_{\pm}(x,t)=\frac{1}{2m}\ev{\delta(\hat x - x)\hat p + \hat p \delta(\hat x - x)}{\phi^{\pm}_\alpha(t)},
\end{equation}
where $\phi^{\pm}_\alpha(t)$ is the time evolution of a coherent state, initially projected on $\theta(\pm \hat x)$ at $t_0$,
\begin{equation}
	\ket{\phi^{\pm}_\alpha(t)} = e^{-iHt}\theta(\pm \hat x)\ket\alpha
\end{equation}
Calculating the current in the position basis, we have
\begin{equation}
J_\pm(x,t)=-\frac{i\hbar}{2m}\left(\phi_\alpha^{\pm*}(x,t)\frac{\partial \phi_\alpha^{\pm}(x,t)}{\partial x}-\frac{\partial \phi_\alpha^{\pm*}(x,t)}{\partial x}\phi_\alpha^{\pm}(x,t)\right),
\end{equation}
equivalent to
\begin{equation}
\label{eqn:Jx}
J_{\pm}(x,t)=\frac{\hbar}{m}\Im\left(\phi_\alpha^{\pm*}(x,t)\frac{\partial \phi_\alpha^{\pm}(x,t)}{\partial x}\right).
\end{equation}
We calculate the evolved chopped state by
\begin{equation}
\label{eq:phipm}
\phi_\alpha^\pm(x,t)=\int_{\Delta(\pm)}\mathop{dy}K(x,y,t)\psi_\alpha(y,t_0),
\end{equation}
where $\Delta(+)=[0,\infty)$, $\Delta(-)=(-\infty,0]$,
\begin{equation}
\psi^{\alpha}(x,t_0)=\frac{1}{\pi^\frac14}\exp(-\frac12(x-x_0)^2 + i p_0 x).
\end{equation}
%\begin{equation}
%\psi_\alpha(x,t)=\left(\frac{m\omega}{\pi\hbar}\right)^\frac14 \exp\left(-\frac{m\omega}{2\hbar}\left(x-\sqrt{\frac{2\hbar}{m\omega}}\Re\alpha(t)\right)^2 + i \sqrt{\frac{2m\omega}{\hbar}}\Im \alpha(t) x\right),
%\end{equation}
is the non-dimenionalized coherent state wave-function, with time evolution $\alpha(t) = e^{-i\omega t}\alpha(0)$, and
\begin{equation}
K(x,y,t)=\left(\frac{1}{2\pi i\sin \omega t}\right)^\frac12 \exp\left(-\frac{1}{2i\sin\omega t}\left((x^2+y^2)\cos\omega t -2 x y\right)\right),
\end{equation}
is the propagator for the harmonic potential.

Inserting the relevant expressions within Eq.~(\ref{eq:phipm}) yields
\begin{multline}
\label{eq:propchop}
\phi_\alpha^\pm(x,t)=\left(\frac{1}{2\pi i \sin \omega t}\right)^\frac12 \left(\frac{1}{\pi}\right)^\frac14 \int_{\Delta(\pm)}\mathop{dy}\exp\left(-\frac12(y-x_0)^2+ i y p_0)\right)\times\\\exp\left(-\frac{x^2 + y^2}{2i \tan \omega t} +\frac{x y}{i \sin \omega t}\right).
\end{multline}
We proceed writing the integral as
\begin{equation}
I_\pm(a,b,c)=\int_{\Delta(\pm)}\mathop{dr}\exp\left(-\frac12(r-a)^2 + i b r+ i c r^2\right),
\end{equation}
where 
\begin{align}
a&=x_0,\\
b&=p_0 - \frac{x}{\sin\omega t},\\
c&=\frac{1}{2\tan\omega t}.
\end{align}
Completing the integration, we have
\begin{equation}
\label{eq:Ipm}
I_\pm(a,b,c)=\frac{\sqrt{\pi } e^{-\frac{2 a^2 c+2 a b+i b^2}{4 c+2 i}} \left(1\pm \erf\left(\frac{a+i b}{\sqrt{2-4 i c}}\right)\right)}{\sqrt{2-4 i c}}.
\end{equation}
We hence can write the chopped wave-function as
\begin{equation}
\label{eq:chopwav}
\phi_\alpha^\pm(x,t)=\left(\frac{1}{2\pi i \sin \omega t}\right)^\frac12 \left(\frac{1}{\pi}\right)^\frac14 e^{-\frac{x^2}{2i \tan \omega t}}I_\pm\left(x_0, p_0-\frac{x}{\sin\omega t},\frac{1}{2\tan\omega t}\right).
\end{equation}
Putting Eq.(\ref{eqn:Jx}) into rescaled units as well, we have
\begin{equation}
\label{eq:Jpmap}
J_{\pm}(x,t)=\omega\Im\left(\phi_\alpha^{\pm*}(x,t)\frac{\partial \phi_\alpha^{\pm}(x,t)}{\partial x}\right).
\end{equation}
To take the derivative we note $I_\pm(a,b,c)$ depends on $s$ only in its second argument, and so we define
\begin{equation}
K_{\pm}(a,b,c)=\frac{\partial b}{\partial s}\frac{\partial}{\partial b}I_\pm(a,b,c),
\end{equation}
which explicitly yields
\begin{equation}
\label{eq:kpm}
K_{\pm}(a,b,c)=\frac{-1}{\sin\omega t}\frac{e^{-\frac{a^2}{2}} \left(-\sqrt{2 \pi } (a+i b) e^{\frac{(a+i b)^2}{2-4 i c}} \left(1\pm \text{erf}\left(\frac{a+i b}{\sqrt{2-4 i c}}\right)\right)\mp 2 \sqrt{1-2 i c}\right)}{2 \sqrt{1-2 i c} (2 c+i)}
\end{equation}

Altogether, this yields the current
\begin{multline}
\label{eq:chopJ}
J_{\pm}(x,t)=\frac{\omega}{2\pi^{\frac32}}\Im\Bigg(\frac{1}{\lvert\sin\omega t\rvert}I_\pm\bigg(x_0,-p_0+\frac{x}{\sin \omega t},-\frac{1}{2\tan \omega t}\bigg)\times\\
\bigg(\frac{ix}{\tan \omega t}I_\pm\bigg(x_0,p_0-\frac{x}{\sin \omega t},\frac{1}{2\tan \omega t}\bigg) +K_\pm\bigg(x_0,p_0-\frac{x}{\sin \omega t},\frac{1}{2\tan \omega t}\bigg)\bigg)
\Bigg).
\end{multline}
We also calculate the current of the original unperturbed coherent state, in these same rescaled units.  Since coherent states are eigenfunctions of the annihilation operator, with eigenvalue $\alpha(t)$, it follows that 
\begin{equation}
\frac{\partial}{\partial x} \psi_\alpha(x,t)=\psi_\alpha(x,t)\left(\sqrt2 \alpha(t) - x\right).
\end{equation}
Hence the current is given by 
\begin{equation}
J(x,t)=\frac{\omega}{\sqrt\pi}\Im\left(e^{-(x-\sqrt{2}\Re \alpha(t))^2}(\sqrt2 \alpha(t) -x)\right),
\end{equation}
where taking the imaginary part, and using Eq.~(\ref{eq:alphatox}), we have
\begin{equation}
	J(x,t)=\frac{\omega}{\sqrt\pi}p_t e^{-(x-x_t)^2}.
\end{equation}
\section{Small Time Current Expansions}
\label{app:tJ}
We are interested in a small time expansion of the chopped current $J_{\pm}(t)$, for any general state $\ket\psi$.  We start by defining the chopped-evolved state,
\begin{equation}
	\ket{\psi^{\pm}(t)}=e^{-iHt}\theta(\pm \hat x)\ket{\psi}.
\end{equation}
 We now follow Appendix~\ref{app:curr} up to Eq.~(\ref{eq:Jpmap}) to calculate the current, however this time using the general $\ket\psi$ state, leading to
\begin{equation}
J_{\pm}(t)=\omega\Im\left(\psi^{\pm*}(0,t)\frac{\partial\psi^\pm(0,t)}{\partial x}\right),
\end{equation}
with $\ket\psi$ represented in the non-dimensional position basis, i.e. $\braket{x}{\psi}=\left(\frac{m\omega}{\hbar}\right)^{\frac{1}{4}}\psi(x)$, with a normalised $\psi(x)$ which is purely a function of $x$.

Using the QHO propagator as in App.~\ref{app:curr} up to Eq.~(\ref{eq:propchop}), now with the $\ket\psi$ state, we have
\begin{equation}
\label{eq:genprop}
\psi^{\pm}(x,t)=\left(\frac{1}{2\pi i \sin\omega t}\right)^\frac12 \int_{\Delta(\pm)}\mathop{dy}\psi(y,0)\exp\left(-\frac{x^2 + y^2}{2i \tan \omega t} +\frac{x y}{i \sin \omega t}\right).
\end{equation}
We interested in the current at $x=0$ which means we only need $\psi(x)$ and its first derivative evaluated at $x=0$, leaving
\begin{equation}
\label{eq:psi0}
	\psi^{\pm}(0,t)=\left(\frac{1}{2\pi i \sin\omega t}\right)^\frac12 \int_{\Delta(\pm)}\mathop{dy}\psi(y,0)\exp\left(-\frac{y^2}{2i \tan \omega t}\right).
\end{equation}
We now argue that for small $t$ the main contribution to the integral will come from near the boundary of the chop,  and hence Taylor expand  $\psi(y,0)$ expand around $y=0$.  By using the parametrisation $y=z(\tan\omega t)^\frac12$, we simplify the exponential part of the integrand.
\begin{equation}
\psi(y,0)=\sum_{n=0}^{\infty}\frac{\psi^{(n)}(0,0)}{n!}z^n(\tan\omega t)^\frac{n}{2}.
\end{equation}
Using this in Eq.~(\ref{eq:psi0}), and using the substitution within the integral, we have
\begin{equation}
	\psi^{\pm}(0,t)=\left(\frac{\tan\omega t}{2\pi i \sin\omega t}\right)^\frac12 \int_{\Delta(\pm)}\mathop{dz}\exp\left(-\frac{z^2}{2i}\right)\sum_{n=0}^{\infty}\frac{\psi^{(n)}(0,0)}{n!}z^n(\tan\omega t)^\frac{n}{2}
\end{equation}
Interchanging the order of summation and integration, we have
\begin{equation}
\psi^\pm(0,t)=\left(\frac{1}{2\pi i \cos\omega t}\right)^\frac12\sum_{n=0}^\infty \frac{\psi^{(n)}(0,0)}{n!}(\tan \omega t)^\frac{n}{2}\int_{\Delta(\pm)}\mathop{dz}e^{-\frac{z^2}{2i}}z^n.
\end{equation}
We now define
\begin{equation}
K_\pm(n)=\int_{\Delta(\pm)}\mathop{dz}e^{-\frac{z^2}{2i}}z^n,
\end{equation}
which can be calculated by taking the Fourier transform of $e^{iz^2}\theta(\pm z)$, taking $n$ derivatives in Fourier space, and then evaluating with the conjugate variable set to $0$, to give the result
\begin{equation}
K_{\pm}(n)=(\pm1)^n 2^{\frac{n-1}{2}} e^{\frac{1}{4} i \pi  (n+1)} \Gamma \left(\frac{n+1}{2}\right),
\end{equation}
were $\Gamma$ is the Gamma function. This gives a final result of
\begin{equation}
\label{eq:psiexp0}
\psi^\pm(0,t)=\left(\frac{1}{2\pi i \cos \omega t}\right)^\frac12 \sum_{n=0}^\infty K_\pm(n)\frac{\psi^{(n)}(0,0)}{n!}(\tan\omega t)^\frac{n}{2}.
\end{equation}
We now calculate the derivative of the chopped wavefunction, by taking the derivative of Eq.~(\ref{eq:genprop}), evaluated at $x=0$ to find
\begin{equation}
\frac{\partial \psi^\pm(x,t)}{\partial x}\eval_{x=0}	=\frac{-i}{\sin \omega t}\left(\frac{1}{2\pi i \sin\omega t}\right)^\frac12 \int_{\Delta(\pm)}\mathop{dy}\psi(y,0)y \exp\left(-\frac{y^2}{2i \tan \omega t}\right).
\end{equation}
We now note, that with the exception of the pre-factor $\frac{-i}{\sin\omega t}$, this is the same result as before, only with $\psi(y,0)$ swapped for $y\psi(y,0)$, leading to the result
\begin{equation}
\frac{\partial \psi^\pm(0,t)}{\partial x}=-\left(\frac{-1}{2\pi i\sin^2\omega t \cos\omega t}\right)^\frac12\sum_{n=0}^\infty K_\pm(n)\frac{\frac{\partial^n}{\partial y^n}(y\psi(y,0))\eval_{y=0}}{n!}(\tan\omega t)^\frac{n}{2}.
\end{equation}
Then since 
\begin{equation}
\frac{\partial^n}{\partial y^n}y\psi(y,0)\eval_{y=0}=n \psi^{(n-1)}(0,0),
\end{equation}
we have as a final result 
\begin{equation}
\label{eq:psipexp0}
\frac{\partial \psi^\pm(0,t)}{\partial x}=-\left(\frac{-1}{2\pi i\sin^2\omega t \cos\omega t}\right)^\frac12\sum_{n=1}^\infty K_\pm(n)\frac{n\psi^{(n-1)}(0,0)}{n!}(\tan\omega t)^\frac{n}{2},
\end{equation}	
noting the change on the sum's lower limit.

We now combine Eq.~(\ref{eq:psiexp0}) and Eq.~(\ref{eq:psipexp0}) to yield the small-time expansion for the chopped current
\begin{equation}
J_\pm(t)=\frac{-\omega}{2\pi\sin\omega t \cos\omega t}\Im \left(i\sum_{\ell=0}^{\infty}\sum_{n=1}^{\infty}K_{\pm}(n)K^*_{\pm}(\ell) \frac{n\psi^{(n-1)}(0,0)\psi^{*(\ell)}(0,0)}{n!\ell!}(\tan\omega t)^\frac{n+\ell}{2}\right)
\end{equation}
We note that this result is in fact trivial to integrate over time by noting the derivative of $\tan(\omega t)$, yielding
\begin{equation}
\int_{0}^{\tau}\mathop{dt}\frac{(\tan\omega t)^\frac{k}{2}}{\sin\omega t\cos\omega t}=\frac{2}{k\omega}(\tan\omega \tau)^{\frac{k}{2}}.
\end{equation}
The time-integral of the chopped current is thus
\begin{equation}
\int_{0}^{\omega \tau}J_\pm(t)\mathop{dt}=-\frac{1}{\pi}\Im\left(i\sum_{\ell=0}^{\infty}\sum_{n=1}^{\infty}K_{\pm}(n)K^*_{\pm}(\ell) \frac{n\psi^{(n-1)}(0,0)\psi^{*(\ell)}(0,0)}{(n+\ell)n!\ell!}(\tan\omega t)^\frac{n+\ell}{2}\right)
\end{equation}
We also note that by defining
\begin{equation}
L(n)=K_+(n)+K_-(n),	
\end{equation}
we may adapt the result to the unchopped current as
\begin{equation}
J(t)=-\frac{\omega}{2\pi\sin\omega t \cos\omega t}\Im \left(i\sum_{\ell=0}^{\infty}\sum_{n=1}^{\infty}L(n)L^*(\ell) \frac{n\psi^{(n-1)}(0,0)\psi^{*(\ell)}(0,0)}{n!\ell!}(\tan\omega t)^\frac{n+\ell}{2}\right),
\end{equation}
with time-integral
\begin{equation}
\int_{0}^{\omega \tau}J(t)\mathop{dt}=-\frac{1}{\pi}\Im\left(i\sum_{\ell=0}^{\infty}\sum_{n=1}^{\infty}L(n)L^*(\ell) \frac{n\psi^{(n-1)}(0,0)\psi^{*(\ell)}(0,0)}{(n+\ell)n!\ell!}(\tan\omega t)^\frac{n+\ell}{2}\right)
\end{equation}
Using Eq.~(\ref{eq:qpJ}) to express the quasi-probability as the time integral of currents, and noting the similarity of the summands, we can write the quasi-probability for as
\begin{equation}
q(-,+)=-\frac{1}{2\pi}\Im\left(i\sum_{\ell=0}^{\infty}\sum_{n=1}^{\infty}\mathcal{Q}(n,\ell) \frac{n\psi^{(n-1)}(0,0)\psi^{*(\ell)}(0,0)}{(n+\ell)n!\ell!}(\tan\omega t)^\frac{n+\ell}{2}\right),
\end{equation}
where
\begin{equation}
\mathcal{Q}(n,\ell)=K_-(n)K_-^*(\ell)-K_+(n)K_+^*(\ell)+L_n L^*_\ell.
\end{equation}
By approximating the infinite sums to finite order, we are able to approximate the quasi-probability.  To get the first three terms of the approximation, we limit both sums to $\ell_{max}=2$ and $n_{max}=3$, yielding
\begin{multline}
\label{eq:tayl3}
q(-,+)=	\frac{1}{2\sqrt{\pi}}\lvert\psi(0,0)\rvert^2 \tan^\frac12\omega\tau + \frac{J(0)}{2}\tan\omega\tau+\\\frac{1}{6\sqrt\pi}\left(\lvert \psi'(0,0)\rvert^2-\left[\frac{1}{4}+\frac{3 i}{4}\right]\psi''^{*}(0,0)\psi(0,0)- \left[\frac{1}{4}-\frac{3 i}{4}\right]\psi''(0,0)\psi^*(0,0)\right)\tan^\frac32\omega t \\+\mathcal{O}(\tan ^2 \omega t)
\end{multline}
Taking the taking the $\omega\rightarrow 0$ expansions of the trigonometric terms recovers the result for the free particle,
\begin{multline}
q(-,+)=\frac{1}{2\sqrt{\pi}}\lvert\psi(0,0)\rvert^2 \tau^\frac12+ \frac{J(0)}{2}\tau+\\\frac{1}{6\sqrt\pi}\left(\lvert \psi'(0,0)\rvert^2-\left[\frac{1}{4}+\frac{3 i}{4}\right]\psi''^{*}(0,0)\psi(0,0)- \left[\frac{1}{4}-\frac{3 i}{4}\right]\psi''(0,0)\psi^*(0,0)\right)\tau^\frac32,
\end{multline}
which we note has a term in $\tau^\frac12$ which was missing from an earlier calculation of this expansion in Ref.~\cite{halliwell2019c}, as well as a different coefficient on the $\tau^\frac32$ term.

The initial divergence of the quantum chopped current is clearly seen. These results also agree with the small time expansion of chopped currents given by Sokolowski \cite{sokolovski2012}, giving another useful check on our calculations. For our gaussian initial state we find agreement with the results above, we plot this expansion alongside our original calculation in Fig.~\ref{fig:tayl}.
\begin{figure}
	\begin{center}
\includegraphics[height=5.3cm]{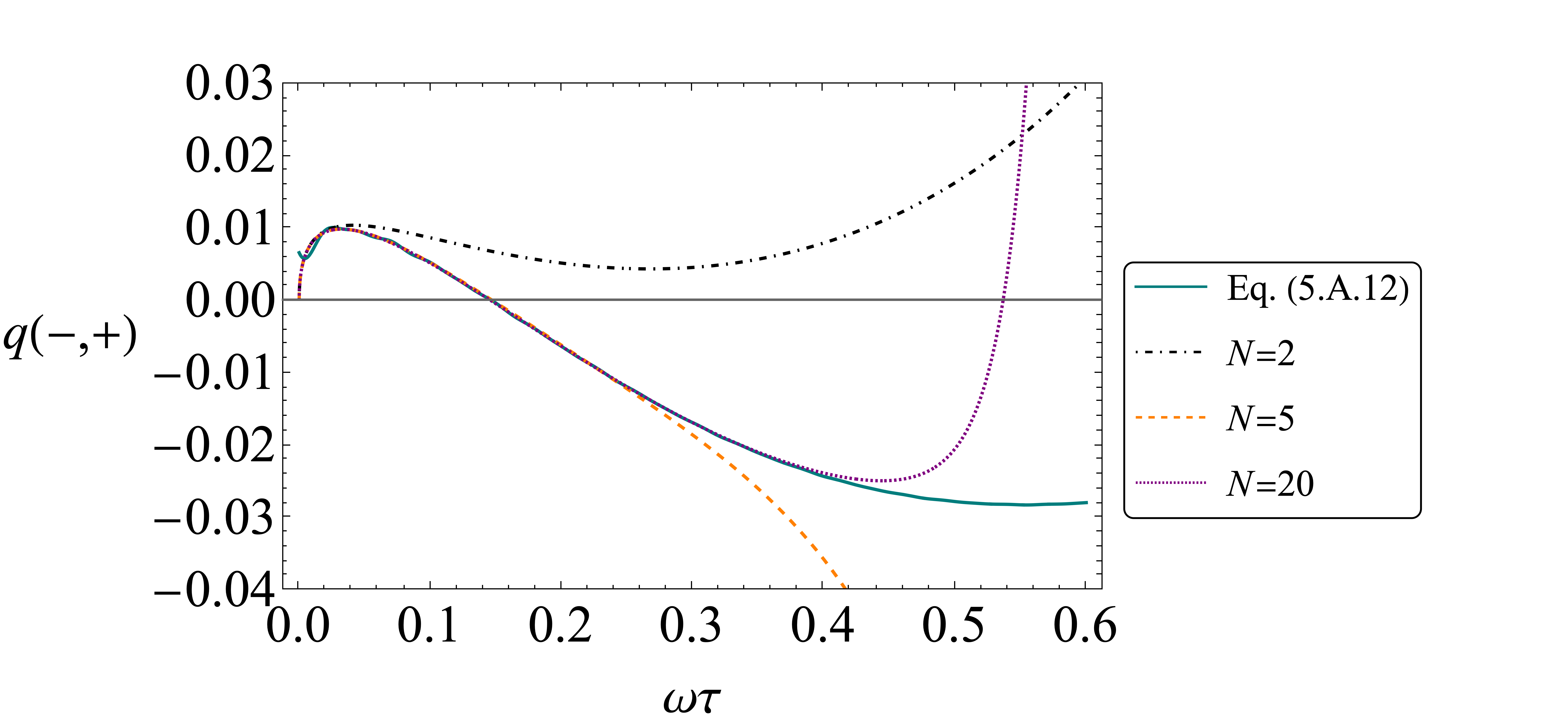}
\end{center}
	\caption[Convergence of small-time expansion valid for generic states]{The small-time expansion of $q(-,+)$ plotted alongside the previous calculation, at varying degrees of truncation, where $N=2$ corresponds to Eq.~(\ref{eq:tayl3}).}%%
	\label{fig:tayl}
\end{figure}

\section{Wigner Function calculational details}
\label{app:wig}
The Wigner-Weyl transform, which maps Hermitian operators to real phase space functions~\cite{hillery1984a, tatarskii1983a, case2008, halliwell1993a}, is defined by
\begin{equation}
\label{eq:wigtx}
	W_A(X,p) = \frac{1}{2\pi}\int_{-\infty}^{\infty}\mathop{d\xi}e^{-ip\xi}\mel{X+\tfrac{\xi}{2}}{A}{X-\tfrac{\xi}{2}}.
\end{equation}
Traces of pairs of operators may be expressed in the Wigner representation as
\begin{equation}
\label{eq:wigexp}
	\Tr(\hat A\hat B) = 2\pi \int_{-\infty}^{\infty}\mathop{dX}\int_{-\infty}^{\infty}\mathop{dp}W_{A}(X,p)W_{B}(X,p).
\end{equation}
To apply this formula to the quasi-probability there are two natural ways to proceed. First, in Ref.~\cite{halliwell2019c}, the free particle quasi-probability was explored in Wigner-Weyl form using $\hat A = \tfrac12\left(P_{s_1}P_{s_2}+P_{s_2}P_{s_1}\right)$ and $\hat B=\rho$. However this was not found to be very useful since $W_A(X,p)$ in this case is highly oscillatory and it was not possible to clearly identify the regions of negativity, hence we proceed with a different approach.

We first write the QP in the form $q(s_1, s_2)=\Tr(\bar{\rho}_{s_1}P_{s_2}(\tau))$, where 
 $\bar{\rho}_{s_1}=\frac12(P_{s_1}\rho+\rho P_{s_1})$, and $t_1=0$ without loss of generality.    
Hence by Eq.~(\ref{eq:wigexp}) we have 
\begin{equation}
\label{eq:wigqp}
	q(s_1,s_2)=2\pi \int_{-\infty}^{\infty}\mathop{dX}\int_{-\infty}^{\infty}W_{\bar\rho_{s_1}}(X,p)W_{P_{s_2}(\tau)}(X,p),
\end{equation}
where $W_{P_{s_2}}(X,p)=\theta(s_2(X \cos \omega t + p \sin\omega t))$.  Using Eq.~(\ref{eq:wigtx}), the transform of $\bar\rho_{s_1}$ is given by
\begin{equation}
\label{eq:wigrhob}
W_{\bar\rho_{s_1}}(X,p)=\frac{1}{4\pi}\int_{-\infty}^{\infty}\mathop{d\xi}e^{-ip\xi}\mel{X+\tfrac{\xi}{2}}{P_{s_1}\rho+\rho P_{s_1}}{X-\tfrac{\xi}{2}},
\end{equation}
We without loss of generality we take $t_1=0$, and so $P_{s_1}=\theta(s_1 \hat x)$, leading to
\begin{equation}
W_{\bar\rho_{s_1}}(X,p)=\frac{1}{4\pi}\int_{-\infty}^{\infty}\mathop{d\xi}e^{-ip\xi}\left(\theta(s_1(X+\tfrac{\xi}{2}))+\theta(s_1(X-\tfrac{\xi}{2}))\right)\mel{X+\tfrac{\xi}{2}}{\rho}{X-\tfrac{\xi}{2}}.
\end{equation}
Since coherent states are pure states, we have simply that $\rho(x,y)=\psi(x)\psi^*(y)$, where we will use natural units with $\psi(x)=\tfrac{1}{\pi^\frac14}\exp(-\tfrac12(x-x_0)^2+ip_0x)$.  This yields the integrand as $I(X,\xi)=e^{-ip\xi}\exp\left(i p_0 \xi-X^2+2 X x_0-x_0^2-\tfrac{\xi^2}{4}\right)$.  

Demonstrating with the $s_1=+1$ case, the theta functions are handled by splitting the integral into two integrals over the regions $[-2X, \infty)$, and $(-\infty, 2X]$, and so we have
\begin{equation}
W_{\bar\rho_{s_1}}(X,p)=\frac{1}{4\pi\sqrt{\pi}}\left(\int_{-2X}^{\infty}\mathop{d\xi}I(X,\xi)+\int_{-\infty}^{2X}\mathop{d\xi}I(X,\xi)\right),
\end{equation}
Computing the integral involved, we reach the result
\begin{equation}
W_{\bar\rho_{s_1}}(X,p)=\frac{1}{2\pi}e^{-(p-p_0)^2- (X-x_0)^2}\left(1+\Re\erf\left(i(p-p_0)+s_1 X\right)\right),
\end{equation}
which may be written as
\begin{equation}
\label{eq:wigint}
W_{\bar\rho_{s_1}}(X,p)=\frac12 W_{\rho}(X,p)\left(1+\Re\erf\left(i(p-p_0)+s_1 X\right)\right).
\end{equation} 
in terms of $W_{\rho}(X,p)$, the Wigner function of the pure coherent state, given by
\begin{equation}
	W_{\rho}(X,p)=\frac{1}{\pi}\exp(-(p-p_0)^2-(X-x_0)^2).
\end{equation}
The classical equivalent for Eq.~(\ref{eq:wigint}) is $\frac12 W_\rho(X,p)(1+\text{sgn}(X))$, which is approached for $p$ close to $p_0$ and for large $\lvert X \rvert$.

The time evolution of the Wigner function in the case of the QHO is given by rigid rotation in accordance with classical paths $X_{\tau}=x_0\cos \omega \tau - p_0 \sin\omega\tau$.  Hence in Eq.~(\ref{eq:wigqp}) $W_{P_{s_2}(\tau)}(X,p)=\theta(s_2 X_{-\tau})$, and the final expression for the quasi-probability is
\begin{equation}
	q(s_1,s_2)=\int_{-\infty}^{\infty}\mathop{dX}\int_{-\infty}^{\infty}\mathop{dp}f_{s_1, s_2}(X,p)
\end{equation}
with the phase-space density $f_{s_1, s_2}(X,p)$ given by
\begin{equation}
\label{eq:wigfin}
f_{s_1, s_2}(X,p)=\frac12 W_{\rho}(X,p)\left(1+\Re\erf\left(i(p-p_0)+s_1 X\right)\right)\theta(s_2 X_{-\tau}).
\end{equation}
Eq. (\ref{eq:wigint}) and Eq. (\ref{eq:wigfin}) are plotted in Fig.~\ref{fig:wig2}.
\begin{figure}[t]
	\subfloat[]{{\includegraphics[width=7.6cm,trim={0 1.5cm 3cm 4cm},clip]{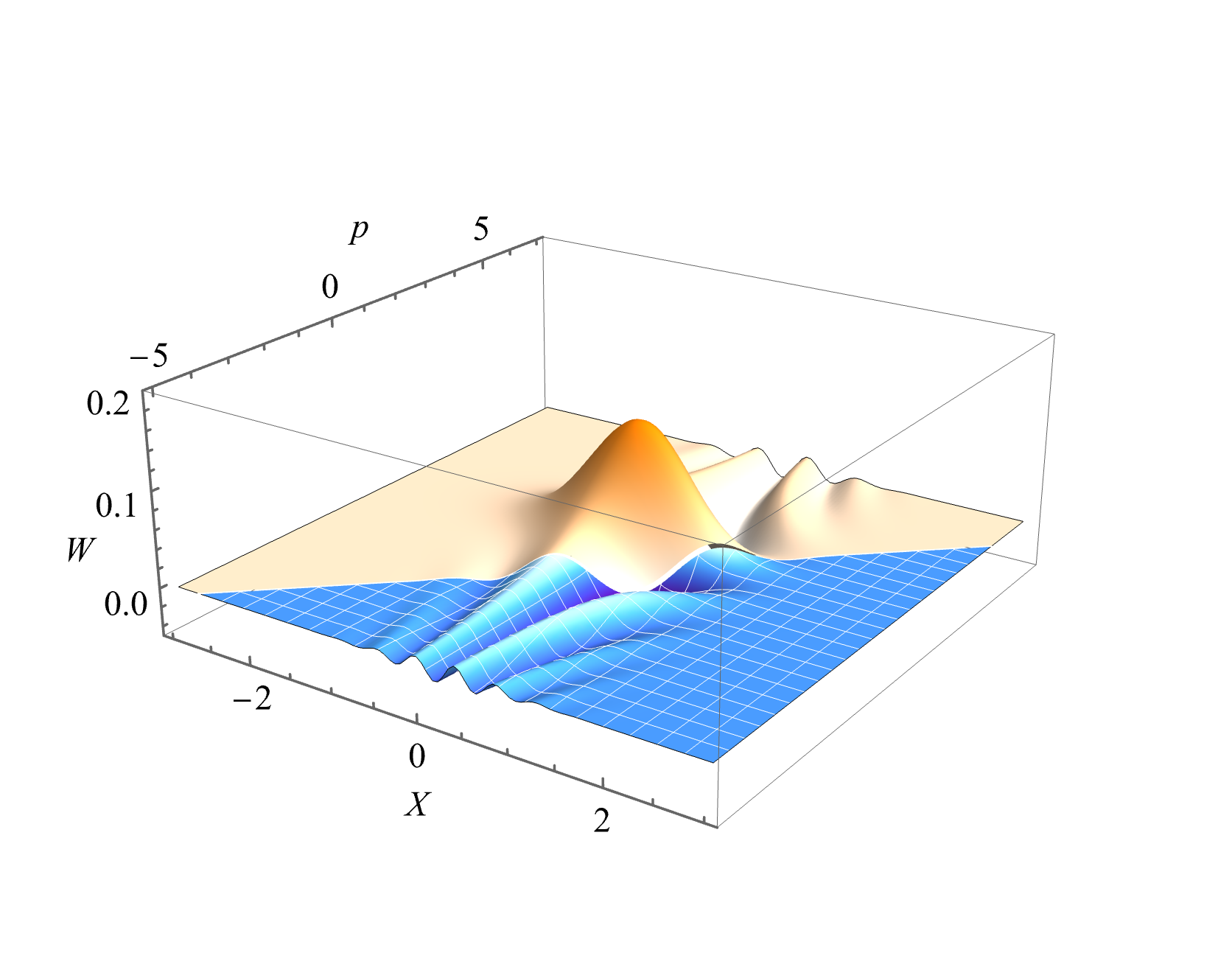}}}%
	%\qquad
	\hspace{5mm}
	\subfloat[]{{\includegraphics[width=7.6cm,trim={3.9cm 0 3.9cm 0},clip]{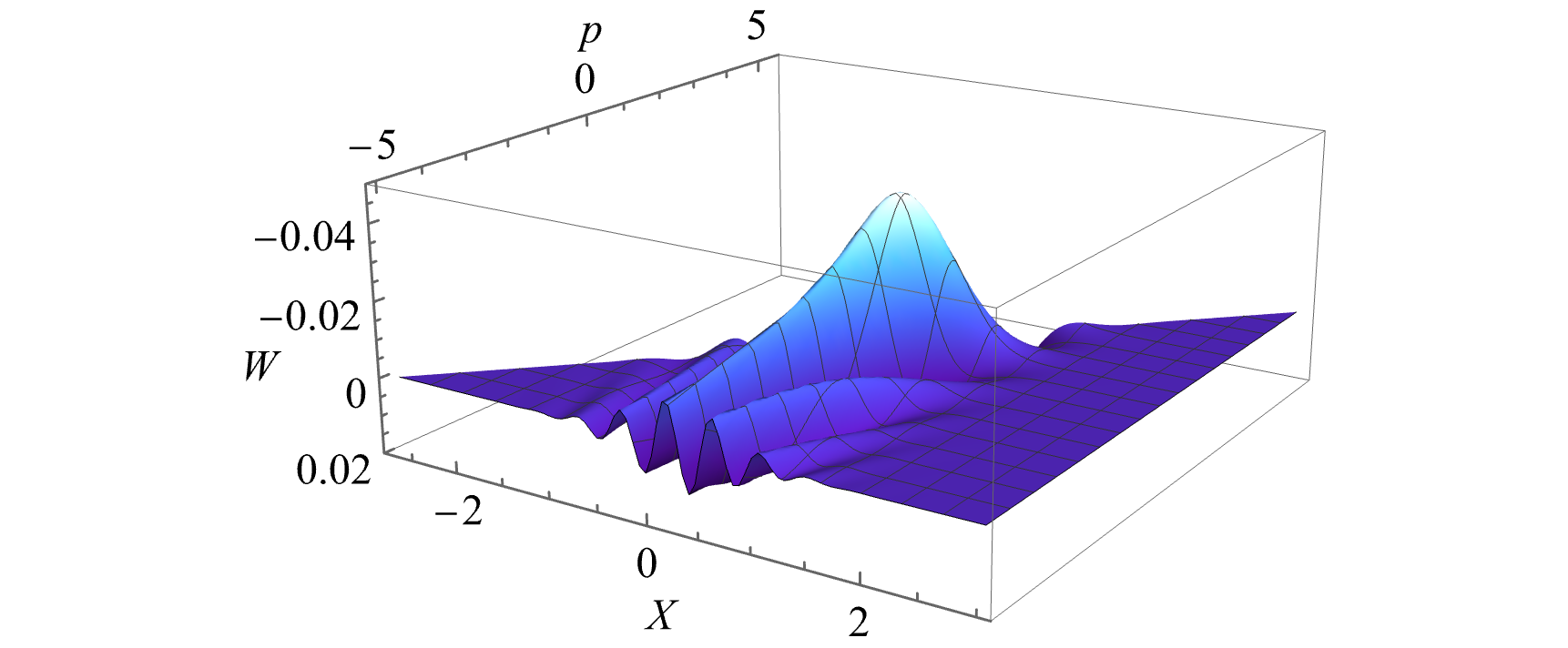}}}
	\caption[Wigner function phase-space negativity associated with LG2 violations]{Plot (a) shows $W_{\bar\rho_{-}}(X,p)$, Eq.(\ref{eq:wigint}) with $s_1=-1$, for an initial coherent state with $x_0=0.55$, $p_0=-1.925$ and $\omega\tau=0.55$. The orange (smooth) region shows the region removed when we go to the phase space density Eq.(\ref{eq:wigfin}), thereby showing how the most significant positive parts are removed. Plot (b) (flipped axis) shows the phase space density Eq.(\ref{eq:wigfin}) which will from visual inspection integrate to a negative number overall.}%
	\label{fig:wig2}%
\end{figure}
\end{subappendices}

\fancyhf{}
\renewcommand{\headrulewidth}{0pt}
\fancyfoot{\makebox[\textwidth][c]{\hyperref[link:7]\thepage}}

\fancyhfoffset[LE,RO, RE, LO]{0cm}
\renewcommand{\chaptermark}[1]{ \markboth{#1}{} }
\renewcommand{\sectionmark}[1]{ \markright{#1}{} }
\fancyhf{}
\fancyhead[L]{\textsl{\thesection~~ \rightmark}}
\fancyhead[R]{\hyperref[link:7]\thepage}
\renewcommand{\headrulewidth}{1pt}

\chapter{Tests for macrorealism for continuous variables using a waiting detector model}
\bookepigraph{3in}{Music is the space between the notes.}{Claude Debussy}{}{0}
\label{chap:wd}
\vspace{-2em}
\section{Introduction}
\label{c6s1}
\graphicspath{{./chapters/WD}}
In this chapter, we investigate an approximately non-invasive measurement protocol which is particularly well-adapted to systems described by continuous variables. It does not involve sequential measurements so is very different to the ideal negative measurement procedure discussed in Section~\ref{subsec:nim}, it may therefore avoid some of its loopholes. It was first proposed for simple spin systems in Ref. \cite{halliwell2016a} and implemented experimentally in Ref. \cite{majidy2019b}, but the implementation considered here for continuous variables is very different.  This chapter addresses the problem of NIM, in the list given in Section~\ref{introsum}. This work is at the time of submission unpublished, and should be considered at the the work in progress stage.

Following the intuitive understanding of the LG inequalities given in Section~\ref{subsec:intuit}, the protocol starts from the observation that the correlators we seek to measure may be written,
\beq
C_{12} = p(S) - p(D),
\eeq
where $p(S) = p(+,+) + p(-,-)$, $p(D) = p(+,-) + p(-,+)$ so $p(S)$ is the probability that $Q_1, Q_2$ have the same sign and $p(D)$ is the probability they have different signs (and clearly $p(S) + p(D) = 1$).
In a macrorealistic view of the process from $t_1$ to $t_2$, we may suppose there is a trajectory $Q(t)$ from initial value $Q_1$ to final value $Q_2$. If $Q_1 = Q_2$, i.e.  they have the same sign, then the trajectory $Q(t)$ must have either zero sign changes or an even number of sign changes. If $Q_1=-Q_2$, then $Q(t)$ must have an odd number of sign changes. However, a reasonable assumption for sufficiently short time intervals is that the motion of $Q(t)$ is sufficiently simple that there is no more than one sign change. The quantities $p(D)$ or $p(S)$ are then readily measured by determining whether $Q(t)$ changes sign, or not, at any time during the time interval $[t_1,t_2]$. The single sign change assumption therefore transforms the measurement of the correlator into a situation involving a {\it single} measurement, which is clearly a desirable situation with regard to non-invasiveness.

Furthermore, for the continuous variable models we are interested in here, the continous variable $x$ is dichotomized by  $Q = {\rm sign} (x)$. Supposing there is a trajectory $x(t)$ for the continuous variable, a sign change of $Q$ simply means the particle crosses the origin at some time. The single sign change assumption is then equivalent to an assumption of a single crossing during the give time interval. All that is required to measure the correlator is then a waiting detector localized around $x=0$ which interacts at most once with the incoming particle -- that is,  a detector acting continuously in time from t1 to t2 which is coupled to a localized function of position. Then
$p(D)$ and hence $C_{12}$, is then readily determined from the fraction of times it clicks over many runs. 

\begin{figure}
\centering
	{\includegraphics[height=7cm]{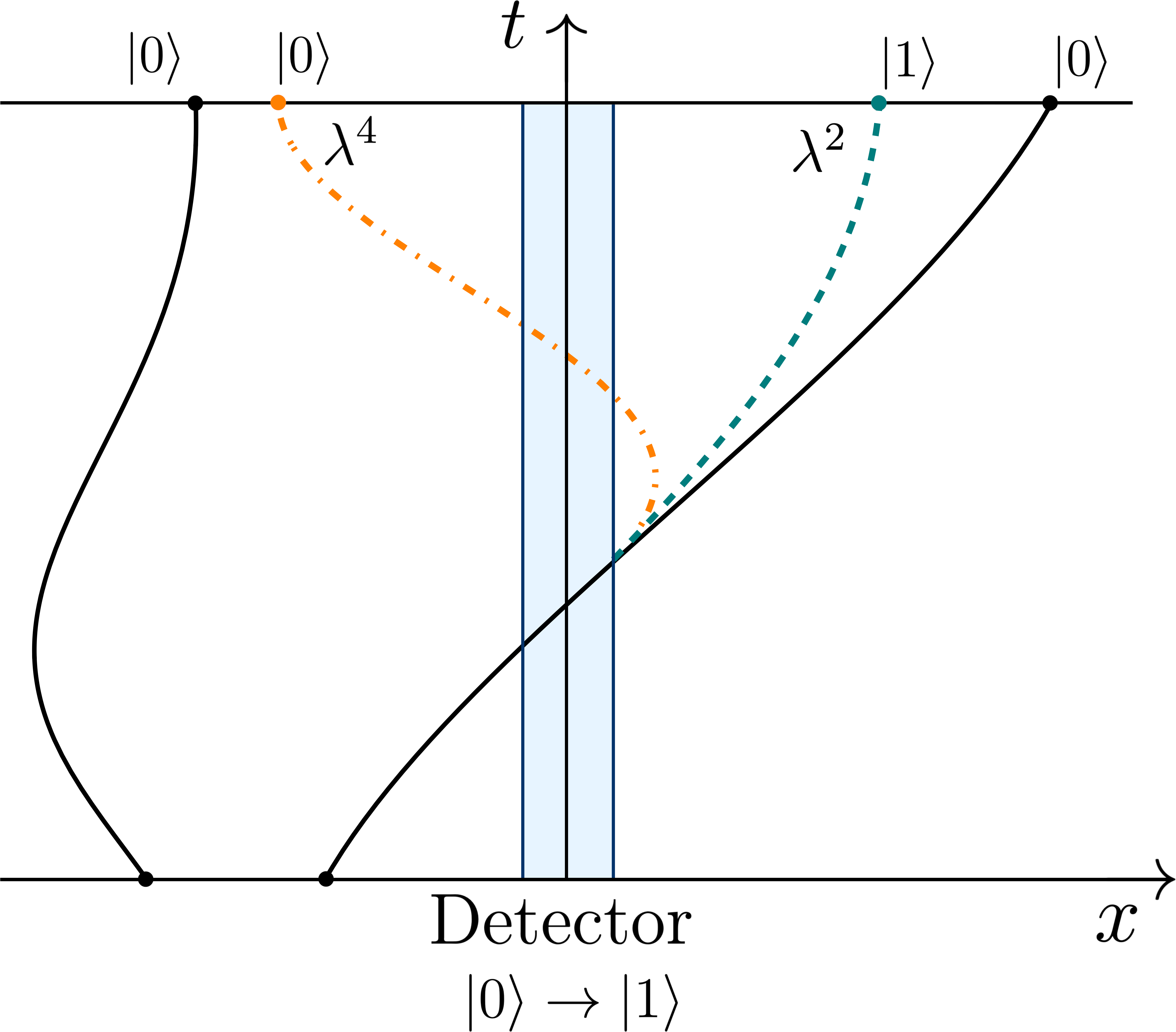}}%
%	%\qquad
%	\hspace{5mm}
%	\subfloat[]{{\includegraphics[height=5.2cm]{fig2a.png}}}
	\caption[The waiting detector model]{The waiting detector model. The detector is localized around the origin initially in the $|0\rangle$ state and weakly coupled to the particle with coupling strength $\lambda$. Most  particles crossing the origin are undetected but some are with probability proportional to $\lambda^2$ and the detector goes to the $|1 \rangle $. The back action of the detector can cause a second crossing and hence the detector returns to the $|0 \rangle $ state, but the probability of this second detection is proportional to $\lambda^4$, which is much smaller than $\lambda^2$ for weak coupling.}%
	\label{fig:waitdet}%
\end{figure}
On the face of it the procedure is non-invasive because, from a macrorealistic point of view, each trajectory interacts only once with the detector (or not at all) and there are no future measurements to disturb.  The interaction with the detector may, however, have an effect on the post-detected evolution of $x(t)$ and in particular, it could cause it recross the origin thereby returning the detector to the undetected state, yielding a false negative. However, this is avoided if the detector coupling is sufficiently weak -- it may be shown that the recrossing detection probability will be an order of magnitude smaller than the effect being measured so may be neglected. Hence the protocol is {\it approximately} non-invasive for sufficiently weak measurements, see Fig.~\ref{fig:waitdet}.  False negatives can also be avoided if the detector is constructed so as to be effectively irreversible.

% what about the case of no detections? IRNM etc. Usual weak measurement story. 

Although the protocol clearly gives information about the sign changes of $x$, once the specific coupling between the particle and detector at the origin is considered
we find that that exact object measured is not in fact ${\rm sign} (x)$, but a more complicated object which also involves the time spent in the region of the detector. However, modified correlators may still be constructed from this object and may be shown to be as useful as the usual LG correlators in terms of detecting macrorealism violations.

In Section \ref{c6s2} we describe the protocol in more detail and exhibit the modified correlators. LG-type inequalities for the modified correlators are derived and we also show that these inequalities have the form of time-averages of conventional LG inequalities.
In Section \ref{sec:sca} we assess the conditions under which the single-crossing assumption is valid.
In Section \ref{c6s4} we compute the modified correlators for the quantum harmonic oscillator and exhibit violations of the modified LG inequalities for a number of different initial states.
%We examine the properties of the modified correlators in the quantum case in Section 4 and in particular show how, for the free particle, they may be transformed into conventional LG correlators by a transformation of the state.  
We summarize in Section \ref{c6s5}.

\section[Measurement Protocol, Modified Correlators and LG Inequalities]{Measurement Protocol, Modified Correlators and \\ Leggett-Garg Inequalities}
\label{c6s2}
The system of interest is a particle in one dimension with Hamiltonian $H$. We will mainly consider the free particle and quantum harmonic oscillator, but the approach will be valid for any system for which the dynamics satisfy the single crossing assumption for reasonable ranges of time.  We suppose that the particle is measured by a weakly coupled detector localized at the origin. We initially consider a detector-particle coupling of the form $ \lambda f(\hat x, \hat p) \otimes H_d$, where  $f( \hat x, \hat p)$ is spatially localized to a small region of size $L$ around $x=0$, $H_d$ is the detector Hamiltonian and $\lambda$ is a suitable small coupling constant.
The detector is allowed to act continuously in time over an interval $[t_1,t_2]$ which is sufficiently short for the single crossing assumption to hold. 
Simple detector models, such as the ones described in Ref.~\cite{halliwell2016a}, show that the detection probability allows us to determine the correlator $ \langle F_{12}^2 \rangle$, where
\beq
F_{12} =  \int_{t_1}^{t_2} dt \ f(\hat x (t), \hat p (t)).
\label{F12}
\eeq
%[MORE HERE? Maybe in appendix].
What follows will apply for many choices of coupling function $f$, but two particular choices are useful.

The first choice is to take $f$ to be the quantum-mechanical current at the origin,
\beq
f  (\hat x, \hat p) =  \frac{1}{2m} \left( \hat p \delta ( \hat x) + \delta (\hat x) \hat p \right),
\eeq
which can be accomplished experimentally for a charged particle in an electromagnetic field localized around the origin. The advantage of this choice is
\beq
f  ( \hat x (t), \hat p (t) ) =  \frac{1}{2} \ \frac{ d \hat Q (t) } {dt}
\eeq
where $ \hat Q = {\rm sign} (\hat x) $, so that as a consequence $F_{12} = (\hat Q_2 - \hat Q_1)/2 $, and 
$  \langle F_{12}^2 \rangle   =  (1 - C_{12})/2  $, from which the standard correlator is obtained. This is clearly the most desirable choice of coupling if it can be arranged experimentally.

The second choice, which is probably easier to arrange experimentally and less dependent on the details of the coupling, is to take $f$ to be a localized function depending only on $\hat x$. This would be the case for a light sheet \cite{lett1988, esslinger1996, bucker2009}, for example, and also a number of other detector models often used in arrival time studies \cite{damborenea2003}.  
We are assuming $L$ is small compared to all other scales, so that we may approximate $f( \hat x) \approx L \delta ( \hat x)$. 
From macrorealistic point of view, we have an ensemble of trajectories $x(t)$. Assuming only single crossings and without at this stage assuming anything about the dynamics, we easily find that $F_{12} = 0$ if the trajectory does not cross the origin, and if it does, we get $ 1 / |\dot x (t_c)|  $, where $t_c$ is the time at which $x(t) = 0$. This is conveniently written
\beq
F_{12} = \frac{L} { \dot x(t_c) } \left(  \theta (x(t_2) ) - \theta (x(t_1) ) \right).
\eeq
For free particle dynamics we get
\beq
F_{12} = \frac{ m L }{p} \left(  \theta (x(t_2) ) - \theta (x(t_1) ) \right).
\label{F12class}
\eeq
For trajectories which cross we get $F_{12} = m L / |p |$, which is the dwell time $\tau_D$  (i.e. the time spent in an interval of size $L$ for a particle of momentum $p$). Noting that 
\beq
\theta (x(t_2) ) - \theta (x(t_1) )  = \frac{1}{2} (Q_2 - Q_1 ),
\eeq
we see that the form of $F_{12}$ is very similar to the one obtained with the coupling to the current except for the overall factor of the dwell time.  Because of this close similarity it is easy to show that the correlators obey the following set of macrorealistic inequalities, which are direct analogues of the LG3 inequalities:
\bea
\langle F_{12}^2 \rangle + \langle F_{13}^2 \rangle - \langle F_{23}^2 \rangle &\ge& 0  
\label{mLG1}
\\
\langle F_{12}^2 \rangle - \langle F_{13}^2 \rangle + \langle F_{23}^2 \rangle
&\ge& 0  
\label{mLG2}
\\
- \langle F_{12}^2 \rangle + \langle F_{13}^2 \rangle + \langle F_{23}^2 \rangle
&\ge& 0  
\label{mLG3}
\\
-\langle F_{12}^2 \rangle - \langle F_{13}^2 \rangle - \langle F_{23}^2 \rangle
+ 2 \langle \tau_D^2  \rangle
&\ge&  0
\label{mLG4}
\eea
We call these the modified LG inequalities (mLG).  The quantity $ \langle \tau_D^2  \rangle $ is measured by measuring $\ev{F_{12}^2}$ and taking $t_1\rightarrow-\infty$ and $t_2\rightarrow\infty$ (where by $\infty$ we mean a sufficiently large time compared to timescales of the system)  It is properly defined only for states which go to zero fast enough at $p=0$, although this is not required for $ \langle F_{12}^2 \rangle $ with finite $t_1$, $t_2$.  These inequalities readily follow using inequalities of the type used to derive standard LG inequalities, such as
\beq
(Q_1 -Q_2)^2 + (Q_2 - Q_3)^2 - (Q_3 - Q_1)^2 \ge  0.
\eeq
There are also analogues of the LG4 inequalities, which are
\beq
\label{F12lg4}
0 \le \langle F_{12}^2 \rangle + \langle F_{23}^2 \rangle + \langle F_{34}^2 \rangle -\langle F_{14}^2 \rangle \le 2
\eeq
plus three more pairs of inequalities with the minus sign in the other three locations.

 In addition there are two-time LG inequalities which follow from the inequality 
\beq
\tau_D^2 (1 -\theta( x(t_1) ) ( 1 - \theta (x(t_2) ) \ge 0 ,
\eeq
and have the form
\beq
\langle \tau_D^2 \rangle -  \langle \tau_D^2 \theta( x(t_1)  \rangle - \langle \tau_D^2 \theta( x(t_2)  \rangle
+ \langle F_{12}^2 \rangle \ge 0.
\eeq
The quantities of the form  $ \langle \tau_D^2 \theta( x(t_2)  \rangle$ can be measured if the dynamics of the system
allows the state to become an approximate eigenstate of $\theta(x)$ for sufficiently large positive or negative times. For example suppose the state starts out localized in $x>0$ as $t_1 \rightarrow - \infty $, then we can measure $\langle F_{12}^2 \rangle $ as usual for large negative $t_1$, and note that it tends to $ \langle \tau_D^2 \theta( x(t_2)  \rangle$. This is clearly a restriction on permissible dynamics and states so these LG2 inequalities may not be as useful. 
A further complication is that in the quantum case the detector model with $f$ proportional to $ \delta(x)$ only picks up symmetric states, which means that $  \langle \tau_D^2 \theta( x(t_2)  \rangle = \frac{1}{2} \langle \tau_D^2   \rangle$ and the LG2 inequality is trivial. However, this simplification is avoided for more general couplings $f$.

Although we have focused on the case where $F_{12}$ is given by Eq.~(\ref{F12class}), actually the general form  of these inequalities does not depend  much on the dynamics and requires only that a regime exists in which the single crossing assumption holds. They are readily generalized to the case of a broader localizing function at the origin in which case the $\theta(x)$ functions become smoothed step functions, but LG type inequalities can still be derived for such functions and still yield interesting MR violations.
In addition, the coupling constant $\lambda$ drops out of the inequalities, as does the length $L$ in the case $ f(x) \approx L \delta (x)$.

%subject only to the existence of a regime in which the single crossing assumption holds.

A precise relationship between the modified and standard correlators in the quantum case is derived in Appendix \ref{app:f12corr} for the free particle (which of course coincides with the harmonic oscillator for sufficiently short times) and $\delta$-function coupling.
Briefly, the modified correlator $\langle F_{12}^2 \rangle$ in a given state $ |\psi \rangle $ is simply related to the standard correlator $C_{12}$ in a different state $|\phi \rangle$,  if we choose $ | \psi \rangle $ proportional to $ \hat p | \phi \rangle $ and take  $ \phi(x) $ to be antisymmetric.
This connection is useful mathematically for linking violations of the modified LG inequalities to  known violations of the usual LG3 inequalities.
For example, in Chapter~\ref{chap:QHO1} it was shown that significant LG3 violations may be obtained for the quantum harmonic oscillator in the first excited state $|1\rangle$. The corresponding $|\phi \rangle$ is proportional to $ \hat p |1 \rangle$, which is a superposition of the $|0\rangle$ and $|2 \rangle $ states, so we expect a violation of the modified LG inequalities in this state.
This connection also allows us to determine the largest possible violations of the modified LG3 inequalities (see Appendix \ref{subsec:mlgluders}) -- the violations can be up to $ -\frac{1}{4} \langle \tau_D^2 \rangle$ on the right-hand side of Eqs.~(\ref{mLG1}--\ref{mLG4}).
%Maximal violations?

%Momentum kernels, Wigner function etc, all in appendices, if they are in fact needed in the end. (We might want to leave this out of the paper, for conciseness).
\section{Assessment of the single crossing assumption}
\label{sec:sca}
%This might be better located before the LG violations section but it seemed more streamlined to get quickly to the LG violations.

The single crossing assumption ensures that in each experimental run there can be at most one interaction with the detector and therefore there is no possibility that any measurement is affected by an earlier measurement.
It is clearly a very reasonable one from a macrorealistic point of view for the free particle and for a particle in a potential which is such that the momentum does not change sign. For bound systems, like the harmonic oscillator, the assumption will be valid for up to half a period and possibly beyond, depending on the initial distribution of positions and momenta.  However, we clearly need to find ways to assess from an experimental point of view whether or not the assumption is valid in a given experiment situation. Since we expect experiments to conform to the laws of quantum theory this also means we need some way to assess the validity of the assumption in a quantum description.

In the spin model considered in Ref. \cite{halliwell2016a}, the analogous assumption is the single sign change of a dichotomic variable $Q = \sigma_z$, where $\sigma_z$ is a Pauli matrix (and the coupling function $f$ is taken to be the time derivative of $Q$). If we do measurements on the system at three times,
the quantum-mechanical probability for two sign changes is $p_{123}(+,-,+) + p_{123}(-,+,-) $, where $ p_{123}(s_1,s_2,s_3)$ is the three-time sequential measurement formula.  This was calculated in the spin model and found to be significantly less than the probabilities for zero or one sign change for sufficiently short times, but for times which could also be sufficiently long to get LG3 violations. Hence there is a regime in which the assumption holds sufficiently well. On general grounds one expects this sort of result to hold for sufficiently short times for many quantum systems, since, as is well known from studies of path integrals, the amplitudes for space time paths tend to diminish as paths have more zig zags. 
Another approach which avoids using three sequential measurements, would be to use two waiting detectors, one active during the interval $[t_1,t_2]$ the other during $[t_2,t_3]$, and then look for the probability that both detectors click. In the quantum case with $  F_{12} = \hat Q_2 - \hat Q_1 $, this can be shown to be proportional to $\langle F_{12} F_{23}^2 F_{12} \rangle$, which can be shown at some length to be proportional to  $p_{123}(+,-,+) + p_{123}(-,+,-) $, hence these two methods coincide.

%There is another way of assessing the single crossing assumption which avoids using sequential measurements and 
%involves examining the time dependence of the correlators obtained from the waiting detector model.
Macrorealistically the SCA means that objects like $\ev{\theta(x(t))\theta(-x))}$ increase monotonically with time and therefore it is useful to consider their time derivatives which are $\ev{J(t) theta(-x)}$. These objects can be measured from time derivatives of the correlator, which may be estimated experimentally through use of a finite difference approach.  For simplicity we focus on the case where the detector coupling is to the current at the origin and so we obtain the usual correlator $C_{12} = {\rm Re} \langle \hat Q_2 \hat Q_1 \rangle $. 
Taking additional correlator measurements at times close to $t_2$, we can experimentally determine the time derivative
\beq
\label{deriv1}
\frac{ \partial}{ \partial t_2}  C_{12}  = 
 2 \ {\rm Re} \  \langle \hat J(t_2)  \left[   \theta( \hat x(t_1) ) -  \theta( - \hat x(t_2)) \right] \rangle
\eeq
Note also that we if we measure $\langle \hat Q(t_2) \rangle$ at a range of times in separate experiments we can determine the derivative
\beq
\label{deriv2}
\frac{ \partial }{ \partial t_2}\langle \hat Q(t_2) \rangle = 2  \langle  \hat J(t_2) \rangle
=2 \ {\rm Re} \  \langle \hat J(t_2)  \left[   \theta( \hat x(t_1) ) +  \theta( - \hat x(t_2)) \right] \rangle
\eeq
\sloppypar{\noindent This means that we can experimentally determine the ``one-sided" chopped currents  $ {\rm Re}  \  \langle \hat J(t_2)   \theta( \pm \hat x(t_1) ) \rangle  $. These are measures of the flow of probability for starting in $x>0$ or $x<0$ and then crossing the origin at some time.  A reasonable assertion is that the single crossing assumption corresponds to the idea that the one-sided chopped current for the left-right crossing is always positive, and always negative for right-left crossings, over a range of times of interest. Hence if this current changes sign the single crossing assumptions fails. These currents are readily related to the fully chopped currents 
$  \langle \theta (\pm \hat x(t_1) )\hat J(t_2)   \theta( \pm \hat x(t_1) ) \rangle  $
considered in detail in Chapter~\ref{chap:QHO2}, using for example identities of the general form Eq.~(\ref{eq:id}), with $\rho$ replaced by $\hat J$. For a harmonic oscillator with initial gaussian state and it can be seen from e.g. Fig~\ref{fig:currs}, that the chopped currents maintain their sign for long periods of time.
}

%[A bit more detail about the behaviour of the different currents may be required here. Although note that if we are only looking at symmetric or antisymmetric states then $\langle \hat Q_2 \rangle =0$.]

A useful elaboration on the use of currents is to replace the $Q = {\rm sign} (\hat x) $ at $t_2$ with a displaced quantity
$Q = {\rm sign} (\hat x - x_0) $, and the above expression for the derivative of $C_{12}$ will then involve the chopped currents  at $x_0$. From these one can easily plot Bohm trajectories and see the flow of probability over a space time region. From the Bohm trajectories plotted for simple states in Section~\ref{subsec:bohm}, it is easily seen that although the Bohm trajectories depart appreciably from the corresponding classical ones (this is what produces the LG violations), there is a large regime in which they remain {\it qualitatively similar} to the classical ones, and in particular the single crossing assumption remains valid.

Another piece of evidence supporting the single crossing assumption in the quantum case comes from the study of backflow \cite{bracken1994, yearsley2013} . This is the phenomenon that a free particle in a state consisting entirely of positive momenta can have negative current, i.e. can flow ``backwards''. Or, in the present language, the state corresponds to trajectories which may cross the origin more than once, unlike the analogous classical paths. What is interesting for our purposes is that this phenomenon is not very large -- it has been shown that it can reduce the probability for a left-right crossing by no more than about $ 4 \%$ (and probably by a lot less for the sorts of states considered here). 
This is to be compared to maximal LG3 violations, which, in Eqs.~(\ref{LG3a}--\ref{LG3d}), can be as much as the L\"uders bound $ - \frac{1}{2}$. However, this violation needs to be compared with the corresponding classical probability, and to this end, note that if an underlying probability $p(s_1,s_2,s_3)$ exists, then it is easily shown that
\beq
p(s_1,s_2,s_3) + p(-s_1,-s_2, -s_3)  = \frac{1}{4} \left(1 + s_1 s_2 C_{12} + s_2 s_3 C_{23} + s_1 s_3 C_{13} 
\label{ppC}
\right),
\eeq
which means that the L\"uders bound corresponds to $ - \frac{1} {8} $, or in other words $ 12.5\%$ of the corresponding classical probability, so is much greater than the backflow effect. This suggest that if the LG violations are sufficiently large, they cannot be explained by failure of the single crossing assumption in quantum mechanics.

%We have focused so far on the case of the usual correlators, but we expect a similar story to also hold for the modified correlators $\langle F_{12}^2 \rangle$. 
%This can be investigated explicitly using the transformation between the modified and usual correlators described in Section 2 but we will not go into that here.

It should be noted that there is an apparent tension between the single crossing assumption and one of the LG3 inequalities.  The single crossing assumption implies that $F_{12}$ and $F_{23}$ cannot both be non-zero in each run, which means that macrorealistically $\langle F_{12} F_{23} \rangle = 0$ and, since $F_{13} = F_{12} + F_{23}$, the LG3 Eq.~(\ref{mLG2}) is then equal to $ -2 \langle F_{12} F_{23} \rangle $ which is zero.  This raises the question as to the significance of a violation of this particular inequality. However, as noted above any measurements to check the single crossing assumption would determine quantum-mechanical analogues of $\langle F_{12} F_{23} \rangle $ which, if zero, would not necessarily make the LG3 inequality trivially zero. This basically comes down to the difference between an LG3 inequality, which, when positive can be written as a probability of certain sign changes as in Eq.~(\ref{ppC}), and the corresponding sequentially measured probability, such as 
$p_{123}(+,-,+) + p_{123}(-,+,-) $. They differ by interference terms as shown in Chapter~\ref{chap:mlev}, Eq~(\ref{corrints}).  Hence, despite the close connection between this particular LG3 and the SCA it is still plausible that we can have an LG3 violation with the SCA arguably satisfied, although this would need to be argued carefully in specific models. And even if it seems likely that an LG3 violation is in fact due to a failure of the SCA this is still an interesting result since it shows the presence of non-classical behaviour.

This issue only arises for the specific LG3 inequality Eq.~(\ref{mLG2}) but as it happens this is the LG3 inequality that is most easily violated \cite{halliwell2016a}. However, the separation between the single crossing conditions and LG violation is much clearer for the other three LG3 inequalities and also for the LG2 and LG4 inequalities, so this could in practice be the place situation to explore in order to find violations when single crossing holds.

In summary, the single crossing assumption can be argued to hold up well in a quantum context and can be checked experimentally by examining the time-dependence of the data obtained by the waiting detector, allowing an experimental finite difference estimation of the derivatives in Eqs.~(\ref{deriv1}-\ref{deriv2}).
 
A different issue which has bearing on the claimed non-invasiveness of the measurement concerns the fact that the detector acts continuously in time. Some LG experiments have used such measurements but questions have been raised as to whether these measurements meet the non-invasiveness requirement \cite{emary2013}. Here, however, the key thing is that we are supposing the particle does not interact for sufficiently long or sufficiently strongly with the detector for there to be any significant disturbance.
For a detector localized to a region $L$, the particle will be in continuous interaction with the detector for the length of time the particle is traversing the region, i.e. the dwell time $\tau_D$. This interaction will probably not affect whether or not the trajectory crosses the region but it will affect the total time spent there. This is because
the interaction will create a spread $\Delta p (t)$ in the momenta of the family of trajectories crossing the region, which will grow with time and will also grow with the coupling strength $\lambda$. For example, in a simple diffusive model of the effects of the interaction, $\Delta p(t)$ will go like $ \lambda t^{\frac{1}{2}}$. (See for example Ref.~\cite{bednorz2012} for a more detailed model of continuous measurement). However, the spread $\Delta p(t) $ is largest once $t$ reaches the dwell time $\tau_D$ and as long as $\Delta p(\tau_D) $ is significantly less than $p$, which can be accomplished for sufficiently small $\lambda$ and $L$,
then the interaction will not have a significant effect on the correlators. Hence we expect there is a regime in which the back action of the continuous measurement is sufficiently small that it can be neglected.

\section{Violations of the modified LG inequalities for the harmonic oscillator}
\label{c6s4}
We now calculate the modified correlators and look for violations of the modified LG inequalities for a number of states of the QHO.  We also check the single crossing assumption, which we expect to be satisfied for sufficiently short times. We initially look at energy eigenstates and coherent states using the simplest coupling $f(\hat x)=L\delta(\hat x)$. We later repeat the calculation for a smoothed gaussian coupling.  We use conventions as defined in Section~\ref{sec:calc}.
%\subsection{Appearance of the Dwell-time}
%As intimated earlier, the $\ev{F_{12}^2}$ quantities are dimensionful, where we expect them to be related to standard correlators through
%\begin{equation}
%\ev{F_{12}^2}\sim \ev{\tau_D^2}\frac{1-C_{12}}{2}.
%\end{equation}
%The presence of the dwell 

\begin{figure}
	\subfloat[]{{\includegraphics[height=4.9cm]{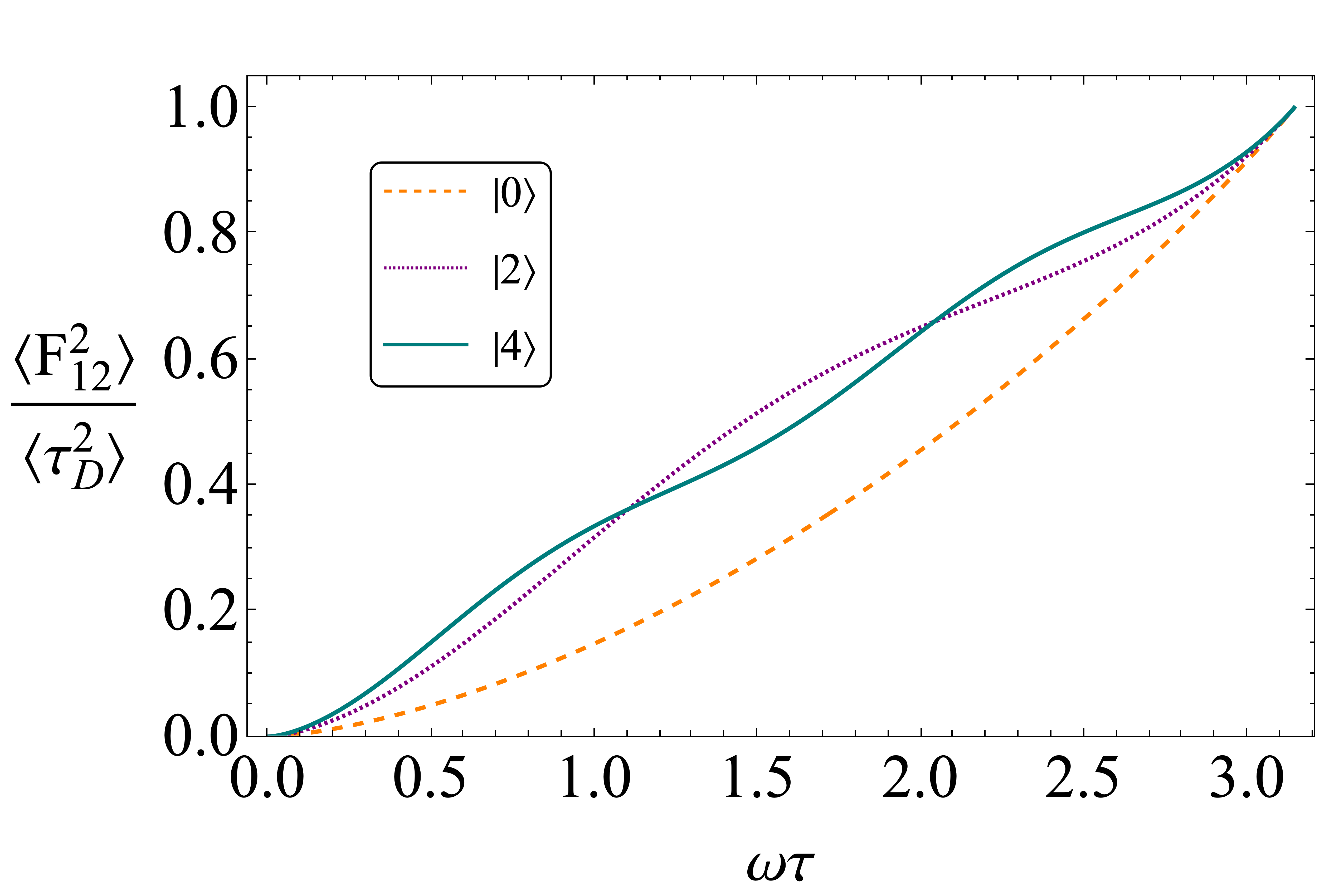}}}%
	%\qquad
	\hspace{5mm}
	\subfloat[]{{\includegraphics[height=5.4cm]{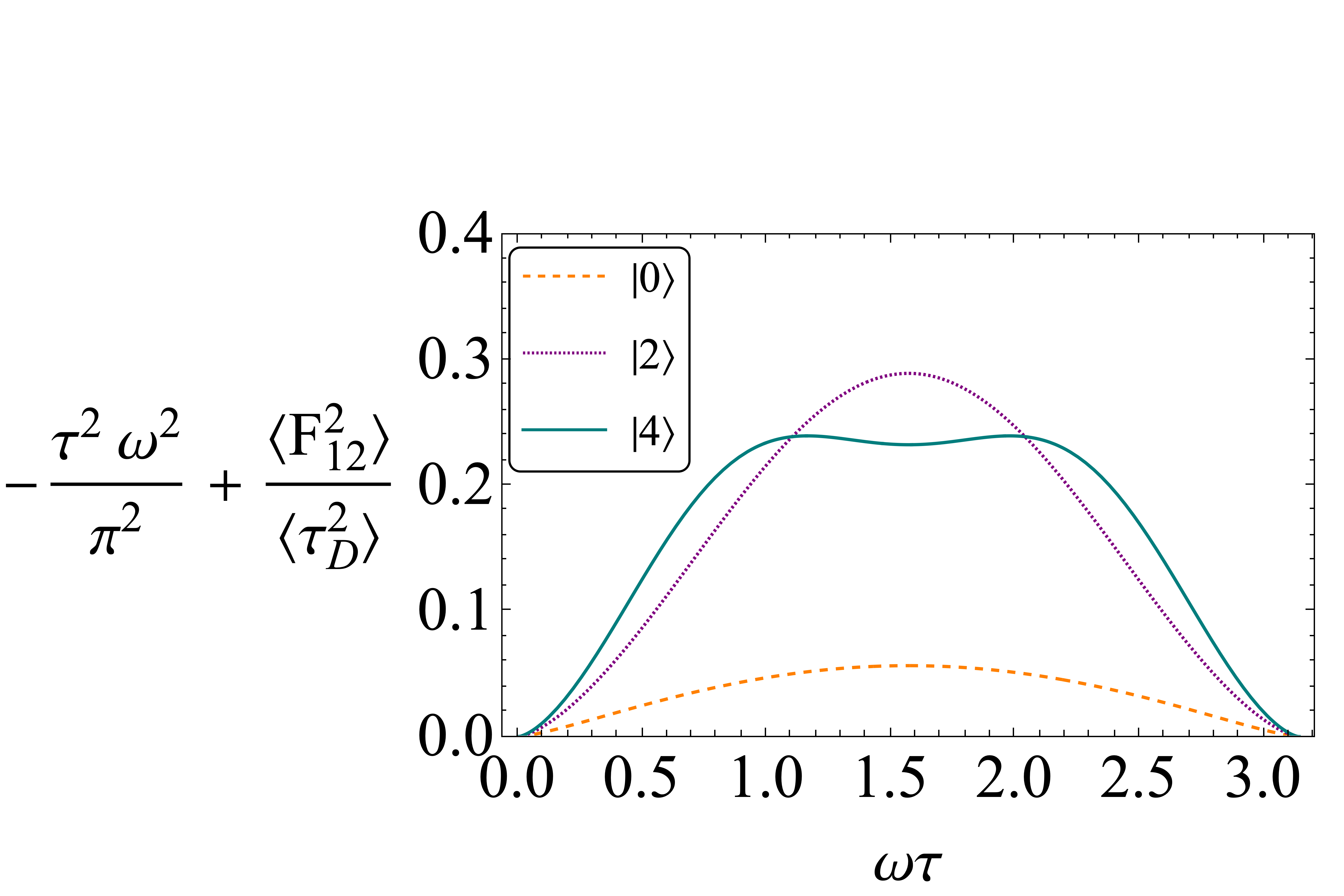}}}
	\caption[$\ev{F_{12}^2}$ for energy eigenstates]{(a) The modified correlators for the energy eigenstates.  In (b) we plot the oscillatory part of the modified correlators.}%
	\label{fig:f12sing}
\end{figure}

\subsection{$\ev*{F_{12}^2}$ for Energy Eigenstates}
%\begin{figure}[b]
%\centering
%	{\subfloat[]{\includegraphics[height=5.4cm]{singlef12.png}}%
%	%\qquad
%%	\hspace{5mm}
%	\subfloat[]{{\includegraphics[height=5.4cm]{singlef12osc.png}}
%	\caption[$\ev{F_{12}^2}$ for energy eigenstates]{(a) The modified correlators for the energy eigenstates.  In (b) we plot the oscillatory part of the modified correlators.}%
%	\label{fig:f12sing}}
%\end{figure}
In Appendix~\ref{app:enmel} we calculate the matrix elements $\mel{m}{F_{12}^2}{n}$.  Taking the case $m=n$ yields the result for an individual energy eigenstate
\begin{equation}
\label{eq:f12eig}
\ev{F_{12}^2}{n}=\tau ^2 M_{nn}^2 +\frac{2}{\omega^2}\sum_{k, k\neq n}^{\infty} \frac{M_{nk}^2 (1-\cos (\omega\tau  (n-k)))}{(n-k)^2},
\end{equation}
with $\tau=t_2-t_1$ the measurement interval.  For the coupling $f(\hat x)=L\delta(\hat x)$, its matrix elements are trivially given by
\begin{equation}
	M_{nk}=L\mel{n}{\delta(\hat x)}{k}=L\psi_n(0)\psi_k(0),
\end{equation}
which for symmetric potentials is non-zero only for even $n$, $k$.  This gives for example summing up to $k=4$ for the ground state
\begin{equation}
\label{gsf12}
	\ev{F_{12}^2}{0}=\frac{19 + 64 \tau^2 \omega^2-16 \cos (2 \omega \tau)-3 \cos (4 
	\omega \tau)}{64 \pi  \omega^2},
\end{equation}
where with the terms in the sum suppressed by $\frac{1}{k^2}$, the next term has coefficient $-0.005$.

We show the behaviour of the modified correlators for the first few even energy eigenstates in Fig.~\ref{fig:f12sing}(a), with just their oscillatory parts singled out in Fig.~\ref{fig:f12sing}(b).

We now assess the single crossing assumption.  Firstly on classical grounds, we would expect the single crossing assumption to hold for as long as $\frac12$ of a cycle.  Hence the mLG3s ought to be valid for up to $\frac14$ of a cycle, and the mLG4s for $\frac16$ of a cycle, since the longest time intervals involved here are $2\omega\tau$ and $3\omega\tau$ respectively.  Within QM higher frequencies are possible, which may indeed be seen in Eq.~(\ref{gsf12}), with the appearance of a $\cos4\omega\tau$ term, however its weighting is smaller than the $\cos2\omega\tau$ term by a factor of more than 5.  The argument in Section~\ref{sec:sca} indicates that the single crossing assumption can be considered to be hold, up until the point where the currents involved turn around, which in Fig~\ref{fig:f12sing}(b) can be seen to occur at $\omega\tau=\pi$ for the $\ket0$ and $\ket2$ states, and for the $\ket4$ state slightly earlier around $\omega\tau\approx \frac{\pi}{3}$, indicating there is a decent range of times for which the assumption holds.

We note for energy eigenstates the modified correlators are dependent only on the time difference, and thus we benefit from the usual simplifications of stationarity.  For energy eigenstates we find violations of the mLG3 inequality Eq.~(\ref{mLG2}) and the mLG4 Eq.~(\ref{F12lg4}), which in the stationary scenario with equal time intervals become
\begin{align}
\label{stat3}
	2\ev{F_{12}^2}-\ev{F_{13}^2}&\geq0,\\
	\label{stat4}
	3\ev{F_{12}^2}-\ev{F_{14}^2}&\geq0.
\end{align}
We plot both of these mLG3 and mLG4 inequalities for the first few even energy eigenstates of the QHO in Fig.~\ref{fig:f12e}, where we find they are readily violated.  Crucially we note these violations occur within time intervals over which we have argued we expect the single crossing assumption to be valid.  In these figures the shaded region represents violations within the L\"uders bound, however as discussed in Appendix~\ref{subsec:mlgluders} we expect this not to be an exact bound but more an indication of the scales involved.  

\begin{figure}
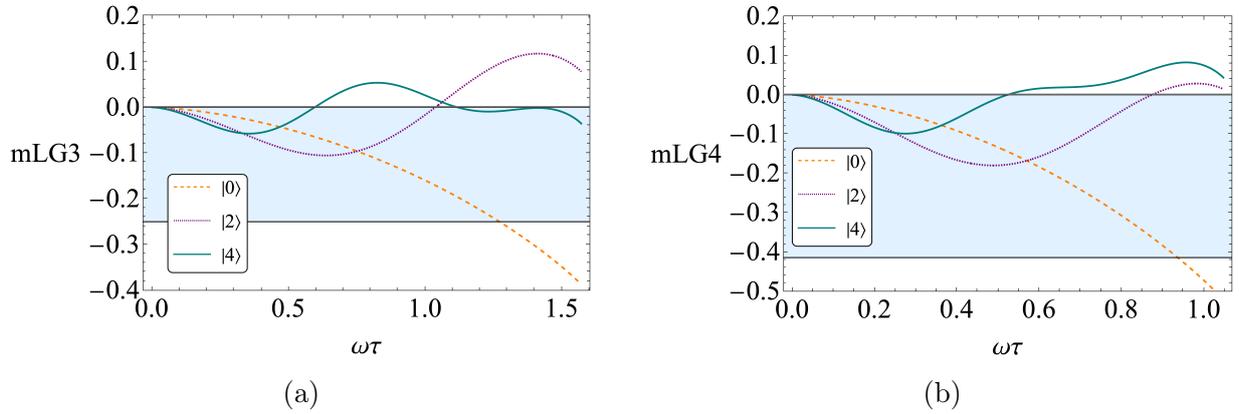

	\subfloat[]{{\includegraphics[height=5.2cm]{fig1a.png}}}%
	%\qquad
	\hspace{5mm}
	\subfloat[]{{\includegraphics[height=5.2cm]{fig1b.png}}}
	\caption[mLG violations for QHO energy eigenstates]{mLG violations for energy eigenstates of the QHO for (a) the three-time inequality Eq.~(\ref{stat3}), and (b) the four-time inequality Eq.~(\ref{stat4}).  The shaded region represents violations with the lower limit the L\"uders bound.}%
	\label{fig:f12e}%
\end{figure}

\subsection{mLG3 violations in the $\hat p\ket1$ state}
As the modified correlator with $\delta$ coupling picks up only the symmetric part of a state, and may potentially have issues with states which do not vanish at $p=0$, the simplest state sidestepping both these issues is the $\hat p\ket1$ state.  As detailed in Appendix~\ref{app:f12corr}, there is a precise correspondence here to the standard correlator in the $\ket1$ state, which we found in Section~\ref{sec:1and0} had near maximal LG3 violations, hence we expect violations here. Further the calculation for this state with details given in Appendix~\ref{secp1app} inherits the simplicity of the temporal correlator for this state, yielding the result
\begin{equation}\ev*{F_{12}^2}{\psi}=\frac{1}{8\omega^2} \left(1+2 \omega^2 \tau ^2  +2 \omega \tau  \sin (2  \omega \tau)-\cos (2  \omega \tau)\right).
\end{equation}
Using the aforementioned correspondence with the conventional correlators, in Fig.~\ref{fig:f12corr}(a) we make a direction comparison between the modified correlator for this state, and the conventional correlator calculated in Chapter~\ref{chap:QHO1}, finding reasonably good agreement for short times.  In Fig.~\ref{fig:f12corr}(b) we show the mLG3 violations for this state.
\begin{figure}
	\subfloat[]{{\includegraphics[height=5.2cm]{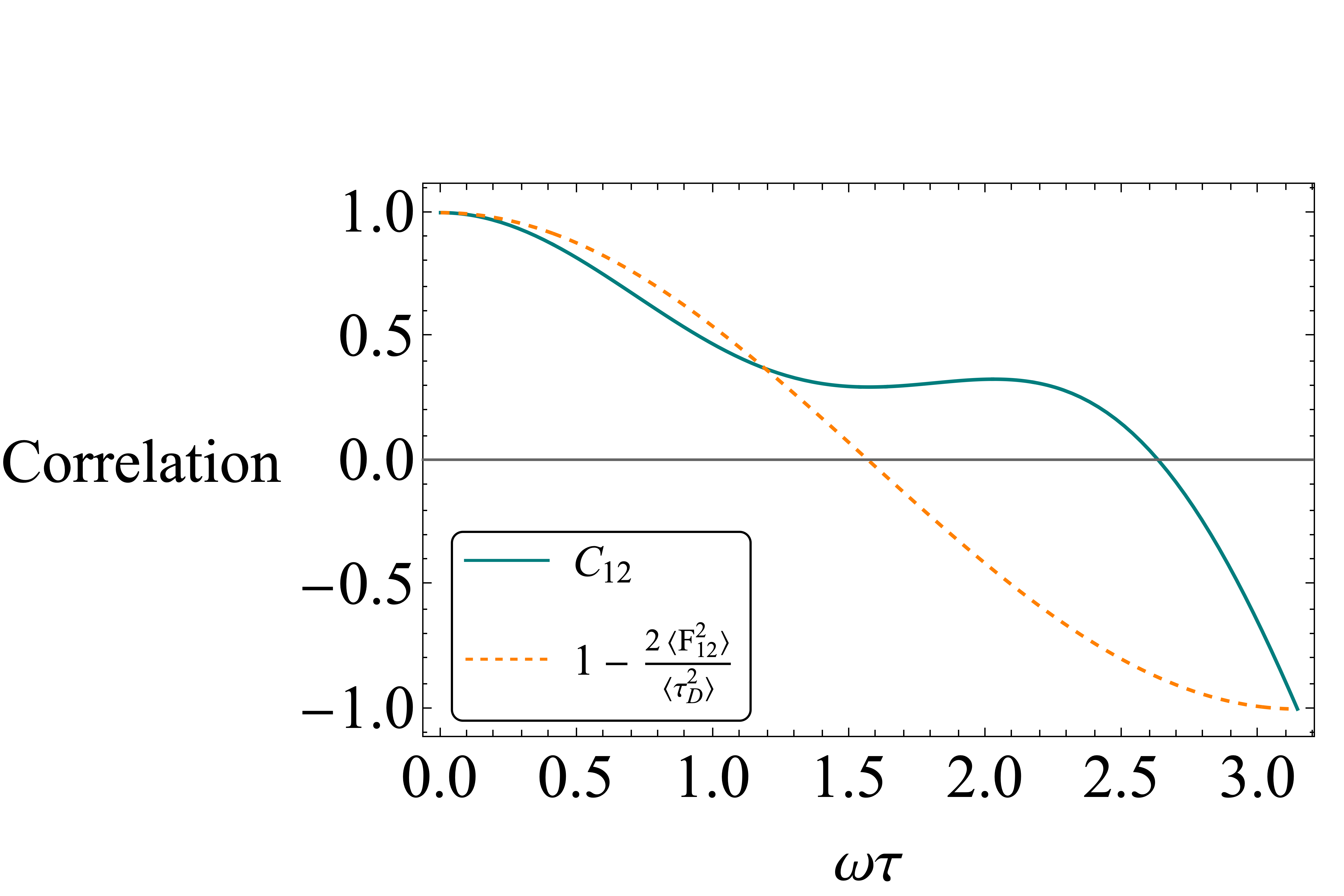}}}%
	%\qquad
	\hspace{5mm}
	\subfloat[]{{\includegraphics[height=5.2cm]{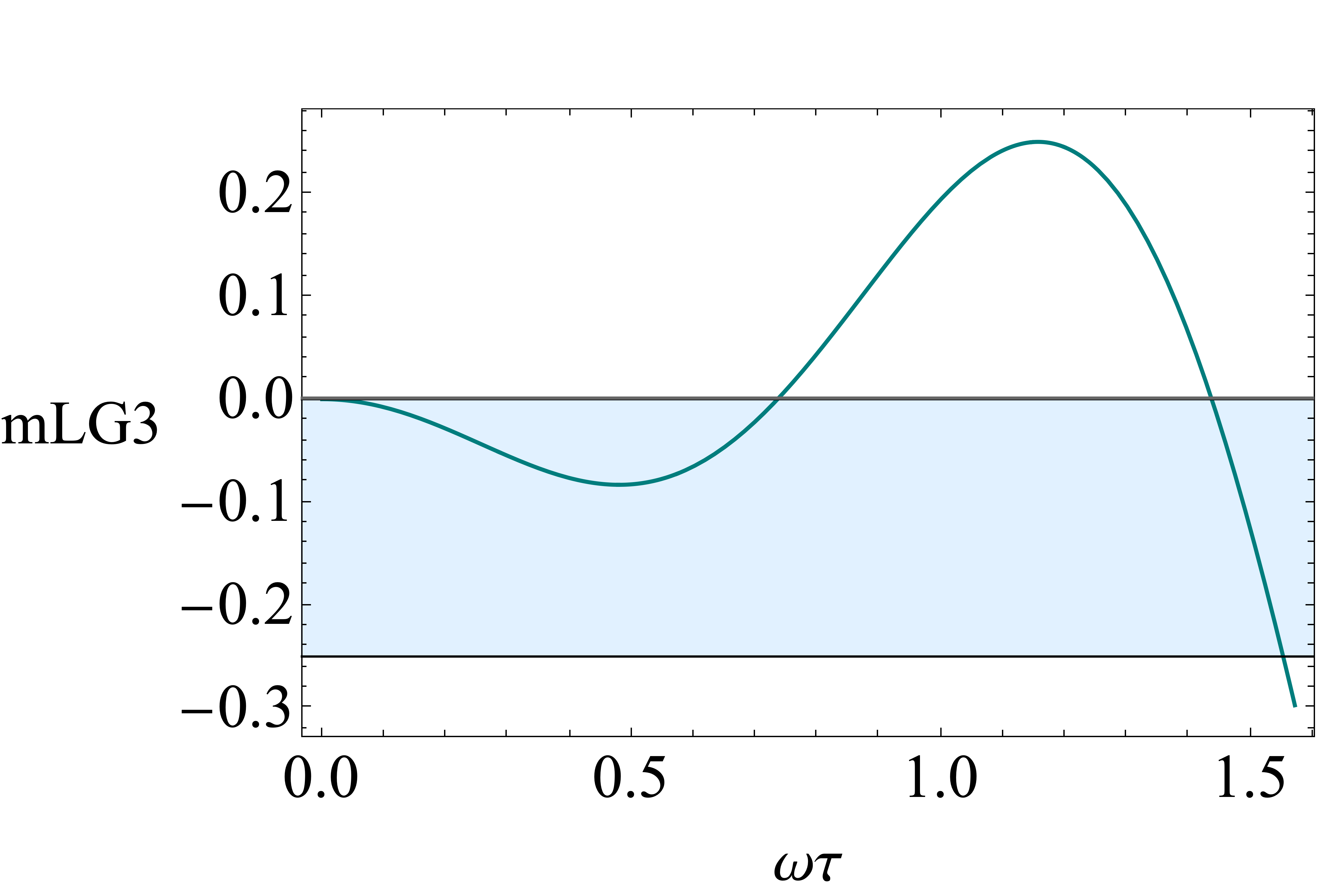}}}
	\caption[$\ev*{F_{12}^2}$ compared to the standard correlator]{(a) Comparison of the standard correlator Eq.~(\ref{cosapprox}), and that found using the waiting detector model. (b) shows the mLG3 violations for the state $\ket\psi \propto \hat p \ket1$}%
	\label{fig:f12corr}
\end{figure}
\subsection{Coherent States}
\begin{figure}
	\subfloat[]{{\includegraphics[height=5.2cm]{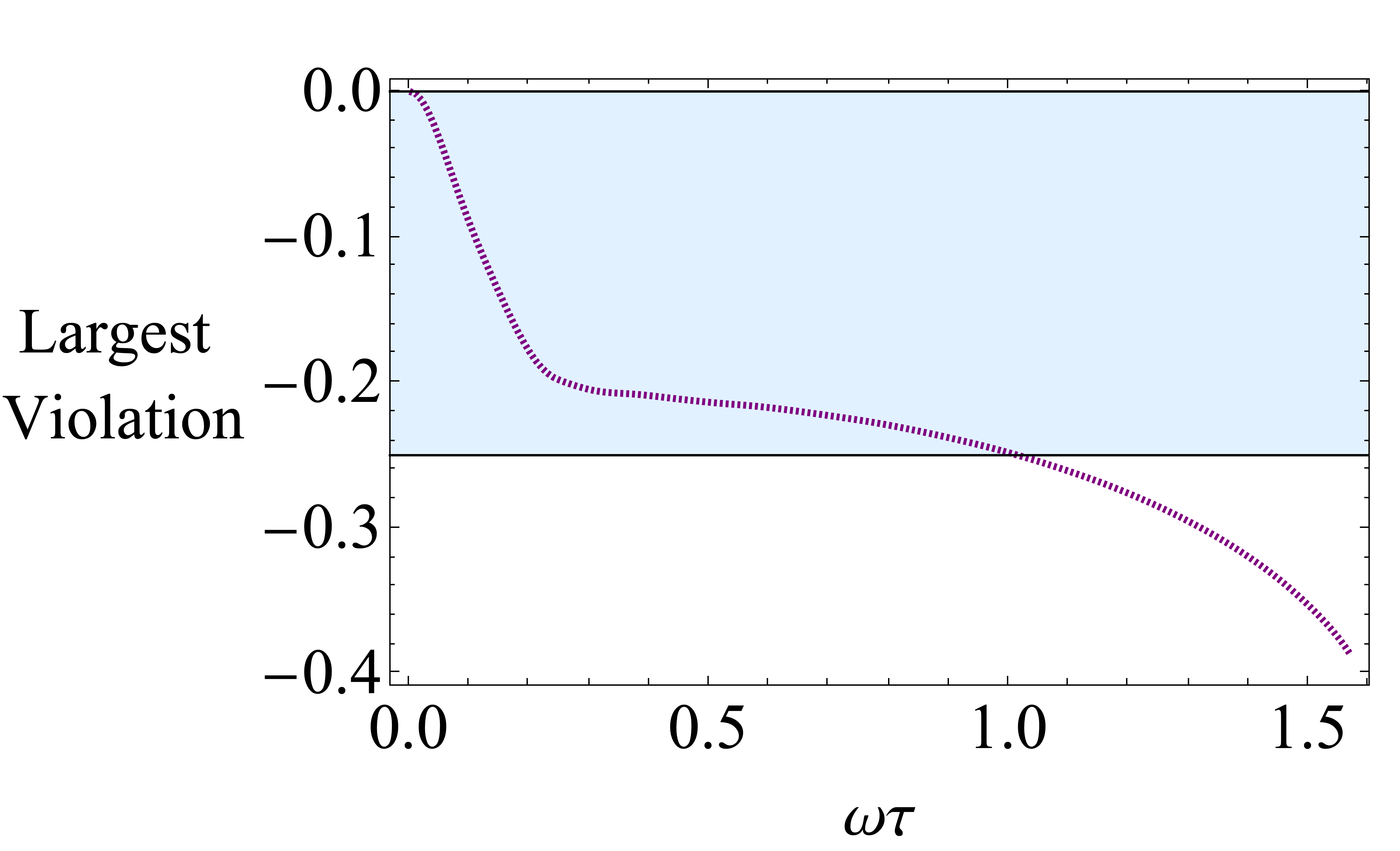}}}%
	%\qquad
	\hspace{5mm}
	\subfloat[]{{\includegraphics[height=5.2cm]{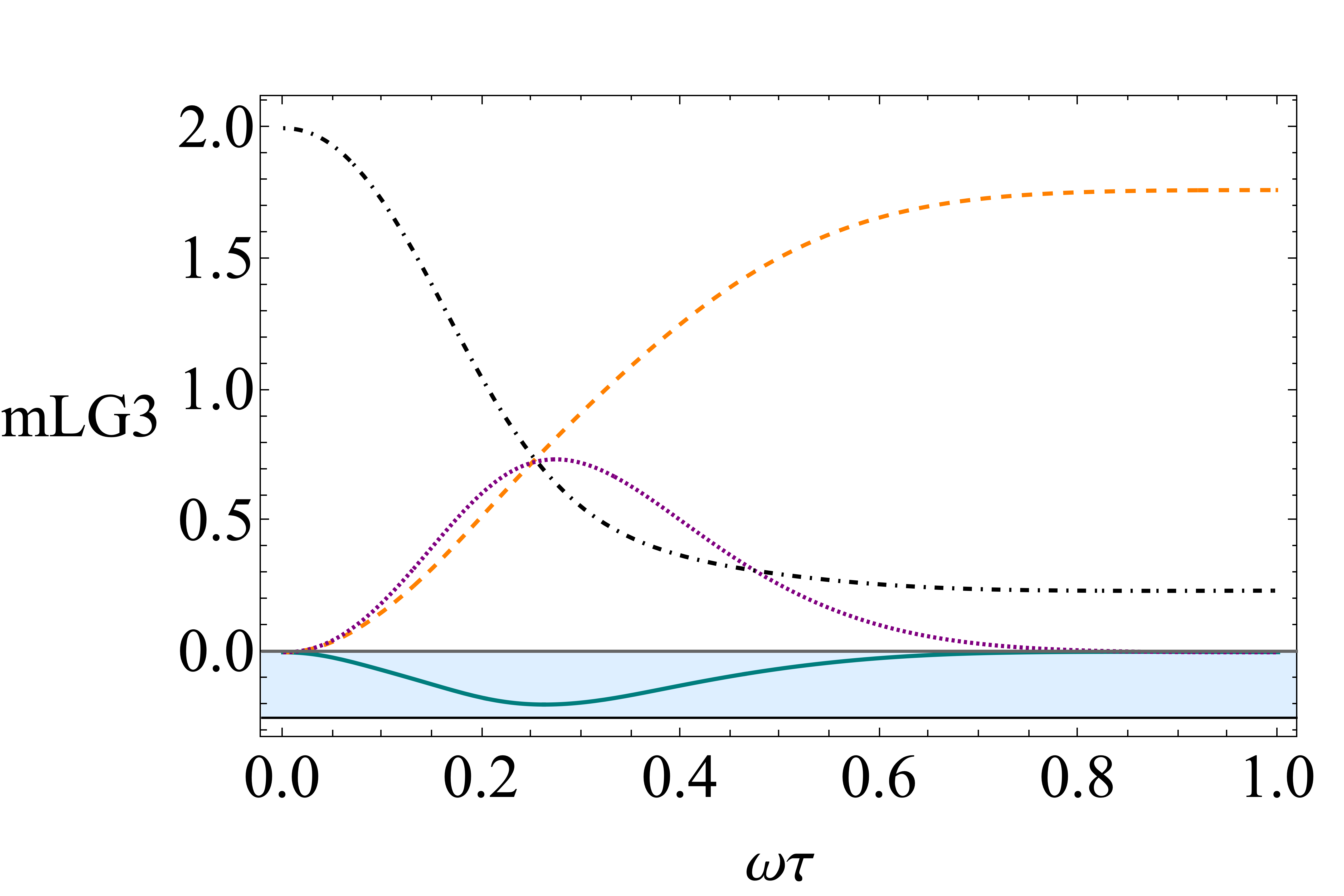}}}
	\\
	\subfloat[]{{\includegraphics[height=5.2cm]{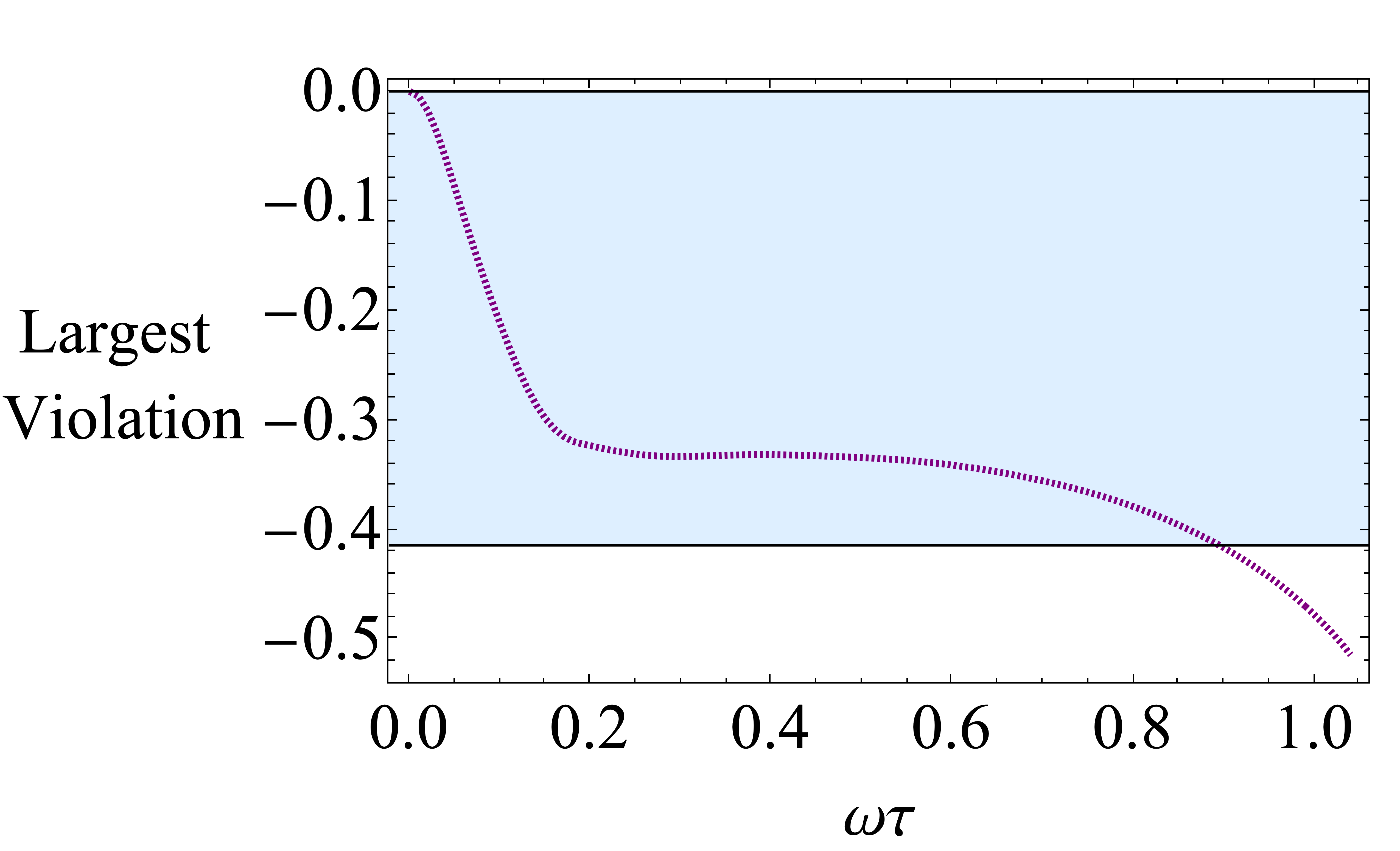}}}%
	%\qquad
	\hspace{5mm}
	\subfloat[]{{\includegraphics[height=5.2cm]{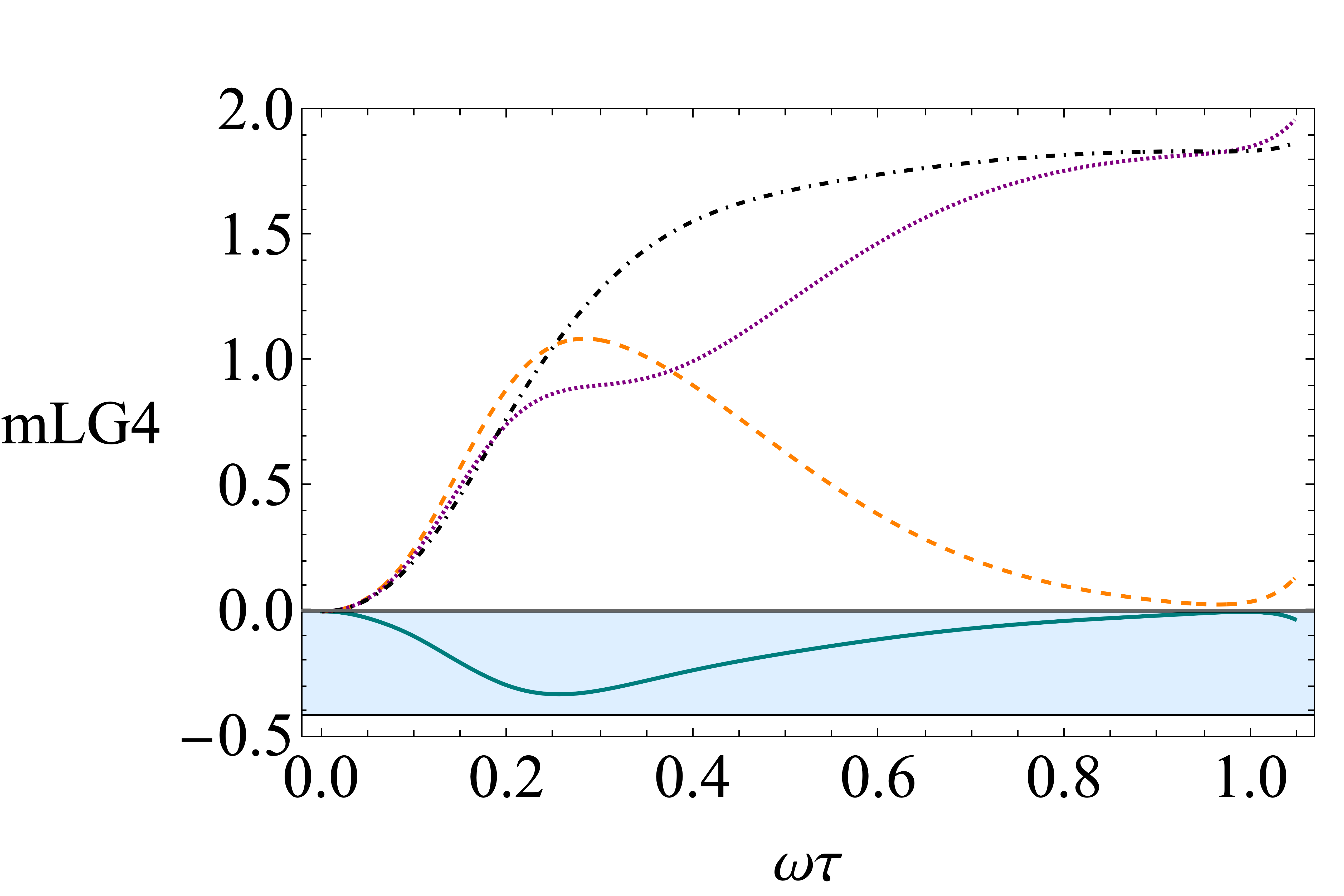}}}
	\caption[mLG violations for coherent states]{
	We show the largest possible mLG3 (a) and mLG4 (c) violations for fixed $\omega\tau$, optimized over all coherent states. We plot the mLG3 (b) and mLG4 (d) inequalities for a coherent state, with violations for Eq.~(\ref{mLG2}) and Eq.~(\ref{F12lg4}) respectively.}%
	\label{fig:f12d}%
\end{figure}
We now calculate the modified correlators for coherent states with details given in Appendix \ref{app:cohf12}.  We find the healthiest violations of the mLG3 inequality Eq.~(\ref{mLG2}) and the mLG4 Eq.~(\ref{F12lg4}) specifically, however the other inequalities do have (smaller) violations for coherent states. 

Since the single time crossing assumption is more plausible for short times, in Figs.~\ref{fig:f12d}(a) and Figs.~\ref{fig:f12d}(c) we numerically optimise over coherent states to find the largest violation for a fixed measurement interval, for the mLG3s and mLG4s respectively. We find significant violations in both for small times.

In Fig.~\ref{fig:f12d}(b), we plot the mLG3s Eqs.~(\ref{mLG1}--\ref{mLG4}), for a coherent state chosen from the point where violations plateau in Fig.~\ref{fig:f12d}(a), meaning these examples are optimised to show the strongest violation over a short time period.  We repeat this for the mLG4s Eq.~(\ref{F12lg4}) in Fig.~\ref{fig:f12d}(d).  The states involved for both these figures, are coherent states with around $\ev{x_0}\sim0.60$, and $\ev{p_0}\sim-2$.

To show that the violations reported here are not the result of the sharpness of the delta function coupling, in Appendix~\ref{app:gausswin}, we repeat the calculation for a gaussian coupling function, finding violations persist for a variety of widths.
%\begin{figure}
%	\subfloat[]{{\includegraphics[height=5.2cm]{mlg3opt.png}}}%
%	%\qquad
%	\hspace{5mm}
%	\subfloat[]{{\includegraphics[height=5.2cm]{figcohMLG3.png}}}
%	\caption[mLG3 violations for coherent states]{
%	We show in (a) the largest possible mLG3 violation for fixed $\omega\tau$, optimized over all coherent states. In (b) we plot the mLG3 inequalities a coherent state, with violations for Eq.~(\ref{mLG2}).}%
%	\label{fig:f12c}%
%\end{figure}

\section{Conclusion}
\label{c6s5}
In this work in progress stage chapter, we have presented an alternative approach to the NIM requirement of LG tests.  By noting that the temporal correlators reflect how a state changes, our protocol attempts to measure the system in these states of change, which for the variable $Q=\textrm{sgn}(x)$ means measuring the particle as it passes the origin.  Heuristically, this allows us to extract what is typically considered two-time correlations, through a single measurement.  In Section~\ref{c6s1} we presented a detector model to implement this, which is non-invasive under the single-time crossing assumption.  The detector model allows us to measure the modified correlators, which in Section~\ref{c6s2} we derived a set of macrorealistic inequalities for.

In Section~\ref{sec:sca} we considered the validity of the single crossing assumption, and gave several arguments to show we can expect the single crossing assumption to be valid within QM for at least a certain period of time.

In Section~\ref{c6s4}, we calculated the modified correlators for the QHO. We found an approximate correspondence between these modified correlators, and the conventional correlators, which allowed us to make connection with the results from Chapters~\ref{chap:QHO1} and \ref{chap:QHO2}, and to establish an estimate of their maximal violation within QM.  We demonstrated significant violations of the mLG3 and mLG4 inequalities for energy eigenstates and coherent states.  We were able to repeat the calculation with a smooth measurement window, for which LG violations persist, showing they are not an artefact of the sharpness of the measurement.

\begin{subappendices}
\section{Matrix elements for $\ev{F^2_{12}}$ in energy eigenbasis}
\label{app:enmel}
We now calculate the matrix elements for $\ev{F_{12}^2}$ in the energy eigenbasis in the QHO. We calculate this for a general coupling $f(\hat x)$, and so have
\begin{equation}
\mel{m}{F_{12}^2}{n}=\iint_{t_1}^{t_2}\mathop{dt}\mathop{ds}\mel{m}{e^{i H t}f(\hat x) e^{-i H (t-s)}f(\hat x)e^{- i H s}}{n}.
\end{equation}
Inserting a resolution of unity in the energy eigenbasis to handle time evolution, we find
\begin{equation}
\mel{m}{F_{12}^2}{n}=\iint_{t_1}^{t_2}\mathop{dt}\mathop{ds}\sum_{k=0}^\infty\mel{m}{f(\hat x)}{k}\!\!\mel{k}{f(\hat x)}{n}e^{i(E_k -E_n)s +i (E_m-E_k)t}.
\end{equation}
We now take the integrals into the sum, which now separated each have the form
\begin{equation}
	\mathcal{I}_{\ell k}(t_1, t_2)=\int_{t_1}^{t_2}\mathop{dt}e^{i\omega(\ell-k)t}= \begin{cases}
 	\frac{i \left(e^{i\omega t_1 (\ell-k)}-e^{i \omega t_2 (\ell-k)}\right)}{\omega(\ell-k)} & \text{for }k\neq \ell, \\
 	t_2-t_1 & \text{otherwise}.\\
 \end{cases}
\end{equation}
We thus find the matrix elements
\begin{equation}
\label{eigmels}
\mel{m}{F_{12}^2}{n}=\sum_{k=0}^{\infty}M_{mk}M_{kn}\mathcal{I}_{kn}(t_1, t_2)\mathcal{I}_{mk}(t_1, t_2).
\end{equation}
where we have shorthanded the matrix elements of the coupling function to
\begin{equation}
M_{mk}=\mel{m}{f(\hat x)}{k}.	
\end{equation}
For the coupling $f(\hat x)=L\delta(\hat x)$, its matrix elements are trivially given by
\begin{equation}
	M_{nk}=L\mel{n}{\delta(\hat x)}{k}=L\psi_n(0)\psi_k(0),
\end{equation}
Note this selects for $n$ and $k$ both to be even, and hence the modified correlators $\ev{F_{12}^2}$ for $\delta$ coupling will only pick up the symmetric part of a state.  With this restriction on $k,\ell$, this also means for $k\neq \ell$, $I_{\ell k}(t_1, t_2)=0$ whenever $\omega(t_2-t_1)=\pi$.
\section{Relation between $\ev*{F_{12}^2}$ and correlators $C_{12}$}
\label{app:f12corr}
We show how to relate $\ev{F_{12}^2}$ in a state $\ket\psi$ to the standard correlator $C_{12}$ in a state $\ket\phi$, related by
\begin{equation}
\label{eq:rship}
\ket\psi =\frac{1}{\ev{p^2}^\frac12}\hat p \ket\phi,
\end{equation}
with $\ev{p^2}=\ev{\hat p^2}{\phi}$. This is hence normalised so $\braket{\psi}=1$.  Working with the coupling $f(\hat x)=L\delta(\hat x)$, which from Eq.~\ref{eigmels} shows $\ev{F_{12}^2}$ will pick up only the symmetric part of the state of $\ket\psi$, and hence we restrict to antisymmetric $\ket\phi$.  We then have 
\begin{equation}
F_{12}\ket\psi=	\frac{1}{\ev{p^2}^\frac12}F_{12} \hat p \ket\phi =\frac{L}{\ev{p^2}^\frac12}\int_{t_1}^{t_2}\mathop{dt} \delta(\hat x (t)) \hat p \ket\phi,
\end{equation}
and noting that
\begin{equation}
\label{eq:ptime}
\delta(\hat x(t))=e^{iHt}\delta(\hat x)e^{-iHt}\hat p=e^{iHt}\delta(\hat x)\hat p(-t)e^{-iHt},
\end{equation}
where with $\hat p(-t)=\hat p \cos \omega t + \hat x \sin \omega t$, and using the simplification $\hat x \delta(\hat x)=0$, we find 
\begin{equation}
F_{12} \hat p \ket \phi=L\int_{t_1}^{t_2}\mathop{dt}e^{iHt}\delta(\hat x)\hat p e^{-iHt}\ket{\phi} \cos\omega t.
\end{equation}
Since $\braket{x}{\phi}$ is antisymmetric, we have $\delta(\hat x)\ket\phi =0$, so may freely include the commuted pair of operators
\begin{equation}
	\delta(\hat x)\hat p \ket \phi = (\delta(\hat x)\hat p + \hat p \delta(\hat x))\ket\phi,
\end{equation}
which represents the current.
\begin{equation}
2m\frac{d \theta(\hat x(t))}{dt}=\delta(\hat x(t))\hat p(t) + \hat p(t) \delta(\hat x(t)),
\end{equation}
Including the time-evolution, we reach
\begin{equation}
\label{inter}
F_{12} \hat p \ket \phi=2mL\int_{t_1}^{t_2}\mathop{dt}\frac{d}{dt}(\theta(\hat x(t))\ket\phi \cos\omega t.
\end{equation}
We now note that for the free particle, in Eq.~(\ref{eq:ptime}) we instead use {$[\hat p, \hat H]=0$}, and so reach 
\begin{equation}
F_{12} \hat p \ket \phi=2mL\int_{t_1}^{t_2}\mathop{dt}\frac{d}{dt}(\theta(\hat x(t))\ket\phi,
\end{equation}
where the integration is now trivially completed yielding
\begin{equation}
F_{12}\hat p \ket\phi = 2mL(\theta(\hat x(t_2)-\hat x(t_1))\ket\phi.
\end{equation}
  Recalling the normalisation through Eq.~(\ref{eq:rship}), this gives us the main result, that
\begin{equation}
\ev{F_{12}^2}{\psi}=\frac{4m^2 L^2}{\ev{p^2}}\ev{(\theta(\hat x (t_2)-\theta (\hat x (t_1)))^2}{\phi},
\end{equation}
where we write
\begin{equation}
\ev{\tau_D^2} = \frac{4m^2L^2}{\ev{p}_\phi^2},
\end{equation}
as the dwell time.  The factor of $4$ difference from the classical dwell time Eq.~(\ref{F12class}) is owing to the presence of a factor of $(1+\Pi)$ unique to the quantum dwell time~\cite{muga2009}, where $\Pi$ is the parity operator introduced in Eq.~(\ref{parityop}), which gives a factor of 2. By relating the theta functions to the dichotomic variable $\hat Q_i=\text{sgn}(\hat x_i)$ through $\theta(\hat x)=\frac{1+\hat Q}{2}$, we have 
\begin{equation}
\label{eq:f12tau}
\frac{\ev{F_{12}^2}}{\ev{\tau_D^2}}=\frac14\expval**{\left(\hat Q_2-\hat Q_1\right)^2}{\phi}.
\end{equation}
Using $Q_i^2=1$, we find
\begin{equation}
\frac{\ev{F_{12}^2}}{\ev{\tau_D^2}}=\frac{1}{2}\expval**{1-\frac12\left(\hat Q_1 \hat Q_2 +\hat Q_2\hat Q_1 \right)}{\varphi}.
\end{equation}
This then yields
\begin{equation}
\label{modcor}
\frac{\ev*{F_{12}^2}}{\ev{\tau_D^2}}=\frac12 (1- C_{12}).
\end{equation}
This gives a precise relationship between $\ev*{F_{12}^2}$ and the standard correlators we used in Chapters~\ref{chap:QHO1} and \ref{chap:QHO2}.  This is exact for the free particle scenario, and approximately correct for small times in the QHO.

To determine $\ev{\tau_D^2}$ in the QHO, we employ the property of harmonicity, which means after a time interval $\omega\tau=\pi$, the entirety of the state will have traversed the origin.  This implies we may take
\begin{equation}
\label{eq:dtdef}
\ev{\tau_D^2}=\ev{F_{12}^2(\omega\tau=\pi)},
\end{equation}
which is given by
\begin{equation}
\label{dtcalc}
\ev{\tau_D^2}=\pi^2 \sum_{n=0}^{\infty}\lvert\braket{\psi}{n}\rvert^2\mel{n}{f(\hat x)}{n}^2
\end{equation}
most generally.
\subsection{$\ev*{F_{12}^2}$ with auxiliary state $\ket1$}
\label{secp1app}
Since in Section \ref{sec:1and0} we found significant LG3 violations for the $\ket1$ state, we expect by using this as the auxiliary state, we find will find decent violations for the mLG inequalities.   

For the QHO, using Eq.~\ref{inter} in $\ev*{F_{12}^2}$, we find
\begin{equation}
\ev*{F_{12}^2}{\psi}=\int_{t_1}^{t_2}\mathop{dt}\int_{t_1}^{t_2}\mathop{ds}\cos\omega t\cos\omega s \frac{\partial^2}{\mathop{\partial t}\mathop{\partial s}}(\Re \ev{\theta(\hat x(t))\theta(\hat x(s))}{\phi}),
\end{equation}
where the last term on the RHS is the QP Eq.~(\ref{qp2}) which considering differentiating the moment expansion Eq.~(\ref{mom}) twice in the antisymmetric $\ev{Q_i}=0$, will leave just $C(t,s)$, the standard correlator evaluated at times $t$ and $s$.

We calculate here the modified correlators where the auxiliary state $\ket\phi=\ket1$.  We may hence use the approximation Eq.~(\ref{cosapprox}) to write $C(s,t)\approx \cos(\omega(t-s))$, and so find
\begin{equation}
\ev*{F_{12}^2}{\psi}= \int_{t_1}^{t_2}\mathop{dt}\int_{t_1}^{t_2}\mathop{ds}\cos\omega t\cos\omega s \frac{\partial^2}{\mathop{\partial t}\mathop{\partial s}}\cos\omega(t-s)).
\end{equation}
Completing the integration, we find
\begin{equation}
\label{f12p1}
\ev*{F_{12}^2}{\psi}=\frac{1}{8\omega^2} \left(1+2 \omega^2 \tau ^2  +2 \omega \tau  \sin (2  \omega \tau)-\cos (2  \omega \tau)\right).
\end{equation}
Using Eq.~\ref{dtcalc} to calculate the dwell time for this state, gives us 
\begin{equation}
\ev{\tau_D^2}=\frac{\pi^2}{4\omega^2}.
\end{equation}
Through Eq.~\ref{modcor} this lets us compare the behaviour of the standard correlator, and that measured using the waiting detector model, which we plot in Fig.~\ref{fig:f12corr}
\section{L\"uders Bound on mLGs}
\label{subsec:mlgluders}
Using Eq.~(\ref{eq:f12tau}) within the mLG inequalities Eqs.~(\ref{mLG1}--\ref{mLG4}), we find for example
\begin{equation}
\frac{\ev{F_{12}^2}}{\ev{\tau^2}}+\frac{\ev{F_{23}^2}}{\ev{\tau^2}}-\frac{\ev{F_{13}^2}}{\ev{\tau^2}}=\frac14\expval**{\left(\hat Q_1-\hat Q_2\right)^2+\left(\hat Q_2-\hat Q_3\right)^2-\left(\hat Q_1-\hat Q_3\right)^2}{\varphi}
\end{equation}
which using Eq.~(\ref{modcor}) indeed yields a typical LG3 kernel Eq.~(\ref{LG3b}),
\begin{equation}
\frac{\ev{F_{12}^2}}{\ev{\tau^2}}+\frac{\ev{F_{23}^2}}{\ev{\tau^2}}-\frac{\ev{F_{13}^2}}{\ev{\tau^2}}=\frac12(1-C_{12}-C_{23}+C_{13}).
\end{equation}
The mLG inequalities hence inherit the standard LG3 L\"uders bound of $-\frac12$, as presented in Section~\ref{subsec:lb}, and hence within QM, the mLG inequalities as presented in Eqs.~(\ref{mLG1}--\ref{mLG4}) have the bound $-\frac14 \ev{\tau_D^2}$ on the right hand side.  Following the same procedure for the mLG4 Eq.~(\ref{F12lg4}) indicates they have a lower bound of $(1-\sqrt{2})\ev{\tau_D^2}$ and upper bound of $(1+\sqrt{2})\ev{\tau_D^2}$

We note some caution on the meaning of the L\"uders bound for the mLGs.  Since the correspondence between $C_{12}$ and $\ev{F_{12}^2}$ holds exactly only for the free particle (and then only for states in which $\ev{\tau_D^2}$ is defined), we do not expect it to be an exact bound on the mLG inequalities, rather an indication of the general scale of violations.  

\section{Calculating $\ev*{F_{12}^2}$ for Coherent states}
\label{app:cohf12}
We now calculate $\ev*{F_{12}^2}$ where the initial state is a coherent state.  Writing the coherent state in the Fock representation,
\begin{equation}
	\ket\alpha = e^{-\frac{\abs{\alpha}^2}{2}}\sum_{n=0}^\infty \frac{\alpha^n}{\sqrt{n!}}\ket{n},
\end{equation}
we find
\begin{equation}
	\ev*{F_{12}^2}{\alpha} = e^{-\abs{\alpha}^2}\sum_n^{\infty}\sum_\ell^{\infty}\frac{\alpha^n\alpha^{*\ell}}{\sqrt{n!\ell!}}\mel{\ell}{F_{12}^2}{n}.
\end{equation}
We may then use Eq.~\ref{eigmels}, whereupon short-handing
\begin{equation}
 G_{n\ell k}(t_1, t_2)=\mathcal{I}_{\ell k}(t_1, t_2)\mathcal{I}_{k n}(t_1, t_2),
\end{equation}
we reach the result
\begin{equation}
\ev*{F_{12}^2}{\alpha}=e^{-\abs{\alpha}^2}\sum_n^{\infty}\sum_\ell^{\infty}\sum_k^{\infty}\frac{\alpha^n\alpha^{*\ell}}{\sqrt{n!\ell!}}G_{n\ell k}(t_1, t_2)M_{\ell k}M_{kn}.
\end{equation}
We hence can write the MLG inequalities here as
\begin{equation}
	e^{-\abs{\alpha}^2}\sum_n^{\infty}\sum_\ell^{\infty}\sum_k^{\infty}\frac{\alpha^n\alpha^{*\ell}}{\sqrt{n!\ell!}}\mathcal{L}_{n\ell k}(t_1, \tau)M_{\ell k}M_{kn}\geq0,
\end{equation}
where for example
\begin{equation}
\mathcal{L}_{n\ell k}(t_1, \tau)=G_{n\ell k}(t_1, t_1+\tau)+G_{n\ell k}(t_1+\tau, t_1+2\tau)-G_{n\ell k}(t_1, t_1+2\tau)
\end{equation}
corresponds to the three-time mLG inequality Eq.~\ref{mLG2}.
\section{Gaussian Window Function}
\label{app:gausswin}
We now repeat the same calculation, but with a gaussian window function.  We hence need the matrix elements 
\begin{equation}
M_{nk}=\frac{1}{\sigma\sqrt \pi}\mel**{n}{\exp(-\frac{\hat x^2}{2\sigma^2})}{k},
\end{equation}
which in the interest of generality we write as
\begin{equation}
M_{nk}=\mel{n}{f(\hat x)}{k}=\int_{-\infty}^\infty \mathop{dx}\psi_n(x)f(x)\psi_k(x).
\end{equation}
We can solve this using the generating function
\begin{equation}
e^{-\frac12 x^2 + 2xq - q^2} = e^{-\frac12 x^2}\sum_{n=0}^\infty H_n(x)\frac{q^n}{n!},
\end{equation}
where $H_n(x)$ are the Hermite polynomials.  Introducing a second generating function, multiplying both sides by $f(x)$ and integrating, we have
\begin{equation}
\int_{-\infty}^{\infty} \mathop{dx}e^{-x^2 + 2x(q_1+q_2) - (q_1^2+q_2^2)}f(x) = \sum_{n=0}^\infty\sum_{m=0}^\infty \frac{q_2^m}{m!}\frac{q_1^n}{n!}\int_{-\infty}^{\infty} \mathop{dx}e^{-x^2}f(x)H_n(x)H_m(x).
\end{equation}
\begin{figure}
\centering
	{\includegraphics[height=5.2cm]{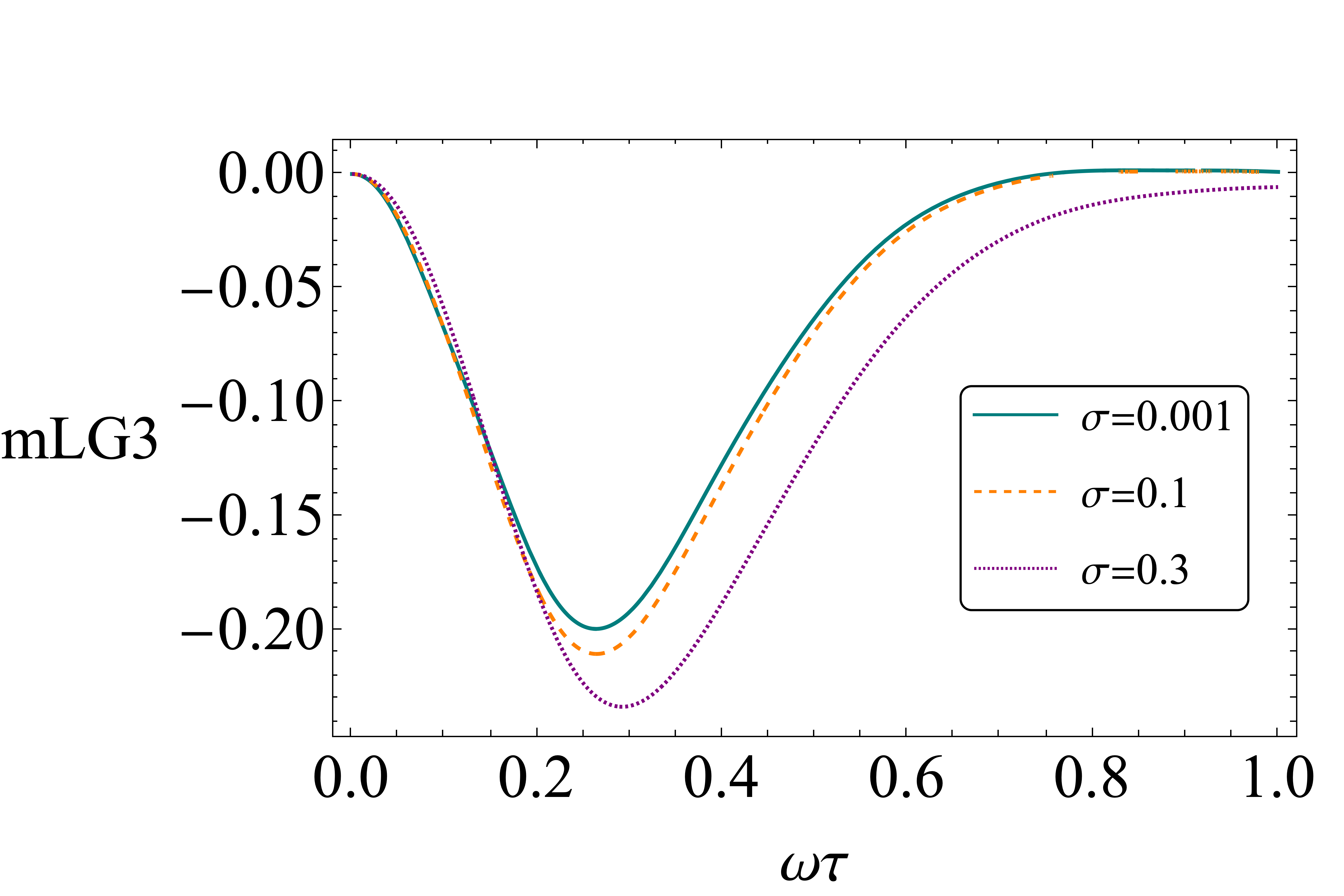}}%
%	%\qquad
%	\hspace{5mm}
%	\subfloat[]{{\includegraphics[height=5.2cm]{fig2a.png}}}
	\caption[mLG violations with a smooth coupling function]{Here we plot the mLG3 inequalities with gaussian coupling of variable width, for the same coherent state as Fig.~\ref{fig:f12d}(b).}%
	\label{fig:f12g}%
\end{figure}
The RHS has the interpretation as a Taylor series, with coefficients proportional to the matrix elements we seek, which may thus be obtained through differentiation of the LHS.  We hence have a procedure for calculating the matrix elements for any $f(x)$ that leaves the integral on the LHS tractable.  We proceed using a Gaussian defined
\begin{equation}
f(x)=\frac{1}{\sigma\sqrt{2\pi}}\exp(-\frac{x^2}{2\sigma^2}),
\end{equation}
where we find
\begin{equation}
\frac{e^{-\frac{q_1^2-4 q_1 q_2 \sigma ^2+q_2^2}{2 \sigma ^2+1}}}{\sigma\sqrt{\frac{1}{\sigma ^2}+2}  }=\sum_{n=0}^\infty\sum_{m=0}^\infty \frac{q_2^m}{m!}\frac{q_1^n}{n!}M_{nm}\sqrt{\pi 2^m 2^n m! n!},
\end{equation}
and hence reach
\begin{equation}
M_{nm}=\frac{1}{\sqrt{\pi 2^n n! 2^m m!}}\pdv[n]{q_1}\pdv[m]{q_2}\left(\frac{\sqrt{2} e^{-\frac{q_1^2-4 q_1 q_2 \sigma ^2+q_2^2}{2 \sigma ^2+1}}}{\sigma\sqrt{\frac{1}{\sigma ^2}+2}  }\right)\eval_{q_1=q_2=0}
\end{equation}
We now use these elements within Eq.~\ref{eq:f12eig} to calculate $\ev*{F_{12}^2}$ for the mLG3 coherent state violation from the previous section, but now with a smooth gaussian coupling.  We plot these inequalities for varying $\sigma$ in Fig.~\ref{fig:f12g}, where violations remain upon broadening.  This is largely left as an indicative result, that the violations observed do not depend upon the sharpness of the $\delta$ function.  

The smoothed coupling function is seen to change the time of the peak violation, and even appears to increase the violation.  This behaviour is markedly different from the smoothed projectors in Section~\ref{subsec:smooth}, which acted only to subdue LG violations.  However we note the gaussian window function is not a projector, and hence the standard simplifications of $\hat Q^2=1$ do not hold, and more work is needed here to establish context for these violations.

\end{subappendices}

\fancyhf{}
\renewcommand{\headrulewidth}{0pt}
\fancyfoot{\makebox[\textwidth][c]{\hyperref[link:7]\thepage}}

\fancyhfoffset[LE,RO, RE, LO]{0cm}
\renewcommand{\chaptermark}[1]{ \markboth{#1}{} }
\renewcommand{\sectionmark}[1]{ \markright{#1}{} }
\fancyhf{}
\fancyhead[L]{\textsl{\thesection~~ \rightmark}}
\fancyhead[R]{\hyperref[link:7]\thepage}
\renewcommand{\headrulewidth}{1pt}

\chapter{Conclusion}
\bookepigraph{3in}{All that you touch\\You Change.\\All that you Change\\Changes you}{Octavia Butler,}{Parable of the Sower}{4.5}
\label{chap:summ}
\vspace{-2em}
\section{Summary of this Thesis}
This thesis was split into three directions.  The first of which, was a mathematical study of generalisations of conditions for MR.  In Chapter~\ref{chap:ntime}, we studied decisive tests for MR in the scenario where measurements are made at $n$ times.  We also looked at conditions involving all possible temporal correlators at $n$ times.  We demonstrated their violation in a simple spin model, and looked at the asymptotic behaviour for large $n$.  In Chapter~\ref{chap:mlev}, we generalised the necessary and sufficient conditions for MR to systems described by many-valued variables. We performed quantum mechanical analysis of the interferences responsible for violating these conditions.  We looked at their inter-relations, where we observed the simple hierarchy associated with the standard dichotomic variable breaks down in this scenario.  There are no new types of interference present here, however we found the existing interferences can be mixed in non-trivial ways, leading to novel behaviour compared to the dichotomic case.

The second direction, corresponds to the implementation of LG tests within continuous variable systems, with variables defined through the coarse-graining of position, as proposed in Ref.~\cite{bose2018}.  In Chapter~\ref{chap:QHO1}, the main result was the calculation of the temporal correlators, over arbitrary spatial coarse-grainings, valid for any exactly soluble potential.  We then used this result to study violations of MR with particular focus on the energy eigenstates of the QHO, where we found an abundance of LG violations.  In Chapter~\ref{chap:QHO2}, we continued this analysis to include coherent states, which are often viewed as one of the most classical-like states of QM.  We performed a parameter-space analysis determining the largest violations possible for any given coherent state.  After demonstrating the violations possible, we sought to understand the physical mechanisms which underpin them, through an analysis of the relevant probability currents, Bohm trajectories, and Wigner transforms.  We found a quantum Zeno effect responsible for the violations, with a connection to diffraction in time.

The third and final direction stays within the realm of continuous variables, where in Chapter~\ref{chap:wd} we presented an alternative measurement protocol to demonstrate violations of MR.  We imagined a waiting detector at the origin, which registers if a particle crosses (and thus transitions states), which yields information related to the standard correlators.  We then argued that implementing this quantum mechanically with a weakly coupled continuous in time measurement, as has been experimentally implemented for spin systems \cite{majidy2019a}, could be considered as approximately non-invasive.  This allowed us to derive a set of modified LG inequalities, which we demonstrated are readily violated in the QHO.  We discussed and gave arguments against some potential criticisms of this approach.

Finally in the remainder of this chapter, I conclude.  In Section~\ref{conc:future}, I survey the state of tests of non-classicality, both for tests of macroscopic coherence, and more generally. With the difficulties around invasiveness, things can sometimes look bleak for LG tests.  I argue instead they still have an important role in our toolkits.  I also outline several areas for future research, and detail some approaches which may get around the invasivity problem.

It is often quite simple for the staunch macrorealist to propose that LG violations could be due to the failure of NIM alone, rather than the much more interesting failure of MRps and NIM.  In Section \ref{sec:nimconc}, I will argue that this may be somewhat of a red herring, and that QM leads to predictions beyond such a simple explanation.  This perhaps touches on the \textsl{je ne sais quoi} of what \textsl{is} quantumness?

Finally, in Section~\ref{sec:end} I close with a broad perspective on our study of the physical world.

\section{LG Tests State of Play \& Future Research}
\label{conc:future}
Classically intuitive notions of reality are well and truly ruled out by Bell tests, which are considered loophole free, braced by the principles of special relativity.  The results of Leggett-Garg tests paint a consistent picture, indicating the inadequacy of non-signalling (in time) hidden variables in explaining observed behaviour. However the loophole of invasivity of measurements remains a significant hurdle to interpreting their violation as absolutely as Bell tests, and through such a fundamental lens.  The recent experiment in Ref. \cite{joarder2022} does make significant progress on the invasivity loophole, where invasiveness beyond that of wavefunction collapse may be ruled out.

If we consider the inadequacy of classical worldviews to describe reality as fact, as demonstrated by Bell tests, then the issues around invasivity seem to just be for argument's sake.  If we instead accept non-classicality, rather than being used to prove something about the universe which we already know, LG tests may be used to quantify and detect how far a system's behaviour departs from a classical understanding.  This is clearly of great importance in the development of technologies which directly exploit non-classicality in their design, and as well in other fields such as quantum biology, where the inequalities may be adapted to identify the scenarios in which living systems utilise quantum mechanics. In this thesis, we have extended the LG formalism to more general situations, with the hope this will better aid their application in novel systems.  The preliminary work for carrying out a number of experimental tests based on the MR conditions proposed in Chapters \ref{chap:ntime} and \ref{chap:mlev} is given in Ref   \cite{majidy2021a}.

This work is far from complete.  Since a quantum computer may ultimately be seen as the application of a given unitary operator followed by a measurement, it seems possible that an LG-like quantity may be derived in this scenario. This could then serve to give a theoretical measure of the degree of quantumness utilised by a given quantum algorithm.  If successful, this could lead to a whole manner of interesting computer science problems -- for example, if Shor's algorithm  were to exploit a different degree of quantumness for the factorising of different prime numbers, what would this mean?  The LG inequalities have been used in benchmarking 

In terms of their original goal of identifying macroscopic coherence, the invasivity loophole is clearly still important, however in the following Section~\ref{sec:nimconc} I put forth an argument that may possibly be developed to make this loophole more far-fetched.  Further if we imagine continuously increasing the degree of macroscopicity in a system, with LG violations maintained at each step, it seems implausible that genuine non-classicality (as bolstered by Bell tests) should at one of these steps seamlessly be replaced by an illusion of non-classicality from invasiveness.

Another approach which should be used in tandem, are the recently unearthed Tsirelson's inequalities, which were first published in a quite technical paper in 2006 \cite{tsirelson2006}, with recent papers developing them in a more accessible way  \cite{zaw2022,plavala2023}.  These inequalities result from using our physical intuitions about harmonic oscillators in place of requirements of non-invasiveness. By making measurements with intervals spanning an oscillator period, non-oscillating trajectories are ruled out as having zero chance of happening.  This places constraints on the moment expansion of the probability, and inequalities may then be constructed out of purely single time averages, which must be satisfied classically.  This ostensibly avoids any issues of invasiveness.  

It would be worthwhile to study the behaviour of the Tsirelson inequalities alongside the LG inequalities, to identify any overlap or disjunction in the non-classicality detected.  A a future calculation would be investigating this using the framework of interference analyses presented in Chapter~\ref{chap:mlev}.  We also note that Tsirelson-like inequalities may be derived in the context of the tests presented in Chapter~\ref{chap:wd}, where there the single-crossing behaviour of classical particles may be implemented as a constraint.  Further there have been other proposals of LG tests based on single time measurements, through placing assumptions on system dynamics \cite{hermens2018}.

For future experimental tests of MR in the harmonic oscillator, the research presented in this thesis, and other papers since \cite{das2022a, hatakeyama2023} indicate there are many circumstances in which violations can be found.  These violations are mass-independent, and so violations will be theoretically present for any size of mass.  Since the LG violations correspond to the way QM handles situations involving uncertain experimental results, at least two of the measurements involved in any test must be made at an instant where we are agnostic to the value of the dichotomic variable.  For variables made through the coarse-graining of position, this means the measurements must significantly chop the wave-function.  For the coherent states presented in this thesis, this means as a general rule of thumb, the implemented measurements must be of a finer resolution than $\sqrt{\frac{\hbar}{m\omega}}$, which likely delineates the feasibility of a given experiment.  Further, the waiting detector model proposed in Chapter~\ref{chap:wd} may be further developed and could be used in future experiments with the QHO.

An important effect we have not analysed in this thesis is the precise effect of decoherence on LG violations within continuous variable systems \cite{zeh1970, zurek1981}.  Experimentally controlling the effects of decoherence is a highly challenging battle, in a war that is always lost.  In earlier work studying the effect of decoherence in spin-$\frac12$ systems, it unsurprisingly leads to a reduction in the magnitude of the correlation functions, which eventually leads to classicalisation of the LG quantities \cite{athalye2011,xu2011,majidy2019a}.  As such, it is important to know how robust the presented LG violations are to decoherence, by modelling the process, and calculating the timescales over which MR violations remain detectable.

With these caveats taken on board, I consider LG tests on macroscopic objects worth experimentally pursuing, as they are capable of giving strong (although not definitive) evidence of non-classical behaviour.  Taken alongside other approaches, for example the results from interference experiments (where such an experiment is feasible), and the Tsirelson inequalities, LG violations will only strengthen the argument that non-classical behaviour has been observed.

\section{Violations of MR Beyond Clumsiness?}
\label{sec:nimconc}
One of the great challenges in interpreting the meaning of LG violations, and their efficacy in refuting a classical worldview lies within the definition of Macrorealism itself.  Since we cannot decouple MRps from NIM, and must always test for both. The possibility our measurements through clumsiness are disturbing a classical system can always serve as an explanation of LG violations, rather than the more radical failing of MRps. This counterargument could well be called the Devil's Advocate (DA) worldview.   Implementation of models consistent with this worldview have been shown to be capable of replicating the canonical spin-$\frac12$ LG violations, where the invasiveness of measurements can be equivalently considered as an MRps system with memory of the measurements made on it \cite{yearsley2013a,montina2012}.

In that vein, we can characterise the nature of this memory.  For simple dichotomic measurements, a single bit is sufficient to replicate MR violations.  However as covered in Chapters \ref{chap:mlev}--\ref{chap:QHO2}, tests of MR are much more general than this, and indeed for continuous variable systems we have a vast choice of possible variables we can define, and subsequently test for MR.  I argue that for the DA position to be truly convincing, we should require it to be capable of replicating all of these violations.

 We now consider how to implement this worldview in our understanding of the motion of a particle.  By MRps, we may understand the ensemble behaviour of a particle by how its individual underlying trajectories change in the face of a disturbing measurement.  We are free to consider an initial ensemble probabilistic through ignorance, whereupon making a disturbing measurement, to be consistent with observation this can only change the momentum distribution.   We may then in principle fine-tune these trajectory changes to mimick the quantum values of $C_{12}$ for a given choice of $Q$. This hence allows the measurement disturbance to replicate an LG violation.
 
 A useful way to characterise what behaviour would be required of such a set of trajectories, is to make a qualitative comparison with the trajectories present in Bohm theory, which we know do match quantum mechanical behaviour \cite{bohm1952a}.  Firstly, the set of trajectories consistent with QMs predictions must be \textsl{guided}, and so a set of ballistic MRps trajectories is not sufficient to emulate quantum statistics most generally.  Secondly, since this guidance depends on the post-measurement wavefunction, the trajectories must differ dependent on the initial measurement made.
  
This has the implication that although clumsy measurements on an MRps system may indeed explain a given MR violation (or equivalently that quantum statistics may be simulated by a classical particle with memory), the measurement disturbances/memory settings must be fine-tuned to each possible initial measurement, each matching the corresponding quantum mechanical behaviour.  This is a challenge to the DA position as the Hilbert space of the system gets larger, and so it is highly problematic in the arena of continuous variables.  It is this requirement for a vast amount of information to simulate a particle's behaviour that I identify as a more accurate notion of the non-classicality inherent in quantum mechanical motion.  This all of course does not rule out the DA worldview, however it results in a requirement for an astounding level of complexity.  This in my opinion makes it much less of a serious counter-argument to violations of macrorealism. 

 The argument presented here aligns with earlier information-theoretic approaches to the nature of sequential measurements in QM \cite{zukowski2014, brierley2015}, with the approach to characterising quantumness through the required classical memory to simulate it well reviewed in Ref \cite{vitagliano2022}, and a very promising future avenue away from the difficulties of countering invasiveness.  This notion of non-classicality is clearly more robust against the invasiveness loophole, and indeed may be more representative of the quantum advantage necessary in the next generation of quantum technology, and hence tests of MR which can incorporate it in some way will be clearly be both more useful and more powerful. 

\section{Closing Note}
\label{sec:end}
As referenced in the start of this thesis, David Bohm speaks of the need for quantum \textsl{non}-mechanics, suggesting that a mechanical picture of the universe may not be sufficient to understand it, leading to the evocative idea of an organic universe, as alive as you, me, or the trees.  This thesis has been dedicated to studying the non-classicalities in QM, and research very strongly suggests we can rule out several classes of hidden variable theories.  We can ask however, if this level of non-classicality is sufficiently \textsl{non-mechanical}, or whether we need to push further with this idea.

In this direction, I believe we as physicists should stop treating consciousness as a dirty word.  The argument against seriously considering consciousness as an ingredient in our theories is largely based on reductionism, the idea that we may eventually understand consciousness from the very first physical principles rigorously applied in full to our brains and bodies.  The problem of bridging between a mechanistic theory and the felt experience of qualia is known as the hard problem of consciousness, and it remains resolutely unsolved.  I hence think we should identify adherence to the reductionist viewpoint as faith-based, and not fundamentally `more scientific' than other perspectives.  

One of the approaches to the hard problem which is gaining in popularity, is the idea of panpsychism, which loosely speaking identifies conscious experience as a fundamental and omnipresent property of the universe.  If this were to be true, then it means that subjective experiences may lie within the motion of all constituent parts of the universe, and therefore have a role to play in our physical theories.  If we deny from the outset the presence of possible non-material elements interacting with the physical world, we may well hamper our ability to understand it -- leaving us wrought with paradoxes and searching in vain for a theory of everything.  Not due to a lack of brilliance on our part, but because the problems may be fundamentally intractable, and even ill-posed -- a mechanistic theory alone may well be incomplete and insufficient to explain the universe in which it was posed.

\fancyhf{}
\renewcommand{\headrulewidth}{0pt}
\fancyfoot{\makebox[\textwidth][c]{\hyperref[link:1]\thepage}}

\fancypagestyle{plain}{%redefine \pagestyle{plain} to add the link <<<
    \fancyhf{} 
\fancyfoot{\makebox[\textwidth][c]{\hyperref[link:1]\thepage}}
    \renewcommand{\headrulewidth}{0pt}
    \renewcommand{\footrulewidth}{0pt}
}%
\newpage
\addcontentsline{toc}{chapter}{Bibliography}
\printbibliography
%\bibliography{lib2}
\end{document}